\documentclass[a4paper,11pt,makeidx]{book}
\pdfoutput=1

\usepackage{a4wide}
\usepackage[utf8]{inputenc}
	\usepackage{draftwatermark}
\usepackage{rotating}
\usepackage{color}
\usepackage{graphicx}
\usepackage{multirow}
\usepackage{amsmath,amssymb}
\usepackage{calc}
\usepackage{feynmp}
\usepackage{simplewick}
\usepackage{subfigure}
\usepackage[inline]{asymptote}
\usepackage[
  backend=bibtex,style=authoryear,sorting=ynt,
  maxcitenames=3,maxbibnames=99,
  firstinits=true,url=false,isbn=false
]{biblatex}
\AtEveryBibitem{\clearlist{language}} 
\AtEveryBibitem{\clearfield{note}}    
\bibliography{Thesis}
\renewbibmacro{in:}{}

\SetWatermarkText{Draft 0.8}
\SetWatermarkScale{2}

\graphicspath{ {../figures/} } 
\begin{asydef}
  import graph;
  defaultpen(fontsize(11));
  real goldenRatio = (1+sqrt(5))/2;
  void sizeRatio(
    picture pic=currentpicture,
    real width=0, real height=0, real ratio=goldenRatio
  ) {
    if (width > 0) {
      size(pic, width, width/ratio, IgnoreAspect);
    } else {
      size(pic, height*ratio, height, IgnoreAspect);
    }
  }
  void plotXY(
    picture pic=currentpicture,
    real[][] data,
    pen p=defaultpen,
    Label l="",
    marker m=nomarker
  ) {
    data = transpose(data);
    draw(pic, graph(data[0],data[1]), p, l, m);
  }
  void plotXYDY(
    picture pic=currentpicture,
    real[][] data,
    pen p=defaultpen,
    Label l="",
    marker m=nomarker
  ) {
    data = transpose(data);
    real[] zeros = array(data[0].length, 0.0);
    errorbars(pic, data[0], data[1], zeros, data[2], p);
    draw(pic, graph(data[0],data[1]), p, l, m);
  }
  void axisXY(
    picture pic=currentpicture,
    Label xLabel="", Label yLabel="",
    pair min=(-infinity,-infinity), pair max=(+infinity,infinity),
    pen p=currentpen,
    ticks xTicks=LeftTicks, ticks yTicks=RightTicks
  ) {
    limits(pic, min, max, crop=true);
    xaxis(pic, xLabel, BottomTop, min.x, max.x, xTicks);
    yaxis(pic, yLabel, LeftRight, min.y, max.y, yTicks);
  }
  void axisY(
    picture pic=currentpicture,
    Label l="", 
    real min=-infinity, real max=+infinity,
    pen p=currentpen
  ) {
    ylimits(pic, min, max, crop=true);
    yaxis(pic, LeftRight, min, max, RightTicks);
  }
\end{asydef}

\newcommand{\der}{\partial}
\newcommand{\bra}[1]{\left\langle #1 \right|}
\newcommand{\ket}[1]{\left| #1 \right\rangle}

\newcommand{\braket}[2]{\left\langle #1 \left| #2 \right. \right\rangle}
\newcommand{\bracket}[3]{\left\langle #1 \left| #2 \right| #3 \right\rangle}
\newcommand{\mbracket}[4][\Big]{#1\langle #2 #1| #3 #1| #4 #1\rangle}

\renewcommand{\vec}[1]{{\bf #1}}

\newcommand{\eps}{\varepsilon}
\renewcommand{\d}{\textrm{d}}

\newcommand{\PDF}{{\operatorname{PDF}}}
\newcommand{\Tr}{{\operatorname{tr}}}
\newcommand{\im}{{\operatorname{i}}}
\newcommand{\diagramBox}[2][0.5]{\raisebox{0.5ex-#1\height}{#2}}
\newcommand{\diagramBoxBorder}[4][0.5]{
  \raisebox{0.5ex-#1\height}[0.5ex+\height-#1\height+#2][#1\height-0.5ex+#3]{#4}
}
\newcommand{\light}[1]{\textcolor[rgb]{0.65,0.65,0.65}{#1}}
\newcommand{\VASP}{\texttt{VASP}}

\makeindex

\begin{document}

\title{
  Density Functional Theory applied to liquid metals and \\
  the Adjacent Pairs Exchange correction to the Random Phase Approximation
  \\[2ex]
  {\Huge Draft 0.8}
}
\author{Felix Hummel \\[1ex] supervisor: Prof. Georg Kresse}

\maketitle 

\tableofcontents

\setcounter{part}{2}
\setcounter{chapter}{4}

\part[The Adjacent Pairs Exchange correction to the RPA]{
  The Adjacent Pairs Exchange correction to the Random Phase Approximation
}

  \chapter{Many Body Perturbation Theory}
\label{cha:MBPT}
Density Functional Theroy (DFT) calculations are sufficiently accurate for many
applications. However its description of electronic correlation is just
approximated locally from the uniform electron gas. It lacks, for instance,
indirect effects of the electron-electron interaction, such as Van-der-Walls
force where electrostatic repulsion deforms the distribution of electrons
and these polarized electron distributions in turn can attract each other.
This effect can not be captured by DFT or HF as they only take the instantaneous
electrostatic interaction into consideration but the formation of the polarized
electron distributions requires time. The missing descriptions of such
dynamical effects is referred to as \emph{dynamical correlation}
\parencite{shavitt_many-body_2009}.

\emph{Static correlation} occurs when the DFT or HF approximation is not
appropriate
for the chemical environment per se. A good example is the the dissociation of 
a hydrogen ion where the protons are already at a great distance. The single
electron should be in a superposition of two states, quite localized at
each respective proton. DFT rather places half an electron on each proton and
yields just one orbital spanning both protons.

Perturbation theory expands properties of the exact solution of the Schrödinger
equation in terms of the orbitals and orbital energies of the corresponding
Hartree-Fock approximation. One can also start from DFT orbitals or from
 other reference which is feasible to solve. 
It is mostly the dynamic correlation that is captured
by the perturbation expansion but it can also include some static correlation
when going to sufficiently many terms. In cases where the static correlation
is large multi reference perturbation theory may be required.
Not unlike a Taylor expansion, a perturbation expansion
is not guaranteed to converge or may converge slowly with the number of terms
included. The convergence also strongly depends on the quality of the reference
state.

We will use time dependent many body perturbation theory following the original
derivation of \parencite{goldstone_derivation_1957}
as it is independent of the reference system (DFT or HF) and it is extensive.
Extensivity means that for two systems $A$ and $B$
the energy of the combined system equals the sum of the energy of the individual
constituents
\[
  E(AB) = E(A) + E(B),
\]
assuming an identical chemical environment.
The time dependent formulation of perturbation theory
also lends itself naturally to Goldstone diagrams to visualize
terms occurring in the perturbation expansion.

Complementary treatments can be found in
\parencite{kutzelnigg_how_2009,lancaster_quantum_2014,shavitt_many-body_2009,fetter_quantum_2003,coleman_introduction_2015}.

    \section{Time dependent perturbation theory}
In perturbation theory the exact Hamiltonian is separated into the unperturbed
part $\hat H_0$, which can be solved, and the perturbation $\hat H_1$, which
contains the full Coulomb electron-electron interaction:
\[
  \hat H =
    \underbrace{\hat T + \hat V_{\rm ne} + \hat V_{\rm eff}}_{\hat H_0} +
    \underbrace{\hat V_{ee} - \hat V_{\rm eff}}_{\hat H_1}.
\]
$\hat V_{\rm eff}$ is the effective interaction employed by the reference, e.g.
the Hartree-Fock approximation.
Note that this effective interaction included in $\hat H_0$
must be subtracted again by the perturbation to arrive at results of the
full Hamiltonian.

Let $\psi_p$ be the spin-orbitals of the unperturbed Hamiltonian of the
HF or DFT reference and let
$\ket\Phi$ denote the Slater determinant of the ground state, where the lowest
$N$ states are occupied by the $N$ electrons present in the system.
These states are called \emph{unexcited states} while the states that
are unoccupied in $\ket\Phi$ are called \emph{excited} states. We will use
the letters $i,j,k,\ldots$ to label unexcited states, $a,b,c,\ldots$ to
label excited states and $p,q,r,\ldots$ to label general states.
We can write $\ket\Phi$ in second quantization as the result of applying the
electron creation operator $\hat c^\dagger_i$ for all unexcited states $i$
on the vacuum state, $\ket{\ }$, without any electrons:
\[
  \ket\Phi = \prod_i \hat c^\dagger_i \ket{\ } =
    \hat c^\dagger_1\ldots \hat c^\dagger_N \ket{\ }.
\]
Note that each application of $\hat c^\dagger_i$ changes the number
of particles and thus the dimensionality of the Hilbert space. The states of
second quantization are elements of the union of
all Hilbert spaces of zero particles, one particle, two particle and so forth,
which is called \emph{Fock-space}. \index{Fock-space}
The beauty of second quantization is that it hides all the tedious footwork of
anti-symmetrization in the algebra of the creation and annihilation operators,
which is completely given by the anti-commutator relations:
\begin{equation}
  \{\hat c^\dagger_p,\hat c^\dagger_q\} = 0  \qquad
    \{\hat c_p,\hat c_q\} = 0  \qquad
    \{\hat c^\dagger_p,\hat c_q\} = \delta_{pq},
  \label{eqn:anti-commutator}
\end{equation}
where $\{\hat A, \hat B\} = \hat A\hat B + \hat B\hat A$. One immediate
consequence of the these relations is the Pauli exclusion principle
disallowing two fermions in the same state:
\[
  \hat c^\dagger_p \hat c^\dagger_p \ket{\ } =
    \frac12\{\hat c^\dagger_p, \hat c^\dagger_p\} \ket{\ } = 0.
\]
Note that the vacuum, $\ket{\ }$, is a state while the number 0 is not.

The Fock-space used for the creation and annihilation operators is
spanned by the Slater determinants of the eigenfunctions $\psi_p$ of
the unperturbed Hamiltonian $\hat H_0$. $\hat H_0$ is therefore diagonal,
counting the eigenenergy $\eps_p$ for each occupied state $p$, irrespective
of whether it is an excited or unexcited state:
\begin{equation}
  \label{eqn:H0Electrons}
  \hat H_0 = \sum_p \eps_p \hat c^\dagger_p \hat c_p.
\end{equation}

\subsection{Particle/hole picture}
Let us now introduce the particle/hole picture where we want to consider
the non-interacting ground state $\ket\Phi$ as the new vacuum state instead
of the true vacuum, $\ket{\ }$, without any electrons.
In this picture we care about the difference to the non-interacting
ground state $\ket\Phi$ and we only count excited states that are
now occupied called \emph{particles}, and unexcited states that are no
longer occupied called \emph{holes}. While $\hat c^\dagger_a$ indeed creates
a particle in an excited state, a hole in an unexcited state has to be created
by annihilating a formerly occupied state by $\hat c_i$.
For unexcited states below the
Fermi energy the meaning of creation and annihilation has to be reversed.
We define the creation and annihilation operators for particles
($\hat a^\dagger, \hat a$) and for holes ($\hat i^\dagger, \hat i$):
\begin{equation}
  \label{eqn:particleHoleOps}
  \begin{array}{cclc}
    \hat a^\dagger = \hat c^\dagger_a \quad &
    \hat a = \hat c_a \quad &
    \textnormal{for} & \eps_a > \eps_{\rm F} \\
    \hat i^\dagger = \hat c_i \quad &
    \hat i = \hat c^\dagger_i \quad &
    \textnormal{for} & \eps_i \leq \eps_{\rm F} \\
  \end{array}
\end{equation}
The anti-commutator relations for these operators follow directly from
(\ref{eqn:anti-commutator}) and the only non-vanishing relations are
\begin{equation}
  \{\hat i^\dagger, \hat j\} = \delta_{ij} \qquad
  \{\hat a^\dagger, \hat b\} = \delta_{ab}.\
\end{equation}
Inserting the particle/hole operators into (\ref{eqn:H0Electrons}) splits the
sum over states $p$ into a sum over holes $i$ and a sum over particles $a$,
giving
\begin{eqnarray}
  \hat H_0 &=&
    \sum_i \eps_i \hat i \hat i^\dagger + \sum_a \eps_a \hat a^\dagger \hat a
  \nonumber \\
  &=&
    \underbrace{\sum_i \eps_i}_{=E_0} -
    \sum_i \eps_i \hat i^\dagger \hat i + \sum_a \eps_a \hat a^\dagger \hat a,
  \label{eqn:H0ParticleHole}
\end{eqnarray}
where we have used the anti-commutator relation
to put the operators into normal order, such that all annihilation
operators appear on the right.

Next, we need to translate the perturbation $\hat H_1$ into the particle/hole
formalism. We start with its second quantized representation using
the electron creation and annihilation operators
$\hat c_p^\dagger$ and $\hat c_p$:
\begin{equation}
  \label{eqn:perturbation2nd}
  \hat H_1 =
    \frac12 \sum_{pqrs} V^{pq}_{sr}
      \hat c^\dagger_p \hat c^\dagger_q \hat c_r \hat c_s
    -\sum_{pq} v^p_q \hat c^\dagger_p \hat c_q,
\end{equation}
where $V^{pq}_{sr}$ and $v^p_q$ are the matrix elements
of the first quantized operators $\hat V_{\rm ee}$ and
 $\hat V_{\rm eff}$, given by
\begin{eqnarray}
  \label{eqn:coulomb}
  V^{pq}_{sr}\ =\ \langle pq|\hat V_{\rm ee}|sr\rangle &=&
    \iint \d{\vec x}\, \d{\vec x'}\,
      \psi_p^\ast(\vec x) \psi_q^\ast(\vec x') \frac1{|\vec r - \vec r'|}
      \psi_r(\vec x') \psi_s(\vec x) \\
  v^p_q\ =\ \langle p|\hat V_{\rm eff}|q\rangle &=&
    \int \d{\vec x}\,
      \psi_p^\ast(\vec x) \left(\hat V_{\rm eff} \psi_q\right)(\vec x),
\end{eqnarray}
with $\int\d\vec x=\sum_\alpha\int\d\vec r$.
The factor $\frac12$ in (\ref{eqn:perturbation2nd}) accounts for double counting
when not restricting the sum over $p,q,r$ and $s$ to distinct elements
of $V^{pq}_{sr}$.

Unfortunately, it is not as straight forward to translate $\hat H_1$ into
the particle/hole picture as it was for $\hat H_0$ since there are now up to
4 different indices which can be either holes or particles. If, for example,
$p,q,r$ are particle indices $a,c,b$ and $s$ is a hole index $i$,
$\hat V_{\rm ee}$ creates the particles $a,c$ and the hole $i$ while it
destroys the particle $b$, as shown in Figure \ref{fig:H1diagrams}.
The electron creation
and annihilation operators $\hat c^\dagger$ and $\hat c$ always occur in pairs
thus keeping the number of electrons constant. However, particle and hole
creation and annihilation operators are in general not normal ordered
and occur in any constellation.
The number of particles and holes is therefore not necessarily
constant.
Figure \ref{fig:H1diagrams} also shows a particle/hole pair created by the
action of $\hat V_{\rm eff}$ indicated by a shaded circle.
\begin{figure}[h]
\[
  V^{ac}_{ib}
    \hat c^\dagger_a \hat c^\dagger_c \hat c_b \hat c_i =
  V^{ac}_{ib}
    \hat a^\dagger \hat c^\dagger \hat b \hat i^\dagger =
  \diagramBox{
    \begin{fmffile}{Vee}
    \begin{fmfgraph*}(60,50)
      \fmfset{arrow_len}{6}
      \fmfstraight
      \fmfleft{v00,v01,v02}
      \fmfright{v60,v61,v62}
      \fmf{phantom}{v00,v10,v20,v30,v40,v50,v60}
      \fmf{phantom}{v01,v11}
        \fmf{photon}{v11,v21}
        \fmf{photon,label=$\hat V_{\rm ee}$,label.side=right,tension=0.5}{v21,v41}
        \fmf{photon}{v41,v51}
        \fmf{phantom}{v51,v61}
      \fmf{phantom}{v02,v12,v22,v32,v42,v52,v62}
      \fmffreeze
      \fmf{fermion,left=0.1,label=$i$,label.dist=4}{v22,v11}
      \fmf{fermion,left=0.1,label=$a$,label.dist=4}{v11,v02}
      \fmf{fermion,label=$b$,label.dist=4,label.side=right}{v60,v51}
      \fmf{fermion,label=$c$,label.dist=4}{v51,v62}
    \end{fmfgraph*}
    \end{fmffile}
  }
  \qquad
  v^a_i
    \hat c^\dagger_a \hat c_i =
  v^a_i
    \hat a^\dagger \hat i^\dagger =
  \diagramBox[0]{
    \begin{fmffile}{Veff}
    \begin{fmfgraph*}(25,25)
      \fmfset{arrow_len}{6}
      \fmfstraight
      \fmftop{v01,v21}
      \fmfbottom{v10}
      \fmf{fermion,label=$a$,left=0.1,label.dist=4,label.side=left}{v10,v01}
      \fmf{fermion,label=$i$,left=0.1,label.dist=4}{v21,v10}
      \fmfv{decor.shape=circle,decor.fill=shaded,decor.size=10,label=$\hat V_{\rm eff}$}{v10}
    \end{fmfgraph*}
    \end{fmffile}
  }
\]
\caption{
  Examples of terms occurring in $\hat H_1$ and their respective
  diagrammatic representation
}
\label{fig:H1diagrams}
\end{figure}

\subsection{Interaction picture}
So far, all operators were given in the Schrödinger picture, where all time
evolution takes place in the states $\psi_p$ and the operators are time
independent. The Heisenberg picture, on the other hand, keeps the states
time independent and all time evolution is put into the operators.
In order to do time dependent perturbation theory, it is convenient to use the
interaction picture, which is a hybrid of the Schrödinger and the Heisenberg
picture. In the interaction picture the operators evolve according to a
time evolution solely based on the unperturbed part of the Hamiltonian
$\hat H_0$, while the states evolve only due to the action of the perturbation
$\hat H_1$.

Given a state in the Schrödinger picture $\ket{\Psi(t)}$ we define the
corresponding state in the interaction picture $\ket{\Psi_I(t)}$ by
``undoing'' the time evolution which originates from $\hat H_0$:
\begin{equation}
  \ket{\Psi_I(t)} = e^{\im \hat H_0 t} \ket{\Psi(t)}.
\end{equation}
To find the equation of motion for states in the interaction picture we
derive with respect to time, giving
\begin{eqnarray}
  \nonumber
  \im \frac\der{\der t} \ket{\Psi_I(t)} &=&
    -\hat H_0 e^{\im \hat H_0 t} \ket{\Psi(t)}
    +e^{\im \hat H_0 t} \im \frac\der{\der t} \ket{\Psi(t)}\\
  \nonumber
  &=&
    -e^{\im \hat H_0 t} \hat H_0  \ket{\Psi(t)}
    +e^{\im \hat H_0 t} (\hat H_0 + \hat H_1) \ket{\Psi(t)}\\
  &=& \underbrace{
    e^{\im \hat H_0 t} \hat H_1 e^{-\im \hat H_0 t}
  }_{= \hat H_{1I}(t)} \ket{\Psi_I(t)}.
  \label{eqn:psiEOM}
\end{eqnarray}
So the time evolution of $\ket{\Psi_I(t)}$ is
determined by the perturbation $\hat H_{1I}(t)$ only.
The time dependence of $\hat H_{1I}(t)$, on the other hand, solely depends
on the unperturbed part of the Hamiltonian $\hat H_0$.
We are now interested in the time evolution operator $\hat U_I(t,t_0)$ that
evolves a state in the interaction picture from the time $t_0$ to the time $t$:
\begin{equation}
  \label{eqn:Udef}
  \hat U_I(t,t_0) \ket{\Psi_I(t_0)} = \ket{\Psi_I(t)}.
\end{equation}
Operating with $\im \der/\der t$ on both sides and using (\ref{eqn:psiEOM}) gives
\[
  \im \frac\der{\der t} \hat U_I(t,t_0) \ket{\Psi_I(t_0)} =
  \hat H_{1I}(t) \hat U_I(t,t_0) \ket{\Psi_I(t_0)},
\]
which must holds for all $\ket{\Psi_I(t_0)}$ thus leading to the equation of
motion for the time evolution operator in the interaction picture
\begin{equation}
  \im \frac\der{\der t} \hat U_I(t,t_0) =
  \hat H_{1I}(t) \hat U_I(t,t_0).
  \label{eqn:UEOM}
\end{equation}
Integrating both sides of the above equation with respect to the first argument
from $t_0$ to $t$ yields
\begin{equation}
  \hat U_I(t,t_0) = 1 - \im \int_{t_0}^t \d t'\,\hat H_{1I}(t') \hat U_I(t',t_0),
\end{equation}
where we have used $\hat U_I(t_0,t_0)=1$, which follows from (\ref{eqn:Udef}).
This integral equation can be solved by iteratively applying the above equation
to each occurrence of $\hat U_I(t',t_0),\hat U_I(t'',t_0)$ and so forth,
giving
\begin{eqnarray}
  \nonumber
  \hat U_I(t,t_0) &=& 1
    -\im \int_{t_0}^t\d t'\,\hat H_{1I}(t')
    +\im^2 \int_{t_0}^t\d t'\int_{t_0}^{t'}\d t''\,
      \hat H_{1I}(t')\hat H_{1I}(t'')
    -\im^3 \ldots \\
  &=&
    \sum_{n=0}^\infty (-\im)^n \int\limits_{t>t_1>\ldots>t_n>t_0}
    \d t_1\ldots \d t_n\,
    \hat H_{1I}(t_1) \ldots \hat H_{1I}(t_n).
  \label{eqn:UExpansion}
\end{eqnarray}
This means that the time evolution operator in the interaction picture can be
constructed by applying the perturbation $\hat H_{1I}$ any number of times and
at all possible times between $t_0$ and $t$ in a time ordered manner. The
number of applications of $\hat H_{1I}$ is called the \emph{order} in the
perturbation expansion.
The perturbation expansion for the time evolution operator can be truncated
at a finite order. However, the truncated expansion is not guaranteed to
converge at any finite order. In metals, for example, no truncation is
convergent beyond first order. This is a considerable drawback for finite order
perturbation methods, such as second order M\o ller--Plesset
perturbation theory (MP2).

    \section{The Gell-Mann--Low theorem}
In the previous section we have seen how to evolve any given state from $t_0$
to $t$. We are, however, interested in eigenstates of the full Hamiltonian,
most importantly in its ground state $\ket{\Psi}$. All we have are
eigenstates of the non-interacting Hamiltonian $\hat H_0$ which we can evolve
from. In general, such an evolved state $\hat U_I(t,t_0) \ket{\Phi(t_0)}$ will
be neither an eigenstate of the non-interacting Hamiltonian $\hat H_0$ nor of
the full Hamiltonian $\hat H$.
We can, however, introduce a time dependent perturbation
\begin{equation}
  \hat H(t) = \hat H_0 + e^{\eta t} \hat H_1
  \label{eqn:AdiabaticSwitching}
\end{equation}
for $t\leq0$, slowly turning on the electron-electron interaction for small
$\eta>0$.
The system starts with the unperturbed Hamiltonian $\hat H_0$ at
$t=-\infty$ and at $t=0$ the interaction is fully turned on,
giving $\hat H$.
According to the adiabatic theorem, the system will stay in an eigenstate
of $\hat H(t)$ at all times when starting from an eigenstate of
$\hat H_0$ at $t=-\infty$ if the transition is sufficiently slow, which holds
for
\[
  \eta \ll \Delta E.
\]
$\Delta E$ is the energy change of the ground state when turning
on the interaction.
Note that starting in the ground state of $\hat H_0$ at $t=-\infty$ does not
guaranteed that the evolved eigenstate at $t=0$ is indeed
the ground state of the full Hamiltonian $\hat H$. Level crossings may occur.
We will, however, assume that they do not occur for the ground state and that
the ground state $\ket\Phi$ of $\hat H_0$ evolves adiabatically into the
ground state $\ket\Psi$ of $\hat H$:
\begin{equation}
  \ket\Psi = \hat U_\eta(0,-\infty) \ket\Phi,
  \label{eqn:psiFromPhi}
\end{equation}
where we drop the explicit denotation of the interaction picture in favor
of denoting the dependence on the parameter $\eta$ of the transition speed,
writing from now on
\begin{equation}
  \hat U_\eta(t,t_0) =
    \sum_{n=0}^\infty (-\im)^n
    \int\limits_{t>t_1>\ldots>t_n>t_0} \d t_1\ldots\d t_n\,
    \hat H_1(t_1)\ldots\hat H_1(t_n)
  \label{eqn:UExpansionNew}
\end{equation}
and
\begin{equation}
  \hat H_1(t) = e^{\im \hat H_0 t} e^{\eta t} \hat H_1 e^{-\im \hat H_0 t}.
\end{equation}
Although $\hat U_\eta$ and $\hat H_1$ depend on $\eta$ we hope that the
results are, in the end, independent of $\eta$ if it is chosen sufficiently
small.
The quantity we are now most interested in is the energy difference
$\Delta E$ between the ground state energy of the non-interacting Hamiltonian
$\hat H_0$ and the ground state energy of the fully interacting Hamiltonian
$\hat H=\hat H_0 + \hat H_1$. The fully interacting ground state $\ket\Psi$
satisfies the Schrödinger equation
\[
  (\hat H_0 + \hat H_1) \ket\Psi = (E_0 + \Delta E) \ket\Psi.
\]
Multiplying both sides with $\bra\Phi$ from the left and using
(\ref{eqn:psiFromPhi}) gives
\begin{eqnarray}
  \nonumber
  E_0 + \Delta E &=&
  \frac{\langle\Phi|\hat H_0|\Psi\rangle}{\braket\Phi\Psi}+
  \frac{\langle\Phi|\hat H_1|\Psi\rangle}{\braket\Phi\Psi}\\
  &=&
  \frac{\langle\Phi|\hat H_0\hat U_\eta(0,-\infty)|\Phi\rangle}{
    \langle\Phi|\hat U_\eta(0,-\infty)|\Phi\rangle
  }+
  \frac{\langle\Phi|\hat H_1(0)\hat U_\eta(0,-\infty)|\Phi\rangle}{
    \langle\Phi|\hat U_\eta(0,-\infty)|\Phi\rangle
  }
  \label{eqn:GellMannLow}
\end{eqnarray}
The last equation now relates the ground state energy of the
fully interacting system to vacuum expectation values (VEV) of the
non-interacting system in the particle/hole picture, where $\ket\Phi$ is the
vacuum state without any particle or hole excitations. In principle, these
expectation values can be evaluated despite the infinite sums hidden
in the time evolution operators $\hat U_\eta$.

    \section{Wick's theorem}
Equation (\ref{eqn:GellMannLow}) still has two practical drawbacks. First,
there is yet no systematic recipe given how to approximate the time evolution
operator $\hat U_\eta$, consisting of an infinite sum, and second, one still
needs to show that $\Delta E$ does not depend on the choice of $\eta$ if
it is chosen sufficiently small. We start with the diagrammatic representation
of the terms occurring in (\ref{eqn:GellMannLow}), where we need to evaluate 
terms of the form
\[
  \langle\Phi|\hat A\hat B\ldots|\Phi\rangle.
\]
$\hat A, \hat B, \ldots$ are arbitrary creation or annihilation operators.
If these operators were normal ordered, such that all annihilation operators
are on the right side of all creation operators, its vacuum expectation value
(VEV) would simply be 0, given there is at least one creation or
annihilation operator. A relation between the VEV of the operators
in the order given and the VEV in normal order is therefore desired.
Let $N[\hat A\hat B\ldots]$ denote normal ordering of the operators where in
the case of fermions the sign changes for each transposition.
For two creation or annihilation operators, there is only one non-trivial case
\[
  N[\hat p\hat q^\dagger] = -\hat q^\dagger\hat p,
\]
in the other three cases the operators are already in normal order. Since the
vacuum expectation value of two operators in normal order is 0, we can rewrite
a given VEV of two operators
\begin{equation}
  \langle \Phi|\hat A\hat B|\Phi\rangle =
    \langle\Phi|\hat A\hat B|\Phi\rangle -
      \langle\Phi|N[\hat A\hat B]|\Phi\rangle =
    \langle\Phi|\hat A\hat B - N[\hat A\hat B]|\Phi\rangle
  =
  \langle\Phi|
  \contraction{}{
    \hat A
  }{
  }{
    \hat B
  }
  \hat A\hat B
  |\Phi\rangle,
  \label{eqn:contractionIntro}
\end{equation}
\index{contraction}
defining the \emph{contraction} of two operators $\hat A$ and $\hat B$
by\footnote{
  Note that contractions are usually defined by
  $\contraction[1ex]{}{\hat A}{}{\hat B}
  \hat A\hat B
  = T[\hat A\hat B] - N[\hat A\hat B]$, where
  $T[\hat A\hat B]$ denotes time ordering of the operators, such
  that the time increases from right to left. In our case, the terms
  occurring in (\ref{eqn:GellMannLow}) are already time ordered by the
  constraints of the integrals so we will drop the time ordering symbol.
}
\[
  \contraction{}{\hat A}{}{\hat B}
  \hat A\hat B := \hat A\hat B - N[\hat A\hat B].
\]
This may seem arbitrary but it turns out that the contraction of two
operators is simply a number.
For two creation or annihilation operators there are four possible contractions
\begin{equation}
  \contraction{}{\hat p}{}{\hat q}
  \hat p\hat q =
  0 \qquad
  \contraction{}{\hat p}{}{\hat q}
  \hat p\hat q^\dagger =
  \{\hat p,\hat q^\dagger\}=\delta_{pq} \qquad
  \contraction{}{\hat p}{^\dagger}{\hat q}
  \hat p^\dagger\hat q =
  0 \qquad
  \contraction{}{\hat p}{^\dagger}{\hat q}
  \hat p^\dagger\hat q^\dagger =
  0.
  \label{eqn:contractionOfTwo}
\end{equation}
Since the contraction of two operators is a number it is unaffected
by normal ordering. This allows us to make the desired connection between the
given order and the normal order of two operators $\hat A\hat B$:
\[
  \hat A\hat B =
  N[\hat A\hat B] +
  \contraction{}{\hat A}{}{\hat B}
  \hat A\hat B
  = 
  N[\hat A\hat B +
  \contraction{}{\hat A}{}{\hat B}
  \hat A\hat B]
\]
This result can be generalized to more than two operators by the virtue of
Wick's theorem \parencite{wick_evaluation_1950,peskin_introduction_1995}:
\index{Wick's theorem}
\begin{equation}
  \hat A\hat B\hat C\ldots = N\left[
    \hat A\hat B\hat C\ldots +
    \textnormal{\emph{all possible contractions of} }\hat A\hat B\hat C\ldots
  \right]
  \label{eqn:wick}
\end{equation}
For a sequence of four operators this gives for instance
\begin{eqnarray}
  \nonumber
  \hat A\hat B\hat C\hat D &=&
  N[\hat A\hat B\hat C\hat D] +
  N[\contraction{}{\hat A}{}{\hat B} \hat A\hat B\hat C\hat D] +
  N[\contraction{}{\hat A}{\hat B}{\hat C} \hat A\hat B\hat C\hat D] +
  N[\contraction{}{\hat A}{\hat B\hat C}{\hat D} \hat A\hat B\hat C\hat D] +
  N[\contraction{\hat A}{\hat B}{}{\hat C} \hat A\hat B\hat C\hat D] + \\
  & &
  N[\contraction{\hat A}{\hat B}{\hat C}{\hat D} \hat A\hat B\hat C\hat D] +
  N[\contraction{\hat A\hat B}{\hat C}{}{\hat D} \hat A\hat B\hat C\hat D] +
  N[\contraction{}{\hat A}{}{\hat B}
    \contraction{\hat A\hat B}{\hat C}{}{\hat D}
    \hat A\hat B\hat C\hat D] +
  N[\contraction[2ex]{}{\hat A}{\hat B}{\hat C}
    \contraction{\hat A}{\hat B}{\hat C}{\hat D}
    \hat A\hat B\hat C\hat D] +
  N[\contraction[2ex]{}{\hat A}{\hat B\hat C}{\hat D}
    \contraction{\hat A}{\hat B}{}{\hat C}
    \hat A\hat B\hat C\hat D],
  \label{eqn:wickFour}
\end{eqnarray}
where we can now reorder the operators, changing the sign appropriately, to
pull contracted operators out of the normal ordering operator. For example
\[
  N[\contraction{}{\hat A}{\hat B}{\hat C} \hat A\hat B\hat C\hat D] =
  -\contraction{}{\hat A}{}{\hat C} \hat A\hat C\times N[\hat B\hat D].
\]
Since vacuum expectation values of normal ordered operators vanish, only
the fully contracted terms survive. The VEV of the four operators in
(\ref{eqn:wickFour}) thus evaluates to
\begin{equation}
  \langle\Phi|\hat A\hat B\hat C\hat D|\Phi\rangle =
  \contraction{}{\hat A}{}{\hat B}
    \contraction{\hat A\hat B}{\hat C}{}{\hat D}
    \hat A\hat B\hat C\hat D -
  \contraction{}{\hat A}{}{\hat C} \hat A\hat C
    \contraction{}{\hat B}{}{\hat D} \hat B\hat D +
  \contraction{}{\hat A}{}{\hat D} \hat A\hat D
    \contraction{}{\hat B}{}{\hat C} \hat B\hat C.
\end{equation}

\subsection{Application to the perturbation}
We can now apply this algebra to a simplified case without an effective
interaction where the perturbation is given by
\begin{equation}
  \hat H_1(t) =
    e^{\im \hat H_0 t} e^{\eta t}
    \frac12 \sum_{pqrs} V_{sr}^{pq}
      \hat c^\dagger_p \hat c^\dagger_q \hat c_r \hat c_s
    e^{-\im \hat H_0 t}.
  \label{eqn:simplePerturbation}
\end{equation}
Note that the electron creation and annihilation operators $\hat c^\dagger_p$
and $\hat c_p$ have to be
expressed in terms of particle and hole creation and annihilation operators
$\hat p^\dagger$ and $\hat p$,
since we want to use the non-interacting ground state $\ket\Phi$ as the vacuum
state. From (\ref{eqn:particleHoleOps}) we get
\begin{equation}
  \hat c^\dagger_p = \left\{
  \begin{array}{ll}
    \hat p^\dagger & \textnormal{for } \eps_p >\eps_{\rm F} \\
    \hat p & \textnormal{otherwise,}
  \end{array}
  \right. \qquad
  \hat c_p = \left\{
  \begin{array}{ll}
    \hat p & \textnormal{for } \eps_p >\eps_{\rm F} \\
    \hat p^\dagger & \textnormal{otherwise.}
  \end{array}
  \right.
\end{equation}
We start with evaluating the numerator of the right term in
(\ref{eqn:GellMannLow}) in zeroth order of the expansion of the time
evolution operator, where $\hat U^{(0)}_\eta(0,-\infty)=1$:
\begin{equation}
  \langle\Phi|\hat H_1(0)|\Phi\rangle =
    \frac12 \sum_{pqrs} V_{sr}^{pq}
    \contraction{}{\hat c}{^\dagger_p}{\hat c}
    \contraction{\hat c^\dagger_p \hat c^\dagger_q}{\hat c}{_r}{\hat c}
      \hat c^\dagger_p \hat c^\dagger_q \hat c_r \hat c_s
  +
    \frac12 \sum_{pqrs} V_{sr}^{pq}
    \contraction[2ex]{}{\hat c}{^\dagger_p \hat c^\dagger_q}{\hat c}
    \contraction{\hat c^\dagger_p}{\hat c}{^\dagger_q\hat c_r}{\hat c}
      \hat c^\dagger_p \hat c^\dagger_q \hat c_r \hat c_s
  +
    \frac12 \sum_{pqrs} V_{sr}^{pq}
    \contraction[2ex]{}{\hat c}{^\dagger_p\hat c^\dagger_q\hat c_r}{\hat c}
    \contraction{\hat c^\dagger_p}{\hat c}{^\dagger_q}{\hat c}
      \hat c^\dagger_p \hat c^\dagger_q \hat c_r \hat c_s
  \label{eqn:firstOrderH1}
\end{equation}
According to (\ref{eqn:contractionOfTwo}) the only non-vanishing contraction
comes from operators of the form $\hat p\hat p^\dagger$, first creating a hole
or a particle in the state $p$ and subsequently destroying it.
Thus, the first term in (\ref{eqn:firstOrderH1}) must vanish.
The second and the third term can only survive if $p,q,r$ and $s$ are hole
indices $i,j,k$ and $l$, giving
\begin{equation}
  \langle\Phi|\hat H_1(0)|\Phi\rangle =
    \frac12 \sum_{ijkl} V_{lk}^{ij}
    \contraction[2ex]{}{\hat i}{\hat j}{\hat k}
    \contraction{\hat i}{\hat j}{\hat k^\dagger}{\hat l}
      \hat i\hat j\hat k^\dagger\hat l^\dagger
  +
    \frac12 \sum_{ijkl} V_{lk}^{ij}
    \contraction[2ex]{}{\hat i}{\hat j\hat k^\dagger}{\hat l}
    \contraction{\hat i}{\hat j}{}{\hat k}
      \hat i\hat j\hat k^\dagger\hat l^\dagger
  =
    -\frac12 \sum_{ij} V_{ji}^{ij}
    +\frac12 \sum_{ij} V_{ij}^{ij}.
  \label{eqn:firstOrderMatrix}
\end{equation}
Note that neither of the two terms individually respects the Pauli exclusion
principle. However, the offending terms, where $i=j$, cancel in the sum of all
terms. By the merit of Wick's theorem it is no longer necessary to
keep track of disallowed states individually. They simply cancel in the sum
of all contractions.

Let us now proceed evaluating the numerator of (\ref{eqn:GellMannLow}) in
the simplified case of above with the perturbation given by
(\ref{eqn:simplePerturbation}).
The next order in the expansion of the time evolution operator is
\[
  \hat U^{(1)}_\eta(0,-\infty) =
  -\im\int\limits_{-\infty}^0\d t_1\,\hat H_1(t_1).
\]
For the application of Wick's theorem at $t\neq0$ it is more convenient to
write the Hamiltonian of the perturbation $\hat H_1(t)$ in terms of time
dependent creation and annihilation operators.
All states in (\ref{eqn:simplePerturbation}) appear in both exponents of
$\pm\im\hat H_0t$ except for $p,q$
and $r,s$ which can only appear in the left or in the right exponent,
respectively. Thus, we get
\[
  \hat H_1(t) = e^{\eta t}\frac12\sum_{pqrs} V_{sr}^{pq}
  \underbrace{
    \left(\hat c^\dagger_p e^{\im\eps_pt}\right)
  }_{=\hat c^\dagger_p(t)}
  \left(\hat c^\dagger_q e^{\im\eps_qt}\right)
  \left(\hat c_r e^{-\im\eps_rt}\right)
  \underbrace{
    \left(\hat c_s e^{-\im\eps_st}\right)
  }_{=\hat c_s(t)},
\]
which can be used to evaluate the next order:
$\langle\Phi|\hat H_1(0)\hat U^{(1)}_\eta(0,-\infty)|\Phi\rangle$
\[
   =
  \bracket\Phi{
    -\im
    \int_{-\infty}^0\d t_1\,e^{\eta t_1}
    \frac14
    \sum_{pqrstuvw}
      V_{sr}^{pq}
      V_{wv}^{tu}\,
      \hat c^\dagger_p\hat c^\dagger_q\hat c_r\hat c_s
      \hat c^\dagger_t(t_1)\hat c^\dagger_u(t_1)\hat c_v(t_1)\hat c_w(t_1)
  }\Phi
\]
There are 4 creation and 4 annihilation operators in this expression, thus
there are $4!=24$ non-vanishing ways of contractions possible. We will, for
now, just look at the following:
\begin{eqnarray*}
  & &  -\im
    \int_{-\infty}^0\d t_1\,e^{\eta t_1}
    \frac14
    \sum_{pqrstuvw}
      V_{sr}^{pq}
      V_{wv}^{tu}\,
    \contraction[2ex]{}{
      \hat c}{^\dagger_p\hat c^\dagger_q\hat c_r\hat c_s
      \hat c^\dagger_t(t_1)\hat c^\dagger_u(t_1)\hat c_v(t_1)}{\hat c}
    \contraction{
      \hat c^\dagger_p\hat c^\dagger_q\hat c_r}{\hat c}{_s}{
      \hat c}
    \bcontraction[2ex]{
      \hat c^\dagger_p}{\hat c}{^\dagger_q\hat c_r\hat c_s
      \hat c^\dagger_t(t_1)\hat c^\dagger_u(t_1)}{\hat c}
    \bcontraction{
      \hat c^\dagger_p\hat c^\dagger_q}{\hat c}{_r\hat c_s
      \hat c^\dagger_t(t_1)}{\hat c}
      \hat c^\dagger_p\hat c^\dagger_q\hat c_r\hat c_s
      \hat c^\dagger_t(t_1)\hat c^\dagger_u(t_1)\hat c_v(t_1)\hat c_w(t_1) \\
  &=&  -\im
    \int_{-\infty}^0\d t_1\,e^{\eta t_1}
    \frac14
    \sum_{pqrstuvw}
      V_{sr}^{pq}
      V_{wv}^{tu}\,
    \contraction{}{\hat c}{^\dagger_p}{\hat c}  \hat c^\dagger_p\hat c_w(t_1)
    \contraction{}{\hat c}{_s}{\hat c}          \hat c_s\hat c^\dagger_t(t_1)
    \contraction{}{\hat c}{^\dagger_q}{\hat c}  \hat c^\dagger_q\hat c_v(t_1)
    \contraction{}{\hat c}{_r}{\hat c}          \hat c_r\hat c^\dagger_u(t_1)
\end{eqnarray*}
For non-vanishing terms, $p,w$ and $q,v$ must be hole indices $i$ and $j$,
respectively. Similarly, $s,t$ and $r,u$ must be particle indices $a$ and $b$,
giving
\begin{equation}
  -\im
    \int_{-\infty}^0\d t_1\,e^{\eta t_1}
    \frac14
    \sum_{ijab}
      V_{ab}^{ij}
      V_{ij}^{ab}\,
      e^{-\im\eps_it_1}
      e^{\im\eps_at_1}
      e^{-\im\eps_jt_1}
      e^{\im\eps_bt_1}
  = 
    \frac14
    \sum_{ijab}
    \frac{
      V_{ab}^{ij}
      V_{ij}^{ab}
    }{
      \eps_i+\eps_j-\eps_a-\eps_b+\im\eta
    }.
  \label{eqn:MP2directMatrix}
\end{equation}
For a system where there exists a finite gap $E_g>0$ such that
$|\eps_i-\eps_a|>E_g$ for
all holes $i$ and particles $a$ we can simply choose $\eta\ll E_g$ to get a
result independent of the choice of $\eta$. For metals with a non-degenerate
ground state there is no finite $E_g>0$ for all $i$ and $a$, however,
$|\eps_i-\eps_a|>0$. In this case the limit
$\eta\rightarrow0$ can only be taken after summing over all states, as
done in Section \ref{sec:RPAUEG} for the Uniform Electron Gas. If the ground
state is degenerate we have to resort to Degenerate State Perturbation Theory,
which is not discussed here.

Wick's theorem provides a systematic way to evaluate the vacuum expectation
values occurring in the Gell-Mann--Low theorem (\ref{eqn:GellMannLow}).
Instead of keeping track of which states are occupied after each interaction
we can simply sum over contractions of the operator matrices $V_{sr}^{pq}$
and $v_q^p$ occurring in the perturbation, as shown in
(\ref{eqn:firstOrderMatrix}) and in (\ref{eqn:MP2directMatrix}).
We do, however, have to sum over all possible contractions.

    \section{Goldstone diagrams}
\label{sec:MBPTGold}
The number of possible contractions in Wick's theorem quickly becomes too
large to evaluate vacuum expectation values as we did in the previous section.
So far, we have used diagrams solely for depicting the action of a second
quantized operator where the initial state was shown below and the final state
above the operator symbol, as shown in Figure \ref{fig:H1diagrams}.
It is time to rigorously introduce the diagrammatic notation employed by
Goldstone in order to enumerate and evaluate all contractions.

\index{vertex}
Each Coulomb interaction $\hat V_{\rm ee}$ is represented by
a horizontal wiggly line between two vertices. Each \emph{vertex} consists of
one electron creation operator and one electron annihilation operator
represented by an outbound and an inbound \emph{leg}, respectively.
Without loss of generality, the outer operators $\hat c^\dagger_p$ and
$\hat c_s$ are associated with the left vertex and
the inner operators $\hat c^\dagger_q$ and $\hat c_r$ with the right vertex.
Each effective interaction $\hat V_{\rm eff}$ is represented by a vertex
in form of a shaded circle.
Therefore, lower indices of the matrices $V_{sr}^{pq}$ and $v_q^p$ are from
inbound legs, upper indices are from outbound legs and left indices are from
left legs.
\begin{figure}[h]
\begin{center}
  \begin{tabular}{cc}
  $V_{sr}^{pq}\hat c_p^\dagger\hat c_q^\dagger\hat c_r\hat c_s=$
  \diagramBox{
    \begin{fmffile}{Vpqrs}
    \begin{fmfgraph*}(60,40)
      \fmfset{arrow_len}{6}
      \fmfstraight
      \fmftop{p,q}
      \fmfbottom{s,r}
      \fmf{fermion,label.side=right,label=$s$}{s,vl}
      \fmf{fermion,label.side=left,label=$r$}{r,vr}
      \fmf{fermion,label.side=right,label=$p$}{vl,p}
      \fmf{fermion,label.side=left,label=$q$}{vr,q}
      \fmf{boson,tension=0.5}{vl,vr}
    \end{fmfgraph*}
    \end{fmffile}
  } & \qquad
  $v_q^p\hat c_p^\dagger\hat c_q=$
  \diagramBox{
    \begin{fmffile}{vpq}
    \begin{fmfgraph*}(25,40)
      \fmfkeep{vpq}
      \fmfset{arrow_len}{6}
      \fmfstraight
      \fmftop{p}
      \fmfbottom{q}
      \fmf{fermion,label.side=left,label=$q$}{q,v}
      \fmf{fermion,label.side=left,label=$p$}{v,p}
      \fmfblob{10}{v}
    \end{fmfgraph*}
    \end{fmffile}
  }    
  \end{tabular}
\end{center}
  \caption{
    Representation of the Coulomb and effective interaction in Goldstone
    diagrams
  }
\end{figure}

Connections of these legs represent the contractions, where only
outbound legs may be connected to inbound legs, as they are the only
non-vanishing operator contractions. The connections are directed from an
outbound to an inbound leg, indicated by an arrow.
Time propagates from bottom to top and the last Coulomb or effective
interaction is usually at $t=0$. Connections directed forwards in time
represent contractions of the form
$\contraction{}{\hat c}{_p\ldots}{\hat c} \hat c_p\ldots\hat c^\dagger_q$
and are hence restricted to particles. Conversely, connections directed
backwards in time are restricted to holes.
Connections between legs of the same Coulomb interaction are called 
\emph{non-propagating} and are also restricted to holes, since they
are of the form
$\contraction{}{\hat c}{^\dagger_p\hat c^\dagger_q\hat c_r}{\hat c}
\hat c^\dagger_p\hat c^\dagger_q\hat c_r\hat c_s$.
Each connection is labeled with a unique index, over which to contract,
$a,b,c,\ldots$ for particles and $i,j,k,\ldots$ for holes.
\begin{figure}[h]
  \[
  \frac12 \sum_{ij} V_{ji}^{ij}
  \contraction[2ex]{}{\hat i}{\hat j}{\hat i}
  \contraction{\hat i}{\hat j}{\hat i^\dagger}{\hat j}
    \hat i\hat j\hat i^\dagger\hat j^\dagger =
  \diagramBox{
    \begin{fmffile}{ExchangeLabel}
    \begin{fmfgraph*}(30,30)
      \fmfkeep{ExchangeLabel}
      \fmfset{arrow_len}{6}
      \fmfleft{v11}
      \fmfright{v21}
      \fmf{boson}{v11,v21}
      \fmf{fermion,left=0.5,label=$i$,label.dist=4,tension=0}{v11,v21}
      \fmf{fermion,left=0.5,label=$j$,label.dist=4,tension=0}{v21,v11}
    \end{fmfgraph*}
    \end{fmffile}
  } \qquad
  \frac12 \sum_{ij} V_{ij}^{ij}
  \contraction[2ex]{}{\hat i}{\hat j\hat j^\dagger}{\hat i}
  \contraction{\hat i}{\hat j}{}{\hat j}
    \hat i\hat j\hat j^\dagger\hat i^\dagger=
  \diagramBox{
    \begin{fmffile}{HartreeLabel}
    \begin{fmfgraph*}(40,20)
      \fmfset{arrow_ang}{25}
      \fmfset{arrow_len}{5pt}
      \fmfleft{v11}
      \fmfright{v51}
      \fmf{phantom}{v11,v21}
        \fmf{boson}{v21,v31,v41} \fmf{phantom}{v41,v51}
      \fmf{plain,left=1.0,tension=0}{v21,v11}
        \fmf{fermion,left=1.0,tension=0}{v11,v21}
      \fmf{fermion,left=1.0,tension=0}{v41,v51}
        \fmf{plain,left=1.0,tension=0}{v51,v41} 
      \fmfv{label=$i$,label.dist=4}{v11}
      \fmfv{label=$j$,label.dist=4}{v51}
    \end{fmfgraph*}
    \end{fmffile}
  }
  \]
  \caption{
    Goldstone diagrams for the terms of (\ref{eqn:firstOrderMatrix})
    with non-propagating connections
  }
  \label{fig:firstOrder}
\end{figure}

\subsection{Symmetries}
\label{ssc:MBPT_GoldstoneSymmetries}
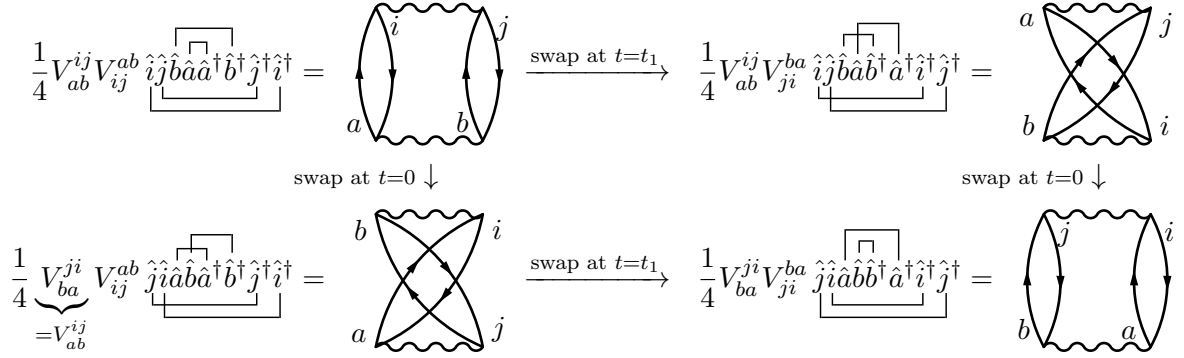
\begin{figure}[h]
  \begin{align*}
    \frac14
      V_{ab}^{ij}
      V_{ij}^{ab}\,
    \bcontraction[2ex]{}{
      \hat i}{\hat j\hat b\hat a
      \hat a^\dagger\hat b^\dagger\hat j^\dagger}{\hat i}
    \bcontraction{\hat i}{
      \hat j}{\hat b\hat a
      \hat a^\dagger\hat b^\dagger}{\hat j}
    \contraction[1.5ex]{
      \hat i\hat j}{\hat b}{\hat a
      \hat a^\dagger}{\hat b}
    \contraction{
      \hat i\hat j \hat b}{\hat a}{}{
      \hat a}
      \hat i\hat j \hat b\hat a
      \hat a^\dagger\hat b^\dagger\hat j^\dagger\hat i^\dagger
    &=
    \diagramBox{
      \begin{fmffile}{MP2d1}
      \begin{fmfgraph*}(50,50)
        \fmfset{arrow_len}{6pt}
        \fmfleft{v00,v01}
        \fmfright{v10,v11}
        \fmf{boson}{v00,v10}
        \fmf{boson}{v01,v11}
        \fmf{fermion,left=0.25}{v00,v01}
        \fmf{fermion,left=0.25}{v01,v00}
        \fmf{fermion,left=0.25}{v10,v11}
        \fmf{fermion,left=0.25}{v11,v10}
        \fmfv{label.angle=150,label.dist=6,label=$a$}{v00}
        \fmfv{label.angle=150,label.dist=6,label=$b$}{v10}
        \fmfv{label.angle=-30,label=$i$}{v01}
        \fmfv{label.angle=-30,label=$j$}{v11}
      \end{fmfgraph*}
      \end{fmffile}
    }
    \xrightarrow{\textnormal{swap at }t=t_1}
    &
    \frac14
      V_{ab}^{ij}
      V_{ji}^{ba}\,
    \bcontraction[2ex]{
      \hat i}{\hat j}{\hat b\hat a
      \hat b^\dagger\hat a^\dagger\hat i^\dagger}{\hat j}
    \bcontraction{}{\hat i}{
      \hat j\hat b\hat a
      \hat b^\dagger\hat a^\dagger}{\hat i}
    \contraction[2.5ex]{
      \hat i\hat j\hat b}{\hat a}{
      \hat b^\dagger}{\hat a}
    \contraction{
      \hat i\hat j}{\hat b}{\hat a}{
      \hat b}
      \hat i\hat j\hat b\hat a
      \hat b^\dagger\hat a^\dagger\hat i^\dagger\hat j^\dagger
    &=
    \diagramBox{
      \begin{fmffile}{MP2d3}
      \begin{fmfgraph*}(50,50)
        \fmfset{arrow_len}{6pt}
        \fmfleft{v00,v01}
        \fmfright{v10,v11}
        \fmf{boson}{v00,v10}
        \fmf{boson}{v01,v11}
        \fmf{fermion,left=0.25}{v00,v11}
        \fmf{fermion,left=0.25}{v11,v00}
        \fmf{fermion,left=0.25}{v10,v01}
        \fmf{fermion,left=0.25}{v01,v10}
        \fmfv{label.angle=150,label.dist=4,label=$b$}{v00}
        \fmfv{label.angle=30,label.dist=4,label=$i$}{v10}
        \fmfv{label.angle=-30,label.dist=4,label=$j$}{v11}
        \fmfv{label.angle=-150,label.dist=4,label=$a$}{v01}
      \end{fmfgraph*}
      \end{fmffile}
    }
    \\[1ex]
    &
    \scriptsize{\textnormal{swap at $t$=0}}\downarrow
    &&
    \scriptsize{\textnormal{swap at $t$=0}}\downarrow
    \\[1ex]
    \frac14
      \underbrace{V_{ba}^{ji}}_{=V_{ab}^{ij}}
      V_{ij}^{ab}\,
    \bcontraction[2ex]{
      \hat j}{\hat i}{\hat a\hat b
      \hat a^\dagger\hat b^\dagger\hat j^\dagger}{\hat i}
    \bcontraction{}{\hat j}{
      \hat i\hat a\hat b
      \hat a^\dagger\hat b^\dagger}{\hat j}
    \contraction[1.5ex]{
      \hat j\hat i\hat a}{\hat b}{
      \hat a^\dagger}{\hat b}
    \contraction{
      \hat j\hat i}{\hat a}{\hat b}{
      \hat a}
      \hat j\hat i\hat a\hat b
      \hat a^\dagger\hat b^\dagger\hat j^\dagger\hat i^\dagger
    &=
    \diagramBox{
      \begin{fmffile}{MP2d2}
      \begin{fmfgraph*}(50,50)
        \fmfset{arrow_len}{6pt}
        \fmfleft{v00,v01}
        \fmfright{v10,v11}
        \fmf{boson}{v00,v10}
        \fmf{boson}{v01,v11}
        \fmf{fermion,left=0.25}{v00,v11}
        \fmf{fermion,left=0.25}{v11,v00}
        \fmf{fermion,left=0.25}{v10,v01}
        \fmf{fermion,left=0.25}{v01,v10}
        \fmfv{label.angle=150,label.dist=4,label=$a$}{v00}
        \fmfv{label.angle=30,label.dist=4,label=$j$}{v10}
        \fmfv{label.angle=-30,label.dist=4,label=$i$}{v11}
        \fmfv{label.angle=-150,label.dist=4,label=$b$}{v01}
      \end{fmfgraph*}
      \end{fmffile}
    }
    \xrightarrow{\textnormal{swap at }t=t_1}
    &
    \frac14
      V_{ba}^{ji}
      V_{ji}^{ba}\,
    \bcontraction[2ex]{}{
      \hat j}{\hat i\hat a\hat b
      \hat b^\dagger\hat a^\dagger\hat i^\dagger}{\hat j}
    \bcontraction{\hat j}{
      \hat i}{\hat a\hat b
      \hat b^\dagger\hat a^\dagger}{\hat i}
    \contraction[2.5ex]{
      \hat j\hat i}{\hat a}{\hat b
      \hat b^\dagger}{\hat a}
    \contraction{
      \hat j\hat i \hat a}{\hat b}{}{
      \hat b}
      \hat j\hat i \hat a\hat b
      \hat b^\dagger\hat a^\dagger\hat i^\dagger\hat j^\dagger
    &=
    \diagramBox{
      \begin{fmffile}{MP2d4}
      \begin{fmfgraph*}(50,50)
        \fmfset{arrow_len}{6pt}
        \fmfleft{v00,v01}
        \fmfright{v10,v11}
        \fmf{boson}{v00,v10}
        \fmf{boson}{v01,v11}
        \fmf{fermion,left=0.25}{v00,v01}
        \fmf{fermion,left=0.25}{v01,v00}
        \fmf{fermion,left=0.25}{v10,v11}
        \fmf{fermion,left=0.25}{v11,v10}
        \fmfv{label.angle=150,label.dist=6,label=$b$}{v00}
        \fmfv{label.angle=150,label.dist=6,label=$a$}{v10}
        \fmfv{label.angle=-30,label=$j$}{v01}
        \fmfv{label.angle=-30,label=$i$}{v11}
      \end{fmfgraph*}
      \end{fmffile}
    }
  \end{align*}
  \caption{
    All possible left/right interchanges of Coulomb interactions for the
    diagram representing the contraction evaluated in
    (\ref{eqn:MP2directMatrix}). The sums and the time arguments of the
    operators are omitted.
  }
  \label{fig:goldstoneSymmetries}
\end{figure}

Interchanging the left and the right side of a Coulomb interaction including
all its connections in the Goldstone diagram leaves the
the respective matrix elements $V^{pq}_{sr}$ invariant. In general, however,
we get a different contraction from the swapped connections. Since we have
to sum over all possible contractions, both contributions have to be counted
if they are distinct. Figure \ref{fig:goldstoneSymmetries} shows all possible
left/right interchanges for the diagram representing the contraction evaluted
in (\ref{eqn:MP2directMatrix}). For the sake of simplicity the time arguments
of the operators are omitted. The two cases on the left hand side have
distinct contractions, so both must be counted. Both evaluate
to the same number since they can be transformed into each other with an
even number of operator transpositions and $V_{sr}^{pq}=V_{rs}^{qp}$.
We can do that
with every Coulomb interaction and from now on identify one Goldstone diagram
with all contractions that arise from interchanging the left and the right
side of all its Coulomb interactions. In general, this cancels the factor
$\frac12$ in all Coulomb operators for evaluating one Goldstone diagram
as opposed to evaluating one single contraction, where this factor is needed
according to (\ref{eqn:coulomb}).

If there is however a global left/right symmetry in the Goldstone diagram,
interchanging all Coulomb interactions simultaneously does not give a distinct
contraction.
It merely interchanges the names of the indices over which to contract,
as shown in the right two cases of Figure \ref{fig:goldstoneSymmetries}.
Thus, for a Goldstone diagram with a global left/right symmetry only half of
the cases arising from interchanging each Coulomb interaction are distinct,
which gives rise to a factor of $\frac12$ for the whole diagram.

\subsection{Time integration}
The number of interactions occurring in the diagram is called its \emph{order},
which is $n+1$, where $n$ is the order of the expansion of the time evolution
operator $\hat U_\eta$ in (\ref{eqn:UExpansionNew}).
According to the expansion of $\hat U_\eta$, we must integrate over all times
$t_1,\ldots,t_n$ of all interactions except the last one, respecting the
order $0>t_1>\ldots>t_n$. We can make the substitutions
$t_{01}=0-t_1,\ldots,t_{n-1n}=t_{n-1}-t_n$
to integrate over the times between each interaction
$t_{01},t_{12},\ldots,t_{n-1n}$.
For a particle state $a$ propagating from an interaction at the time $t_n$ to
the time $t=0$ this gives for instance
\begin{multline}
  (-\im)^n\int\limits_{0>t_1>\ldots>t_n}
  \hat ae^{-\im\eps_a0}\ldots e^{\eta t_1}
  \ldots e^{\eta t_n}\hat a^\dagger e^{\im\eps_at_n} \ldots
  =\\
  (-\im)^n
  \int\limits_0^\infty\d t_{01} e^{-\im(\eps_a+\ldots-n\im\eta)t_{01}}
    \int\limits_0^\infty\d t_{12} e^{-\im(\eps_a+\ldots-(n-1)\im\eta)t_{12}}
    \ldots
    \int\limits_0^\infty\d t_{n-1n} e^{-\im(\eps_a+\ldots-\im\eta)t_{n-1n}}\\
  \ldots\hat a\ldots\hat a^\dagger\ldots
\end{multline}
So the eigenenergy $\eps_a$ appears in the exponential of each
time interval where the state propagates. For hole states $i$ the sign
is inverted.
Integrating out all time intervals gives an energy denominator for each
interval between two interactions of the form
\[
  \frac1{
    \displaystyle
    \sum\limits_{i\in H_k} \eps_i -
    \sum\limits_{a\in P_k} \eps_a
    +k\im\eta
  },
\]
where $H_k$ and $P_k$ are the sets of the holes and particles propagating in the
respective $k$-th time interval. $k$ is counted from the bottom.
Figure \ref{fig:MP2d} shows an example with two holes and two particles.
\begin{figure}[h]
  \begin{align*}
    \frac12
    \sum_{ijab}
    \frac{
      V_{ab}^{ij}
      V_{ij}^{ab}
    }{
      \eps_i+\eps_j-\eps_a-\eps_b+\im\eta
    }=
  \diagramBox{
    \begin{fmffile}{MP2dEnergy}
    \begin{fmfgraph*}(80,60)
      \fmfset{arrow_len}{6}
      \fmfstraight
      \fmfleft{v00,v01,v02,v03}
      \fmfright{v60,v61,v62,v63}
      \fmf{dots}{v01,v61}
      \fmf{phantom,tension=4}{v00,v10}
        \fmf{photon}{v10,v50}
        \fmf{phantom,tension=4}{v50,v60}
      \fmf{phantom,tension=4}{v03,v13}
        \fmf{photon}{v13,v53}
        \fmf{phantom,tension=4}{v53,v63}
      \fmffreeze
      \fmf{fermion,left=0.25,label=$a$,label.dist=4}{v10,v13}
      \fmf{fermion,left=0.25,label=$i$,label.dist=4}{v13,v10}
      \fmf{fermion,left=0.25,label=$b$,label.dist=4}{v50,v53}
      \fmf{fermion,left=0.25,label=$j$,label.dist=4}{v53,v50}
      \fmfv{
        label.angle=0,label.dist=8,
        label=$\footnotesize{H=\{i,,j\},, P=\{a,,b\}}$
      }{v61}
    \end{fmfgraph*}
    \end{fmffile}
  }
  \end{align*}
  \caption{
    Evaluation of all contractions represented by the given Goldstone
    diagram. There is one time interval where the holes $i,j$ and the
    particles $a,b$ are propagating. The symmetry factor is $\frac12$.
  }
  \label{fig:MP2d}
\end{figure}

For finite orders $n$ of the diagrams and systems with a finite energy gap
$E_g$ one can choose $n\eta \ll E_g$ to retrieve results independent of $\eta$.
When going to infinite orders in the diagrams, as it is done in the Random
Phase Approximation, one has to care of the limit of $\eta\rightarrow0$.

\subsection{Fermion sign}
\label{ssc:MBPT_Goldstone_FermionSign}
Finally, we need a rule to determine the sign of all contractions
arising from one Goldstone diagram in a graphical way. The sign of a contraction
depends on whether an even or an odd number $P$ of transpositions is required
to reorder the contractions in pairs:
\[
  \contraction[2ex]{}{\hat A}{\hat B}{\hat C}
  \contraction{\hat A}{\hat B}{\hat C\ldots}{\hat Z}
    \hat A\hat B\hat C\ldots\hat Z =
  (-1)^P
    \contraction{}{\hat A}{}{\hat C}\hat A\hat C
    \contraction{}{\hat B}{}{\hat Z}\hat B\hat Z
    \ldots
\]
Note that contracted operators must not be swapped since that
turns a particle connection into a hole connection or vice versa,
representing a different diagram. In the upper example, the following reordering
is therefore not allowed:
$
    \contraction{}{\hat A}{}{\hat C}\hat A\hat C
    \contraction{}{\hat Z}{}{\hat B}\hat Z\hat B
    \ldots
$
In Goldstone diagrams the contracted operators are the operators occurring in
the perturbation. They are of the general form
\[
  \ldots V_{sr}^{pq}\hat c^\dagger_p\hat c^\dagger_q\hat c_r\hat c_s \ldots
  v_{q'}^{p'}\hat c^\dagger_{p'}\hat c_{q'} \ldots
\]
omiting the sums for brevity.
Their respective time arguments can be integrated out according to the last
section, the order of the interactions from right to left is, however, still
relevant.

To determine the sign of all contractions of one Goldstone diagram, we first
reorder the operators such that operators of the same vertex of an
interaction are grouped together:
\[
  \ldots V_{sr}^{pq}
  \underbrace{\hat c^\dagger_p\hat c_s}_{\textnormal{vertex}}
  \hat c^\dagger_q\hat c_r \ldots
  v_{q'}^{p'}\hat c^\dagger_{p'}\hat c_{q'} \ldots
\]
This only affects the Coulomb interactions and involves an even number of
transpositions leaving the sign invariant.
The matrix elements $V_{sr}^{pq}$ and $v_q^p$ are complex numbers. Their
order is therefore irrelevant and we will omit them here.

\begin{figure}[h]
\[
  \diagramBox[0.25]{
    \begin{fmffile}{vertexLoop}
    \begin{fmfgraph*}(60,60)
      \fmfset{arrow_len}{6}
      \fmfstraight
      \fmfleft{v00,v01,v02,v03,v04}
      \fmfright{v40,v41,v42,v43,v44}
      \fmf{phantom}{v04,v14,v24,v34,v44}
      \fmf{phantom}{v03,v13,v23,v33,v43}
      \fmf{phantom,tension=2}{v02,l} \fmf{dots,tension=2}{l,v12}
        \fmf{dots}{v12,v22,v32}
        \fmf{dots,tension=2}{v32,r} \fmf{phantom,tension=2}{r,v42}
      \fmf{phantom}{v01,v11,v21,v31,v41}
      \fmf{phantom}{v00,v10,v20,v30,v40}
      \fmffreeze
      \fmf{fermion}{v30,v22,v11,v03,v34,v41,v30}
      \fmfv{
        decor.shape=circle,decor.size=4,label.dist=4,label.angle=0,label=$\hat A_1$
      }{v30}
      \fmfv{
        decor.shape=circle,decor.size=4,label.dist=4,label.angle=30,label=$\hat A_2$
      }{v22}
      \fmfv{
        decor.shape=circle,decor.size=4,label.dist=4,label.angle=-90,label=$\hat A_3$
      }{v11}
      \fmfv{
        decor.shape=circle,decor.size=4,label.dist=4,label.angle=180,label=$\hat A_4$
      }{v03}
      \fmfv{
        decor.shape=circle,decor.size=4,label.dist=4,label.angle=0,label=$\hat A_5$
      }{v34}
      \fmfv{
        decor.shape=circle,decor.size=4,label.dist=4,label.angle=0,label=$\hat A_6$
      }{v41}
    \end{fmfgraph*}
    \end{fmffile}
  }
  \qquad\qquad
  \diagramBox[0]{
    \begin{fmffile}{vertexCC}
    \begin{fmfgraph*}(20,20)
      \fmfkeep{vertexCC}
      \fmfset{arrow_len}{6}
      \fmfstraight
      \fmfbottom{v10}
      \fmftop{v01,v21}
      \fmf{fermion,label.side=left,label=$i$}{v21,v10}
      \fmf{fermion,label.side=left,label=$a$}{v10,v01}
      \fmfv{
        decor.shape=circle,decor.size=4,
        label.dist=6,label.angle=-90,label=$\hat a^\dagger\hat i^\dagger$
      }{v10}
    \end{fmfgraph*}
    \end{fmffile}
  }
  \qquad
  \diagramBox[0.5]{
    \begin{fmffile}{vertexCA}
    \begin{fmfgraph*}(10,30)
      \fmfset{arrow_len}{6}
      \fmfstraight
      \fmfbottom{v00,v10}
      \fmftop{v01,v11}
      \fmf{fermion}{v10,m,v01}
      \fmfv{
        decor.shape=circle,decor.size=4,
        label.dist=6,label.angle=0,label=$\hat b^\dagger\hat a$
      }{m}
    \end{fmfgraph*}
    \end{fmffile}
  }
  \qquad
  \diagramBox[0.5]{
    \begin{fmffile}{vertexAC}
    \begin{fmfgraph*}(10,30)
      \fmfset{arrow_len}{6}
      \fmfstraight
      \fmfbottom{v00,v10}
      \fmftop{v01,v11}
      \fmf{fermion}{v11,m,v00}
      \fmfv{
        decor.shape=circle,decor.size=4,
        label.dist=6,label.angle=0,label=$\hat i\hat j^\dagger$
      }{m}
    \end{fmfgraph*}
    \end{fmffile}
  }
  \qquad
  \diagramBox[1]{
    \begin{fmffile}{vertexAA}
    \begin{fmfgraph*}(20,20)
      \fmfset{arrow_len}{6}
      \fmfstraight
      \fmftop{v11}
      \fmfbottom{v00,v20}
      \fmf{fermion}{v00,v11,v20}
      \fmfv{
        decor.shape=circle,decor.size=4,
        label.dist=6,label.angle=0,label=$\hat i\hat a$
      }{v11}
    \end{fmfgraph*}
    \end{fmffile}
  }
\]
\caption{
  A closed loop of vertices $\hat A_1, \ldots, \hat A_6$, where each vertex
  operator $\hat A$ is of one of the four possible forms shown.
}
\label{fig:vertexLoop}
\end{figure}
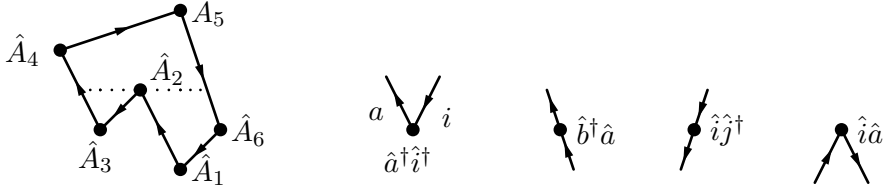

The finite set of vertices and connections form a directed graph, where
each vertex has exactly one outgoing and one incoming connection. Thus,
the graph consists of disconnected loops of vertices allowing us to
group the vertices of each loop together as long as we do not change
the order of the vertices within each loop.
The sign will not be affected since we only move vertices.
For the diagram in Figure
\ref{fig:MP2d} this gives for instance
\begin{equation}
  \contraction[2ex]{}{
    \hat i}{\hat a\hat j\hat b
    \hat a^\dagger}{\hat i}
  \contraction[1.5ex]{
    \hat i}{\hat a}{\hat j\hat b}{
    \hat a}
  \contraction[4ex]{
    \hat i\hat a}{\hat j}{\hat b
    \hat a^\dagger\hat i^\dagger\hat b^\dagger}{\hat j}
  \contraction[3ex]{
    \hat i\hat a\hat j}{\hat b}{
    \hat a^\dagger\hat i^\dagger}{\hat b}
  \hat i\hat a\hat j\hat b
  \hat a^\dagger\hat i^\dagger\hat b^\dagger \hat j^\dagger
  =
  (+1)\,
  \contraction[2ex]{}{
    \hat i}{\hat a\hat a^\dagger}{\hat i}
  \contraction[1.5ex]{
    \hat i}{\hat a}{}{\hat a}
  \hat i\hat a\hat a^\dagger\hat i^\dagger
  \contraction[2ex]{}{
    \hat j}{\hat b\hat b^\dagger}{\hat j}
  \contraction[1ex]{
    \hat j}{\hat b}{}{\hat b}
  \hat j\hat b\hat b^\dagger\hat j^\dagger
  .
  \label{eqn:loops}
\end{equation}
The symmetry of the Coulomb interaction allows us to freely move the vertices
horizontally such that each loop forms a simple polygon where the connections
do not intersect each other. Without loss of generality, we choose a clockwise
orientation of the directed connections in order to resemble the order of the
creation operators in a vertex of the form $\hat a^\dagger\hat i^\dagger$
as shown in Figure \ref{fig:vertexLoop}. 

Simple polygons can be decomposed into truncated monotone polygons 
\parencite{preparata_computational_2008}.
In a monotone polygon a horizontal line intersects the edges at most
twice, such that its left side consists only of particle connections and its
right side only of hole connections.
In the loop in Figure \ref{fig:vertexLoop} there are
two monotone polygons starting at the vertices $\hat A_1$ and $\hat A_3$.
After merging a particle and a hole connection from two different monotone
polygons at vertex $\hat A_2$ the rest of the polygon above the dotted line is
also monotone ending at vertex $\hat A_5$.

We have already reordered the operators to group vertices and loops together,
as shown in (\ref{eqn:loops}). As a last preparatory step we will group
vertices of monotone polygons together, keeping the time order within each
monotone polygon. For the loop in Figure \ref{fig:vertexLoop} this
rearrangement from the initial time order in the gives
\[
  \hat A_5\hat A_4\hat A_2  \hat A_6\hat A_3\hat A_1 = (+1)\,
  \hat A_5\hat A_4\hat A_2 (\hat A_3)  (\hat A_6\hat A_1),
\]
where the operators in the left and in the right parenthesis respectively
belong to the left and right monotone polygon below the dotted line.

With the operators ordered this way it is now sufficient to treat monotone
polygons only. We will follow the vertices within a monotone polygon
starting with a vertex of the form $\hat a^\dagger\hat i^\dagger$.
Then, vertices of the form $\hat b^\dagger\hat a$ add particle connections
on the left side of the monotone polygon under consideration
while vertices of the form $\hat j\hat i^\dagger$
add hole connections on the right side.
A vertex of the form $\hat i\hat a$ can either join two different monotone
polygons, as $\hat A_2$ does in Figure \ref{fig:vertexLoop}, or it can
close the loop.
At each vertex we reorder the involved contractions into the general
form
\begin{equation}
  \contraction[2ex]{}{
    \hat i}{\hat a....\hat a^\dagger}{\hat i}
  \contraction[1.5ex]{
    \hat i}{\hat a}{....}{\hat a}
  \light{\hat i\hat a....} \hat a^\dagger\hat i^\dagger
  \hat P
  \qquad
  \diagramBox[0.25]{
    \begin{fmffile}{vertexai}
    \begin{fmfgraph*}(30,30)
      \fmfkeep{vertexai}
      \fmfset{arrow_len}{6}
      \fmfstraight
      \fmfbottom{v10}
      \fmftop{v01,v21}
      \fmf{fermion,label.side=left,label.dist=4,label=$i$}{v21,v10}
      \fmf{fermion,label.side=left,label.dist=4,label=$a$}{v10,v01}
      \fmfv{
        decor.shape=circle,decor.fill=empty,decor.size=10,
        label.dist=6,label.angle=0,label=$\hat P$
      }{v10}
    \end{fmfgraph*}
    \end{fmffile}
  }  
  \label{eqn:polygonForm}
\end{equation}
such that $\hat P$ consists only of contracted pairs
$
  \contraction{}{\hat b}{}{\hat b}
  \contraction{\hat b\hat b^\dagger}{\hat j}{}{\hat j}
  \hat b\hat b^\dagger\hat j\hat j^\dagger\ldots
$
and the creation operators $\hat a^\dagger\hat i^\dagger$ of the open particle
and hole connections are to the left of $\hat P$.
All operators
to the right of $\hat a^\dagger\hat i^\dagger$ are then in the desired order,
denoted here by a circle in the diagram.
Note that the order of the
creation operators $\hat a^\dagger\hat i^\dagger$ is relevant and the
particle
creation operator must stand to the left of the hole creation operator.

A monotone polygon starts with a vertex of the form
$\hat a^\dagger\hat i^\dagger$. This is trivially in the above form
\begin{equation}
  \diagramBox[0.25]{
    \begin{fmffile}{vertexStart}
    \begin{fmfgraph*}(30,30)
      \fmfset{arrow_len}{6}
      \fmfstraight
      \fmfbottom{v10}
      \fmftop{v01,v21}
      \fmf{fermion,label.side=left,label.dist=4,label=$i$}{v21,v10}
      \fmf{fermion,label.side=left,label.dist=4,label=$a$}{v10,v01}
      \fmfdot{v10}
    \end{fmfgraph*}
    \end{fmffile}
  }
  \qquad
  \contraction[2ex]{}{
    \hat i}{\hat a....\hat a^\dagger}{\hat i}
  \contraction[1.5ex]{
    \hat i}{\hat a}{....}{\hat a}
  \light{\hat i\hat a....} \hat a^\dagger\hat i^\dagger
  =
  (+1)\,
  \contraction[2ex]{}{
    \hat i}{\hat a....\hat a^\dagger}{\hat i}
  \contraction[1.5ex]{
    \hat i}{\hat a}{....}{\hat a}
  \light{\hat i\hat a....} \hat a^\dagger\hat i^\dagger
  \underbrace{1}_{\hat P}
  \qquad
  \diagramBox[0.25]{
    \fmfreuse{vertexai}
  }
\end{equation}
with no change of sign.
A vertex of the form $\hat b^\dagger\hat a$ adds
a particle connection on the left side of the polygon.
Using two transpositions we can bring the operators into the desired form
\begin{equation}
  \diagramBox[0.25]{
    \begin{fmffile}{vertexParticle}
    \begin{fmfgraph*}(40,40)
      \fmfset{arrow_len}{6}
      \fmfstraight
      \fmfbottom{v10}
      \fmftop{v01,v21}
      \fmf{fermion,label.side=left,label.dist=4,label=$i$}{v21,v10}
      \fmf{fermion,label.side=left,label.dist=4,label=$a$}{v10,m}
      \fmf{fermion,label.side=left,label.dist=4,label=$b$}{m,v01}
      \fmfv{
        decor.shape=circle,decor.fill=empty,decor.size=10,
        label.dist=6,label.angle=0,label=$\hat P$
      }{v10}
      \fmfdot{m}
    \end{fmfgraph*}
    \end{fmffile}
  }
  \qquad
  \contraction[2ex]{}{
    \hat i}{\hat b.... \hat b^\dagger\hat a \hat a^\dagger}{\hat i}
  \contraction{
    \hat i}{\hat b}{....}{\hat b}
  \contraction[1.5ex]{
    \hat i\hat b.... \hat b^\dagger}{\hat a}{}{\hat a}
  \light{\hat i\hat b....} \hat b^\dagger\hat a \hat a^\dagger\hat i^\dagger
    \hat P
  =
  (+1)\,
  \contraction[2ex]{}{
    \hat i}{\hat b.... \hat b^\dagger}{\hat i}
  \contraction{
    \hat i}{\hat b}{....}{\hat b}
  \light{\hat i\hat b....} \hat b^\dagger\hat i^\dagger
  \underbrace{
    \contraction[1.5ex]{}{\hat a}{}{\hat a}
    \hat a\hat a^\dagger\hat P
  }_{\hat P'}
  \qquad
  \diagramBox[0.25]{
    \begin{fmffile}{vertexbi}
    \begin{fmfgraph*}(40,40)
      \fmfkeep{vertexbi}
      \fmfset{arrow_len}{6}
      \fmfstraight
      \fmfbottom{v10}
      \fmftop{v01,v21}
      \fmf{fermion,label.side=left,label.dist=4,label=$i$}{v21,v10}
      \fmf{fermion,label.side=left,label.dist=4,label=$b$}{v10,v01}
      \fmfv{
        decor.shape=circle,decor.fill=empty,decor.size=10,
        label.dist=6,label.angle=0,label=$\hat P'$
      }{v10}
    \end{fmfgraph*}
    \end{fmffile}
  }
  \label{eqn:vertexParticle}
\end{equation}
such that
$\hat a\hat a^\dagger$ can be
absorbed into the set of paired operators $\hat P'$ to continue with.
The sign does not change in this case.
A vertex of the form $\hat i\hat j^\dagger$ adds a hole connection on the
right side of the polygon. Now, three transpositions are required to
bring the operators into the form 
\begin{equation}
  \diagramBox[0.25]{
    \begin{fmffile}{vertexHole}
    \begin{fmfgraph*}(40,40)
      \fmfset{arrow_len}{6}
      \fmfstraight
      \fmfbottom{v10}
      \fmftop{v01,v21}
      \fmf{fermion,label.side=left,label.dist=4,label=$j$}{v21,m}
      \fmf{fermion,label.side=left,label.dist=4,label=$i$}{m,v10}
      \fmf{fermion,label.side=left,label.dist=4,label=$a$}{v10,v01}
      \fmfv{
        decor.shape=circle,decor.fill=empty,decor.size=10,
        label.dist=6,label.angle=180,label=$\hat P$
      }{v10}
      \fmfdot{m}
    \end{fmfgraph*}
    \end{fmffile}
  }
  \qquad
  \contraction[2ex]{}{
    \hat j}{\hat a.... \hat i}{\hat j}
  \contraction[1.5ex]{
    \hat j}{\hat a}{.... \hat i\hat j^\dagger}{\hat a}
  \contraction[3ex]{
    \hat j\hat a....}{\hat i}{\hat j^\dagger \hat a^\dagger}{\hat i}
  \light{\hat j\hat a....} \hat i\hat j^\dagger \hat a^\dagger\hat i^\dagger
    \hat P
  =
  (-1)\,
  \contraction[2ex]{}{
    \hat j}{\hat a.... \hat a^\dagger}{\hat j}
  \contraction[1.5ex]{
    \hat j}{\hat a}{....}{\hat a}
  \light{\hat j\hat a....}\hat a^\dagger\hat j^\dagger
  \underbrace{
    \contraction{}{\hat i}{}{\hat i}
    \hat i\hat i^\dagger\hat P
  }_{\hat P'}
  \qquad
  \diagramBox[0.25]{
    \begin{fmffile}{vertexaj}
    \begin{fmfgraph*}(40,40)
      \fmfkeep{vertexaj}
      \fmfset{arrow_len}{6}
      \fmfstraight
      \fmfbottom{v10}
      \fmftop{v01,v21}
      \fmf{fermion,label.side=left,label.dist=4,label=$j$}{v21,v10}
      \fmf{fermion,label.side=left,label.dist=4,label=$a$}{v10,v01}
      \fmfv{
        decor.shape=circle,decor.fill=empty,decor.size=10,
        label.dist=6,label.angle=0,label=$\hat P'$
      }{v10}
    \end{fmfgraph*}
    \end{fmffile}
  }
  \label{eqn:vertexHole}
\end{equation}
to absorb $\hat i\hat i^\dagger$ into $\hat P'$ and to bring the hole creation
operator $\hat j^\dagger$ of the open connection to the right side of the
particle creation operator $\hat a^\dagger$, as required by
(\ref{eqn:polygonForm}). In this case the sign changes.

At a vertex of the form $\hat j\hat a$ two different monotone polygons can
be joined to a polygon that is monotone after this vertex. We assume that the
right monotone polygon starts earlier in time, such that all its operators
$\hat a^\dagger\hat i^\dagger\hat P$
are to the right of all operators of the left monotone polygon
$\hat b^\dagger\hat j^\dagger\hat Q$. The converse case can be treated
analogously. Bringing the operators into the desired form
\begin{equation}
  \diagramBox[0.25]{
    \begin{fmffile}{vertexParticleHole}
    \begin{fmfgraph*}(60,60)
      \fmfset{arrow_len}{6}
      \fmfstraight
      \fmfleft{v00,v01,v02,v03}
      \fmfright{v40,v41,v42,v43}
      \fmf{phantom}{v00,v10,v20,v30,v40}
      \fmf{phantom}{v01,v11,v21,v31,v41}
      \fmf{phantom}{v02,v12,v22,v32,v42}
      \fmf{phantom}{v03,v13,v23,v33,v43}
      \fmffreeze
      \fmf{fermion,label.side=left,label.dist=4,label=$i$}{v43,v30}
      \fmf{fermion,label.side=right,label.dist=4,label=$a$}{v30,v22}
      \fmf{fermion,label.side=right,label.dist=4,label=$j$}{v22,v11}
      \fmf{fermion,label.side=left,label.dist=4,label=$b$}{v11,v03}
      \fmfv{
        decor.shape=circle,decor.fill=empty,decor.size=10,
        label.dist=6,label.angle=0,label=$\hat P$
      }{v30}
      \fmfv{
        decor.shape=circle,decor.fill=empty,decor.size=10,
        label.dist=6,label.angle=180,label=$\hat Q$
      }{v11}
      \fmfdot{v22}
    \end{fmfgraph*}
    \end{fmffile}
  }
  \qquad
  \contraction[4ex]{}{
    \hat i}{\hat b.... \hat j\hat a
    \hat b^\dagger\hat j^\dagger\hat Q \hat a^\dagger}{\hat i}
  \contraction{
    \hat i}{\hat b}{.... \hat j\hat a}{
    \hat b}
  \contraction[2ex]{
    \hat i\hat b....}{\hat j}{\hat a
    \hat b^\dagger}{\hat j}
  \contraction[3.5ex]{
    \hat i\hat b.... \hat j}{\hat a}{
    \hat b^\dagger\hat j^\dagger\hat Q}{\hat a}
  \light{\hat i\hat b....} \hat j\hat a
  \hat b^\dagger\hat j^\dagger\hat Q \hat a^\dagger\hat i^\dagger\hat P
  =
  (-1)\,
  \contraction[2ex]{}{
    \hat i}{\hat b.... \hat b^\dagger}{\hat i}
  \contraction{
    \hat i}{\hat b}{....}{\hat b}
  \light{\hat i\hat b....} \hat b^\dagger\hat i^\dagger
  \underbrace{
    \contraction{}{
      \hat j}{}{\hat j}
    \contraction[1.5ex]{
      \hat j\hat j^\dagger\hat Q}{\hat a}{}{\hat a}
    \hat j\hat j^\dagger\hat Q \hat a\hat a^\dagger\hat P
  }_{\hat P'}
  \qquad
  \diagramBox[0.25]{
    \fmfreuse{vertexbi}
  }
  \label{eqn:vertexParticleHole}
\end{equation}
requires an odd number of transpositions and changes the sign.
The last vertex of a loop is of the form $\hat i\hat a$. It closes the
monotone polygon without change of sign
\begin{equation}
  \diagramBox[0.25]{
    \begin{fmffile}{vertexEnd}
    \begin{fmfgraph*}(30,30)
      \fmfset{arrow_len}{6}
      \fmfstraight
      \fmfbottom{v10}
      \fmftop{v11}
      \fmf{fermion,left=0.4,label.side=left,label.dist=4,label=$a$}{v10,v11}
      \fmf{fermion,left=0.4,label.side=left,label.dist=4,label=$i$}{v11,v10}
      \fmfv{
        decor.shape=circle,decor.fill=empty,decor.size=10,
        label.dist=6,label.angle=0,label=$\hat P$
      }{v10}
      \fmfdot{v11}
    \end{fmfgraph*}
    \end{fmffile}
  }
  \qquad
  \contraction[2ex]{}{
    \hat i}{\hat a\hat a^\dagger}{\hat i}
  \contraction[1.5ex]{
    \hat i}{\hat a}{}{\hat a}
  \hat i\hat a \hat a^\dagger\hat i^\dagger
  \hat P
  =
  (+1)\,
  \underbrace{
    \contraction[1.5ex]{}{\hat a}{}{\hat a}\hat a\hat a^\dagger
    \contraction{}{\hat i}{}{\hat i}\hat i\hat i^\dagger
    \hat P
  }_{\hat P'}.
  \label{eqn:vertexEnd}
\end{equation}

In summary, the sign changes when a contraction of hole operators
$\contraction{}{\hat i}{}{\hat i}\hat i\hat i^\dagger$
is absorbed into the set of paired operators $\hat P'$,
as in (\ref{eqn:vertexHole}) and (\ref{eqn:vertexParticleHole}).
When closing the loop in (\ref{eqn:vertexEnd}) there is however no change of
sign although there is a pair of hole operators absorbed in $\hat P'$.
The Fermion sign of a single loop is therefore
$
  (-1)^{h-1} = (-1)\,(-1)^h
$,
where $h$ is the number of hole connections in the loop. For a Goldstone
diagram consisting of $l$ loops and $h$ hole connections in total the Fermion
sign is thus
\begin{equation}
  (-1)^{l+h},
  \label{eqn:MBPTGoldstoneSign}
\end{equation}
which can easily be determined graphically. Note that non-propagating
connections, as shown in Figure \ref{fig:firstOrder} also count as holes.

    \section{The Linked-Cluster theorem}
We can now employ the framework of Goldstone diagrams to systematically
evaluate the terms of the Gell-Mann--Low theorem (\ref{eqn:GellMannLow}):
\[
  E_0 + \Delta E =
  \frac{
    \langle\Phi|\hat H_0\hat U_\eta(0,-\infty)|\Phi\rangle
  }{
    \langle\Phi|\hat U_\eta(0,-\infty)|\Phi\rangle
  }
  +
  \frac{
    \langle\Phi|\hat H_1(0)\hat U_\eta(0,-\infty)|\Phi\rangle
  }{
    \langle\Phi|\hat U_\eta(0,-\infty)|\Phi\rangle
  }
\]
Only fully contracted operators contribute to the vacuum expectation values
which requires that all occurring diagrams are closed with no dangling
connection left.
In Figure \ref{fig:firstOrder} and \ref{fig:MP2d} we have already evaluated
selected diagrams of the numerator containing the final interaction
$\hat H_1(0)$.
Diagrams that are connected to the final interaction at $t=0$
are called \emph{connected} or, historically, \emph{linked} diagrams.
In second order there are already disconnected diagrams, such as
\index{linked}\index{connected}
\[
  \diagramBox{
    \begin{fmffile}{disconnected2}
    \begin{fmfgraph*}(40,30)
      \fmfset{arrow_ang}{25}
      \fmfset{arrow_len}{5pt}
      \fmfleft{v10,v11}
      \fmfright{v50,v51}
      \fmf{phantom}{v10,v20}
        \fmf{boson,tension=0.5}{v20,v40}
        \fmf{phantom}{v40,v50}
      \fmf{phantom}{v11,v21}
        \fmf{boson,tension=0.5}{v21,v41}
        \fmf{phantom}{v41,v51}
      \fmf{plain,left=1.0,tension=0}{v20,v10}
        \fmf{fermion,left=1.0,tension=0}{v10,v20}
      \fmf{fermion,left=1.0,tension=0}{v40,v50}
        \fmf{plain,left=1.0,tension=0}{v50,v40}
      \fmf{plain,left=1.0,tension=0}{v21,v11}
        \fmf{fermion,left=1.0,tension=0}{v11,v21}
      \fmf{fermion,left=1.0,tension=0}{v41,v51}
        \fmf{plain,left=1.0,tension=0}{v51,v41}
      \fmfv{label.ang=180,label=$\hat H_1(0)$}{v11}
      \fmfv{label.ang=180,label=$\hat H_1(t_1)$}{v10}
    \end{fmfgraph*}
    \end{fmffile}
  }
  =
  \quad
  -\im\int\limits_{-\infty}^0\d t_1
  \sum_{ijkl}
    V_{ij}^{ij}
    \contraction[2ex]{}{\hat i}{\hat j\hat j^\dagger}{\hat i}  
    \contraction{\hat i}{\hat j}{}{\hat j}
      \hat i\hat j\hat j^\dagger \hat i^\dagger
    \,
    e^{\eta t_1}
    \,
    V_{kl}^{kl}
    \contraction[2ex]{}{\hat k}{\hat l\hat l^\dagger}{\hat k}  
    \contraction{\hat k}{\hat l}{}{\hat l}
      \hat k\hat l\hat l^\dagger \hat k^\dagger
  =
  \left(\sum_{ij} V_{ij}^{ij}\right)
    \left(\sum_{ij}
    \frac{V_{ij}^{ij}}{\im\eta}\right),
\]
where the lower diagram is disconnected.
Since there are no contractions between disconnected diagrams their vacuum
expectation value decomposes into factors which can be evaluated
independently. Note that each disconnected diagram comes with an additional
factor of $1/\im\eta$ which diverges in the limit of $\eta\rightarrow0$.
In general, we get the following form for the numerator
of the Gell-Mann--Low energy expression containing the final interaction:
\[
  \langle\Phi|\hat H_1(0)\hat U_\eta(0,-\infty)|\Phi\rangle =
  \bigg(\sum\textnormal{\emph{connected diagrams}}\bigg)
  \bigg(\sum\textnormal{\emph{disconnected diagrams}}\bigg).
\]
Let us now look at the denominator
$\langle\Phi|\hat U_\eta(0,-\infty)|\Phi\rangle$, where there is no
operator at $t=0$. Since the diagrams must be closed we simply get the
same diagrams as the disconnected diagrams in the previous case, 
as for instance in
\[
  \diagramBox{
    \begin{fmffile}{disconnected1}
    \begin{fmfgraph*}(40,30)
      \fmfset{arrow_ang}{25}
      \fmfset{arrow_len}{5pt}
      \fmfleft{v10,v11}
      \fmfright{v50,v51}
      \fmf{phantom}{v10,v20}
        \fmf{boson,tension=0.5}{v20,v40}
        \fmf{phantom}{v40,v50}
      \fmf{dots}{v11,v51}
      \fmf{plain,left=1.0,tension=0}{v20,v10}
        \fmf{fermion,left=1.0,tension=0}{v10,v20}
      \fmf{fermion,left=1.0,tension=0}{v40,v50}
        \fmf{plain,left=1.0,tension=0}{v50,v40}
      \fmfv{label.ang=180,label=$t=0$}{v11}
      \fmfv{label.ang=180,label=$\hat H_1(t_1)$}{v10}
    \end{fmfgraph*}
    \end{fmffile}
  }
  =
  \quad
  \left(\sum_{ij}
    \frac{V_{ij}^{ij}}{\im\eta}
  \right).
\]
Thus, the denominator has the following general form
\[
  \langle\Phi|\hat U_\eta(0,-\infty)|\Phi\rangle =
  \bigg(\sum\textnormal{{disconnected diagrams}}\bigg).
\]

Finally, we need to evaluate the numerator containing $\hat H_0$, which
is by construction diagonal and from (\ref{eqn:H0ParticleHole}) given by
$
  \hat H_0 =
    E_0 - \sum_i\eps_i\hat i^\dagger\hat i + \sum_a\eps_a\hat a^\dagger\hat a.
$
Therefore, all diagrams from $\hat U_\eta$ must already be closed at
$t=0$ for non-vanishing contributions, just like in the denominator.
In general we get
\[
  \langle\Phi|\hat H_0\hat U_\eta(0,-\infty)|\Phi\rangle =
  E_0\bigg(\sum\textnormal{\emph{disconnected diagrams}}\bigg).
\]
Thus, all disconnected diagrams, containing diverging factors $1/\im\eta$,
cancel and we finally arrive at the expression for the correlation energy
in many-body perturbation theory
\[
  \Delta E = \bigg(\sum\textnormal{\emph{connected diagrams}}\bigg).
\]

    \section{Hartree-Fock reference}
\label{sec:MBPT_HartreeFock}
For the sake of simplicity we have not yet taken the effective interaction
$\hat V_{\rm eff}$ into consideration although it is an integral part of the
perturbation $\hat H_1$. Including the effective interaction in first order
gives just one additional diagram leading to three non-vanishing
contributions:
\begin{equation}
  +\diagramBox{
    \begin{fmffile}{Hartree}
    \begin{fmfgraph}(40,20)
      \fmfkeep{Hartree}
      \fmfset{arrow_ang}{25}
      \fmfset{arrow_len}{5pt}
      \fmfleft{v11}
      \fmfright{v51}
      \fmf{phantom}{v11,v21}
        \fmf{boson}{v21,v31,v41} \fmf{phantom}{v41,v51}
      \fmf{plain,left=1.0,tension=0}{v21,v11}
        \fmf{fermion,left=1.0,tension=0}{v11,v21}
      \fmf{fermion,left=1.0,tension=0}{v41,v51}
        \fmf{plain,left=1.0,tension=0}{v51,v41} 
    \end{fmfgraph}
    \end{fmffile}
  }
  \quad \quad
  +\diagramBox{
    \begin{fmffile}{Exchange}
    \begin{fmfgraph}(30,30)
      \fmfkeep{Exchange}
      \fmfset{arrow_len}{6}
      \fmfleft{v11}
      \fmfright{v21}
      \fmf{boson}{v11,v21}
      \fmf{fermion,left=0.5}{v11,v21}
      \fmf{fermion,left=0.5}{v21,v11}
    \end{fmfgraph}
    \end{fmffile}
  }
  \quad \quad
  -\diagramBox{
    \begin{fmffile}{Effective}
    \begin{fmfgraph}(30,30)
      \fmfset{arrow_ang}{25}
      \fmfset{arrow_len}{5pt}
      \fmftop{v12}
      \fmfbottom{v10}
      \fmf{phantom}{v10,v11,v12}
      \fmffreeze
      \fmfblob{10}{v11}
      \fmf{plain,right=1.0}{v11,v12}
      \fmf{fermion,right=1.0}{v12,v11}
    \end{fmfgraph}
    \end{fmffile}
  }
  \label{eqn:first}
\end{equation}
The perturbation $\hat H_1$
contains $\hat V_{\rm eff}$ with a negative sign which has to be taken into
account additionally to the sign of the Goldstone diagram originating from
the number its loops and holes. The former is explicitly given here, the
latter not.

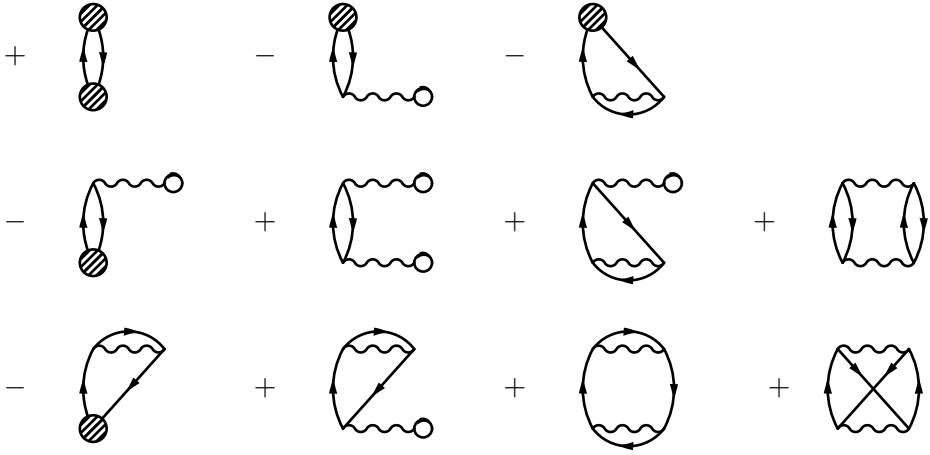
\begin{figure}
\begin{center}
\begin{tabular}{cccc}
  $+$
  \diagramBoxBorder{2ex}{2ex}{
    \begin{fmffile}{second11}
    \begin{fmfgraph}(40,30)
      \fmfset{arrow_len}{6}
      \fmfstraight
      \fmfbottom{llb,lb,b1,b2,b3,rb,rrb}
      \fmftop{llt,lt,t1,t2,t3,rt,rrt}
      \fmfblob{10}{lt,lb}
      \fmf{fermion,left=0.25}{lb,lt,lb}
    \end{fmfgraph}
    \end{fmffile}
  }
  &
  $-$
  \diagramBoxBorder{2ex}{2ex}{
    \begin{fmffile}{second21}
    \begin{fmfgraph}(40,30)
      \fmfset{arrow_len}{6}
      \fmfstraight
      \fmfbottom{llb,lb,b1,b2,b3,rb,rrb}
      \fmftop{llt,lt,t1,t2,t3,rt,rrt}
      \fmfblob{10}{lt}
      \fmf{boson}{lb,rb}
      \fmf{fermion,left=0.25}{lb,lt,lb}
      \fmf{plain,left=1.0}{rrb,rb}
      \fmf{fermion,left=1.0}{rb,rrb}
    \end{fmfgraph}
    \end{fmffile}
  }
  &
  $-$
  \diagramBoxBorder{2ex}{2ex}{
    \begin{fmffile}{second31}
    \begin{fmfgraph}(40,30)
      \fmfset{arrow_len}{6}
      \fmfstraight
      \fmfbottom{llb,lb,b1,b2,b3,rb,rrb}
      \fmftop{llt,lt,t1,t2,t3,rt,rrt}
      \fmfblob{10}{lt}
      \fmf{boson}{lb,rb}
      \fmf{fermion,left=0.25}{lb,lt}
      \fmf{fermion,left=0.5}{rb,lb}
      \fmf{fermion}{lt,rb}
    \end{fmfgraph}
    \end{fmffile}
  }
  \\
  \\
  $-$
  \diagramBoxBorder{2ex}{2ex}{
    \begin{fmffile}{second12}
    \begin{fmfgraph}(40,30)
      \fmfset{arrow_len}{6}
      \fmfstraight
      \fmfbottom{llb,lb,b1,b2,b3,rb,rrb}
      \fmftop{llt,lt,t1,t2,t3,rt,rrt}
      \fmfblob{10}{lb}
      \fmf{boson}{lt,rt}
      \fmf{fermion,left=0.25}{lb,lt,lb}
      \fmf{plain,left=1.0}{rrt,rt}
      \fmf{fermion,left=1.0}{rt,rrt}
    \end{fmfgraph}
    \end{fmffile}
  }
  &
  $+$
  \diagramBoxBorder{2ex}{2ex}{
    \begin{fmffile}{second22}
    \begin{fmfgraph}(40,30)
      \fmfset{arrow_len}{6}
      \fmfstraight
      \fmfbottom{llb,lb,b1,b2,b3,rb,rrb}
      \fmftop{llt,lt,t1,t2,t3,rt,rrt}
      \fmf{boson}{lb,rb}
      \fmf{boson}{lt,rt}
      \fmf{fermion,left=0.25}{lb,lt,lb}
      \fmf{plain,left=1.0}{rrb,rb}
      \fmf{fermion,left=1.0}{rb,rrb}
      \fmf{plain,left=1.0}{rrt,rt}
      \fmf{fermion,left=1.0}{rt,rrt}
    \end{fmfgraph}
    \end{fmffile}
  }
  &
  $+$
  \diagramBoxBorder{2ex}{2ex}{
    \begin{fmffile}{second32}
    \begin{fmfgraph}(40,30)
      \fmfset{arrow_len}{6}
      \fmfstraight
      \fmfbottom{llb,lb,b1,b2,b3,rb,rrb}
      \fmftop{llt,lt,t1,t2,t3,rt,rrt}
      \fmf{boson}{lb,rb}
      \fmf{boson}{lt,rt}
      \fmf{fermion,left=0.25}{lb,lt}
      \fmf{fermion,left=0.5}{rb,lb}
      \fmf{fermion}{lt,rb}
      \fmf{plain,left=1.0}{rrt,rt}
      \fmf{fermion,left=1.0}{rt,rrt}
    \end{fmfgraph}
    \end{fmffile}
  }
  &
    $+$
    \diagramBoxBorder{2ex}{2ex}{
      \begin{fmffile}{MP2d}
      \begin{fmfgraph}(40,30)
        \fmfkeep{MP2d}
        \fmfset{arrow_len}{6}
        \fmfstraight
        \fmfbottom{llb,lb,b1,b2,b3,rb,rrb}
        \fmftop{llt,lt,t1,t2,t3,rt,rrt}
        \fmf{boson}{lb,rb}
        \fmf{boson}{lt,rt}
        \fmf{fermion,left=0.25}{lb,lt,lb}
        \fmf{fermion,left=0.25}{rb,rt,rb}
      \end{fmfgraph}
      \end{fmffile}
    }
  \\
  \\
  $-$
  \diagramBoxBorder{2ex}{2ex}{
      \begin{fmffile}{second13}
    \begin{fmfgraph}(40,30)
      \fmfset{arrow_len}{6}
      \fmfstraight
      \fmfbottom{llb,lb,b1,b2,b3,rb,rrb}
      \fmftop{llt,lt,t1,t2,t3,rt,rrt}
      \fmfblob{10}{lb}
      \fmf{boson}{lt,rt}
      \fmf{fermion,left=0.25}{lb,lt}
      \fmf{fermion,left=0.5}{lt,rt}
      \fmf{fermion}{rt,lb}
    \end{fmfgraph}
    \end{fmffile}
  }
  &
  $+$
  \diagramBoxBorder{2ex}{2ex}{
    \begin{fmffile}{second23}
    \begin{fmfgraph}(40,30)
      \fmfset{arrow_len}{6}
      \fmfstraight
      \fmfbottom{llb,lb,b1,b2,b3,rb,rrb}
      \fmftop{llt,lt,t1,t2,t3,rt,rrt}
      \fmf{boson}{lb,rb}
      \fmf{boson}{lt,rt}
      \fmf{fermion,left=0.25}{lb,lt}
      \fmf{fermion}{rt,lb}
      \fmf{plain,left=1.0}{rrb,rb}
      \fmf{fermion,left=1.0}{rb,rrb}
      \fmf{fermion,left=0.5}{lt,rt}
    \end{fmfgraph}
    \end{fmffile}
  }
  &
  $+$
  \diagramBoxBorder{2ex}{2ex}{
    \begin{fmffile}{second33}
    \begin{fmfgraph}(40,30)
      \fmfset{arrow_len}{6}
      \fmfstraight
      \fmfbottom{llb,lb,b1,b2,b3,rb,rrb}
      \fmftop{llt,lt,t1,t2,t3,rt,rrt}
      \fmf{boson}{lb,rb}
      \fmf{boson}{lt,rt}
      \fmf{fermion,left=0.25}{lb,lt}
      \fmf{fermion,left=0.5}{rb,lb}
      \fmf{fermion,left=0.25}{rt,rb}
      \fmf{fermion,left=0.5}{lt,rt}
    \end{fmfgraph}
    \end{fmffile}
  }
  &
    $+$\diagramBox{
      \begin{fmffile}{MP2x}
      \begin{fmfgraph}(40,30)
        \fmfkeep{MP2x}
        \fmfset{arrow_len}{6}
        \fmfstraight
        \fmfbottom{llb,lb,b1,b2,b3,rb,rrb}
        \fmftop{llt,lt,t1,t2,t3,rt,rrt}
        \fmf{boson}{lb,rb}
        \fmf{boson}{lt,rt}
        \fmf{fermion,left=0.25}{lb,lt}
        \fmf{fermion,right=0.25}{rb,rt}
        \fmf{fermion}{lt,lm} \fmf{plain}{lm,rb}
        \fmf{fermion}{rt,rm} \fmf{plain}{rm,lb}
      \end{fmfgraph}
      \end{fmffile}
    }
  \\
\end{tabular}
\end{center}
\caption{
  All connected Goldstone diagrams of second order.
  In a Hartree-Fock reference the left three columns cancel leaving only
  the two contributions of second order M\o ller--Plesset perturbation theory
  (MP2).
}
\label{fig:second}
\end{figure}

For a Hartree-Fock Hamiltonian $\hat H_0$, $\hat V_{\rm eff}$ can
be explicitly given
\[
  v_q^p\hat c^\dagger_p \hat c_q\ =
  \ \sum_i V_{qi}^{pi} \hat c^\dagger_p\hat c_q -
  \sum_i V_{qi}^{ip} \hat c^\dagger_p\hat c_q,
\]
or in diagrams:
\begin{equation}
  \diagramBox{
    \fmfreuse{vpq}
  }
  =
  \hspace*{1ex}
  \diagramBox{
    \begin{fmffile}{HartreeSelf}
    \begin{fmfgraph*}(40,40)
      \fmfkeep{HartreeSelf}
      \fmfstraight
      \fmfset{arrow_len}{6}
      \fmfleft{v00,v01,v02}
      \fmfright{v30,v31,v32}
      \fmf{phantom}{v21,v31}
      \fmf{boson,tension=0.5}{v01,v21}
      \fmffreeze
      \fmf{plain,right=1.0}{v21,v31}
      \fmf{fermion,right=1.0,label.dist=4,label=$i$}{v31,v21}
      \fmf{fermion,label.side=left,label.dist=4,label=$q$}{v00,v01}
      \fmf{fermion,label.side=left,label.dist=4,label=$p$}{v01,v02}
    \end{fmfgraph*}
    \end{fmffile}
  }
  +
  \diagramBox{
    \begin{fmffile}{ExchangeSelf}
    \begin{fmfgraph*}(40,40)
      \fmfstraight
      \fmfset{arrow_len}{6}
      \fmfleft{v00,v01,v02}
      \fmfright{v30,v31,v32}
      \fmf{phantom}{v21,v31}
      \fmf{boson,tension=0.5}{v01,v21}
      \fmffreeze
      \fmf{fermion,right=0.5,label.dist=4,label=$i$}{v01,v21}
      \fmf{fermion,label.side=left,label.dist=4,label=$q$}{v00,v01}
      \fmf{fermion,label.side=right,label.dist=4,label=$p$}{v21,v02}
    \end{fmfgraph*}
    \end{fmffile}
  }
  \label{eqn:effectiveFock}
\end{equation}
We can insert (\ref{eqn:effectiveFock}) into (\ref{eqn:first}) giving
\[
  \diagramBox{
    \fmfreuse{Hartree}
  }
  +
  \diagramBox{
    \fmfreuse{Exchange}
  }
  -
  \diagramBox{
    \fmfreuse{Hartree}
  }
  -
  \diagramBox{
    \fmfreuse{Exchange}
  }
  =\ 0.
\]
In the case of the Hartree-Fock reference it turns out that
the effective interaction contained in the perturbation exactly cancels with
all Coulomb interactions containing non-\-propagating connections.
This greatly
simplifies the set of Goldstone diagrams to consider. In second order there
are already 11 connected Goldstone diagrams, which are shown in Figure
\ref{fig:second}, and only two of them do not cancel in Hartree-Fock.
Finite order many-body perturbation theory based on Hartree-Fock
is referred to as M\o ller--Plesset perturbation theory: MP2, MP3 or MP4.
The maximum order is given as suffix and rarely exceeds four.
M\o ller--Plesset perturbation theory was developed by
\parencite{moller_note_1934}
well before Goldstone introduced the diagrammatic treatment of perturbation
theory discussed here.

For a non-Hartree-Fock reference, such as Density Functional Theory, one
must take all diagrams into consideration that contain either the effective
interaction or non-propagating connections.
There are nine such diagrams in second order, shown on the left in Figure
\ref{fig:second}. In case of Density Functional Theory, where the effective
interaction consists of the Hartree contribution and an exchange-correlation
interaction
\[
  \diagramBox{
    \fmfreuse{vpq}
  }
  =
  \hspace*{2ex}
  \diagramBox{
    \fmfreuse{HartreeSelf}
  }
  \hspace*{1ex}
  +
  \diagramBox{
    \begin{fmffile}{vxcpq}
    \begin{fmfgraph*}(25,40)
      \fmfset{arrow_len}{6}
      \fmfstraight
      \fmftop{p}
      \fmfbottom{q}
      \fmf{fermion,label.side=left,label=$q$}{q,v}
      \fmf{fermion,label.side=left,label=$p$}{v,p}
      \fmfv{
        decor.shape=circle,decor.size=10,decor.filled=empty,
        label.dist=8, label=$\hat V_{\rm xc}$
      }{v}
    \end{fmfgraph*}
    \end{fmffile}
  }    
\]
all diagrams containing the Hartree contribution cancel. This leaves only 4 of
the 9 additional diagrams and they are to be evaluated with $\hat V_{\rm xc}$
instead of $\hat V_{\rm eff}$. Note that these diagrams are convergent in
second order even for metals, offering an alternative to
renormalization\index{renormalization}
as done for instance by \parencite{ren_renormalized_2013}.
Unless explicitly stated, the diagrams containing the effective or the
exchange-correlation interaction are only taken into account to first
order according to (\ref{eqn:first}) computing the Hartree-Fock interaction
energy with the DFT orbitals.

    \section{Propagators}
\label{sec:MBPTpropagators}
In the current approach to many-body perturbation theory we use the matrix
elements $V_{sr}^{pq}$ and $v_q^p$ from the second quantized
representation of the interactions present in the perturbation $\hat H_1$.
However, storing the Coulomb integrals $V_{sr}^{pq}$ on the computer
requires a large amount of memory scaling like $\mathcal O(N^4)$ with
the size of the system since this matrix has four indices. It is also
time consuming to calculate all elements of $V_{sr}^{pq}$ scaling like
$\mathcal O(N^5)$ with system size. 
$V_{sr}^{pq}$ must be computed before we can even begin to evaluate any
of its contractions for the diagrams in the perturbation expansion.

It can be beneficial to defer the calculation of the Coulomb integrals
to a later stage, especially for diagrams where the sums of the Coulomb
integrals can be factored into independent contributions, such as in the
Hartree term
\begin{align}
  \nonumber
  \diagramBox{
    \fmfreuse{Hartree}
  }
  =\ \frac12\sum_{ij}V_{ij}^{ij}\,
  &=\frac12\sum_{ij}\iint \d\vec x_1\d\vec x_2\,
    \psi_i^\ast(\vec x_1)\psi_j^\ast(\vec x_2)
    \frac1{|\vec r_1-\vec r_2|}
    \psi_j(\vec x_2)\psi_i(\vec x_1) \\
  \nonumber
  &=\frac12\iint \d\vec x_1\d\vec x_2
    \left(\sum_i \psi^\ast_i(\vec x_1)\psi_i(\vec x_1)\right)
    \frac1{|\vec r_1-\vec r_2|}
    \left(\sum_i \psi^\ast_i(\vec x_2)\psi_i(\vec x_2)\right)
\end{align}
which can even be computed in $\mathcal O(N^2)$. We can rewrite the above
expression to
\[
 \frac12 \iint \d \vec x_1\d \vec x_2\,
  G_0(\vec x_10,\vec x_10)\,
    \frac1{|\vec r_1-\vec r_2|}\,G_0(\vec x_20,\vec x_20)\,,
\]
defining the complex valued \emph{propagator} or
\emph{Green's function} from the spin-position $\vec x'$ to the
spin-position $\vec x$ at the same instance in time $t$
\index{Green's function}\index{propagator}
\begin{equation}
  G_0(\vec xt,\vec x't) := -\sum_i \psi_i(\vec x)\psi_i^\ast(\vec x')\,.
  \label{eqn:greensNonPropagating}
\end{equation}
Note that the sign in the non-propagating case for equal times is negative in
accordance with the Fermion sign rules discussed in Subsection
\ref{ssc:MBPT_Goldstone_FermionSign}.

For propagating states we can use the time dependent creation and
annihilation operators to express contractions over $V_{sr}^{pq}$ and $v_q^p$
in terms of complex valued functions
of two spin, space and time coordinates. In case of the direct MP2 diagram
this gives
\begin{align}
  \nonumber
  \diagramBox{\fmfreuse{MP2d}}
  =\ \frac12
  (-\im)\int_{-\infty}^0\d t\,e^{\eta t}\,
  &\sum_{ijab}V_{ij}^{ab} V_{ab}^{ij}
  \\
  \nonumber
  =\ \frac12
  (-\im)\int_{-\infty}^0\d t\,e^{\eta t}
  &\iiiint \d\vec x_1\,\d\vec x_2\,\d\vec x_3\,\d\vec x_4\,
  \frac1{|\vec r_1-\vec r_3|}\,
  \frac1{|\vec r_2-\vec r_4|}\,
  \\
  \nonumber
  &
  \sum_i
    \psi^\ast_i(\vec x_1)
    \psi_i(\vec x_3)
      e^{-\im\eps_it}
  \sum_a
    \psi_a(\vec x_1)
    \psi^\ast_a(\vec x_3)
      e^{\im\eps_at}
  \\
  \nonumber
  &
  \sum_j
    \psi^\ast_j(\vec x_2)
    \psi_j(\vec x_4)
      e^{-\im\eps_jt}
  \sum_b
    \psi_b(\vec x_2)
    \psi^\ast_b(\vec x_4)
      e^{\im\eps_bt}\,.
\end{align}
According to the Linked-Cluster theorem we can choose $\eta$ arbitrarily
small since the terms diverging with $\eta\rightarrow0$ cancel. Therefore,
the adiabatic switching function $e^{\eta t}$ has no physical effect and
rather serves to make the integrals convergent in the considered interval.
Thus, we can as well absorb the switching function into the time dependent
creation and annihilation operators. We get
\begin{align}
  \nonumber
  \diagramBox{\fmfreuse{MP2d}}
  =\ \frac12
  (-\im)\int_{-\infty}^0\d t\,
  &\iiiint \d\vec x_1\,\d\vec x_2\,\d\vec x_3\,\d\vec x_4\,
  \frac1{|\vec r_1-\vec r_3|}\,
  \frac1{|\vec r_2-\vec r_4|}\,
  \\
  \nonumber
  &
  \sum_i
    \psi^\ast_i(\vec x_1)
    \psi_i(\vec x_3)
      e^{(-\im\eps_i+\eta)t}
  \sum_a
    \psi_a(\vec x_1)
    \psi^\ast_a(\vec x_3)
      e^{(\im\eps_a+\eta)t}
  \\
  &
  \underbrace{
    \sum_j
      \psi^\ast_j(\vec x_2)
      \psi_j(\vec x_4)
        e^{(-\im\eps_j+\eta)t}
  }_{=-G_0(\vec x_4t,\,\vec x_20)}
  \underbrace{
    \sum_b
      \psi_b(\vec x_2)
      \psi^\ast_b(\vec x_4)
        e^{(\im\eps_b+\eta)t}
  }_{=+G_0(\vec x_20,\,\vec x_4t)}\,,
  \label{eqn:MP2directGreens}
\end{align}
extending the definition of the propagator\footnote{
Note that it is often $\im G_0$ which is defined by the right
hand side of (\ref{eqn:Greens}). This factor is normally introduced such that
the time evolution of the propagator is compatible to that of the Hamiltonian.
We omit this factor here for brevity and in accordance with
\parencite{lancaster_quantum_2014}.
}
to all cases:
\index{Green's function}\index{propagator}
\begin{equation}
  G_0(\vec xt,\vec x't') :=
  \left\{
  \begin{array}{ll}
    \displaystyle
    -\sum_i
      \psi_i(\vec x)
      \psi^\ast_i(\vec x')
        e^{(-\im\eps_i+\eta)(t-t')} & \textnormal{for } t\leq t' \\
    \displaystyle
    +\sum_a
      \psi_a(\vec x)
      \psi^\ast_a(\vec x')
        e^{(-\im\eps_a-\eta)(t-t')} & \textnormal{otherwise.}
  \end{array}
  \right.
  \label{eqn:Greens}
\end{equation}
In other words, particle states $a$ propagate forwards in time form $t'$ to $t$,
while hole states $i$ propagate backwards in time from $t'$ to $t$. Note that
we have to use hole states $i$ in the non-propagating case, where $t'=t$, to be
consistent with definition (\ref{eqn:greensNonPropagating}).
The adiabatic switching function is now part the propagator and its sole
purpose is to make positive time intervals convergent for particles and
negative time intervals convergent for holes.

    \section{Feynman diagrams}
\label{sec:MBPTFeyn}
Eventually, it is desirable to have a diagrammatic framework that
treats space and time on equal footing. Here, we do not
need it for a relativistic treatment of the many-body system, however
we want to be able to work in the frequency domain. This requires
independent integrals over the whole time domain of the form
$
  \int_{-\infty}^\infty\d t_1 \ldots \int_{-\infty}^\infty\d t_n
$
rather than the dependent integrals
$
  \int_{0>t_1>\ldots>t_n}\d t_1\ldots\d t_n
$
we have from the expansion
of the time evolution operator $\hat U_\eta(0,-\infty)$ according to
(\ref{eqn:UExpansionNew}):
\[
  \hat U_\eta(t,t_0) =
    \sum_{n=0}^\infty (-\im)^n
    \int\limits_{t>t_1>\ldots>t_n>t_0} \d t_1\ldots\d t_n\,
    \hat H_1(t_1)\ldots\hat H_1(t_n)\,.
\]
Each interaction that comes from the time evolution operator introduces a time
variable to integrate over and a factor of $(-\im)$.
In connected diagrams that applies to all interactions except the last one.
We can, however, introduce the same for the
last interaction and rewrite (\ref{eqn:MP2directGreens}) to
\begin{align}
  \nonumber
  (-\im)\diagramBox{\fmfreuse{MP2d}}
  =\ \frac12
  &\int\limits_{t_1>t_3}\d t_1\d t_3\,
  \iiiint \d\vec x_1\,\d\vec x_2\,\d\vec x_3\,\d\vec x_4\,
  \delta(t_1)\,
  \frac{(-\im)}{|\vec r_1-\vec r_2|}\,
  \frac{(-\im)}{|\vec r_3-\vec r_4|}
  \\
  \nonumber
  &
  G_0(x_1t_1,x_3t_3)\, G_0(x_3t_3,x_1t_1)\,
  G_0(x_2t_1,x_4t_3)\, G_0(x_4t_3,x_2t_1)
\end{align}
Since the propagators $G_0$ only depend on time differences we no longer
explicitly require the times to be negative and we can integrate over the whole
domain instead. However, we still require the times to be ordered and we also
need to anchor one of the interaction times for a convergent result.
Without loss of generality, we choose to fix the last interaction at $t_1=0$
by including $\delta(t_1)$.

In the case of the MP2 direct diagram, we can now simply drop the constraints
on the time variables taking double counting into consideration:
\begin{align}
  \nonumber
  (-\im)\diagramBox{\fmfreuse{MP2d}}
  =\ \frac14
  &\iint_{-\infty}^\infty \d t_1\d t_3\,
  \iiiint \d\vec x_1\,\d\vec x_2\,\d\vec x_3\,\d\vec x_4\,
  \delta(t_1)\,
  \frac{(-\im)}{|\vec r_1-\vec r_2|}\,
  \frac{(-\im)}{|\vec r_3-\vec r_4|}
  \\
  \nonumber
  &
  G_0(x_1t_1,x_3t_3)\, G_0(x_3t_3,x_1t_1)\,
  G_0(x_2t_1,x_4t_3)\, G_0(x_4t_3,x_2t_1)
\end{align}
In general, most permutations of the order of the time variables will lead to
distinct Goldstone diagrams that we all want to include. This will be discussed
in Subsection \ref{ssc:MBPT_FeynmanSymmetries}.

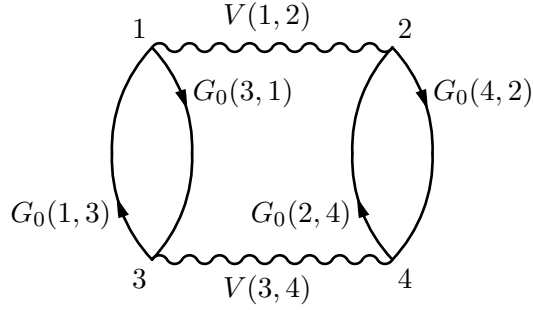
\begin{figure}[h]
\begin{center}
  \begin{tabular}{cc}
  \diagramBoxBorder{3ex}{3ex}{
    \begin{fmffile}{MP2dGreens}
    \begin{fmfgraph*}(120,80)
      \fmfset{arrow_len}{8}
      \fmfstraight
      \fmfleft{v00,v01,v02}
      \fmfright{v80,v81,v82}
      \fmf{phantom}{v01,v11,v21,v31,v41,v51,v61,v71,v81}
      \fmf{phantom,tension=6}{v00,v10}
        \fmf{
          photon,label.side=right,label=$V(3,,4)$
        }{v10,v70}
        \fmf{phantom,tension=6}{v70,v80}
      \fmf{phantom,tension=6}{v02,v12}
        \fmf{
          photon,label.side=left,label=$V(1,,2)$
        }{v12,v72}
        \fmf{phantom,tension=6}{v72,v82}
      \fmffreeze
      \fmf{
        fermion,left=0.2,label=$G_0(1,,3)$,label.dist=4
      }{v10,v01}
      \fmf{plain,left=0.2}{v01,v12}
      \fmf{
        fermion,left=0.2,label=$G_0(3,,1)$,label.dist=4
      }{v12,v21}
      \fmf{plain,left=0.2}{v21,v10}
      \fmf{
        fermion,left=0.2,label=$G_0(2,,4)$,label.dist=4
      }{v70,v61}
      \fmf{plain,left=0.2}{v61,v72}
      \fmf{
        fermion,left=0.2,label=$G_0(4,,2)$,label.dist=4
      }{v72,v81}
      \fmf{plain,left=0.2}{v81,v70}
      \fmfv{label=$1$,label.dist=4,label.angle=120}{v12}
      \fmfv{label=$2$,label.dist=4,label.angle=60}{v72}
      \fmfv{label=$3$,label.dist=4,label.angle=-120}{v10}
      \fmfv{label=$4$,label.dist=4,label.angle=-60}{v70}
    \end{fmfgraph*}
    \end{fmffile}
  }
  \end{tabular}
\end{center}
  \caption{
    In Feynman diagrams space and time coordinates are integrated over
    the entire domain and there is no required time order of the Coulomb
    interactions. Each vertex is given a label denoting its space
    and time coordinates. The propagator $G_0$ represents particles or
    holes propagating between the respective vertices and the propagator
    $V$ represents Coulomb interactions between vertices.
    The depicted diagram is the one evaluated in (\ref{eqn:MP2dFeyn}).
  }
  \label{fig:MP2dFeyn}
\end{figure}

Finally, we introduce time variables on every vertex to arrive at an expression
where time and space coordinates are treated on equal footing
\begin{align}
  \nonumber
  (-\im)\diagramBox{\fmfreuse{MP2d}}
  =\ \frac14
  &\iiiint_{-\infty}^{\infty}\d t_1\d t_2\d t_3\d t_4\,
  \iiiint \d\vec x_1\,\d\vec x_2\,\d\vec x_3\,\d\vec x_4\,
  \\
  \nonumber
  &
  G_0(x_1t_1,x_3t_3)\, G_0(x_3t_3,x_1t_1)\,
  G_0(x_4t_4,x_2t_2)\, G_0(x_2t_2,x_4t_4)
  \\
  \nonumber
  &
  \delta(t_1)\,
  \underbrace{
    \frac{(-\im)}{|\vec r_1-\vec r_2|}\,\delta(t_1-t_2)
  }_{=V(\vec x_1t_1,\vec x_2t_2)}\,
  \frac{(-\im)}{|\vec r_3-\vec r_4|}\,\delta(t_3-t_4)
  \\[1ex]
  \nonumber
  =\ \frac14
  &\iiiint \d1\,\d2\,\d3\,\d4\,\delta(t_1)
  \\
  &
  G_0(1,3)\, G_0(3,1)\, V(3,4)\,
  G_0(4,2)\, G_0(2,4)\, V(2,1)
  \label{eqn:MP2dFeyn}
\end{align}
defining the propagator for the Coulomb interaction
\begin{equation}
  V(\vec xt,\vec x't') := \frac{(-\im)}{|\vec r-\vec r'|}\,\delta(t-t')
  \label{eqn:CoulombPropagator}
\end{equation}
and the short forms
$G_0(1,2) = G_0(\vec x_1t_1,\vec x_2t_2)$ and
$\int\d1 = \int_{-\infty}^\infty\d t_1 \int\d\vec r_1 \sum_{\alpha_1}$.
Figure \ref{fig:MP2dFeyn} shows the coordinate labels used in
(\ref{eqn:MP2dFeyn}) and it is
called \emph{Feynman diagram}\index{Feynman diagram}.

\subsection{Effective interaction}
If the effective interaction is a multiplicative effective potential
$\left(\hat V_{\rm eff}\psi_q\right)(\vec x) =
  v_{\rm eff}(\vec x) \psi_q(\vec x)$
we can simply use the potential $v_{\rm eff}(\vec x)$ to evaluate Feynman
diagrams containing effective interactions, respecting its negative sign.
For a process involving a single effective interaction to be inserted
between the two space-time coordinates $1$ and $2$ we get for instance
\[
  \diagramBoxBorder{2ex}{2ex}{
    \begin{fmffile}{EffPropagator}
    \begin{fmfgraph*}(15,40)
      \fmfset{arrow_len}{6}
      \fmfstraight
      \fmfleft{v00,v02}
      \fmfright{v11}
      \fmfv{label=$1$}{v00}
      \fmfv{label=$2$}{v02}
      \fmfv{decor.shape=circle,decor.fill=shaded,decor.size=10,label=$3$}{v11}
      \fmf{fermion}{v00,v11,v02}
    \end{fmfgraph*}
    \end{fmffile}
  }
  \ \ =\ -\int\d3\, G_0(2,3)\,v_{\rm eff}(3)\,G_0(3,1)\,.
\]

\subsection{Symmetries}
\label{ssc:MBPT_FeynmanSymmetries}
\begin{figure}
\begin{center}
  \begin{tabular}{ccccc}
    &
    &
    \quad
    \diagramBox{
      \begin{fmffile}{MP3a1}
      \begin{fmfgraph*}(50,50)
        \fmfkeep{MP3a1}
        \fmfset{arrow_len}{6}
        \fmfstraight
        \fmfleft{v00,v01,v02}
        \fmfright{v20,v21,v22}
        \fmf{boson}{v00,v10,v20}
        \fmf{phantom}{v01,v11} \fmf{boson}{v11,v21}
        \fmf{boson}{v02,v12} \fmf{phantom}{v12,v22}
        \fmffreeze
        \fmf{fermion,left=0.25}{v00,v02}
        \fmf{fermion}{v02,v11}
        \fmf{fermion,right=0.25}{v11,v12}
        \fmf{fermion}{v12,v00}
        \fmf{fermion,left=0.25}{v20,v21,v20}
        \fmfv{label=$t_1$,label.angle=180,label.dist=9}{v00}
        \fmfv{label=$t_2$,label.angle=180,label.dist=9}{v01}
        \fmfv{label=$t_3$,label.angle=180,label.dist=9}{v02}
      \end{fmfgraph*}
      \end{fmffile}
    }
    \quad
  &
    $=$
  &
    \quad
    \diagramBox{
      \begin{fmffile}{MP3a2}
      \begin{fmfgraph*}(50,50)
        \fmfset{arrow_len}{6}
        \fmfstraight
        \fmfleft{v01,v00,v02}
        \fmfright{v21,v20,v22}
        \fmf{boson,tension=3}{v00,v10}
          \fmf{phantom,tension=12}{v10,c}
          \fmf{phantom,tension=4}{c,v20}
        \fmf{phantom}{v01,v11} \fmf{boson}{v11,v21}
        \fmf{boson}{v02,v12} \fmf{phantom}{v12,v22}
        \fmffreeze
        \fmf{boson}{v10,v20}
        \fmf{fermion,left=0.25}{v00,v02}
        \fmf{plain}{v02,m}\fmf{fermion}{m,v11}
        \fmf{fermion,right=0.125}{v11,c}\fmf{plain,right=0.125}{c,v12}
        \fmf{fermion}{v12,v00}
        \fmf{fermion,left=0.25}{v20,v21,v20}
        \fmfv{label=$t_1$,label.angle=180,label.dist=9}{v00}
        \fmfv{label=$t_2$,label.angle=180,label.dist=9}{v01}
        \fmfv{label=$t_3$,label.angle=180,label.dist=9}{v02}
      \end{fmfgraph*}
      \end{fmffile}
    }
    \\
    &
      $\nearrow$
    \\
      \diagramBoxBorder{3ex}{3ex}{
        \begin{fmffile}{Exchange3Feyn}
        \begin{fmfgraph*}(60,60)
          \fmfkeep{Exchange3Feyn}
          \fmfset{arrow_len}{6}
          \fmfsurround{d1,v01,d2,v11,d3,v02,d4,v12,d5,v00,d0,v10}
          \fmf{boson,label.side=right,label=$t_1$}{v00,v10}
          \fmf{boson,label.side=right,label=$t_2$}{v01,v11}
          \fmf{boson,label.side=right,label=$t_3$}{v02,v12}
          \fmf{plain}{v02,m1} \fmf{fermion}{m1,v00}
            \fmf{fermion,left=0.25}{v00,v12}
          \fmf{plain}{v12,m2} \fmf{fermion}{m2,v11}
            \fmf{fermion,right=0.25}{v11,v02}
          \fmf{fermion,left=0.3}{v10,v01,v10}
          \fmf{dots}{d1,d4}
        \end{fmfgraph*}
        \end{fmffile}
      }
      \quad
    &
      $\rightarrow$
    &
    \quad
    \diagramBox{
      \begin{fmffile}{MP3b1}
      \begin{fmfgraph*}(50,50)
        \fmfkeep{MP3b1}
        \fmfset{arrow_len}{6}
        \fmfstraight
        \fmfleft{v00,v02,v01}
        \fmfright{v20,v22,v21}
        \fmf{boson}{v00,v10,v20}
        \fmf{phantom}{v01,v11} \fmf{boson}{v11,v21}
        \fmf{boson}{v02,v12} \fmf{phantom}{v12,v22}
        \fmffreeze
        \fmf{fermion,left=0.25}{v00,v02}
        \fmf{fermion}{v02,v11}
        \fmf{fermion,left=0.25}{v11,v12}
        \fmf{fermion}{v12,v00}
        \fmf{fermion,left=0.25}{v20,v21,v20}
        \fmfv{label=$t_1$,label.angle=180,label.dist=9}{v00}
        \fmfv{label=$t_2$,label.angle=180,label.dist=9}{v01}
        \fmfv{label=$t_3$,label.angle=180,label.dist=9}{v02}
      \end{fmfgraph*}
      \end{fmffile}
    }
    \quad
  &
    $=$
  &
    \quad
    \diagramBox{
      \begin{fmffile}{MP3b2}
      \begin{fmfgraph*}(50,50)
        \fmfset{arrow_len}{6}
        \fmfstraight
        \fmfleft{v01,v02,v00}
        \fmfright{v21,v22,v20}
        \fmf{boson}{v00,v10,v20}
        \fmf{phantom}{v01,v11} \fmf{boson}{v11,v21}
        \fmf{boson}{v02,v12} \fmf{phantom}{v12,v22}
        \fmffreeze
        \fmf{fermion,right=0.25}{v00,v02}
        \fmf{fermion}{v02,v11}
        \fmf{fermion,right=0.25}{v11,v12}
        \fmf{fermion}{v12,v00}
        \fmf{fermion,left=0.25}{v20,v21,v20}
        \fmfv{label=$t_1$,label.angle=180,label.dist=9}{v00}
        \fmfv{label=$t_2$,label.angle=180,label.dist=9}{v01}
        \fmfv{label=$t_3$,label.angle=180,label.dist=9}{v02}
      \end{fmfgraph*}
      \end{fmffile}
    }
    \\
    &
      $\searrow$
    \\
  &
  &
    \quad
    \diagramBox{
      \begin{fmffile}{MP3c1}
      \begin{fmfgraph*}(50,50)
        \fmfkeep{MP3c1}
        \fmfset{arrow_len}{6}
        \fmfstraight
        \fmfleft{v02,v01,v00}
        \fmfright{v22,v21,v20}
        \fmf{boson}{v00,v10,v20}
        \fmf{phantom}{v01,v11} \fmf{boson}{v11,v21}
        \fmf{boson}{v02,v12} \fmf{phantom}{v12,v22}
        \fmffreeze
        \fmf{fermion,right=0.25}{v00,v02}
        \fmf{fermion}{v02,v11}
        \fmf{fermion,left=0.25}{v11,v12}
        \fmf{fermion}{v12,v00}
        \fmf{fermion,left=0.25}{v20,v21,v20}
        \fmfv{label=$t_1$,label.angle=180,label.dist=9}{v00}
        \fmfv{label=$t_2$,label.angle=180,label.dist=9}{v01}
        \fmfv{label=$t_3$,label.angle=180,label.dist=9}{v02}
      \end{fmfgraph*}
      \end{fmffile}
    }
    \quad
  &
    $=$
  &
    \quad
    \diagramBox{
      \begin{fmffile}{MP3c2}
      \begin{fmfgraph*}(50,50)
        \fmfset{arrow_len}{6}
        \fmfstraight
        \fmfleft{v02,v00,v01}
        \fmfright{v22,v20,v21}
        \fmf{boson,tension=3}{v00,v10}
          \fmf{phantom,tension=12}{v10,c}
          \fmf{phantom,tension=4}{c,v20}
        \fmf{phantom}{v01,v11} \fmf{boson}{v11,v21}
        \fmf{boson}{v02,v12} \fmf{phantom}{v12,v22}
        \fmffreeze
        \fmf{boson}{v10,v20}
        \fmf{fermion,right=0.25}{v00,v02}
        \fmf{plain}{v02,m}\fmf{fermion}{m,v11}
        \fmf{fermion,left=0.125}{v11,c}\fmf{plain,left=0.125}{c,v12}
        \fmf{fermion}{v12,v00}
        \fmf{fermion,left=0.25}{v20,v21,v20}
        \fmfv{label=$t_1$,label.angle=180,label.dist=9}{v00}
        \fmfv{label=$t_2$,label.angle=180,label.dist=9}{v01}
        \fmfv{label=$t_3$,label.angle=180,label.dist=9}{v02}
      \end{fmfgraph*}
      \end{fmffile}
    }
  \end{tabular}
\end{center}
  \caption{
    One Feynman diagram on the left represents all Goldstone diagrams
    originating from permutations of its interaction times
    $t_1,t_2$ and $t_3$. Due to the reflection symmetry when swapping $t_1$ and
    $t_2$, indicated by the dotted line, only three of the six permutations
    give rise to distinct Goldstone diagrams. The symmetry factor is 2
    and the diagram must be divided by that upon evaluation.
  }
  \label{fig:MP3Feyn}
\end{figure}
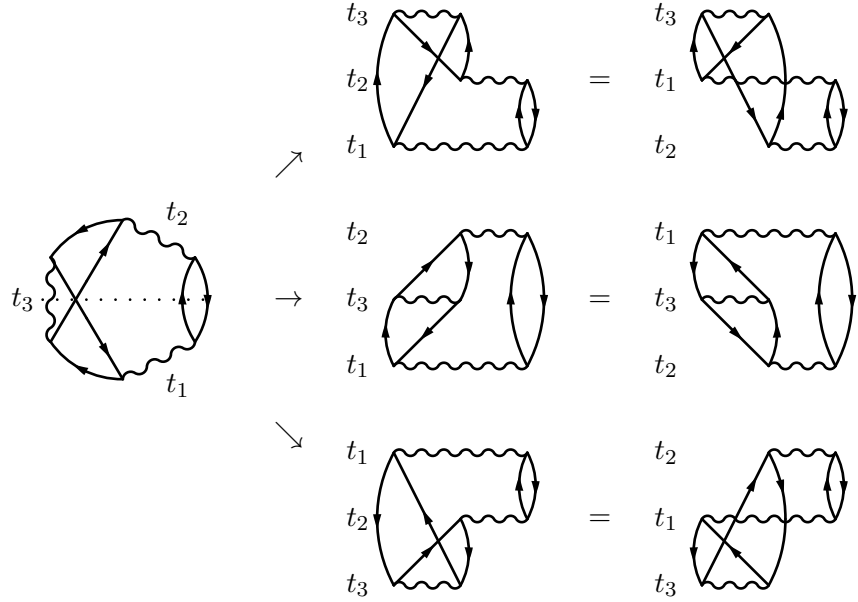

When dropping the constraints on the order of the interaction times we
get $n!$ permutations of the initial order. We now have to consider, how
many of them lead to distinct Goldstone diagrams and in turn to distinct
contractions, which we all have to count. Due to symmetries some permutations
may lead to the same Goldstone diagram and must not be counted more than once.
Figure \ref{fig:MP3Feyn} shows the case for a third order diagram, where
three of the six possible permutations lead to distinct Goldstone diagrams.
This is due to the reflection symmetry when swapping $t_1$ and $t_2$, indicated by
the dotted line.

The product of the order of all symmetry operations on a certain Feynman
diagram is called its \emph{symmetry factor}.\index{Symmetry factor}
In the example in Figure \ref{fig:MP3Feyn} there is only one reflection symmetry
with the order 2. Thus, the symmetry factor of the considered diagram is 2,
which means that only half of all the six permutations are distinct.
The symmetry factor can be determined graphically or computer aided,
considering all permutations of the vertices.
For the diagram in Figure \ref{fig:MP2dFeyn} there are 4 vertices, 2 Bosonic
edges $B$ and 4 Fermionic edges $F$:
\begin{align}
  \nonumber
  B &= \{\{1,2\},\{3,4\}\} \\
  \nonumber
  F &= \{(1,3),(3,1),(2,4),(4,2)\}\,.
\end{align}
The Bosonic edges are undirected since swapping the left and the right
side of a Coulomb interaction leads to the same Goldstone diagram.
The Fermionic edges are directed.
The permutation
$\tau=\left(\begin{array}{cccc}1&2&3&4\\4&3&2&1\end{array}\right)$
is for instance one of four permutations leaving the sets $B$ and $F$
unaltered:
\begin{align}
  \nonumber
  \tau(B) &= \{\{4,3\},\{2,1\}\} = B\\
  \nonumber
  \tau(F) &= \{(4,2),(2,4),(3,1),(1,3) \} = F\,,
\end{align}
thus being a symmetry operation. For this diagram, there are two reflection
symmetries of order 2 resulting in a symmetry factor of 4. Therefore, the
diagram has to be divided by 4 as done in (\ref{eqn:MP2dFeyn}).

\subsection{Fermion sign}
The Fermion sign of a Goldstone diagram is $(-1)^{l+h}$, where $l$ is the
number of Fermion loops and $h$ is the number of hole connections in the
entire diagram. In Feynman diagrams, the negative sign for each hole connection
is already contained in the definition of the propagator $G_0$
in (\ref{eqn:Greens}) for the hole case. Thus, only the number of loops $l$
still needs to be taken into consideration and the Fermion sign of a given
Feynman diagram is
\[
  (-1)^l\,.
\]

\section*{Summary}
Many-body perturbation theory provides a recipe to approximate the ground
state energy of the fully interacting Hamiltonian given the spin orbitals
$\psi_p(\vec x)$ and their eigenenergies $\eps_p$ of a single body reference,
such as Hartree-Fock or DFT. It is derived from time dependent
Rayleigh-Schr\"odinger perturbation theory which makes it extensive and thus
applicable to molecules as well as solids.
We can write the expansion of the perturbation series leading to the
ground state in terms of connected diagrams and we can use either
the rules of Goldstone diagrams or the ones of Feynman diagrams to evaluate the
individual terms in the expansion.

One Goldstone diagram of order $n$ expands in general to $2^n$ different
contractions originating from swapping the two vertices at each Coulomb
interaction.
The time order of the interactions is fixed indicated by drawing the
Coulomb interactions as parallel wiggly lines. Goldstone diagrams are evaluated
by contracting the occurring matrix elements of the Coulomb integrals
$V_{sr}^{pq}$.
One Feynman diagram of order $n$ expands in general to $n!$ different
Goldstone diagrams arising from all permutations of the order of the
interactions. The order of the interactions is not fixed and Coulomb
interactions are normally not drawn as parallel lines. Feynman diagrams are
evaluated by integrating the spin, space and time coordinates of all its
vertices which are arguments to complex valued functions, the propagators,
defining its connections.

The symmetry of a diagram determines how many distinct contractions arise
from evaluating a single Goldstone or Feynman diagram. In a Goldstone diagram
there can only be a global left/right symmetry with a symmetry factor of 2,
while in a Feynman diagram more complex symmetries are possible.
In both approaches we can define diagrams with open legs as building blocks
to be inserted in larger diagrams. In the Goldstone approach this has to be
done in a time ordered fashion while in Feynman diagrams it cannot be done
in a time ordered way.

Although, many-body perturbation theory is a valuable framework for
approximating the ground state energy it also has several drawbacks.
The number of diagrams is still infinite and one can only evaluate a small
subset. Using building blocks iteratively allows for evaluating an infinite
number of diagrams of a certain class. This improves the results in many
cases but there are still infinitely many diagrams neglected that can not be
constructed by iterating building blocks. The Random Phase Approximation
is a prominent example for this procedure.
Another drawback of MBPT is that it requires not only the reference solutions
of the unexcited states $\psi_i(x)$ and $\eps_i$ but also of the
theoretically infinite number of excited states $\psi_a(x)$ and $\eps_a$,
called virtual orbitals. The convergence with respect to the number
of virtual orbitals if often slow and it easilly exceeds twice the number
of electrons.
Finally, MBPT is not variational and one cannot give an upper bound for the
ground state energy as it is given by the Hartree-Fock approximation.
The non-variational nature of MBPT also considerably complicates the evaluation
of analytic gradients, for instance with respect to the atom positions, i.e.
forces.

  \chapter{The Random Phase Approximation}
\label{cha:RPA}
The Random Phase Approximation (RPA) is one of the most prominent methods
beyond Hartree-Fock or DFT. Historically, it has been derived within two
different frameworks rather independently. Within Rayleigh-Schr\"odinger
perturbation theory, Heisenberg already noticed that certain processes are
diverging in the
uniform electron gas due to the vanishing band gap and the sign of the
divergence is alternating with the order.
In the diagrammatic notation
introduced later by Feynman and Goldstone these processes are
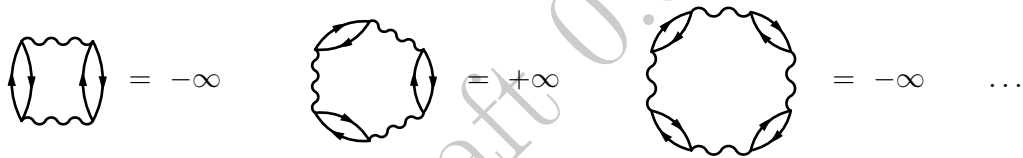
\[
  \diagramBox{\fmfreuse{MP2d}} =\ -\infty
  \qquad
  \diagramBox{
    \begin{fmffile}{Bubble3Feyn}
    \begin{fmfgraph}(45,45)
      \fmfset{arrow_len}{6}
      \fmfsurround{d1,v01,d2,v11,d3,v02,d4,v12,d5,v00,d0,v10}
      \fmf{boson,label.side=right}{v00,v10}
      \fmf{boson,label.side=right}{v01,v11}
      \fmf{boson,label.side=right}{v02,v12}
      \fmf{fermion,left=0.3}{v10,v01,v10}
      \fmf{fermion,left=0.3}{v11,v02,v11}
      \fmf{fermion,left=0.3}{v12,v00,v12}
    \end{fmfgraph}
    \end{fmffile}
  }
  =\ +\infty
  \qquad
  \diagramBox{
    \begin{fmffile}{Bubble4Feyn}
    \begin{fmfgraph}(55,55)
      \fmfset{arrow_len}{6}
      \fmfsurround{d2,v11,d3,v02,d4,v12,d5,v03,d6,v13,d7,v00,d0,v10,d1,v01}
      \fmf{boson,label.side=right}{v00,v10}
      \fmf{boson,label.side=right}{v01,v11}
      \fmf{boson,label.side=right}{v02,v12}
      \fmf{boson,label.side=right}{v03,v13}
      \fmf{fermion,left=0.3}{v10,v01,v10}
      \fmf{fermion,left=0.3}{v11,v02,v11}
      \fmf{fermion,left=0.3}{v12,v03,v12}
      \fmf{fermion,left=0.3}{v13,v00,v13}
    \end{fmfgraph}
    \end{fmffile}
  }
  =\ -\infty
  \qquad
  \ldots
\]
and they are referred to as \emph{ring diagrams}.\index{ring diagrams}
These divergencies pose serious problems as they render any finite order
perturbation theory useless for metals. 
However, \parencite{macke_uber_1950},
a student of Heisenberg, found a finite sum of all
such diagrams - later to be termed RPA - if one carries out the summation over
the perturbation orders
before summing over the states in the perturbation expression of each
term.
This reconciles the use of perturbation theory for metals again.
The summation over the perturbation orders can be done by iterating the diagram
  \hspace*{-2ex}
  \diagramBox{
    \begin{fmffile}{PolarizationV}
    \begin{fmfgraph}(20,20)
      \fmfkeep{PolarizationV}
      \fmfset{arrow_len}{6}
      \fmfleft{v00,v01}
      \fmfright{v10,v11}
      \fmf{fermion,right=0.3}{v00,v01,v00}
      \fmf{boson}{v00,v10}
    \end{fmfgraph}
    \end{fmffile}
  }
  \hspace*{-4ex}
as a building block like in a geometric series.

With the advent of Quantum Field Theory the procedure of redefining summation
orders became more common and was termed \emph{resummation} or
\emph{renormalization}.\index{resummation}\index{renormalization}
We will not argue in depth whether this procedure is justified. After all, the
sum of a conditionally convergent series depends crucially on the order
in which the individual terms are summed and - even worse - any result can be
achieved just by choosing an appropriate order. However, we can argue that
the notion of perturbation order is in a way arbitrary regarding that
it solely originates from iteratively solving the equation
of motion for the time evolution operator $\hat U(t,t_0)$
in (\ref{eqn:UExpansion}). Therefore, it seems a natural choice to sum over
the perturbation order first.

In 1953, Bohm and Pines developed the RPA independenlty in the framework
of the adiabatic connection (AC) and coined the term Random Phase Approximation.
Both frameworks arrive at the same result in case of the RPA but despite the common
use of diagrams in the adiabatic connection their meaning differs from
Goldstone and Feynman diagrams of many-body perturbation theory discussed in
Chapter \ref{cha:MBPT}. 

From a more applied point of view, RPA poses an important improvement over
Hartree-Fock and DFT. Just like finite order perturbation theories, such
as M\o ller--Plesset PT, it can describe van der Waals interaction but unlike
its finite order counterparts it can also be applied to metals. Recent
developments allow the RPA to be calculated in $\mathcal O(N^3)$ steps, just as
DFT but with a considerably higher prefactor compared to DFT
\parencite{kaltak_low_2014}.

    \section{RPA in the frequency domain}
\label{sec:RPAFreq}
We will first derive the Random Phase Approximation using many-body
perturbation theory in the frequency domain. Given the propagators
$G_0(\vec x t,\vec x't')$ and $V(\vec x t,\vec x't')$
of a system according to (\ref{eqn:Greens}) and (\ref{eqn:CoulombPropagator}),
we introduce the matrix notation\footnote{
  A matrix with continuous indices is actually an operator.
  However, for numerical evaluation the coordinates will be discretized
  justifying a matrix notation for the practical application.
}
\begin{equation}
  {\vec G_0}_{\vec x\vec x'}(t-t') = G_0(\vec x t,\vec x't')
  \quad
  \textnormal{and}
  \quad
  \vec V_{\vec x\vec x'}(t-t') = \frac{(-\im)}{|\vec r-\vec r'|}\,\delta(t-t')
\end{equation}
noting that the propagators only depend on the time difference. We define
the trace and the matrix product of two propagators $\vec A$ and $\vec B$ by
\[
  \Tr\big\{\vec A\big\} = \int\d\vec x\,\vec A_{\vec x\vec x}\,,
  \qquad
  (\vec A\vec B)_{\vec x\vec x''} =
    \int\d\vec x'\,\vec A_{\vec x\vec x'}\vec B_{\vec x'\vec x''}\,.
\]
Next, we define the \emph{independent particle polarizability}
\index{independent particle polarizability}
$\chi_0$
as the first building block of the ring diagrams and we let $\vec X_0$ denote
its matrix representation
\begin{equation}
  \hspace*{-2ex}
  \diagramBox{
    \begin{fmffile}{Polarization}
    \begin{fmfgraph}(2,20)
      \fmfkeep{Polarization}
      \fmfset{arrow_len}{6}
      \fmfbottom{v00}
      \fmftop{v01}
      \fmf{fermion,right=0.3}{v00,v01,v00}
    \end{fmfgraph}
    \end{fmffile}
  }
  \hspace*{-2ex}\
  =\ \chi_0(\vec xt,\vec x't')\ =
  \ -G_0(\vec xt,\vec x't')\,G_0(\vec x't',\vec xt)\,,
  \quad
  {\vec X_0}_{\vec x\vec x'}(t-t') = \chi_0(\vec xt,\vec x't')
  \label{eqn:RPAPolReal}
\end{equation}
The negative sign is required according to the Fermion sign rule of Feynman
diagrams since $\vec X_0$ is one closed Fermion loop.

From now on, we will connect the building blocks $\vec X_0$ and $\vec V$
in series. In the time domain this corresponds to convolutions while
it corresponds to simple products in the frequency domain.
Thus, we transform $\vec X_0$ into the frequency domain
with respect to the time difference, giving
\[
  \vec X_0(\omega) = \int\d(t-t')\,e^{+\im\omega(t-t')}\,\vec X_0(t-t')\,.
\]
We use $e^{+\im\omega(t-t')}$ for the forward Fourier transform
into the frequency domain. Using this convention, the poles of the
polarizability as a function of $\omega$ coincide with the positive
elementary excitation energies $\eps_a-\eps_i$ of $\hat H_0$.
See (\ref{eqn:UEGRealPropagator}) for more details in the case of the uniform
electron gas. The Coulomb propagator $\vec V$ is independent of the frequency.

We can now evaluate the diagrams of the Random Phase Approximation.
In (\ref{eqn:MP2dFeyn}) we already derived an expression for the second
order ring diagram in the time domain where
$\d i=\d\vec x_i\,\d t_i$:
\[
  (-\im)\diagramBox{\fmfreuse{MP2d}} =\ \frac14
  \iiiint \d1\,\d2\,\d3\,\d4\,\delta(t_1)\,
  \underbrace{G_0(1,3)\, G_0(3,1)}_{=-\chi_0(1,3)}\, V(3,4)\,
  \underbrace{G_0(4,2)\, G_0(2,4)}_{=-\chi_0(4,2)}\, V(2,1)
\]
In the frequency domain there is only one frequency $\omega$ to integrate
over, since we have only one loop and the frequency is conserved at every
vertex. The integral over all frequencies corresponds to a convolution of all
time differences rather than absolute times. Therefore, we do not need to
anchor the diagram at a certain time anymore, here done by $\delta(t_1)$.
Using the matrix notation for the integrals over space we get
\begin{equation}
  \diagramBox{\fmfreuse{MP2d}}
  =\ \frac\im4 \int\frac{\d\omega}{2\pi}\,
    \Tr\Big\{\vec X_0(\omega)\vec V\vec X_0(\omega)\vec V\Big\}
  \ =\ \frac\im4
    \int\frac{\d\omega}{2\pi}\,
    \Tr\Big\{(\vec X_0(\omega)\vec V)^2\Big\}\,.
  \label{eqn:RPARing2}
\end{equation}
This diagram has two reflection symmetries of order 2. Thus, the symmetry
factor of this diagram is 4 giving rise to the factor $1/4$ as discussed
in Section \ref{sec:MBPTFeyn}. The next diagram in the RPA is the third order
ring diagram. It has one reflection symmetry of order 2 and one rotational
symmetry of order 3 indicated by the dotted lines. It is therefore given by
\begin{equation}
  \diagramBox{
    \begin{fmffile}{Bubble3FeynSym}
    \begin{fmfgraph}(55,55)
      \fmfset{arrow_len}{6}
      \fmfsurround{d1,v01,d2,v11,d3,v02,d4,v12,d5,v00,d0,v10}
      \fmf{boson,label.side=right}{v00,v10}
      \fmf{boson,label.side=right}{v01,v11}
      \fmf{boson,label.side=right}{v02,v12}
      \fmf{fermion,left=0.3}{v10,v01,v10}
      \fmf{fermion,left=0.3}{v11,v02,v11}
      \fmf{fermion,left=0.3}{v12,v00,v12}
      \fmf{dots}{d0,c}\fmf{dots}{d2,c}\fmf{dots}{d4,c}
    \end{fmfgraph}
    \end{fmffile}
  }
  =\ \frac\im{2\cdot3}
    \int\frac{\d\omega}{2\pi}\,
    \Tr\Big\{(\vec X_0(\omega)\vec V)^3\Big\}\,.
  \label{eqn:RPARing3}
\end{equation}
All ring diagrams have a reflection symmetry due the symmetry of the independent
particle polarizability $\vec X_0$. Additionally, each ring diagram of order
$n$ has a rotational symmetry of order $n$, allowing us to evaluate any
given order
\begin{equation}
  \diagramBox{
    \begin{fmffile}{BubbleNFeynSym}
    \begin{fmfgraph*}(65,65)
      \fmfset{arrow_len}{6}
      \fmfsurround{d7,v04,d8,v14,d9,v00,d0,v10,d1,v01,d2,v11,d3,v02,d4,v12,d5,v03,d6,v13}
      \fmf{boson,label.side=right}{v00,v10}
      \fmf{boson,label.side=right}{v01,v11}
      \fmf{boson,label.side=right}{v02,v12}
      \fmf{boson,label.side=right}{v03,b1} \fmf{phantom}{b1,v13}
      \fmf{boson,label.side=right}{b2,v14} \fmf{phantom}{v04,b2}
      \fmf{fermion,left=0.3}{v10,v01,v10}
      \fmf{fermion,left=0.3}{v11,v02,v11}
      \fmf{fermion,left=0.3}{v12,v03,v12}
      \fmf{fermion,left=0.3}{v14,v00,v14}
      \fmf{dots}{d0,c}\fmf{dots}{d2,c}\fmf{dots}{d4,c}
        \fmf{dots}{d6,c}
        \fmf{dots}{d8,c}
      \fmfv{label=$n$,label.angle=180}{d7}
    \end{fmfgraph*}
    \end{fmffile}
  }
  =\ \frac\im{2n}
    \int\frac{\d\omega}{2\pi}\,
    \Tr\Big\{(\vec X_0(\omega)\vec V)^n\Big\}\,.
  \label{eqn:RPARingN}
\end{equation}
We can now do the resummation, summing over the orders
before evaluating the frequency integration and the trace:
\begin{equation}
  \hspace*{-2ex}
    \diagramBox{
      \begin{fmffile}{RPA}
      \begin{fmfgraph}(40,30)
        \fmfkeep{RPA}
        \fmfset{arrow_len}{6}
        \fmfstraight
        \fmfbottom{llb,lb,b1,b2,b3,rb,rrb}
        \fmftop{llt,lt,t1,t2,t3,rt,rrt}
        \fmf{dbl_wiggly}{lb,rb}
        \fmf{boson}{lt,rt}
        \fmf{fermion,left=0.25}{lb,lt,lb}
        \fmf{fermion,left=0.25}{rb,rt,rb}
      \end{fmfgraph}
      \end{fmffile}
    }
  \hspace*{-2ex}
  =\ \frac\im2
    \int\frac{\d\omega}{2\pi}\,
    \Tr\left\{
      \sum_{n=2}^\infty
      \frac1n(\vec X_0(\omega)\vec V)^n
    \right\}\,.
  \label{eqn:RPAByOrder}
\end{equation}
For the diagrammatic notation of the series we use the
\emph{screened interaction}\index{screened interaction}
$W$ in RPA given by
\begin{equation}
  \diagramBox{
    \begin{fmffile}{ScreenedW}
    \begin{fmfgraph*}(26,10)
      \fmfkeep{ScreenedW}
      \fmfstraight
      \fmfleft{v11}
      \fmfright{v21}
      \fmf{dbl_wiggly,label=$W$,label.side=left}{v11,v21}
    \end{fmfgraph*}
    \end{fmffile}
  }
  =
  \hspace*{-1ex}
  \diagramBox{
    \begin{fmffile}{CoulombInteraction}
    \begin{fmfgraph}(26,10)
      \fmfkeep{CoulombInteraction}
      \fmfstraight
      \fmfleft{v11}
      \fmfright{v21}
      \fmf{photon}{v11,v21}
    \end{fmfgraph}
    \end{fmffile}
  }
  \hspace*{-1ex}
  +
  \diagramBox{
    \begin{fmffile}{VChiV}
    \begin{fmfgraph}(52,20)
      \fmfkeep{VChiV}
      \fmfstraight
      \fmfset{arrow_len}{6}
      \fmfleft{v00,v01}
      \fmfright{v20,v21}
      \fmf{photon}{v00,v10} \fmf{phantom}{v10,v20}
      \fmf{phantom}{v01,v11} \fmf{photon}{v11,v21}
      \fmffreeze
      \fmf{fermion,right=0.3}{v10,v11,v10}
    \end{fmfgraph}
    \end{fmffile}
  }
  +
  \diagramBox{
    \begin{fmffile}{VChiVChiV}
    \begin{fmfgraph}(64,32)
      \fmfkeep{VChiVChiV}
      \fmfstraight
      \fmfset{arrow_len}{6}
      \fmfleft{v00,v01,v02}
      \fmfright{v30,v31,v32}
      \fmf{photon}{v00,v10} \fmf{phantom}{v10,v20,v30}
      \fmf{phantom}{v01,v11} \fmf{photon}{v11,v21} \fmf{phantom}{v21,v31}
      \fmf{phantom}{v02,v12,v22} \fmf{photon}{v22,v32}
      \fmffreeze
      \fmf{fermion,right=0.3}{v10,v11,v10}
      \fmf{fermion,right=0.3}{v21,v22,v21}
    \end{fmfgraph}
    \end{fmffile}
  }
  +\ \ \ldots
  \label{eqn:RPAScreenedW}
\end{equation}
Note that the graphical appearance of a diagram containing the screened
interaction $W$ might be deceptive. The eye suggests a simple reflection
symmetry of this diagram but in fact
the symmetry factor must be considered for each order separately, as we have
done it here.

Instead of carrying out the matrix products in (\ref{eqn:RPAByOrder}) order
by order we search for a matrix function having the same series expansion.
The function $\log(1-x)$ has the power series $-x-x^2/2-x^3/3-\ldots$ so
we can write the RPA energy as
\begin{equation}
    \diagramBox{
      \fmfreuse{RPA}
    }
  =\ -\,\frac\im2
    \int_{-\infty}^\infty\frac{\d\omega}{2\pi}\,
    \Tr\Big\{
      \log\Big(\vec 1-\vec X_0(\omega)\vec V\Big)+\vec X_0(\omega)\vec V
    \Big\}\,.
  \label{eqn:RPAReal}
\end{equation}
$\vec X_0(\omega)$ has poles along the real frequency axis making a numerical
integration difficult. We can rotate the integration contour as long as we
do not cross any poles and use the imaginary frequency instead.
This rotation is called \emph{Wick rotation}\index{Wick rotation} and it is
discussed in more detail for the uniform electron gas in Section
\ref{sec:RPAUEG}. Given $\vec X_0(\im\nu)$ in imaginary frequency we can
substitute $\omega = \im\nu$ in (\ref{eqn:RPAReal}) and finally get
\begin{equation}
    \diagramBoxBorder{0ex}{1ex}{
      \fmfreuse{RPA}
    }
  =\ \frac12
    \int_{-\infty}^\infty\frac{\d\nu}{2\pi}\,
    \Tr\Big\{
      \log\Big(\vec 1-\vec X_0(\im\nu)\vec V\Big)+\vec X_0(\im\nu)\vec V
    \Big\}\,.
  \label{eqn:RPAImag}
\end{equation}

For the uniform electron gas (UEG) we can evaluate $\vec X_0(\im\nu)$ and
$\vec V$ analytically and use (\ref{eqn:RPAImag}) to evaluate the RPA energy
for the UEG numerically. This is done in Section \ref{sec:RPAUEG}. For
a molecule or a solid $\vec X_0(\im\nu)$ has to be computed from the
Hartree-Fock or DFT spin-orbitals $\psi_p(\vec x)$. Instead of calculating
$\vec X_0(t-t')$ in real time according to (\ref{eqn:RPAPolReal}), one can
already perform the Wick rotation in the time domain and evaluate
\begin{equation}
  {\vec X_0}_{\vec x\vec x'}(\im\tau) =
  -{\vec G_0}_{\vec x\vec x'}(\im\tau){\vec G_0}_{\vec x'\vec x}(-\im\tau)
  \label{eqn:RPAChiImag}
\end{equation}
in imaginary time $\im\tau$. Note that the matrices are multiplied elementwise.
The particle/hole propagator in imaginary time is given by
\begin{equation}
  {\vec G_0}_{\vec x\vec x'}(\im\tau) =
  \left\{
  \begin{array}{ll}
    \displaystyle
    -\sum_i \psi_i(\vec x)\psi_i^\ast(\vec x') e^{-(\eps_i-\mu)\tau} &
    \textnormal{for }\tau\leq 0 \\[3ex]
    \displaystyle
    +\sum_a \psi_a(\vec x)\psi_a^\ast(\vec x') e^{-(\eps_a-\mu)\tau} &
    \textnormal{otherwise,}
  \end{array}
  \right.
  \label{eqn:RPAGreenImag}
\end{equation}
where $\mu$ is the Fermi energy.
Evaluating the energies of second order M\o ller--Plesset Perturbation Theory
using the imaginary time propagators is equivalent to the Laplace transformed
MP2 approach proposed by \parencite{almlof_elimination_1991}.
To evaluate the Random Phase Approximation $\vec X_0(\im\tau)$ needs to be
Fourier transformed with respect to $\tau$ to arrive at the independent
particle polarizability in imaginary frequency $\vec X_0(\im\nu)$ employed by
(\ref{eqn:RPAImag}).
The Fourier transform from imaginary time to imaginary frequency, as well
as the imaginary frequency integration in (\ref{eqn:RPAImag}) can be done
numerically on a non-equidistant grid to high accuracy with a only few
integration points \parencite{kaltak_low_2014,kaltak_cubic_2014}.
To determine the employed quadrature frequencies and weights
a function is chosen that resembles the RPA energy function and whose exact
frequency integral is known. The proposed function of imaginary time is the
direct MP2 term
\[
  \frac12\int\frac{\d\nu}{2\pi}\,
    \frac{
      2(\eps_a-\eps_i)
    }{
      (\eps_a-\eps_i)^2+\nu^2
    }\,
    \frac{
      2(\eps_b-\eps_j)
    }{
      (\eps_b-\eps_j)^2+\nu^2
    }
  =\frac1{\eps_i+\eps_j-\eps_a-\eps_b}
\]
which is the lowest order of the RPA expansion.
The quadrature frequencies $\nu_k$ and weights $w_k$ can then be fit
such that the dominant terms with $a=b$ and $i=j$
are best reproduced by the numeric integral
\begin{equation}
  \frac1{\eps_i+\eps_i-\eps_a-\eps_a}
  \approx
    \frac12 \sum_k w_k
    \left(
    \frac{
      2(\eps_a-\eps_i)
    }{
      (\eps_a-\eps_i)^2+\nu_k^2
    }
    \right)^2
  \,.
  \label{eqn:RPA_NuFit}
\end{equation}
This fit is done for all single particle excitation energies $\eps_a-\eps_i$
and the quality of the fit depends on the ratio of the largest and the smallest
excitation energy $\max(\eps_a-\eps_i)/\min(\eps_a-\eps_i)$.
For a non-metallic system the number of frequency points is negligible
compared to the number of possible excitations, required in the
conventional approach to calculate $\vec X_0$ from the Adler-Wiser formula
\parencite{adler_quantum_1962,wiser_dielectric_1963}.
The above frequency grid allows the RPA energy to be evaluated in
$\mathcal O(N^3)$ steps, just like DFT but with a considerably higher
prefactor.
For a metallic system the behavior of the RPA energy for large imaginary
frequencies is important for an accurate numerical quadrature. This is
discussed in the end of Section \ref{sec:RPAUEG}.

    \section{Direct Ring Coupled Cluster Doubles}
\label{sec:RPAdrCCD}
The Random Phase Approximation can also be evaluated using the matrix
elements of the Coulomb integral
\[
  V_{sr}^{pq} = \iint\d\vec x\d\vec x'\,
    \psi_p^\ast(\vec x)\psi_q^\ast(\vec x')\,\frac1{|\vec r-\vec r'|}\,
    \psi_r(\vec x')\psi_s(\vec x)
\]
rather than the propagators $G_0$ and $V$. This approach is not as efficient,
however an important correction to the error remaining in the RPA is based on
this approach.
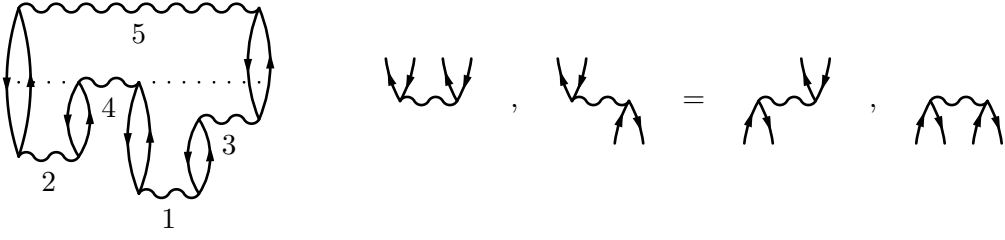
\begin{figure}
\begin{center}
  \diagramBoxBorder{0ex}{2ex}{
    \begin{fmffile}{Bubbles6}
    \begin{fmfgraph*}(90,70)
      \fmfstraight
      \fmfset{arrow_len}{6}
      \fmfleft{v00,v01,v02,v03,v04,v05}
      \fmfright{v40,v41,v42,v43,v44,v45}
      \fmf{phantom,tension=0.5}{v00,v20} \fmf{boson,label=$1$}{v20,v30}
        \fmf{phantom}{v30,v40}
      \fmf{boson,label=$2$,tension=3}{v01,v11} \fmf{phantom}{v11,v41}
      \fmf{phantom}{v02,v32} \fmf{boson,label=$3$,tension=3}{v32,v42}
      \fmf{dots}{v03,v13} \fmf{boson,label=$4$}{v13,v23}
        \fmf{dots,tension=0.5}{v23,v43}
      \fmf{boson,label=$5$}{v05,v45}
      \fmffreeze
      \fmf{fermion,right=0.16}{v01,v05,v01}
      \fmf{fermion,right=0.3}{v11,v13,v11}
      \fmf{fermion,right=0.2}{v20,v23,v20}
      \fmf{fermion,right=0.3}{v30,v32,v30}
      \fmf{fermion,right=0.2}{v42,v45,v42}
    \end{fmfgraph*}
    \end{fmffile}
  }
  \qquad
  \diagramBox{
    \begin{fmffile}{CCDvUpUp}
    \begin{fmfgraph}(32,32)
      \fmfset{arrow_len}{6}
      \fmfstraight
      \fmfleft{v00,v01,v02}
      \fmfright{v60,v61,v62}
      \fmf{phantom,tension=4}{v01,v11} \fmf{boson}{v11,v51}
        \fmf{phantom,tension=4}{v51,v61}
      \fmf{phantom}{v02,v22,v42,v62}
      \fmffreeze
      \fmf{fermion,left=0.1}{v22,v11,v02}
      \fmf{fermion,left=0.1}{v62,v51,v42}
    \end{fmfgraph}
    \end{fmffile}
  }
  ,
  \diagramBox{
    \begin{fmffile}{CCDvUpDown}
    \begin{fmfgraph}(32,32)
      \fmfset{arrow_len}{6}
      \fmfstraight
      \fmfleft{v00,v01,v02}
      \fmfright{v60,v61,v62}
      \fmf{phantom,tension=4}{v01,v11} \fmf{boson}{v11,v51}
        \fmf{phantom,tension=4}{v51,v61}
      \fmf{phantom}{v02,v22,v42,v62}
      \fmf{phantom}{v00,v20,v40,v60}
      \fmffreeze
      \fmf{fermion,left=0.1}{v22,v11,v02}
      \fmf{fermion,left=0.1}{v40,v51,v60}
    \end{fmfgraph}
    \end{fmffile}
  }
  =
  \diagramBox{
    \begin{fmffile}{CCDvDownUp}
    \begin{fmfgraph}(32,32)
      \fmfset{arrow_len}{6}
      \fmfstraight
      \fmfleft{v00,v01,v02}
      \fmfright{v60,v61,v62}
      \fmf{phantom,tension=4}{v01,v11} \fmf{boson}{v11,v51}
        \fmf{phantom,tension=4}{v51,v61}
      \fmf{phantom}{v02,v22,v42,v62}
      \fmf{phantom}{v00,v20,v40,v60}
      \fmffreeze
      \fmf{fermion,left=0.1}{v00,v11,v20}
      \fmf{fermion,left=0.1}{v62,v51,v42}
    \end{fmfgraph}
    \end{fmffile}
  }
  ,
  \diagramBox{
    \begin{fmffile}{CCDvDownDown}
    \begin{fmfgraph}(32,32)
      \fmfset{arrow_len}{6}
      \fmfstraight
      \fmfleft{v00,v01,v02}
      \fmfright{v60,v61,v62}
      \fmf{phantom,tension=4}{v01,v11} \fmf{boson}{v11,v51}
        \fmf{phantom,tension=4}{v51,v61}
      \fmf{phantom}{v00,v20,v40,v60}
      \fmffreeze
      \fmf{fermion,left=0.1}{v00,v11,v20}
      \fmf{fermion,left=0.1}{v40,v51,v60}
    \end{fmfgraph}
    \end{fmffile}
  }
  \caption{
    Goldstone diagram of an RPA ring diagram forming a closed loop.
    Each Coulomb interaction is connected to the ring in one of four
    possible ways of which three are distinct.
  }
  \label{fig:RPAGold}
\end{center}
\end{figure}

The RPA ring diagrams form a closed loop and Figure
\ref{fig:RPAGold} shows the Goldstone diagram of one such ring diagram.
As discussed in Section \ref{sec:MBPTGold}, connected vertices represent
contractions and we need to integrate over the time interval between Coulomb
interactions. We will do that from bottom to top following the left and
the right particle/hole pairs along the loop. In the diagram given as
example in Figure \ref{fig:RPAGold} we start with interaction $1$
following the left and the right particle/hole pairs to interaction $3$.
There, the right pair is contracted at one vertex and a new pair to follow
on the right side emerges on the other vertex of interaction 3.
Independently, we also follow the left and the right particle/hole pairs
starting at interaction $2$. At interaction $4$ the two processes merge
such that there is only one left pair and one right pair remaining after
the time indicated by the dotted line. At interaction $5$ both pairs
are finally contracted and the loop is closed.
This procedure is analogous to the one
employed for deriving the Fermion sign of Goldstone diagrams in 
Subsection \ref{ssc:MBPT_Goldstone_FermionSign}.

While following the left and the right particle/hole pairs we perform all
occurring contractions and integrals over the respective time intervals
and we keep the intermediate results in a matrix $t_{ij}^{ab}$ depending on
the states of two particle/hole pairs, diagrammatically denoted by
\[
  t_{ij}^{ab}\ =
  \diagramBoxBorder{2ex}{0ex}{
    \begin{fmffile}{CCDtLabel}
    \begin{fmfgraph*}(32,16)
      \fmfkeep{CCDtLabel}
      \fmfset{arrow_len}{6}
      \fmfstraight
      \fmfleft{v00,v01}
      \fmfright{v60,v61}
      \fmf{phantom,tension=4}{v00,v10} \fmf{dashes}{v10,v50}
        \fmf{phantom,tension=4}{v50,v60}
      \fmf{phantom}{v01,v21,v41,v61}
      \fmffreeze
      \fmf{fermion,left=0.1}{v21,v10,v01}
      \fmf{fermion,left=0.1}{v61,v50,v41}
      \fmfv{label=$a$,label.angle=90}{v01}
      \fmfv{label=$i$,label.angle=90}{v21}
      \fmfv{label=$b$,label.angle=90}{v41}
      \fmfv{label=$j$,label.angle=90}{v61}
    \end{fmfgraph*}
    \end{fmffile}
  }\,.
\]
$t_{ij}^{ab}$ is called \emph{direct ring Coupled Cluster Doubles
(drCCD) amplitudes}
\index{direct ring Coupled Cluster} and it contains the probability amplitude
to arrive at two particle/hole pairs in the given states in any of the possible
ways forming the ring diagrams. We will now go through all cases that can
occur following the left and the right particle/hole pair from bottom to
top indicated in Figure \ref{fig:RPAGold}.

In the first case, a Coulomb interaction creates two new particle/hole pairs
as in interaction $1$ in Figure \ref{fig:RPAGold}.
This can occur at any time in the past so we need to integrate over the time
interval between the Coulomb interaction and the time where we want to
use the probability amplitudes $t_{ij}^{ab}$, which we always move at $t=0$.
One contribution to the drCCD amplitudes is thus
\begin{align}
  \nonumber
  \hspace*{-2ex}
  \diagramBoxBorder{1ex}{0ex}{
    \begin{fmffile}{CCDtLeft}
    \begin{fmfgraph*}(48,48)
      \fmfkeep{CCDtLeft}
      \fmfset{arrow_len}{6}
      \fmfstraight
      \fmfleft{v00,v01,v02,v03}
      \fmfright{v60,v61,v62,v63}
      \fmf{phantom,tension=4}{v02,v12} \fmf{dashes}{v12,v52}
        \fmf{phantom,tension=4}{v52,v62}
      \fmf{dots}{v03,v23,v43,v63}
      \fmffreeze
      \fmf{phantom}{v03,v13,v23}
      \fmf{fermion,left=0.1}{v23,v12,v03}
      \fmf{fermion,left=0.1}{v63,v52,v43}
      \fmfv{label=$a$,label.angle=90}{v03}
      \fmfv{label=$i$,label.angle=90}{v23}
      \fmfv{label=$b$,label.angle=90}{v43}
      \fmfv{label=$j$,label.angle=90}{v63}
      \fmfv{label=$t=0$,label.angle=180,label.dist=16}{v13}
    \end{fmfgraph*}
    \end{fmffile}
  }
  \hspace*{-1ex}
  \quad
  t_{ij}^{ab} &=
  (-\im)\int_{-\infty}^0 \d t\,e^{\eta t}\,
  e^{\eps_a t}\,e^{\eps_b t}\,e^{-\eps_i t}\,e^{-\eps_j t}\,V_{ij}^{ab} +
  \ldots
  \quad
    \diagramBox{
    \begin{fmffile}{CCDv}
    \begin{fmfgraph*}(48,48)
      \fmfset{arrow_len}{6}
      \fmfstraight
      \fmfleft{v00,v01,v02,v03}
      \fmfright{v60,v61,v62,v63}
      \fmf{dots,tension=4}{v01,v11} \fmf{boson}{v11,v51}
        \fmf{dots,tension=4}{v51,v61}
      \fmf{dots}{v03,v23,v43,v63}
      \fmffreeze
      \fmf{phantom}{v43,v53,v63}
      \fmf{fermion,left=0.1}{v23,v11,v03}
      \fmf{fermion,left=0.1}{v63,v51,v43}
      \fmfv{label=$a$,label.angle=90}{v03}
      \fmfv{label=$i$,label.angle=90}{v23}
      \fmfv{label=$b$,label.angle=90}{v43}
      \fmfv{label=$j$,label.angle=90}{v63}
      \fmfv{label=$t=0$,label.angle=0,label.dist=16}{v53}
      \fmfv{label=$t$,label.angle=0,label.dist=16}{v51}
    \end{fmfgraph*}
    \end{fmffile}
  }
  \\
  &=\frac{V_{ij}^{ab}}{\eps_i+\eps_j-\eps_a-\eps_b+\im\eta}\,+\ldots
  \label{eqn:drCCDV}
\end{align}
In the next case, a Coulomb interaction contracts the right particle/hole
pair creating a new one. This is the case for interaction $3$ in Figure
\ref{fig:RPAGold}. We can now use the drCCD amplitudes recursively containing
all processes until the time of the interaction. The time interval between
the interaction and the time where we want to use the new drCCD amplitudes
still needs to be integrated as before. The second contribution to the drCCD
amplitudes is thus
\begin{equation}
  \diagramBoxBorder{2ex}{0ex}{
    \fmfreuse{CCDtLeft}
  }
  \quad
  t_{ij}^{ab} = \ldots +
  \,\frac{\sum_{kc}t_{ik}^{ac}V_{cj}^{kb}}{
    \eps_i+\eps_j-\eps_a-\eps_b+\im\eta
  }\,+\ldots
  \quad
    \diagramBox{
    \begin{fmffile}{CCDtv}
    \begin{fmfgraph*}(64,48)
      \fmfset{arrow_len}{6}
      \fmfstraight
      \fmfleft{v00,v01,v02,v03}
      \fmfright{v80,v81,v82,v83}
      \fmf{phantom,tension=12}{v00,v10} \fmf{dashes,tension=4}{v10,v40}
        \fmf{phantom,tension=3}{v40,v80}
      \fmf{dots,tension=3}{v01,v41} \fmf{boson,tension=4}{v41,v71}
        \fmf{dots,tension=12}{v71,v81}
      \fmf{dots}{v03,v23,v43,v63,v83}
      \fmffreeze
      \fmf{phantom}{v63,v73,v83}
      \fmf{fermion,left=0.1}{v23,v10,v03}
      \fmf{fermion,left=0.1}{v83,v71,v63}
      \fmf{fermion,left=0.4,label=$c$,label.dist=4}{v40,v41}
        \fmf{fermion,left=0.4,label=$k$,label.dist=4}{v41,v40}
      \fmfv{label=$a$,label.angle=90}{v03}
      \fmfv{label=$i$,label.angle=90}{v23}
      \fmfv{label=$b$,label.angle=90}{v63}
      \fmfv{label=$j$,label.angle=90}{v83}
      \fmfv{label=$t=0$,label.angle=0,label.dist=16}{v73}
      \fmfv{label=$t$,label.angle=0,label.dist=16}{v71}
    \end{fmfgraph*}
    \end{fmffile}
  }
  \label{eqn:drCCDtv}
\end{equation}
The Fermion sign of this contribution is positive since there is one more hole
and one more closed Fermion loop. Note however, that the denominator is negative
giving rise to the alternating sign when evaluating the RPA order by order.
The same case can occur for the left particle/hole pair giving
\begin{equation}
  \diagramBoxBorder{2ex}{0ex}{
    \fmfreuse{CCDtLeft}
  }
  \quad
  t_{ij}^{ab} = \ldots +
  \,\frac{\sum_{kc}V_{ic}^{ak}t_{kj}^{cb}}{
    \eps_i+\eps_j-\eps_a-\eps_b+\im\eta
  }\,+\ldots
  \quad
  \diagramBox{
    \begin{fmffile}{CCDvt}
    \begin{fmfgraph*}(64,48)
      \fmfset{arrow_len}{6}
      \fmfstraight
      \fmfleft{v00,v01,v02,v03}
      \fmfright{v80,v81,v82,v83}
      \fmf{dots,tension=12}{v01,v11} \fmf{boson,tension=4}{v11,v41}
        \fmf{dots,tension=3}{v41,v81}
      \fmf{phantom,tension=3}{v00,v40} \fmf{dashes,tension=4}{v40,v70}
        \fmf{phantom,tension=12}{v70,v80}
      \fmf{dots}{v03,v23,v43,v63,v83}
      \fmffreeze
      \fmf{phantom}{v41,v51,v61,v71,v81}
      \fmf{phantom}{v63,v73,v83}
      \fmf{fermion,left=0.1}{v23,v11,v03}
      \fmf{fermion,left=0.1}{v83,v70,v63}
      \fmf{fermion,left=0.4,label=$c$,label.dist=4}{v40,v41}
        \fmf{fermion,left=0.4,label=$k$,label.dist=4}{v41,v40}
      \fmfv{label=$a$,label.angle=90}{v03}
      \fmfv{label=$i$,label.angle=90}{v23}
      \fmfv{label=$b$,label.angle=90}{v63}
      \fmfv{label=$j$,label.angle=90}{v83}
      \fmfv{label=$t=0$,label.angle=0,label.dist=16}{v73}
      \fmfv{label=$t$,label.angle=0,label.dist=16}{v71}
    \end{fmfgraph*}
    \end{fmffile}
  }
\end{equation}
The last case occurs when two independent processes merge at a Coulomb
interaction, as in interaction $4$ of Figure \ref{fig:RPAGold}. In this
contribution, the drCCD amplitudes occur in a quadratic form on the right hand
side
\begin{equation}
  \diagramBoxBorder{2ex}{0ex}{
    \fmfreuse{CCDtLeft}
  }
  \quad
  t_{ij}^{ab} = \ldots +
  \,\frac{\sum_{klcd}t_{ik}^{ac}V_{cd}^{kl}t_{lj}^{db}}{
    \eps_i+\eps_j-\eps_a-\eps_b+\im\eta
  }
  \,.
  \quad
  \diagramBox{
    \begin{fmffile}{CCDtvt}
    \begin{fmfgraph*}(96,48)
      \fmfset{arrow_len}{6}
      \fmfstraight
      \fmfleft{v00,v01,v02,v03}
      \fmfright{v120,v121,v122,v123}
      \fmf{dots}{v01,v41} \fmf{boson}{v41,v81} \fmf{dots}{v81,v121}
      \fmf{phantom,tension=12}{v00,v10} \fmf{dashes,tension=4}{v10,v40}
        \fmf{phantom,tension=3}{v40,v80} \fmf{dashes,tension=4}{v80,v110}
        \fmf{phantom,tension=12}{v110,v120}
      \fmf{dots}{v03,v23,v43,v63,v83,v103,v123}
      \fmffreeze
      \fmf{phantom}{v81,v91,v101,v111,v121}
      \fmf{phantom}{v103,v113,v123}
      \fmf{fermion,left=0.1}{v23,v10,v03}
      \fmf{fermion,left=0.1}{v123,v110,v103}
      \fmf{fermion,left=0.4,label=$c$,label.dist=4}{v40,v41}
        \fmf{fermion,left=0.4,label=$k$,label.dist=4}{v41,v40}
      \fmf{fermion,left=0.4,label=$d$,label.dist=4}{v80,v81}
        \fmf{fermion,left=0.4,label=$l$,label.dist=4}{v81,v80}
      \fmfv{label=$a$,label.angle=90}{v03}
      \fmfv{label=$i$,label.angle=90}{v23}
      \fmfv{label=$b$,label.angle=90}{v103}
      \fmfv{label=$j$,label.angle=90}{v123}
      \fmfv{label=$t=0$,label.angle=0,label.dist=16}{v113}
      \fmfv{label=$t$,label.angle=0,label.dist=16}{v111}
    \end{fmfgraph*}
    \end{fmffile}
  }
  \label{eqn:drCCDtVt}
\end{equation}
Taking all contribution (\ref{eqn:drCCDV}) to (\ref{eqn:drCCDtVt}) into
consideration and taking the limit $\eta\rightarrow0$ yields the drCCD
amplitudes equation
\begin{equation}
  (\eps_i+\eps_j-\eps_a-\eps_b)\,t_{ij}^{ab} =
    V_{ij}^{ab}
    + \sum_{kc}t_{ik}^{ac}V_{cj}^{kb} + \sum_{kc}V_{ic}^{ak}t_{kj}^{cb}
    + \sum_{klcd}t_{ik}^{ac}V_{cd}^{kl}t_{lj}^{db}\,.
  \label{eqn:drCCDAmplitudes}
\end{equation}
This equation is quadratic and can only be solved by iteration. The convergence
with respect to the number of iterations is, however, fast. 
Employing a Shanks transform, 8 iterations are sufficient to yield a converged
RPA energy to 5 significant digits of precision even for a
system with a low band gap, such as a finite size uniform electron gas
\parencite{freeman_coupled-cluster_1977,shanks_non-linear_1955}.
Each iteration is still costly requiring $\mathcal O(N^5)$ steps.
The iteration process also requires the amplitudes to be stored demanding
$\mathcal O(N^4)$ of memory.

Given the drCCD amplitudes, we can evaluate the RPA energy by contracting
both particle/hole pairs with the last Coulomb interaction, corresponding
to interaction $5$ in Figure \ref{fig:RPAGold}. The drCCD amplitudes have
a left/right reflection symmetry. The RPA energy is thus
\begin{equation}
  \hspace*{-2ex}
  \diagramBoxBorder{2ex}{2ex}{
    \begin{fmffile}{drCCDd}
    \begin{fmfgraph}(30,30)
      \fmfkeep{drCCDd}
      \fmfset{arrow_len}{6}
      \fmfstraight
      \fmfbottom{lb,rb}
      \fmftop{lt,rt}
      \fmf{dashes}{lb,rb}
      \fmf{boson}{lt,rt}
      \fmf{fermion,left=0.25}{lb,lt,lb}
      \fmf{fermion,left=0.25}{rb,rt,rb}
    \end{fmfgraph}
    \end{fmffile}
  }
  \hspace*{-2ex}
  \ =\ \frac12 \sum_{ijab}t_{ij}^{ab}V_{ab}^{ij}\,,
\end{equation}
where the two closing Fermion loops result in a positive Fermion sign.

In this approach the time order is always maintained by the way the
drCCD amplitudes are recursively defined. Therefore, no particular symmetries
have to be considered apart from the reflection symmetry.
Ring diagrams that have more than two
particle/hole pairs at some instances in time arise from the quadratic
contribution (\ref{eqn:drCCDtVt}). This is the case for the diagram in
Figure \ref{fig:RPAGold} between interaction $2$ and $4$. 
Excluding the quadratic contribution (\ref{eqn:drCCDtVt}) gives the
\emph{Tamm-Dancoff Approximation (TDA)}\index{Tamm-Dancoff approximation},
which is the subset of all RPA ring diagrams
where there are exactly two particle/hole pairs at all times between the first
and the last interaction.
From (\ref{eqn:drCCDtVt}) and (\ref{eqn:drCCDV}) follows that the lowest order
diagram of RPA that is not part of TDA is of fourth order.

On the other hand, the drCCD amplitude equations are a subset of the amplitude
equations including all possible ways to arrive at two particle/hole pairs.
These amplitude equations are called
\emph{Coupled Cluster Singles Doubles (CCSD) amplitudes}\index{Coupled Cluster Singles Doubles}
and they contain processes such as
\[
  \diagramBox{
    \begin{fmffile}{CCDtvxt}
    \begin{fmfgraph}(96,48)
      \fmfkeep{CCDtvxt}
      \fmfset{arrow_len}{6}
      \fmfstraight
      \fmfleft{v00,v01,v02,v03}
      \fmfright{v120,v121,v122,v123}
      \fmf{phantom}{v01,v41} \fmf{boson}{v41,v81} \fmf{phantom}{v81,v121}
      \fmf{phantom,tension=12}{v00,v10} \fmf{dashes,tension=4}{v10,v40}
        \fmf{phantom,tension=3}{v40,v80} \fmf{dashes,tension=4}{v80,v110}
        \fmf{phantom,tension=12}{v110,v120}
      \fmf{phantom}{v03,v23,v43,v63,v83,v103,v123}
      \fmffreeze
      \fmf{fermion,left=0.1}{v23,v10,v03}
      \fmf{fermion,left=0.1}{v123,v110,v103}
      \fmf{fermion,left=0.4}{v40,v41}
        \fmf{plain}{v41,m1} \fmf{fermion}{m1,v80}
      \fmf{fermion,right=0.4}{v80,v81}
        \fmf{plain}{v81,m2} \fmf{fermion}{m2,v40}
    \end{fmfgraph}
    \end{fmffile}
  }
  \hspace*{-1ex}.
\]
Unfortunately, it is computationally more time consuming to calculate the
full CCSD amplitudes, scaling like $\mathcal O(N^6)$, since the employed
tensor multiplications cannot be split into pieces, involving no more than
5 indices. For the calculation of the drCCD amplitudes this can be done.
since the matrix of the Coulomb interaction $V^{pq}_{sr}$ can be decomposed
into a product of two tensors with 3 indices in the momentum basis:
\[
  V^{pq}_{sr} =
    \int\frac{\d\vec G}{(2\pi)^3}\,
      \chi_s^p(\vec G)\,
      {\chi_q^r}^\ast(\vec G)\,,
  \quad
  \textnormal{with }
  \chi_q^p(\vec G)=\int\d\vec x\,\frac{\sqrt{4\pi}}{|\vec G|}\,
    \psi_p^\ast(\vec x)e^{\im\vec r\cdot\vec G}\psi_q(\vec x)\,.
\]
This allows tensor products involving $V_{sr}^{pq}$ to be ``cut'' at the
Coulomb line, accelerating the evaluation of the direct ring Coupled Cluster
Doubles amplitudes to $\mathcal O(N^5)$. For the first
non-trivial case in (\ref{eqn:drCCDAmplitudes}), this is done, for instance, by
\[
  \nonumber
  \sigma_i^a(\vec G) = \sum_{ck} t_{ik}^{ac}\chi_c^k(\vec G)\,,
  \qquad
  \nonumber
  (\eps_i+\eps_j-\eps_a-\eps_b)\, t_{ij}^{ab} =
    \ldots +
    \int\frac{\d\vec G}{(2\pi)^3}\,\sigma_i^a(\vec G){\chi_b^j}^\ast(\vec G) +
    \ldots
\]

The Coupled Cluster method was developed by
\parencite{coester_short-range_1960} for
the atomic nucleus and later adopted for electronic correlation
by \parencite{cizek_use_1969}.

    \section{RPA from the Adiabatic Connection}
\label{sec:RPAAC}
In the framework of many-body perturbation theory the derivation of the
Random Phase Approximation is straight forward. According to the last two
sections it can be done either in the frequency domain employing Feynman
diagrams and regarding the symmetries of the ring diagrams or in the time
domain employing Goldstone diagrams.
In Chapter \ref{cha:MBPT} we discussed the equivalence of the two approaches.

Despite the straight forward derivations and the general applicability of
many-body perturbation theory to arbitrary
reference systems, neither of the two previously discussed derivations are
considered standard approaches to the Random Phase Approximation.
In this section we discuss the framework of the Adiabatic Connection (AC)
employed by Bohm and Pines, which is considered the standard approach to
the RPA, at least in solid state physics.

In the Adiabatic Connection (AC) we define a Hamiltonian $\hat H(\lambda)$
depending on a coupling constant $\lambda$ specifying the strength
of the full electron-electron Coulomb interaction
\begin{equation}
  \hat H(\lambda) = \hat T + \hat V_{\rm ne} + \hat V_{\rm eff}(\lambda)
    + \lambda \hat V_{\rm ee}\,,\quad \lambda\in[0,1]\,.
  \label{eqn:RPAACHamiltonian}
\end{equation}
The effective interaction $\hat V_{\rm eff}(\lambda)$ also depends on
the coupling constant and we choose $\hat V_{\rm eff}(\lambda)$ such
that the density of the system with the Hamiltonian $\hat H(\lambda)$
is for all $\lambda$ equivalent to the density of the fully interacting system
with the Hamiltonian $\hat H(1)$. This is in contrast to many-body perturbation
theory where the effective interaction is simply scaled by the factor
$1-\lambda$ according to (\ref{eqn:AdiabaticSwitching}) and where $\lambda$ is
time dependent.
Requiring a constant density for all $\lambda$ is a strong condition implying
that the density of the reference system with the Hamiltonian
$\hat H(0)$ is equivalent to the density of the fully interacting system.
Although this condition is met to a good degree by using a DFT Hamiltonian 
for $\hat H(0)$, the Hohenberg-Kohn theorem only states the existence of such an
effective potential $\hat V_{\rm eff}(\lambda)$. For practical considerations
it is known that in general no DFT density fully agrees with the density of the
respective fully interacting system.

Let now $\Psi(\lambda)$ denote the normalized ground state
of the respective Hamiltonian $\hat H(\lambda)$, being the solution of
\[
  \hat H(\lambda)\ket{\Psi(\lambda)} = E \ket{\Psi(\lambda)}\,.
\]
The ground state energy $E$ is equivalent for all $\lambda$ since $E$ is
a functional of the density according to the Hohenberg-Kohn theorem and
the density is the same for all $\lambda$.
Even assuming that the density $n(\vec r)$ of the DFT reference is exact we can
only directly evaluate the nuclei-electron potential energy $E_{\rm ne}$ as the
functionals for the kinetic energy $T$ and for the electron-electron potential
energy $E_{\rm ee}$ are unknown:
\[
  E[n(\vec r)] = T[n(\vec r)] + E_{\rm ne}[n(\vec r)] + E_{\rm ee}[n(\vec r)]\,.
\]
However, we are only interested in the sum of the kinetic and the potential
energy so we may as well use the known kinetic energy of the Kohn-Sham system
\[
  T_{\rm S} = -\mbracket{\Psi(0)}{\sum_{n=1}^N \frac{\nabla_n^2}2}{\Psi(0)}
\]
and ask for the energy to be added to $T_{\rm S}$ and $E_{\rm ne}$ to arrive at
the same total energy $E$. This energy is called
\emph{Hartree-exchange-correlation energy}
\index{Hartree-exchange-correlation energy} and it is given by
\[
  E_{\rm Hxc} = E - E_{\rm ne} - T_{\rm S}\,.
\]
Using (\ref{eqn:RPAACHamiltonian}) we can express the terms above by
expectation values of the fully and the non-interacting Hamiltonian:
\begin{align}
  \nonumber
  E-E_{\rm ne} =&\ \langle\Psi(1)|\hat H(1)|\Psi(1)\rangle
    - \langle\Psi(1)|\hat V_{\rm ne} + \hat V_{\rm eff}(1)|\Psi(1)\rangle \\
  \nonumber
  T_{\rm S} =&\ \langle\Psi(0)|\hat H(0)|\Psi(0)\rangle
    - \langle\Psi(0)|\hat V_{\rm ne} + \hat V_{\rm eff}(0)|\Psi(0)\rangle\,,
\end{align}
noting that $\hat V_{\rm eff}(1) = \hat 0$ and $\hat V_{\rm eff}(0)$ is the
effective potential of the Kohn-Sham system. This allows us to make the
connection between the fully interacting and the non-interacting
system
\begin{equation}
  E_{\rm Hxc} = \int_0^1 \d\lambda\, \frac\d{\d\lambda}\,
    \langle\Psi(\lambda)|
      \hat H(\lambda) - \hat V_{\rm ne} - \hat V_{\rm eff}(\lambda)|
      \Psi(\lambda)\rangle
  \label{eqn:RPAAC}
\end{equation}
$\Psi(\lambda)$ is the ground state of $\hat H(\lambda)$. As a consequence,
we can use the G\"uttinger theorem \parencite{guttinger_verhalten_1932},
also known as Hellman-Feynman theorem, to evaluate the total derivative of
the expectation values involving the Hamiltonian:
\begin{equation}
  \frac\d{\d\lambda}\,
    \langle \Psi(\lambda)|\hat H(\lambda)|\Psi(\lambda)\rangle=
  \mbracket{\Psi(\lambda)}{\frac\d{\d\lambda}\,\hat H(\lambda)}{\Psi(\lambda)}
  \label{eqn:RPAGuettinger}
\end{equation}
Furthermore, $\hat V_{\rm ne}$ and $\hat V_{\rm eff}(\lambda)$ are local
potentials and their expectation value only depends on the electron density
$n(\vec r)$. Since the density is constant for all $\lambda$ the total derivate
of their respective expectation value simplifies to
\begin{align}
  \frac\d{\d\lambda}\,
    \langle \Psi(\lambda)|\hat V_{\rm ne}|\Psi(\lambda)\rangle
  &= 0 \\
  \frac\d{\d\lambda}\,
    \langle \Psi(\lambda)|\hat V_{\rm eff}(\lambda)|\Psi(\lambda)\rangle
  &=
  \mbracket{\Psi(\lambda)}{
    \frac\d{\d\lambda}\,\hat V_{\rm eff}(\lambda)
  }{\Psi(\lambda)}\,.
  \label{eqn:RPALocalDerivative}
\end{align}
Inserting the total derivatives into (\ref{eqn:RPAAC}) yields the final
expression for the Hartree-exchange-correlation energy in the adiabatic
connection:
\begin{equation}
  E_{\rm Hxc} = \int_0^1\d\lambda\,
  \langle\Psi(\lambda)|\hat V_{\rm ee}|\Psi(\lambda)\rangle\,.
  \label{eqn:RPAACFinal}
\end{equation}
In other words, the Hartree-exchange-correlation energy is the coupling
strength averaged potential electron-electron energy.

\subsection{Fluctuation dissipation theorem}
Equation (\ref{eqn:RPAACFinal}) cannot be evaluated directly since the
ground states $\Psi(\lambda)$ are not available for any $\lambda$ except 0.
However, the whole information of the many-body wavefunction $\Psi(\lambda)$ is
not required to evaluate the Coulomb operator. The pair density
$n^2(\vec x,\vec x')$ is sufficient:
\begin{align}
  \nonumber
  \langle\Psi|\hat V_{\rm ee}|\Psi\rangle =&
    \int\d\vec x_1\ldots\int\d\vec x_N\,
      \Psi^\ast(\vec x_1,\ldots,\vec x_N)
      \,\frac12\sum_{n\neq m}\frac1{|\vec r_n-\vec r_m|}\,
      \Psi(\vec x_1,\ldots,\vec x_N) \\
  \nonumber
    =&\ \frac12\int\d\vec x\int\d\vec x'\,\frac1{|\vec r-\vec r'|}\,
      \underbrace{
        \int\d\vec x_1\ldots\int\d\vec x_N\,
        \Psi^\ast(\ldots)\,
        \sum_{n\neq m}\delta(\vec x-\vec x_n)\delta(\vec x'-x_m)\,
        \Psi(\ldots)
      }_{=\,n^2(\vec x,\vec x')}
\end{align}
The pair density can be written in terms of unrestricted summations
\begin{align}
  \nonumber
  n^2(\vec x, \vec x')
  =&\
    \mbracket{\Psi}{
      \sum_{n\neq m}\delta(\vec x-\vec x_n)\delta(\vec x'-\vec x_m)
    }{\Psi} \\
  \nonumber
  =&\
    \mbracket{\Psi}{
      \sum_{nm}\delta(\vec x-\vec x_n)\delta(\vec x'-\vec x_m)
    }{\Psi} -
    \underbrace{
      \mbracket{\Psi}{
        \sum_{n}\delta(\vec x-\vec x')\delta(\vec x-\vec x_n)
      }{\Psi}
    }_{=\,\delta(\vec x-\vec x') n(\vec x)}\,,
\end{align}
which allows us to relate it to the density fluctuation operator
$
  \delta\hat n(x) = \sum_n \delta(\vec x-\vec x_n) - n(x)
$, since
\begin{align}
  \nonumber
  \mbracket[]{\Psi}{
    \delta\hat n(\vec x) \delta\hat n(\vec x')
  }{\Psi}
  =&\
    \mbracket{\Psi}{
      \sum_n\delta(\vec x-\vec x_n)
      \sum_m\delta(\vec x'-\vec x_m)
    }{\Psi}
    +n(\vec x)n(\vec x') \\
  \nonumber
  &\
    -\underbrace{
      \mbracket{\Psi}{
        \sum_n\delta(\vec x-\vec x_n)
      }{\Psi}
    }_{=\,n(\vec x)}n(\vec x')
    -n(\vec x)n(\vec x')\,.
\end{align}
The pair density and the density fluctuation operator are thus related by
\begin{equation}
  n^2(\vec x,\vec x') =\,
  \mbracket[]{\Psi}{\delta\hat n(\vec x)\delta\hat n(\vec x')}{\Psi}
  + n(\vec x)n(\vec x') - \delta(\vec x-\vec x')n(\vec x)\,.
  \label{eqn:RPADensityFluctuations}
\end{equation}
Note that although $\mbracket[]{\Psi}{\delta\hat n(\vec x)}{\Psi}$ is
zero, the expectation value of a quadratic form of the density fluctuation
operator is in general non-zero. The fluctuation dissipation theorem
links the density-density fluctuations at the two sites $\vec x$ and $\vec x'$
occurring in (\ref{eqn:RPADensityFluctuations}) to the density-density response
function $\chi_\lambda(\vec x, \vec x', \im\nu)$ of the system with the coupling
strength $\lambda$, here given in imaginary frequency:
\begin{equation}
  -\int\frac{\d\nu}{2\pi}\,\chi_\lambda(\vec x,\vec x',\im\nu) =
    \mbracket[]{\Psi(\lambda)}{
      \delta\hat n(\vec x)\delta\hat n(\vec x')
    }{\Psi(\lambda)}\,.
  \label{eqn:RPAFDT}
\end{equation}
With time dependent density functional theory (TDDFT) we can finally
connect the density-density response function $\chi_\lambda$ of the system
with a coupling strength $\lambda$ to the density-density response function
$\chi_0$ of the DFT reference system
\begin{multline}
  \chi_\lambda(\vec x_1,\vec x_4,\im\nu) =\
    \chi_0(\vec x_1,\vec x_4,\im\nu) \\
  +
    \iint\d\vec x_2\d\vec x_3\,
      \chi_0(\vec x_1,\vec x_2,\im\nu)
      \left(
        \frac\lambda{|\vec r_2-\vec r_3|} +
        f^\lambda_{\rm xc}(\vec x_2,\vec x_3,\im\nu)
      \right)
      \chi_\lambda(\vec x_3,\vec x_4,\im\nu)\,,
  \label{eqn:RPAChiDyson}
\end{multline}
where the Coulomb interaction is scaled by $\lambda$. $f^\lambda_{\rm xc}$
is called \emph{exchange-correlation kernel}\index{exchange-correlation kernel}
and it contains the change of the DFT effective potential with respect
to a change in the electron density. $f^\lambda_{\rm xc}$ is not explicitly
known and it is coupling strength dependent.
(\ref{eqn:RPAChiDyson}) is an implicit integral equation for $\chi_\lambda$
called \emph{Dyson-like equation}\index{Dyson-like equation}.
It requires that the
density response of the interacting system to a change of the external
potential is the same as the density response of the DFT reference system to a
change of the effective potential. This assumption is the
TDDFT counterpart of the assumption that the DFT density is exact.

The response function of the DFT reference system $\chi_0$ can be evaluated
from the DFT spin-orbitals $\psi_p(\vec x)$ with first order perturbation
theory:
\begin{align}
  \chi_0(\vec x,\vec x',\im\nu)
  =&
  \,-\sum_{ia}
  \left(
  \frac{
    \psi_a(\vec x)\psi^\ast_a(\vec x')\psi_i(\vec x')\psi^\ast_i(\vec x)
  }{\eps_a-\eps_i-\im\nu}
  +
  \frac{
    \psi_i(\vec x)\psi^\ast_i(\vec x')\psi_a(\vec x')\psi^\ast_a(\vec x)
  }{\eps_a-\eps_i+\im\nu}
  \right)
  \label{eqn:RPAPolImag}
\end{align}
See (\ref{eqn:RPAChiFreqInt}) for an analogous derivation of the
above result for the uniform electron gas. Note that the definition of
$\chi_0$ given in this section expects the Coulomb propagator to be
$1/|\vec r-\vec r'|$. This differs from the convention used elsewhere
in this work and has been chosen for consistency of this section with
literature.
In (\ref{eqn:RPAChiDifference}) this is discussed in detail.
The above definition further assumes a spin-unrestricted reference and an
imaginary frequency integration over the entire domain from $-\infty$ to
$\infty$.

Using (\ref{eqn:RPAACFinal}), (\ref{eqn:RPADensityFluctuations}) and
(\ref{eqn:RPAFDT}) the
Hartree-exchange-correlation energy is given by
\begin{align}
  \nonumber
  E_{\rm Hxc} =&
  \ \frac12\int_0^1\d\lambda\int\d\vec x\int\d\vec x'
    \frac1{|\vec r-\vec r'|}
  \left(
    n(\vec x)n(\vec x')
    - \delta(\vec x-\vec x')n(\vec x)
    -\int\frac{\d\nu}{2\pi}\chi_\lambda(\vec x,\vec x',\im\nu)
  \right) \\
  =&\ E_{\rm H} + 
  \frac12\int_0^1\d\lambda\int\d\vec x\int\d\vec x'
    \frac1{|\vec r-\vec r'|}
  \left(
    -\delta(\vec x-\vec x')n(\vec x)
    -\int\frac{\d\nu}{2\pi}\chi_\lambda(\vec x,\vec x',\im\nu)
  \right)\,,
  \label{eqn:RPAExc}
\end{align}
where the Hartree energy $E_{\rm H}$ is readily factored out.
To tackle the remaining Dirac delta function we rewrite
$n(\vec x)=\sum_i\psi_i(\vec x)\psi^\ast_i(\vec x)$ in the context of the
delta function and then use the completeness of the eigenstates
$\psi_p(\vec x)$:
\begin{align}
  \nonumber
  \delta(\vec x-\vec x') n(\vec x) =&\,
    \delta(\vec x-\vec x')\sum_i\psi_i(\vec x')\psi^\ast_i(\vec x)\,
  =\,
    \sum_{pi}
      \psi_p(\vec x)\psi^\ast_p(\vec x')\psi_i(\vec x')\psi^\ast_i(\vec x) \\
  =&\,
    \sum_{ji}
      \psi_j(\vec x)\psi^\ast_j(\vec x')\psi_i(\vec x')\psi^\ast_i(\vec x)
    +
    \sum_{ai}
      \psi_a(\vec x)\psi^\ast_a(\vec x')\psi_i(\vec x')\psi^\ast_i(\vec x)\,.
  \label{eqn:RPADelta}
\end{align}
The first sum over unexcited states can be inserted into (\ref{eqn:RPAExc})
and factored out giving the \emph{exact exchange}\index{exact exchange}
energy
\[
  \diagramBox{\fmfreuse{Exchange}}
  \quad
  E_{\rm x} =\,
  -\frac12\int\d\vec x\int\d\vec x'\,
  \sum_{ij}
  \frac{
      \psi_j(\vec x)\psi^\ast_j(\vec x')\psi_i(\vec x')\psi^\ast_i(\vec x)
  }
  {|\vec r-\vec r'|}\,.
\]
The exact exchange energy corresponds to the given first order diagram in
many-body perturbation theory with the Kohn-Sham spin-orbitals
$\psi_p(\vec x)$ as reference system.

The second sum in (\ref{eqn:RPADelta}) over excited and unexcited states
is the integral of $\chi_0(\vec x,\vec x',\im\nu)$ given in
(\ref{eqn:RPAPolImag}) over all imaginary frequencies.
Note that there is
only one pole at $\nu=\mp\im(\eps_a-\eps_i)$ requiring a convergence factor
of $e^{-\im\nu\tau}$ with $0<|\tau|\ll1$ for a contour enclosing the
pole with vanishing contribution at infinity. We choose to take the limit
$\tau\rightarrow0^+$ from above resulting in a clockwise contour.
The sum over the residue gives
\[
  \int\frac{\d\nu}{2\pi}\,\chi_0(\vec x,\vec x',\im\nu) =
    -\sum_{ia}
      \psi_a(\vec x)\psi^\ast_a(\vec x')\psi_i(\vec x')\psi^\ast_i(\vec x)\,.
\]
The above result can also be obtained evaluating
$\chi_0(\vec x,\vec x',\im\tau)$ using imaginary time propagators according
to (\ref{eqn:RPAChiImag}) and (\ref{eqn:RPAGreenImag}) and then taking
the limit $\tau\rightarrow0^+$.
Finally, we can insert the exact exchange energy and the above expression
into (\ref{eqn:RPAExc}) to arrive at an expression for the
correlation energy:
\begin{equation}
  E_{\rm c} =
  -\frac12\int_0^1\d\lambda\int\d\vec x\int\d\vec x'
    \frac1{|\vec r-\vec r'|}
    \int\frac{\d\nu}{2\pi}
    \left(
      \chi_\lambda(\vec x,\vec x',\im\nu) -
      \chi_0(\vec x,\vec x',\im\nu) 
    \right)\,,
  \label{eqn:RPAACFDT}
\end{equation}
where $E_{\rm Hxc} = E_{\rm H} + E_{\rm x} + E_{\rm c}$.

\subsection{Random Phase Approximation of the polarizability}
The Dyson-like equation (\ref{eqn:RPAChiDyson}) for the polarizability
$\chi_\lambda(\vec x,\vec x',\im\nu)$ can be written in the matrix notation
introduced in Section \ref{sec:RPAFreq}
\begin{equation}
  \vec X_\lambda(\im\nu) = \vec X_0(\im\nu) +
    \vec X_0(\im\nu) \left(
      \lambda\vec V +
      \vec F^\lambda_{\rm xc}(\im\nu)
    \right)
    \vec X_\lambda(\im\nu)\,.
  \label{ref:RPAChiDysonMatrix}
\end{equation}
In the Random Phase Approximation one considers only the Hartree term in the
time dependent density functional theory derivation of the above Dyson-like
equation, neglecting the exchange correlation kernel $\vec F^\lambda_{\rm xc}$.
The polarizability can then be expanded to
\[
  \vec X_\lambda(\im\nu) = \vec X_0(\im\nu)
    + \lambda\vec X_0(\im\nu)\vec V\vec X_0(\im\nu)
    + \lambda^2\vec X_0(\im\nu)\vec V\vec X_0(\im\nu)\vec V\vec X_0(\im\nu)
    + \ldots
\]
Writing (\ref{eqn:RPAACFDT}) in the matrix notation and inserting the
above expansion gives
\begin{align}
  \nonumber
  E_{\rm c} &=\,
  -\frac12\int_0^1\d\lambda
    \int\frac{\d\nu}{2\pi}\,
    \Tr\big\{
      \vec X_\lambda(\im\nu)\vec V -
      \vec X_0(\im\nu)\vec V
    \big\} \\
  &=\,
  -\frac12\int_0^1\d\lambda
    \int\frac{\d\nu}{2\pi}\,
    \Tr\left\{
      \lambda\big(\vec X_0(\im\nu)\vec V\big)^2 +
      \lambda^2\big(\vec X_0(\im\nu)\vec V\big)^3 + \ldots
    \right\}\,.
\end{align}
The explicitly known $\lambda$ dependency allows us
to integrate $\lambda$ analytically, which leads to the final expression
of the correlation energy in the Random Phase Approximation:
\begin{align}
  \nonumber
  E_{\rm c}^{\rm RPA} &=\,
  -\frac12
    \int\frac{\d\nu}{2\pi}\,
    \Tr\left\{
      \frac12\big(\vec X_0(\im\nu)\vec V\big)^2 +
      \frac13\big(\vec X_0(\im\nu)\vec V\big)^3 + \ldots
    \right\} \\[1ex]
  &=\,
    +\frac12
    \int\frac{\d\nu}{2\pi}\,
    \Tr\Big\{
      \log\Big(\vec 1-\vec X_0(\im\nu)\vec V\Big)+\vec X_0(\im\nu)\vec V
    \Big\}\,.
  \label{eqn:RPAACFDTFinal}
\end{align}

This result is formally equivalent to (\ref{eqn:RPAImag}) although two
very different approaches have been used in the respective derivations.
The prescribed Hartree $E_{\rm H}$ and exact-exchange term $E_{\rm x}$ also
correspond to the two first order diagrams in many-body perturbation theory
using a non-Hartree-Fock reference.
Note however, that the factors in the series expansion
\[
      \frac12\big(\vec X_0(\im\nu)\vec V\big)^2 +
      \frac13\big(\vec X_0(\im\nu)\vec V\big)^3 + \ldots
\]
have entirely different reasons in the two approaches. In the approach using
many-body perturbation theory they originate from the rotational symmetry of
the ring diagrams. In the Adiabatic Connection, on the other side, they stem
from averaging the potential energy over all $\lambda$ to arrive at the total
energy. Thus, the two approaches can very well disagree when different classes
of diagrams are considered.


    \section{RPA for the uniform electron gas}
\label{sec:RPAUEG}
In this section we apply the framework of many-body perturbation theory
as discussed in Chapter \ref{cha:MBPT} to the uniform electron gas (UEG) and
and use resulting propagators to calculate the Random Phase Approximation for
this system.

We choose the free Hamiltonian as a reference only containing the kinetic
energy
\[
  \hat H_0 = -\sum_{n=1}^N \frac{\nabla_n^2}2
\]
for a system with $N$ electrons in a cubic box of length $a$ and volume
$\Omega=a^3$. The solutions of the non-interacting Hamiltonian are plane
waves commensurate with the box. The normalized spin orbitals and eigenenergies
are
\begin{equation}
  \psi_{\sigma\vec k}(\alpha\vec r) =
    \delta_{\sigma\alpha}
    \frac1{\sqrt\Omega} e^{\im\vec k\cdot\vec r}\,,
  \qquad
  \eps_{\sigma\vec k} = \frac{\vec k^2}2\,,
  \qquad
  \textnormal{where }
  \vec k\in \frac{2\pi}a\mathbb{Z}^3
  \textnormal{, and }
  \sigma,\alpha\in\{\uparrow,\downarrow\}\,.
  \label{eqn:UEGSpinOrbitals}
\end{equation}
The uniform electron gas is the limit of this system taking its volume
$\Omega$ to infinity while keeping the electron density $N/\Omega$
fixed. The average volume per electron is usually given by specifying the
the radius $r_s$ of a sphere of equal volume
\begin{equation}
  \Omega/N = \frac{4\pi r_s^3}3\,.
  \label{eqn:WignerSeitz}
\end{equation}
$r_s$ is called \emph{Wigner--Seitz radius}.\index{Wigner-Seitz radius}
In the non-interacting ground state the lowest $N$ spin-orbitals are occupied
and the wave numbers $\vec k$ of these states lie inside a sphere
called \emph{Fermi sphere}\index{Fermi sphere} with radius $k_{\rm F}$:
\begin{equation}
  \sum_{\sigma,|\vec k| < k_{\rm F}} = N\,.
  \label{eqn:KFermiCondition}
\end{equation}
The density of the wave vectors $\vec k\in2\pi\mathbb Z^3/a$ increases with
increasing box volume so we can approximate the above sum by an appropriate
integral
\begin{equation}
  \sum_{\sigma,\vec k} =
  \sum_\sigma \int\frac\Omega{(2\pi)^3}\,\d\vec k
  \label{eqn:UEGSumOverStates}
\end{equation}
and use it together with (\ref{eqn:KFermiCondition}) and (\ref{eqn:WignerSeitz})
to find $k_{\rm F}$ as a function of the Wigner--Seitz radius $r_s$:
\begin{equation}
  k_{\rm F}^3 = \frac{9\pi}{2\sum_\sigma}\, \frac1{r_s^3}\,.
  \label{eqn:RPA_UegKfOfRs}
\end{equation}
In the non-spin polarized case $\sum_\sigma=2$ and in the spin polarized
case $\sum_\sigma=1$.

\subsection{Propagators}
We can now evaluate the
propagators for the uniform electron gas according to (\ref{eqn:Greens}):
\[
  G_0(\alpha\vec rt,\alpha'\vec r't') =
  \left\{
  \begin{array}{ll}
    \displaystyle
    -\sum_{\sigma,|\vec k|<k_{\rm F}}
      \psi_{\sigma\vec k}(\alpha\vec r)
      \psi_{\sigma\vec k}^\ast(\alpha'\vec r')
      e^{(-\im\eps_{\sigma\vec k} + \eta)(t-t')} &
    \textnormal{for }t\leq t'\\
    \displaystyle
    +\sum_{\sigma,|\vec k|\geq k_{\rm F}}
      \psi_{\sigma\vec k}(\alpha\vec r)
      \psi_{\sigma\vec k}^\ast(\alpha'\vec r')
      e^{(-\im\eps_{\sigma\vec k} - \eta)(t-t')} &
    \textnormal{otherwise.}
  \end{array}
  \right.
\]
Inserting the spin orbitals and eigenenergies from (\ref{eqn:UEGSpinOrbitals})
and approximating the sum over states by the appropriate integral
from (\ref{eqn:UEGSumOverStates}) gives
\[
  G_0(\alpha\vec rt,\alpha'\vec r't') =
  \left\{
  \begin{array}{ll}
    \displaystyle
    -\int\limits_{|\vec k'|<k_{\rm F}} \frac{\d\vec k'}{(2\pi)^3}\,
      \delta_{\alpha\alpha'}\,e^{\im\vec k'\cdot(\vec r-\vec r')}
      e^{(-\im\vec k'^2/2 + \eta)(t-t')} &
    \textnormal{for }t\leq t'\\
    \displaystyle
    +\int\limits_{|\vec k'|\geq k_{\rm F}} \frac{\d\vec k'}{(2\pi)^3}\,
      \delta_{\alpha\alpha'}\,e^{\im\vec k'\cdot(\vec r-\vec r')}
      e^{(-\im\vec k'^2/2 - \eta)(t-t')} &
    \textnormal{otherwise,}
  \end{array}
  \right.
\]
which we can transform into momentum space $\vec k$ with respect to
$(\vec r-\vec r')$. For the case $t\leq t'$ this yields
\begin{align}
  \nonumber
    &\
      -\int\d(\vec r-\vec r') e^{-\im\vec k\cdot(\vec r-\vec r')}
      \int\limits_{|\vec k'|<k_{\rm F}} \frac{\d\vec k'}{(2\pi)^3}\,
      \delta_{\alpha\alpha'}\,e^{\im\vec k'\cdot(\vec r-\vec r')}
      e^{(-\im\vec k'^2/2 + \eta)(t-t')} \\
  \nonumber
  =&\
      -\int\limits_{|\vec k'|<k_{\rm F}}
      \frac{\d\vec k'}{(2\pi)^3}\,
      (2\pi)^3\,\delta(\vec k'-\vec k)\,
      \delta_{\alpha\alpha'}\,
      e^{(-\im\vec k'^2/2 + \eta)(t-t')} \\
  \nonumber
  =&\
      -\theta(k_{\rm F}-|\vec k|)\,\delta_{\alpha\alpha'}\,
      e^{(-\im\vec k^2/2 + \eta)(t-t')}\,,
\end{align}
where $\theta(x)$ is the Heaviside step function. In the general case this gives
\begin{equation}
  G_0(\vec k, t-t') =
  \left\{
  \begin{array}{ll}
    \displaystyle
      -\theta(k_{\rm F}-|\vec k|)\,
      e^{(-\im\vec k^2/2 + \eta)(t-t')} &
    \textnormal{for }t\leq t'\\[1ex]
    \displaystyle
      +\theta(|\vec k|-k_{\rm F})\,
      e^{(-\im\vec k^2/2 - \eta)(t-t')} &
    \textnormal{otherwise,}\\
  \end{array}
  \right.
\end{equation}
as a function of $\vec k$ and the time difference $(t-t')$,
omitting the electron spins.
Finally, we can transform the propagator into frequency space $\omega$
with respect to $(t-t')$. For the case $t\leq t'$ this gives
\begin{align}
  \nonumber
    &\
      -\int_{-\infty}^0\d(t-t')\,e^{+\im\omega(t-t')}\,
      \theta(k_{\rm F}-|\vec k|)\,
      e^{(-\im\vec k^2/2 + \eta)(t-t')} \\
  \nonumber
    =&\
      -\frac{(1-0)\,\theta(k_{\rm F}-|\vec k|)}{\im(\omega-\vec k^2/2) + \eta}
    \ =\
      \frac{\im\,\theta(k_{\rm F}-|\vec k|)}{\omega-\vec k^2/2-\im\eta}\,.
\end{align}
Note that, unlike for the momentum, we use $e^{+\im\omega(t-t')}$ for the
forward Fourier transform into frequency space so that the poles of
the propagator coincide with the positive eigenenergies.
The particle/hole propagator is now quite compact, even for the general case:
\begin{equation}
  G_0(\vec k,\omega) = 
    \frac\im{\omega-\vec k^2/2+\im\eta_{\vec k}}\,,
  \qquad
  \textnormal{where }
  \eta_{\vec k} =
    \left\{
    \begin{array}{ll}
      -\eta & \textnormal{for }|\vec k|<k_{\rm F} \\
      +\eta & \textnormal{otherwise.}
    \end{array}
    \right.
  \label{eqn:UEGRealPropagator}
\end{equation}
In the space and time domain we have to integrate over all spins,
space and time coordinates of each vertex:
\[
  \sum_\alpha\int\limits_\Omega\d\vec r\int \d t\,.
\]
In the momentum and frequency domain we integrate over all spins,
momenta and frequencies of all propagators connecting the vertices:
\begin{equation}
  \sum_\alpha\int\frac{\Omega}{(2\pi)^3}\,\d\vec k\int\frac1{2\pi}\,\d \omega\,.
\end{equation}
Next, we also want to transform the propagator for the Coulomb interaction
\[
  V(\alpha\vec r t,\alpha'\vec r't') =
    \frac{-\im}{|\vec r-\vec r'|}\,\delta(t-t')
\]
in the momentum and frequency domain. The transform with respect to
$(t-t')$ is straight forward and simply gives a frequency independent
propagator
\[
  V(\vec r-\vec r',\omega) = \frac{-\im}{|\vec r-\vec r'|}\,,
\]
omitting the spin coordinates. The transform into momentum space
does not converge, however, the propagator for the
Yukawa potential can be expressed by a Fourier integral
for any parameter $m>0$:
\[
  \frac{-\im e^{-m|\vec r-\vec r'|}}{|\vec r-\vec r'|} =
    \int\frac1{(2\pi)^3}\,\d \vec q\,e^{\im\vec q\cdot(\vec r-\vec r')}\,
    \frac{-4\pi\im}{\vec q^2 + m^2}
  =
    \int\frac\Omega{(2\pi)^3}\,\d \vec q\,e^{\im\vec q\cdot(\vec r-\vec r')}\,
    \frac{-4\pi\im}{\Omega(\vec q^2 + m^2)}\,.
\]
Taking the limit $m\rightarrow0$ gives the propagator for the Coulomb
interaction in momentum space:
\begin{equation}
  V(\vec q, \omega) = -\,\frac{4\pi\im}{\Omega\vec q^2}\,.
\end{equation}
Note that we use $\vec k$ for particle/hole propagators while
we use $\vec q$ for Coulomb propagators. The spin coordinates
remain omitted for brevity as the behavior of spin is intuitive in the UEG.
A particle or a hole does not change spin during propagation while the
Coulomb interaction mediates between particles or holes irrespective of the
their spin.

\subsection{Imaginary frequencies}
\begin{figure}
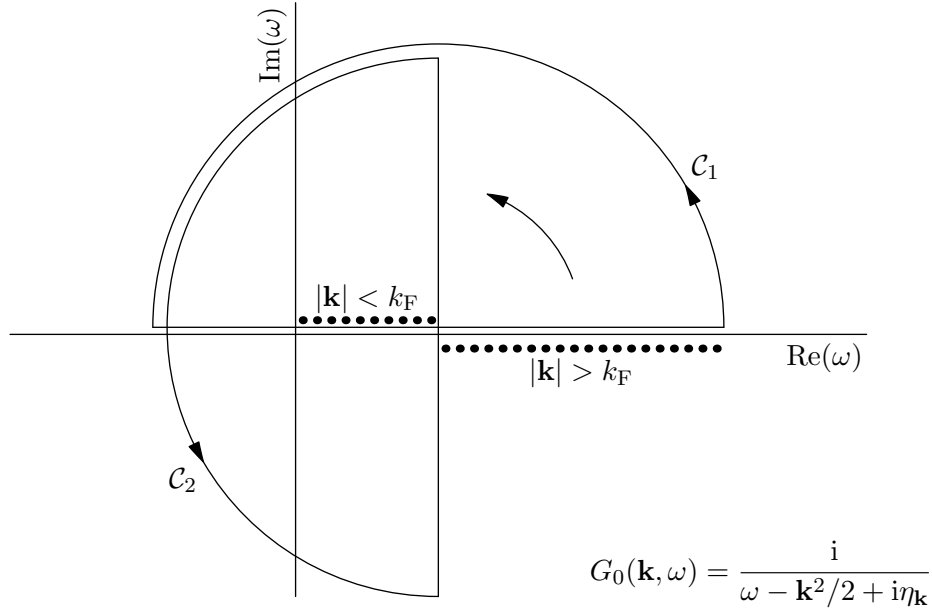

\begin{center}
  \begin{asy}
    import graph;
    size(320,0);

    real d=0.1;
    pair c=(1,d/2);
    real rr=2,ir=1.9;
    real ra=30,ia=210;
    for (real x=0.05; x<0.99; x+=0.1) {
      dot((x,+d),black);
    }
    for (real x=1.05; x<2.99; x+=0.1) {
      dot((x,-d),black);
    }
    draw(arc(c,rr,0,ra),black,Arrow);   draw(arc(c,rr,ra,180),black);
    draw((c-(rr,0))--(c+(rr,0)));
    draw(arc(c,ir,90,ia),black,Arrow); draw(arc(c,ir,ia,270));
    draw((c-(0,ir))--(c+(0,ir)));
    draw(arc(c,1,20,70),black,Arrow);
    xaxis(Label("${\rm Re}(\omega)$"),-2,4);
    yaxis("${\rm Im}(\omega)$");
    label("$|\vec k|>k_{\rm F}$",(2,-d),S);
    label("$|\vec k|<k_{\rm F}$",(0.5,d),N);
    label("$\mathcal C_1$", c+rr*(cos(radians(ra)),sin(radians(ra))),NE);
    label("$\mathcal C_2$", c+ir*(cos(radians(ia)),sin(radians(ia))),SW);
    label(
      "$\displaystyle G_0(\vec k,\omega)=\frac{\im}{\omega-\vec k^2/2+\im\eta_{\vec k}}$",
      (2,-1.5),E
    );
  \end{asy}
\end{center}
\caption{
  $G_0(\omega)$ has one pole above the real axis for $|\vec k|<k_{\rm F}$ or
  below the real axis for $|\vec k|>k_{\rm F}$. We can smoothly rotate
  the integration contour from the real axis to a contour parallel to the
  imaginary axis without crossing any pole.
  The arcs indicate the contour at infinity.
}
\label{fig:PropagatorPole}
\end{figure}
$G_0$ has one pole on the positive real frequency axis. The pole is
infinitesimally below or above the real axis depending on whether it is
a particle, having $|\vec k|>k_{\rm F}$, or a hole.
On which side of the real axis the pole lies determines whether there
is forwards or backwards propagation in time.
Figure \ref{fig:PropagatorPole} shows the pole of $G_0(\vec k,\omega)$ as
a function of $\omega$ with $\vec k$ as parameter.
The poles of the propagators make a numeric frequency integration
along the real axis difficult. We can, however, smoothly deform the integration
contour as long as we do not cross any poles. As indicated in Figure
\ref{fig:PropagatorPole} we can rotate the contour $\mathcal C_1$
counterclockwise around the Fermi energy $\mu=k_{\rm F}^2/2$ to arrive at
the contour $\mathcal C_2$.
Substituting $\omega=\mu+\im\nu$ we get
\[
  \int_{-\infty}^\infty \d\omega = \oint\limits_{\mathcal C_1} \d\omega =
  \oint\limits_{\mathcal C_2} \d\omega = \int_{-\infty}^\infty \im\,\d\nu\,,
\]
where we assume that the contribution of a contour at infinity vanishes.
This rotation of the integral contour is called \emph{Wick} rotation
to \emph{imaginary frequency}.\index{Wick rotation}\index{imaginary frequency}
When integrating in imaginary frequency the side of the poles relative to the
contour
no longer depend on $\eta$ and we can as well take the limit $\eta\rightarrow0$
before evaluating the imaginary frequency integrals. The particle/hole
propagator in imaginary frequency is then given by
\begin{equation}
  G_0(\vec k,\im\nu)=\frac{\im}{\im\nu-(\vec k^2/2-\mu)}=
  \frac1{\nu+\im\Delta\eps_{\vec k}}\,,
  \quad
  \textnormal{where }\Delta\eps_{\vec k}=\vec k^2/2-k_{\rm F}^2/2\,.
  \label{eqn:RPAUEGPropagatorImag}
\end{equation}

\subsection{Polarizability}
Next, we can evaluate the independent particle polarizability.
The polarizability diagram has two open vertices. Let $\vec q$ and $\nu$
denote the momentum and the imaginary frequency incident at the lower vertex
shown in the diagram below.
From the momentum and frequency conservation at
every vertex follows that there is one pair of momentum $\vec k$ and
frequency $\eta$ to integrate over and that the outgoing momentum and
frequency is equal to the incoming ones.
In the context of this diagram part
$\vec k$ and $\eta$ are called \emph{internal}\index{internal} momentum and frequency,
respectively, while $\vec q$ and $\nu$ are connected to other diagram parts and
are therefore referred to as \emph{external}\index{external} momentum and frequency,
respectively. We evaluate the diagram part by integrating over all internal
momenta and frequencies starting with the analytic $\eta$ integration:
\begin{align}
  \nonumber
  \diagramBoxBorder{2ex}{3ex}{
    \begin{fmffile}{PolarizationLabled}
    \begin{fmfgraph*}(40,40)
      \fmfstraight
      \fmfset{arrow_len}{6}
      \fmfbottom{v00,v10,v20}
      \fmftop{v01,v11,v21}
      \fmf{fermion,left=0.3,label=$
        \begin{array}{l}\vec k+\vec q,,\\\eta+\nu\end{array}
      $,label.side=left,label.dist=2}{v10,v11}
      \fmf{fermion,left=0.3,label=$
        \begin{array}{l}\vec k,,\\\eta\end{array}
      $,label.side=left,label.dist=2}{v11,v10}
      \fmf{boson,label=$\vec q,,\nu$,label.side=right,label.dist=6}{v00,v10}
      \fmf{boson,label=$\vec q,,\nu$,label.side=left,label.dist=6}{v11,v21}
    \end{fmfgraph*}
    \end{fmffile}
  }
  \quad
  \chi_0(\vec q,\im\nu) =\,
   -\sum_\sigma\int\frac{\Omega\,\d\vec k}{(2\pi)^3}\,&
    \int\frac{\im\,\d\eta}{2\pi}\,
    G_0(\vec k+\vec q, \im\eta+\im\nu)\,G_0(\vec k,\im\eta)\\[-3ex]
  \nonumber
  =\,
   -\sum_\sigma\int\frac{\Omega\,\d\vec k}{(2\pi)^3}\,&
    \int\frac{\im\,\d\eta}{2\pi}\,
    \frac1{\eta+\nu+\im\Delta\eps_{\vec k+\vec q}}\,
    \frac1{\eta+\im\Delta\eps_{\vec k}} \\
  \nonumber
  =\,
   -\im\,\sum_\sigma\int\frac{\Omega\,\d\vec k}{(2\pi)^3}\,&
    \Bigg(
    \frac{\theta(|\vec k+\vec q|-k_{\rm F})\theta(k_{\rm F}-|\vec k|)}{
      \eps_{\vec k+\vec q} - \eps_{\vec k} - \im\nu
    }
    \\
    &
    +
    \frac{\theta(|\vec k|-k_{\rm F})\theta(k_{\rm F}-|\vec k+\vec q|)}{
      \eps_{\vec k} - \eps_{\vec k+\vec q} + \im\nu
    }
    \Bigg)
  \label{eqn:RPAChiFreqInt}
\end{align}
The conditions on $\vec k$ expressed by the Heaviside theta functions
arise from the necessity to have one pole on either
side of the integration contour for a non-vanishing result. The conditions
in (\ref{eqn:RPAChiFreqInt}) can be unified by substituting
$-\vec k'=\vec k+\vec q$ in the second sum. This gives
\begin{equation}
  \chi_0(\vec q,\im\nu) =\,
    -\im\,\sum_\sigma\int\limits_{|\vec k|<k_{\rm F}<|\vec k+\vec q|}
      \frac{\Omega\,\d\vec k}{(2\pi)^3}\,
    \Bigg(
    \frac1{
      \eps_{\vec k+\vec q} - \eps_{\vec k} - \im\nu
    }
    +
    \frac1{
      \eps_{\vec k+\vec q} - \eps_{\vec k} + \im\nu
    }
    \Bigg)
  \label{eqn:RPAChiUEG}
\end{equation}
With effort this expression can also be integrated analytically with respect
to the momentum $\vec k$ and expressed in terms of a real valued function
$R(q,u)$ where the Fermi momentum and the Fermi energy are normalized to 1 and
1/2, respectively \parencite{ziesche_high-density_2010}:
\begin{align}
  \chi_0(\vec q,\im\nu) \,=&
    -\im\,\frac{k_{\rm F}}{\pi^2}\, R\left(
      \frac{|\vec q|}{k_{\rm F}}, \frac\nu{|\vec q|k_{\rm F}}
   \right)
   \\[2ex]
\label{eqn:RPA_Chi0}
  2R(q,u) \,=&
        1 - u\left(
          \arctan\left(\frac{1+q/2}u\right)+\arctan\left(\frac{1-q/2}u\right)
        \right)\\
    &
        + \frac{1+u^2-q^2/4}{2q}\,\log\left(
            \frac{u^2+(q/2+1)^2}{u^2+(q/2-1)^2}
          \right) 
\label{eqn:RPA_Chi0R}
\end{align}
Note that the imaginary units are often traded between the Coulomb kernel
$V(\vec q)$ and the polarizability $\chi_0(\vec q,\im\nu)$ since only their
product occurs in the expression for the RPA correlation energy.
Throughout this work, except Section \ref{sec:RPAAC}, imaginary units are used
exactly as they emerge from Wick rotated imaginary frequency integrations:
\begin{align}
  V(\vec q) &= -\frac{4\pi\im}{\Omega\,\vec q^2} &
  \chi(\vec q,\im\nu) &= -\im\,\sum_{\sigma\vec k} \left(\ldots\right)
  &&\textnormal{used in this work,}\\
  V(\vec q) &= \frac{4\pi}{\Omega\,\vec q^2} &
  \chi(\vec q,\im\nu) &= -\,\sum_{\sigma\vec k} (\ldots)
  &&\textnormal{also common in literature.}
  \label{eqn:RPAChiDifference}
\end{align}

\subsection{Correlation energy}
Finally, we can evaluate the correlation energy in the Random Phase
Approximation for the uniform electron gas according to (\ref{eqn:RPAImag}).
In the space domain we needed to convolve the propagators connected in series
which we denoted by the matrix product. In the momentum domain of a homogeneous
system this simplifies to a product. The RPA energy per electron in the uniform
electron gas is thus
\begin{align}
  \nonumber
  E_{\rm c}^{\rm RPA}/N \,=& \frac\Omega N\,\frac12 \int\frac{\d\vec q}{(2\pi)^3}\,
    \int_{-\infty}^\infty\frac{\d\nu}{2\pi}\,
    \Big\{
      \log\Big(1-\chi_0(\vec q,\im\nu)V(\vec q)\Big)+\chi_0(\vec q\im\nu)V(\vec q)
    \Big\}
    \\
   =&\, \frac{4\pi r_s^3}3\,
   \int_0^\infty\frac{4\pi q^2\d q}{(2\pi)^3}\,
   \int_0^\infty\frac{\d\nu}{2\pi}\,
    \Big\{
      \log\Big(1-\chi_0(q,\im\nu)V(q)\Big)+\chi_0(q,\im\nu)V(q)
    \Big\}
  \label{eqn:RPA_UEG}
\end{align}
Note that the sum over all momenta $\vec q$ of the Coulomb propagator does not
include a sum over spins. The given integral can been evaluated using a
Gauss--Kronrod rule first in $q$ then in $\nu$ to yield the correlation
energy in the Random Phase Approximation to 5 significant digits of precision
for the non-spin polarized, paramagnetic case with $\sum_\sigma=2$ and for the
spin polarized, ferromagnetic case
with $\sum_\sigma=1$. The results are listed in Table \ref{tab:RPAUEG}
for different densities and they include Quantum Monte Carlo results for
comparison. Although the RPA systematically overestimates the correlation
energy it lacks only about 1/3 of the correlation energy at $r_s=1$.
\parencite{gell-mann_correlation_1957}
have shown that in the limit of $r_s\rightarrow0$ the
entire correlation energy is contained in the RPA ring diagrams. For low
densities, however, the relative error becomes larger and other diagrams
become important.
\begin{table}
\begin{center}
\begin{tabular}{|c|rr|r||c|rr|r|}
\hline
  \multicolumn{4}{|c||}{paramagnetic} & \multicolumn{4}{c|}{ferromagnetic} \\
\hline
&&&&&&& \\[-2.5ex]
  $r_s$ & \multicolumn{2}{c|}{$E_{\rm c}^{\rm RPA}$} &
    \multicolumn{1}{c||}{$E_{\rm c}$} &
  $r_s$ & \multicolumn{2}{c|}{$E_{\rm c}^{\rm RPA}$} &
    \multicolumn{1}{c|}{$E_{\rm c}$} \\[0ex]
  [a.u.] & [m$E_h\,N$] & \multicolumn{1}{c|}{$\pm$} &
    [m$E_h\,N$] &
  [a.u.] & [m$E_h\,N$] & \multicolumn{1}{c|}{$\pm$} &
    [m$E_h\,N$] \\
&&&&&&& \\[-2.6ex]
\hline
&&&&&&& \\[-2.5ex]
 1 & -78.799 & 0.001 & -59.632  &   1 & -51.893 & 0.002 & -31.701 \\
 2 & -61.801 & 0.001 & -45.091  &   2 & -42.416 & 0.001 & -24.090 \\
 3 & -52.759 & $<$0.001 & -37.214  &   3 & -37.179 & 0.001 & -20.048 \\
 4 & -46.806 & $<$0.001 & -32.054  &   4 & -33.633 & $<$0.001 & -17.415 \\
 5 & -42.470 & $<$0.001 & -28.339  &   5 & -30.992 & $<$0.001 & -15.520 \\
 6 & -39.117 & $<$0.001 & -25.504  &   6 & -28.911 & $<$0.001 & -14.071 \\
 7 & -36.418 & $<$0.001 & -23.253  &   7 & -27.209 & $<$0.001 & -12.916 \\
 8 & -34.182 & $<$0.001 & -21.414  &   8 & -25.778 & $<$0.001 & -11.969 \\
 9 & -32.289 & $<$0.001 & -19.876  &   9 & -24.551 & $<$0.001 & -11.174 \\
10 & -30.658 & $<$0.001 & -18.568  &  10 & -23.482 & $<$0.001 & -10.495 \\
12 & -27.975 & $<$0.001 & -16.454  &  12 & -21.698 & $<$0.001 & -9.391 \\
15 & -24.929 & $<$0.001 & -14.119  &  15 & -19.629 & $<$0.001 & -8.160 \\
20 & -21.381 & $<$0.001 & -11.497  &  20 & -17.156 & $<$0.001 & -6.758 \\
30 & -17.068 & $<$0.001 & -8.486  &  30 & -14.044 & $<$0.001 & -5.112 \\
40 & -14.463 & $<$0.001 & -6.778  &  40 & -12.099 & $<$0.001 & -4.156 \\
50 & -12.680 & $<$0.001 & -5.666  &  50 & -10.736 & $<$0.001 & -3.521 \\
\hline
\end{tabular}
\end{center}
\caption{
  Correlation energy in the Random Phase Approximation for the uniform
  electron gas at different densities compared to Quantum Monte-Carlo
  (QMC) results obtained by \parencite{ceperley_ground_1980} fitted
  by \parencite{perdew_self-interaction_1981}.
  For RPA energies at intermediate spin polarizations see
  \parencite{vosko_accurate_1980}.
}
\label{tab:RPAUEG}
\end{table}

  \subsection{Large momentum behavior}
\label{ssc:RPA_Ueg_LargeMomentum}
In case of the uniform electron gas the polarizability can be evaluated for
an arbitrary magnitude of the momentum transfer $q$ allowing for an accurate
numerical integration. For solids or molecules the finite resolution used
for finding the Hartree-Fock or DFT oribitals imposes however an upper
limit $G_{\rm max}$ up to where $\chi_0$ can be evaluated. Therefore, we need
to know the asymptotic behavior of the RPA correlation energy for sufficiently
large values of $G_{\rm max}$ in order to extrapolate numerical results to
$G_{\rm max} \rightarrow \infty$. Assuming that the system is sufficiently
homogeneous at the resolution corresponding to a given $G_{\rm max}$, the
asymptotic behavior of the RPA in a solid or in a molecule is the same as
in the uniform electron gas. The latter can be derived analytically and will
be outlined here.

Given the expression for the independent particle polarizability $\chi_0$
of the uniform electron gas from (\ref{eqn:RPAChiUEG})
\begin{align}
  \nonumber
  \chi_0(\vec q,\im\nu) &=\,
    -\im\,\sum_\sigma\int\limits_{|\vec k|<k_{\rm F}<|\vec k+\vec q|}
      \frac{\Omega\,\d\vec k}{(2\pi)^3}\,
    \Bigg(
    \frac1{
      \eps_{\vec k+\vec q} - \eps_{\vec k} - \im\nu
    }
    +
    \frac1{
      \eps_{\vec k+\vec q} - \eps_{\vec k} + \im\nu
    }
    \Bigg)
  \\
  &=\,
    -\im\,\sum_\sigma\int\limits_{|\vec k|<k_{\rm F}<|\vec k+\vec q|}
      \frac{\Omega\,\d\vec k}{(2\pi)^3}\,
    \frac {\Delta\eps_{\vec k,\vec q}}{
      \Delta\eps_{\vec k,\vec q}^2 + \nu^2
    }
  \,,
  \label{eqn:RPA_ChiCollected}
\end{align}
where
$
  \Delta\eps_{\vec k,\vec q} = \eps_{\vec k+\vec q}-\eps_{\vec k} =
  \vec k\cdot\vec q + \vec q^2/2
$, we can trivially integrate out $\vec k$ for large enough magnitudes of
$\vec q$ since $|\vec k|<k_{\rm F}\ll |\vec q|$. This yields
\begin{equation}
  \chi_0(q, \im \nu) \sim
    -\im\,k_{\rm F}^3\,\frac {q^2} {
      \left(q^2/2\right)^2 + \nu^2
    }
    \,.
  \label{eqn:RPA_ChiLargeQ}
\end{equation}
Next, we can insert this approximation into the RPA energy expression
(\ref{eqn:RPA_UEG})
for a given, large $q$ and integrate out the imaginary frequency
$\nu$, getting
\begin{align}
  \nonumber
  \int_0^\infty\frac{\d\nu}{2\pi}\,
    \Big\{
      \log\Big(1-\chi_0(q,\im\nu)V(q)\Big)+\chi_0(q,\im\nu)V(q)
    \Big\}
  &\sim
  \frac{
    \left(q^2/2\right)\sqrt{\left(q^2/2\right)^2+1}
    -\left(q^2/2\right)^2-1/2
  }{q^2}
  \\
  \nonumber
  &= -\frac1{q^6} + \frac1{q^{10}} + \mathcal O\left(\frac1{q^{14}}\right)
  \,,
\end{align}
where we expanded in the variable $u=1/q$ at $u=0$. The leading order term
can finally be used to estimate the missing RPA energy per electron
$\Delta E_{\rm c}^{\rm RPA}(G_{\rm max})$ when truncating the momentum integration
at a finite momentum $G_{\rm max}$:
\begin{equation}
  \Delta E_{\rm c}^{\rm RPA}(G_{\rm max}) \sim
  \int_{G_{\rm max}}^\infty\d q\,q^2\,\left(\frac1{q^6} +
    \mathcal O\left(\frac1{q^{10}}\right)
  \right)
  \sim \frac1{G_{\rm max}^3} + \mathcal O\left(\frac1{G_{\rm max}^7}\right)\,.
  \label{eqn:RPA_RpaLargeQ}
\end{equation}

  \subsection{Large imaginary frequency behavior}
\begin{figure}
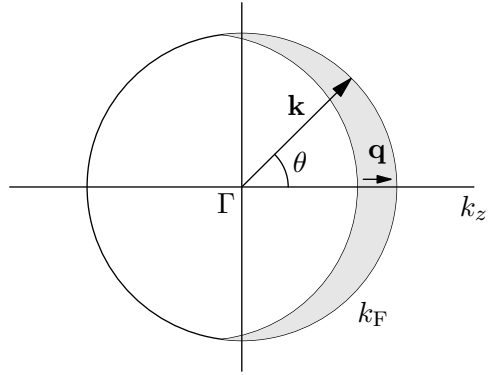

\begin{center}
\begin{asy}
    import graph;
    size(180,0);

    real dy=0.05;
    real dx=0.035;
    real q=0.25;
    real z=-q/2;
    real x=sqrt(1-z^2);
    pair gamma=(0,0);
    draw(circle(gamma,1),black);
    draw(arc((-q,0),(z,-x),(z,+x)),black);
    fill(arc(gamma,(z,-x),(z,+x))--arc((-q,0),(z,+x),(z,-x),CW)--cycle, lightgrey);
    draw(Label("$\vec q$",align=2N),(1-q+dx,dy)--(1-dx,dy),EndArrow);
    draw(Label("$\vec k$",align=2N),gamma--0.5*(sqrt(2),sqrt(2)),EndArrow);
    draw(Label("$\theta$"), arc(gamma,0.3,0,45));
    label("$\Gamma$",gamma,SW);
    label("$k_{\rm F}$",(1/sqrt(2),-1/sqrt(2)),SE);
    draw(Label("$k_z$",1,align=S), (-1.5,0)--(1.5,0));
    draw((0,-1.2)--(0,1.2));
\end{asy}
\end{center}
\caption{
  Cross section of the set of momenta $\vec k$ such that
  $|\vec k|<k_{\rm F}<|\vec k+\vec q|$ for small magnitudes of the excitation
  momentum $\vec q$. In this limit, the set covers half the surface of the
  Fermi sphere and its maximum thickness is $|\vec q|$.
}
\label{fig:RPA_SmallQ}
\end{figure}
Although most implementations of the RPA can evaluate the independent
particle polarizability $\chi_0$ at arbitrary imaginary frequencies, knowledge
of the asymptotic behavior of the RPA energy for large frequencies
$\nu$ is useful for choosing an appropriate variable transform for integrating
the tail.
This is relevant for metallic systems, where we can assume that the system
behaves like a Uniform Electron Gas at times short enough.

Unlike in the case of large momenta $q$, $\vec k$ cannot
be trivially integrated out. However, we can separate the momentum integration
of $q$ into three regions and show that they all come to an analogous form
of the integrand for that momentum integration. In Subsection
\ref{ssc:RPA_Ueg_LargeMomentum} we already have gotten an approximation
of $\chi_0(q,\im\nu)$ for large $q$ in (\ref{eqn:RPA_ChiLargeQ}).
For small $q$, the volume of the momenta $\vec k$, such that
$|\vec k|<k_{\rm F}<|\vec k+\vec q|$, is proportional to $q$,
as shown in Figure \ref{fig:RPA_SmallQ}.
Therefore. the integral in (\ref{eqn:RPA_ChiCollected}) transforms as follows
\begin{align}
  \nonumber
    -\im\,\sum_\sigma\int\limits_{|\vec k|<k_{\rm F}<|\vec k+\vec q|}
      \frac{\Omega\,\d\vec k}{(2\pi)^3}\,
    \frac {\Delta\eps_{\vec k,\vec q}}{
      \Delta\eps_{\vec k,\vec q}^2 + \nu^2
    }
  \sim&
  -\im\,4\pi k_{\rm F}^2\,q\int_0^1\d\cos\theta\,\cos\theta\,
    \frac {q^2/2+k_{\rm F}q\cos\theta}{
      \left(q^2/2+k_{\rm F}q\cos\theta\right)^2 + \nu^2
    }
  \\
  \sim&
  -\im\,k_{\rm F}^2\,\frac {
    q^3/4 + k_{\rm F}q^2/6
  }{
    \left(q^2/2\right)^2 + \nu^2
  }
  \sim
  -\im\,k_{\rm F}^3\,\frac {
    q^2 + \mathcal O\left(q^3\right)
  }{
    \left(q^2/2\right)^2 + \nu^2
  }\,,
  \label{eqn:RPA_LargeFreqSmallQ}
\end{align}
where we used that $\Delta\eps_{\vec k,\vec q}=q^2/2+k_{\rm F}q\cos\theta$
and that $\nu$ is large compared to $k_{\rm F}q$. This is the same
behavior of $\chi_0(q,\im\nu)$ as for large momenta $q$.
For intermediate $q$, where the integration volume for $\vec k$ is roughly
independent of $\vec q$, the $\vec k$ integration merely averages the
contributions to $\chi_0$. For large $\nu$ we retrieve
\begin{align}
  \nonumber
    -\im\,\sum_\sigma\int\limits_{|\vec k|<k_{\rm F}<|\vec k+\vec q|}
      \frac{\Omega\,\d\vec k}{(2\pi)^3}\,
    \frac {\Delta\eps_{\vec k,\vec q}}{
      \Delta\eps_{\vec k,\vec q}^2 + \nu^2
    }
  \sim&
  -\im\,\frac{4\pi k_{\rm F}^3}3\,\int_{-1}^1\d\cos\theta\,
    \frac {q^2/2+k_{\rm F}q\cos\theta}{
      \left(q^2/2+k_{\rm F}q\cos\theta\right)^2 + \nu^2
    }
  \\
  \sim&
  -\im\,k_{\rm F}^3\,\frac {
    q^2 + \mathcal O(q)
  }{
    \left(q^2/2\right)^2 + \nu^2
  }\,,
  \label{eqn:RPA_LargeFreqIntermediateQ}
\end{align}
Since the integrand has the same form for large $\nu$ in all cases, we can
use it in the integral over the whole domain of $q$, getting
\begin{multline}
  \nonumber
  \int_0^\infty\frac{4\pi q^2\d q}{(2\pi)^3}\,
    \Big\{
      \log\Big(1-\chi_0(q,\im\nu)V(q)\Big)+\chi_0(q,\im\nu)V(q)
    \Big\}
  \\
  \sim
  \frac{4\sqrt\nu\left(\nu^2+1\right)^{3/4}-4\nu^2-3}{6\sqrt\nu}
  = -\frac1{16}\frac1{\nu^{5/2}} + \frac5{192}\frac1{\nu^{9/2}} +
    \mathcal O\left(\frac1{\nu^{13/2}}\right)\,,
\end{multline}
which is expanded in the variable $u=1/\nu$ at $u=0$.
The other orders in the terms (\ref{eqn:RPA_LargeFreqSmallQ}) and
 (\ref{eqn:RPA_LargeFreqIntermediateQ}) yield a leading order term
of the form $\mathcal O(1/\nu^4)$ for the respective integration region of
$q$ and can therefore be neglected for sufficiently large frequencies $\nu$.
These terms can, however, contribute to the next-to-leading order.
Finally, we can insert this expansion in the imaginary frequency integration
to estimate the missing RPA energy per electron $\Delta E_{\rm c}^{\rm RPA}$
when truncating the frequency integration at some finite but large
$\nu_{\rm max}$:
\begin{equation}
  \Delta E_{\rm c}^{\rm RPA} \sim \int_{\nu_{\rm max}}^\infty\d\nu
  \left(
    \frac1{\nu^{5/2}} + \mathcal O\left(\frac1{\nu^{9/2}}\right)
  \right)
  \sim \frac1{\nu_{\rm max}^{3/2}} +
    \mathcal O\left(\frac1{\nu_{\rm max}^{7/2}}\right)\,.
  \label{eqn:RPA_LargeNu}
\end{equation}

\section*{Summary}
The Random Phase Approximation is the sum of all ring diagrams to infinite
order. Momentum conservation dictates that every Coulomb interaction in a ring
diagram mediates the same momentum giving rise to a $1/q^{2n}$ divergence
in $n$-th order. This is the strongest divergence possible in $n$-th order
rendering the ring diagrams the most important contribution for low momenta,
i.e. at long distances or in the high density regime. The divergence of
each diagram when integrating over low momenta is referred to as
\emph{infrared catastrophe}\index{infrared catastrophe}. Evaluating the sum
over all orders before integrating over the mediated momenta turns the
$1/q^{2n}$ divergence of each order into a $\log(1+1/q^2)$ divergence,
which yields a finite result in the subsequent momentum integration and
solves the infrared catastrophe.

Within the framework of many-body perturbation theory, discussed in
Chapter \ref{cha:MBPT}, the Random Phase Approximation can be readily derived
using the independent particle polarizability
$\diagramBox{\fmfreuse{Polarization}}=\,\chi_0$ as a building block. In the
frequency domain this can be done using Feynman diagrams where the rotational
symmetry of the ring diagrams gives rise to the factors in the expansion
of the RPA energy
\[
  \frac12\left(\diagramBox{\fmfreuse{PolarizationV}}\right)^2 +
  \frac13\left(\diagramBox{\fmfreuse{PolarizationV}}\right)^3 + \ \ldots
\]

The RPA can also be derived in the frequency domain within the Adiabatic
Connection (AC) arriving at a formally equivalent result for very different
reasons. In many-body
perturbation theory the perturbation is slowly introduced to the system
leaving it in its ground state. The sum over all connected diagrams,
respecting their symmetry, yields
the total correlation energy. In the Adiabatic Connection the correlation
energy is retrieved from averaging the potential energy over the coupling
strength $\lambda$:
\[
  \int_0^1\d\lambda\,\left(
    \lambda\left(\diagramBox{\fmfreuse{PolarizationV}}\right)^2 +
    \lambda^2\left(\diagramBox{\fmfreuse{PolarizationV}}\right)^3 +
    \ \ldots
  \right)
\]
In the AC the polarizability $\chi_\lambda$ is the key quantity of interest
rather than connected diagrams. Therefore, there are no symmetries to consider
and connected (closed) diagrams should be avoided for depicting the RPA within
the Adiabatic Connection.
It is important to remark that the AC derivation is tailored to a DFT reference
system assuming an exact density of the reference for the Adiabatic Connection
and an exact density response for the Dyson-like equation of the polarizability.

Finally, the RPA can also be derived in the time domain using Goldstone
diagrams and the direct ring Coupled Cluster Doubles amplitudes
\[
  \diagramBox{
    \begin{fmffile}{CCDt}
    \begin{fmfgraph}(32,16)
      \fmfkeep{CCDt}
      \fmfset{arrow_len}{6}
      \fmfstraight
      \fmfleft{v00,v01}
      \fmfright{v60,v61}
      \fmf{phantom,tension=4}{v00,v10} \fmf{dashes}{v10,v50}
        \fmf{phantom,tension=4}{v50,v60}
      \fmf{phantom}{v01,v21,v41,v61}
      \fmffreeze
      \fmf{fermion,left=0.1}{v21,v10,v01}
      \fmf{fermion,left=0.1}{v61,v50,v41}
    \end{fmfgraph}
    \end{fmffile}
  }
  \hspace*{-2ex}
  \,.
\]
In each iteration of the amplitude equation ring diagrams are added in all
possible ways in a time ordered manner. The simple fact that rings have a left
and a right side when building them bottom to top requires 4 open connections
of this building block according to the left and the right particle/hole pair.
Although this approach is not as efficient as calculating the RPA in the
imaginary frequency domain, it is easy to include a larger set of diagrams
once the amplitudes are found.
After all, Wick's theorem requires all contractions to be considered
not just the ring diagrams.
The additional set of diagrams available given the direct ring Coupled Cluster
Doubles amplitudes is called \emph{Second Order Screened Exchange} diagrams.
They represent the lowest order correction to the Random Phase Approximation.

  \chapter{Second Order Screened Exchange}
\label{cha:SOSEX}
The Random Phase Approximation (RPA) improves considerably on Hartree-Fock
or Density Functional Theory results. It is capable of describing van
der Waals interactions and works well in a large variety of chemical
environments.  
However, it is biased and tends to overestimate the negative correlation
energy. In the
previous chapter we have seen that for the uniform electron gas but this
has also been shown for various solids and molecules.
It is not surprising that the Random Phase Approximation shows an error.
Wick's theorem
states that all contractions should be considered not just those forming
the ring diagrams. That the RPA exhibits a systematic error, however,
indicates a general reason behind this error which can serve
as a guide to the next class of diagrams partially correcting RPA's bias.

In second order the overestimation of the correlation energy is more evident.
According Figure \ref{fig:MP2d} the second order (MP2) direct term is
negative and reads
\[
  (-1)^{(2+2)}\, \frac12
  \sum_{ijab}\frac{V_{ij}^{ab} V_{ab}^{ij}}{\eps_i+\eps_j-\eps_a-\eps_b}\,,
\]
where the Fermion sign is explicitly given depending on the number of loops and
holes according to (\ref{eqn:MBPTGoldstoneSign}).
This term contains contributions in violations of the Pauli exclusion principle,
for instance as illustrated below on the left, where the state $i$
occurs twice at the same instance in time.
Such contributions are canceled
exactly by the respective \emph{exchange diagram}\index{exchange diagram} where
the two offending states are crossed by anti-symmetrizing the affected
interaction as shown on the right. The resulting diagram has one loop less
giving an opposite sign:
\begin{align}
  \nonumber
  \diagramBox{
    \begin{fmffile}{Bubbles2Pauli}
    \begin{fmfgraph*}(50,50)
      \fmfkeep{Bubbles2Pauli}
      \fmfset{arrow_len}{6}
      \fmfleft{v11,v12}
      \fmfright{v21,v22}
      \fmf{photon}{v11,v21}
      \fmf{photon}{v12,v22}
      \fmf{fermion,left=0.3,tension=0,label=$a$,l.dist=3}{v11,v12}
      \fmf{fermion,left=0.3,tension=0,label=$\textcolor{red}{i}$,l.dist=3}
        {v12,v11}
      \fmf{fermion,right=0.3,tension=0,label=$b$,l.dist=3}{v21,v22}
      \fmf{fermion,right=0.3,tension=0,label=$\textcolor{red}{i}$,l.dist=3}
        {v22,v21}
    \end{fmfgraph*}
    \end{fmffile}
  }\
  =& -\left\{
  \diagramBox{
    \begin{fmffile}{Exchange2Pauli}
    \begin{fmfgraph*}(50,50)
      \fmfset{arrow_len}{6}
      \fmfleft{v11,v12}
      \fmfright{v21,v22}
      \fmf{photon}{v11,v21}
      \fmf{photon}{v12,v22}
      \fmf{fermion,left=0.3,label=$a$,l.dist=3,l.side=left}{v11,v12}
      \fmf{fermion,label=$\textcolor{blue}{i}$,l.dist=2,l.side=left}{v12,vm}
      \fmf{plain}{vm,v21}
      \fmf{fermion,right=0.3,label=$b$,l.dist=3}{v21,v22}
      \fmf{fermion,label=$\textcolor{blue}{i}$,l.dist=2}{v22,vm}
      \fmf{plain}{vm,v11}
    \end{fmfgraph*}
    \end{fmffile}
  }
  \ \right\}
  \\[2ex]
  \nonumber
  (-1)^{(2+2)}\, \frac12\,
  \frac{V_{ii}^{ab} V_{ab}^{ii}}{\eps_i+\eps_i-\eps_a-\eps_b}
  =&
  -\left\{(-1)^{(1+2)}\, \frac12\,
  \frac{V_{ii}^{ab} V_{ab}^{ii}}{\eps_i+\eps_i-\eps_a-\eps_b}
  \right\}
\end{align}
As a consequence, violating contributions would be canceled if all
contractions were considered. Ignoring this exchange diagram leaves
the violating negative contributions uncanceled resulting in a negative error.
For the RPA the sign of the ring diagrams alternates with the order. However,
the contributions of each order decay such that the error of the lowest order
dominates, which is negative.

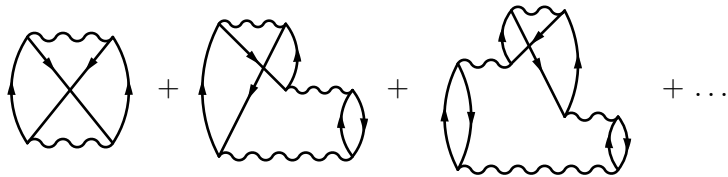
\begin{figure}
\begin{center}
  \diagramBox{
    \begin{fmffile}{Exchange2}
    \begin{fmfgraph}(40,40)
      \fmfkeep{Exchange2}
      \fmfset{arrow_len}{6}
      \fmfleft{v11,v12}
      \fmfright{v21,v22}
      \fmf{photon}{v11,v21}
      \fmf{photon}{v12,v22}
      \fmf{fermion,left=0.3}{v11,v12} \fmf{fermion}{v12,vm} \fmf{plain}{vm,v21}
      \fmf{fermion,right=0.3}{v21,v22} \fmf{fermion}{v22,vm} \fmf{plain}{vm,v11}
    \end{fmfgraph}
    \end{fmffile}
  }
  \hspace*{-2ex}
  +
  \diagramBox{
    \begin{fmffile}{BubblesExchange3}
    \begin{fmfgraph}(50,50)
      \fmfkeep{BubblesExchange3}
      \fmfset{arrow_len}{6}
      \fmfstraight
      \fmfleft{v11,v12,v13}
      \fmfright{v31,v32,v33}
      \fmf{photon}{v11,v31}
      \fmf{phantom}{v12,v22} \fmf{photon}{v22,v32}
      \fmf{photon}{v13,v23} \fmf{phantom}{v23,v33}
      \fmffreeze
      \fmf{fermion,left=0.35}{v31,v32,v31}
      \fmf{fermion,left=0.25}{v11,v13}
      \fmf{fermion}{v13,v22}
      \fmf{fermion,right=0.35}{v22,v23}
      \fmf{fermion}{v23,v11}
    \end{fmfgraph}
    \end{fmffile}
  }
  \hspace*{-2ex}
  +
  \diagramBoxBorder{0ex}{2ex}{
    \begin{fmffile}{BubblesExchange4}
    \begin{fmfgraph}(60,60)
      \fmfkeep{BubblesExchange4}
      \fmfset{arrow_len}{6}
      \fmfstraight
      \fmfleft{v11,v12,v13,v14}
      \fmfright{v41,v42,v43,v44}
      \fmf{photon}{v11,v41}
      \fmf{phantom}{v12,v22,v32} \fmf{photon}{v32,v42}
      \fmf{photon}{v13,v23} \fmf{phantom}{v23,v33,v43}
      \fmf{phantom}{v14,v24} \fmf{photon}{v24,v34} \fmf{phantom}{v34,v44}
      \fmffreeze
      \fmf{fermion,left=0.25}{v11,v13,v11}
      \fmf{fermion,left=0.35}{v41,v42,v41}
      \fmf{fermion,left=0.35}{v23,v24}
      \fmf{fermion}{v24,v32}
      \fmf{fermion,right=0.25}{v32,v34}
      \fmf{fermion}{v34,v23}
    \end{fmfgraph}
    \end{fmffile}
  }
  \hspace*{-2ex}
  +\ \ldots
\end{center}
  \caption{
    Exchanging only the last interaction in the ring
    diagrams leads to the Second Order Screened Exchange (SOSEX) correction to
    the Random Phase Approximation.
  }
  \label{fig:SOSEXDiagrams}
\end{figure}

The systematic error can be alleviated by including exchange diagrams
where violations of the Pauli exclusion principle can occur.
The lowest order correction to
the Random Phase Approximation anti-symmetrizes only one interaction of the RPA
ring diagrams. The correction is termed
\emph{Second Order Screened Exchange (SOSEX)}
\index{Second Order Screened Exchange (SOSEX)}if only the
last interaction in time is anti-symmetrized.
Figure \ref{fig:SOSEXDiagrams} shows the additional diagrams included in the
SOSEX.

    \section{SOSEX from Direct Ring Coupled Cluster Doubles}
The lowest order correction anti-symmetrizes only one interaction of the
RPA ring diagrams. This could be any interaction and not necessarily just
the last one. However, when calculating the Random Phase Approximation using
the direct ring Coupled Cluster Doubles amplitudes $t_{ij}^{ab}$, as discussed
in Section \ref{sec:RPAdrCCD}, calculating the SOSEX corrections comes with
hardly any additional costs. We simply close the amplitudes once with a 
direct interaction $V_{ab}^{ij}$ and once with two indices swapped.
Respecting the Fermion sign the RPA+SOSEX energy is then given by
\begin{equation}
  \diagramBoxBorder{2ex}{2ex}{
    \fmfreuse{drCCDd}
  }
  \ +
  \diagramBox{
    \begin{fmffile}{drCCDx}
    \begin{fmfgraph}(30,30)
      \fmfkeep{drCCDx}
      \fmfset{arrow_len}{6}
      \fmfstraight
      \fmfbottom{lb,rb}
      \fmftop{lt,rt}
      \fmf{dashes}{lb,rb}
      \fmf{boson}{lt,rt}
      \fmf{fermion,left=0.25}{lb,lt}
        \fmf{fermion}{lt,m1} \fmf{plain}{m1,rb}
      \fmf{fermion,right=0.25}{rb,rt}
        \fmf{fermion}{rt,m2} \fmf{plain}{m2,lb}
    \end{fmfgraph}
    \end{fmffile}
  }
  \hspace*{-2ex}
  \ =\ \frac12\,\sum_{ijab}t_{ij}^{ab}V_{ab}^{ij} - t_{ij}^{ab}V_{ab}^{ji}\,.
  \label{eqn:SOSEXdrCCD}
\end{equation}

\begin{figure}[p]
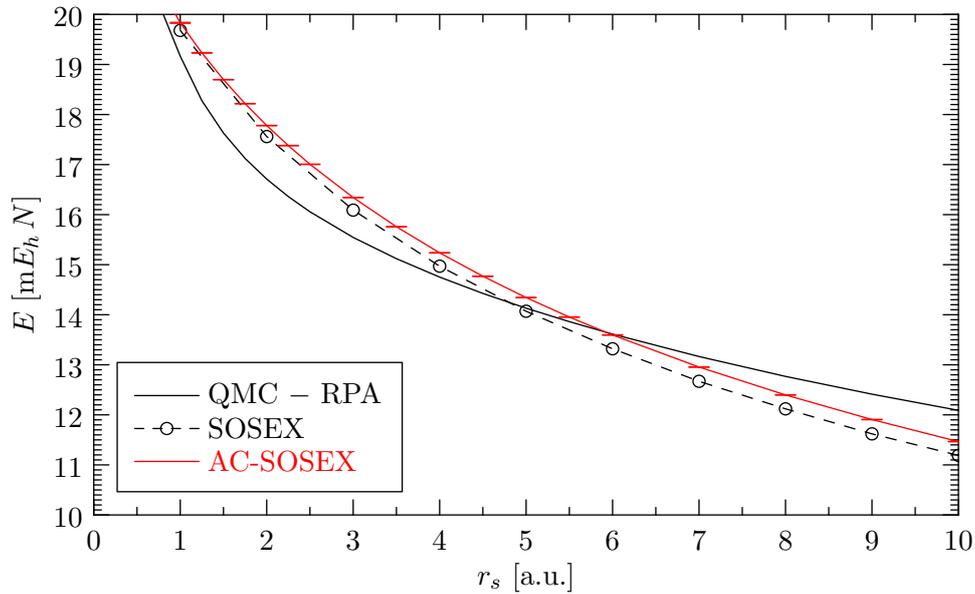

\begin{center}
\begin{asy}
  real[][] eps = input("EpsPara.dat").line().dimension(0,0);
  real[][] sosex = input("SosexPara.dat").line().dimension(0,0);
  real[][] acSosex = input("AcSosexPara.dat").line().dimension(0,0);

  sizeRatio(width=360);
  plotXY(eps, "QMC $-$ RPA");
  plotXY(sosex, dashed, "SOSEX", marker(scale(0.8mm)*unitcircle,black));
  plotXYDY(acSosex, red, "AC-SOSEX");
  axisXY(
    Label("$r_s$ [a.u.]",MidPoint), Label("$E$ [m$E_h\,N$]",MidPoint),
    (0,10), (10,20),
    yTicks=RightTicks(Step=1,n=10)
  );
  attach(legend(1,6,6,24,vskip=1),(0,10),12NE,UnFill);
\end{asy}
\end{center}
\vspace*{-2ex}
\caption{
  The Second Order Screened Exchange energy per electron obtained by
  \parencite{freeman_coupled-cluster_1977} for the uniform
  electron gas. It is compared to the error of the Random Phase Approximation
  with respect to Quantum Monte-Carlo (QMC)
  calculations by \parencite{ceperley_ground_1980}
  fitted by \parencite{perdew_self-interaction_1981}.
  At $r_s\approx 5$ RPA+SOSEX fortuitously matches QMC results.
  The Adiabatic Connection-SOSEX according to (\ref{eqn:SOSEX_ACUEG}),
  which is discussed later, is also shown here.
}
\label{fig:SOSEXUEG}
\end{figure}
\begin{table}[p]
\begin{center}
\begin{tabular}{|c|r|r|rr|}
  \hline
  &&&& \\[-2.5ex]
  \multicolumn{1}{|c|}{$r_s$} &
    \multicolumn{1}{|c|}{$(E_{\rm c}-E_{\rm c}^{\rm RPA})$} &
    \multicolumn{1}{|c|}{$E_{\rm c}^{\rm SOSEX}$} &
    \multicolumn{2}{c|}{$E_{\rm c}^{\rm AC-SOSEX}$}
    \\
  \multicolumn{1}{|c|}{$[$a.u.$]$} &
    \multicolumn{1}{|c}{[m$E_h\,N$]} &
    \multicolumn{1}{|c|}{[m$E_h\,N$]} &
    \multicolumn{1}{|c}{[m$E_h\,N$]} &
    \multicolumn{1}{c|}{$\pm$}
     \\
 &&&& \\[-2.6ex]
\hline
 &&&& \\[-2.5ex]
 1 & 19.167 & 19.680 &  19.832 & 0.009 \\
 2 & 16.710 & 17.560 &  17.780 & 0.003 \\
 3 & 15.545 & 16.090 &  16.342 & 0.003 \\
 4 & 14.752 & 14.970 &  15.237 & 0.003 \\
 5 & 14.131 & 14.070 &  14.343 & 0.003 \\
 6 & 13.613 & 13.320 &  13.595 & 0.003 \\
 7 & 13.165 & 12.670 &  12.955 & 0.003 \\
 8 & 12.768 & 12.120 &  12.398 & 0.003 \\
 9 & 12.413 & 11.620 &  11.906 & 0.003 \\
10 & 12.090 & 11.190 &  11.466 & 0.003 \\
12 & 11.521 & \multicolumn{1}{c|}{---} &  10.712 & 0.003 \\
15 & 10.810 & \multicolumn{1}{c|}{---} &   9.806 & 0.003 \\
20 &  9.884 & \multicolumn{1}{c|}{---} &   8.679 & 0.003 \\
30 &  8.582 & \multicolumn{1}{c|}{---} &   7.201 & 0.003 \\
40 &  7.685 & \multicolumn{1}{c|}{---} &   6.246 & 0.003 \\
50 &  7.014 & \multicolumn{1}{c|}{---} &   5.564 & 0.003 \\
  \hline
\end{tabular}
\end{center}
\caption{
  Data of Figure \ref{fig:SOSEXUEG} including low densities.
}
\label{tab:SOSEX_UEG}
\end{table}
The Second Order Screened Exchange correction to the RPA was introduced by
\parencite{freeman_coupled-cluster_1977}
who applied it to the uniform electron gas.
The term SOSEX was later coined by 
\parencite{gruneis_making_2009} who studied this correction also for solids.
Figure \ref{fig:SOSEXUEG} shows the Second Order Screened Exchange correction
per electron for the uniform electron gas as calculated by Freeman. The error
of the Random Phase Approximation with respect to Quantum Monte-Carlo (QMC)
results of \parencite{ceperley_ground_1980}
fitted by \parencite{perdew_self-interaction_1981}
is also given. Remarkably
and certainly fortuitously RPA+SOSEX matches the QMC results at
$r_s\approx 5$, which is in the density region of real metals.
Anti-symmetrizing only the last interaction seems gives just the right 
correction to the RPA in the uniform electron gas. 
However, from a strictly \textit{ab-initio} point of view, there is no reason
to restrict the exchange diagrams to those where only the last interaction is
anti-symmetrized, except technical convenience when having the drCCD amplitudes
at hand.

Anti-symmetrizing each but still only one interaction in the ring diagrams
gives worse agreement with QMC in the high density regime $r_s\leq 8$ but
better agreement for low densities, where correlation effects are stronger.
This is shown in Section \ref{sec:APX_UEG}.
Note that in a spin-polarized system SOSEX is less fortunate. While still
improving on RPA in the density range of interest, it cancels the RPA energy
in the low density limit, as discussed in Subsection
\ref{ssc:APX_UegFerromagnetic}.

One can also include full anti-symmetrization in the amplitude equation,
including terms such as
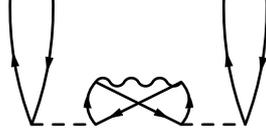
\[
  \diagramBox{\fmfreuse{CCDtvxt}}\,
\]
which lead to the full Coupled Cluster Doubles (CCD) amplitude equation.
However, calculating the CCD amplitudes requires $\mathcal O(N^6)$ operations
while the direct ring CCD amplitudes of RPA and SOSEX can be computed
in $\mathcal O(N^5)$ steps
\parencite{scuseria_is_1989}.
A major drawback of either method is the large memory requirement
scaling like $\mathcal O(N^4)$ since the amplitudes $t_{ij}^{ab}$ need to be
stored for iterating the amplitude equation.

    \section{Adiabatic Connection-SOSEX}
\label{sec:SOSEX_ACSOSEX}
In contrast to calculating the drCCD amplitudes, evaluating the Random Phase
Approximation in the frequency domain only requires
$\mathcal O(N^2)$ memory. Calculating an exchange correction based on
two point quantities, such as the independent particle polarizability
$\diagramBox{\fmfreuse{Polarization}}=\,\chi_0$
would therefore be favorable.
\'Ang\'yan \emph{et al.} suggested an approximation to the drCCD SOSEX
within the framework of the Adiabatic Connection that can be implemented
with a memory usage of $\mathcal O(N^2)$, which we will outline here.

Within the Adiabatic Connection one can define a screened interaction
similar to the one defined in many-body perturbation theory in
(\ref{eqn:RPAScreenedW})
\begin{equation}
\begin{array}{ccc ccc ccc}
  \diagramBox{
    \begin{fmffile}{ScreenedWLambda}
    \begin{fmfgraph*}(26,10)
      \fmfstraight
      \fmfleft{v11}
      \fmfright{v21}
      \fmf{dbl_wiggly,label=$W_\lambda$,label.side=left}{v11,v21}
    \end{fmfgraph*}
    \end{fmffile}
  }
  &=&
  \diagramBox{
    \begin{fmffile}{CoulombInteractionLambda}
    \begin{fmfgraph*}(26,10)
      \fmfstraight
      \fmfleft{v11}
      \fmfright{v21}
      \fmf{photon,label=$\lambda$,label.side=left}{v11,v21}
    \end{fmfgraph*}
    \end{fmffile}
  }
  &+&
  \diagramBox{
    \begin{fmffile}{VChiVLambda}
    \begin{fmfgraph*}(52,20)
      \fmfstraight
      \fmfset{arrow_len}{6}
      \fmfleft{v00,v01}
      \fmfright{v20,v21}
      \fmf{photon,label=$\lambda$,label.side=left}{v00,v10}
        \fmf{phantom}{v10,v20}
      \fmf{phantom}{v01,v11}
        \fmf{photon,label=$\lambda$,label.side=left}{v11,v21}
      \fmffreeze
      \fmf{fermion,right=0.3}{v10,v11,v10}
    \end{fmfgraph*}
    \end{fmffile}
  }
  &+&
  \diagramBoxBorder{1ex}{2ex}{
    \begin{fmffile}{VChiVChiVLambda}
    \begin{fmfgraph*}(64,32)
      \fmfstraight
      \fmfset{arrow_len}{6}
      \fmfleft{v00,v01,v02}
      \fmfright{v30,v31,v32}
      \fmf{photon,label=$\lambda$,label.side=left}{v00,v10} \fmf{phantom}{v10,v20,v30}
      \fmf{phantom}{v01,v11}
        \fmf{photon,label=$\lambda$,label.side=left}{v11,v21}
        \fmf{phantom}{v21,v31}
      \fmf{phantom}{v02,v12,v22}
        \fmf{photon,label=$\lambda$,label.side=left}{v22,v32}
      \fmffreeze
      \fmf{fermion,right=0.3}{v10,v11,v10}
      \fmf{fermion,right=0.3}{v21,v22,v21}
    \end{fmfgraph*}
    \end{fmffile}
  }
  &+& \ldots
  \\[2ex]
  \vec W_\lambda(\im\nu)
  &=&
    \lambda \vec V
  &+& \lambda^2 \vec V\vec X_0(\im\nu)\vec V
  &+& \lambda^3 \vec V\vec X_0(\im\nu)\vec V\vec X_0(\im\nu)\vec V
  &+& \ldots
\end{array}
  \label{eqn:SOSEXScreenedW}
\end{equation}
using the matrix notation introduced in Section \ref{sec:RPAFreq}.
The only difference to (\ref{eqn:RPAScreenedW}) is the dependence on
the coupling strength $\lambda$. Note that we explicitly write the
coupling strength in all diagrams within the Adiabatic Connection to make
a clear distinction from the diagrams within many-body perturbation theory
discussed in Chapter \ref{cha:MBPT}.

We can now define the coupling strength averaged screened interaction
\[
  \overline{\vec W}(\im\nu) = \int_0^1\d\lambda\,\vec W_\lambda(\im\nu)
    = \frac12\vec V + \frac13\vec V\vec X_0(\im\nu)\vec V + \ldots 
\]
and write the RPA correlation energy found in (\ref{eqn:RPAACFDTFinal}) in
terms of $\overline{\vec W}$:
\begin{equation}
  E_{\rm c}^{\rm RPA} = -\frac12\int\frac{\d\nu}{2\pi}\,
    \Tr\Big\{\vec X_0(\im\nu)\vec V\vec X_0(\im\nu)\overline{\vec W}(\im\nu)\Big\}
  \,.
  \label{eqn:SOSEXScreenedRPA}
\end{equation}
Next, we insert the independent particle polarizability given in
(\ref{eqn:RPAPolImag}),
\[
  {\vec X_0}_{\vec x\vec x'}(\im\nu) =
  -\sum_{ia}
  \left(
  \frac{
    \psi_a(\vec x)\psi^\ast_a(\vec x')\psi_i(\vec x')\psi^\ast_i(\vec x)
  }{\eps_a-\eps_i-\im\nu}
  +
  \frac{
    \psi_i(\vec x)\psi^\ast_i(\vec x')\psi_a(\vec x')\psi^\ast_a(\vec x)
  }{\eps_a-\eps_i+\im\nu}
  \right)\,,
\]
into (\ref{eqn:SOSEXScreenedRPA}), giving
\begin{align}
  \nonumber
  E_{\rm c}^{\rm RPA} =
    -\frac12\int\frac{\d\nu}{2\pi}\,
    \sum_{ijab}
      \Bigg(&
        \frac{
          V^{aj}_{ib}\,{\overline W}^{bi}_{ja}(\im\nu)
        }{
          (\eps_a-\eps_i-\im\nu)(\eps_b-\eps_j-\im\nu)
        }
        +
        \frac{
          V^{ab}_{ij}\,\overline W^{ji}_{ba}(\im\nu)
        }{
          (\eps_a-\eps_i-\im\nu)(\eps_b-\eps_j+\im\nu)
        }+
        \\ &
        \frac{
          V^{ij}_{ab}\,\overline W^{ba}_{ji}(\im\nu)
        }{
          (\eps_a-\eps_i+\im\nu)(\eps_b-\eps_j-\im\nu)
        }
        +
        \frac{
          V^{ib}_{aj}\,\overline W^{ja}_{bi}(\im\nu)
        }{
          (\eps_a-\eps_i+\im\nu)(\eps_b-\eps_j+\im\nu)
        }
      \Bigg)\,,
  \label{eqn:SOSEXRPA}
\end{align}
where we write $\overline W^{pq}_{sr}$ analogous to the matrix elements
$V^{pq}_{sr}$ of the Coulomb operator:
\[
  \overline W^{pq}_{sr}(\im\nu) = \iint\d\vec x_1\,\d\vec x_2\,
    \psi_p^\ast(\vec x_1)\psi_q^\ast(\vec x_2)\,
    \overline{\vec W}_{\vec x_1\vec x_2}(\im\nu)\,
    \psi_r(\vec x_2)\psi_s(\vec x_1)\,.
\]
Although in the Adiabatic Connection the polarizability is the central
quantity of interest rather than connected diagrams, we can use similar
diagrams to depict the terms in (\ref{eqn:SOSEXRPA}):
\begin{equation}
  E_{\rm c}^{\rm RPA} = -\frac12\,
  \left(
    \hspace*{4ex}
    \diagramBox[0.5]{
      \begin{fmffile}{ACRPA1}
      \begin{fmfgraph*}(32,64)
        \fmfkeep{ACRPA1}
        \fmfstraight
        \fmfset{arrow_len}{6}
        \fmfleft{v00,v01,v02}
        \fmfright{v10,v11,v12}
        \fmf{photon,label=$\leftarrow$,label.dist=3}{v11,v01}
        \fmf{
          dbl_wiggly,left=1.25,label=$\overline W$,label.side=right,label.dist=3
        }{v10,v02}
        \fmffreeze
        \fmf{fermion,left=0.3,label=$a$,label.dist=3}{v01,v02}
        \fmf{fermion,left=0.3,label=$i$,label.dist=3}{v02,v01}
        \fmf{fermion,left=0.3,label=$b$,label.dist=3}{v10,v11}
        \fmf{fermion,left=0.3,label=$j$,label.dist=3}{v11,v10}
      \end{fmfgraph*}
      \end{fmffile}
    }
    \hspace*{1ex}
    +
    \hspace*{1ex}
    \diagramBox[0.0]{
      \begin{fmffile}{ACRPA2}
      \begin{fmfgraph*}(32,32)
        \fmfstraight
        \fmfset{arrow_len}{6}
        \fmfleft{v00,v01}
        \fmfright{v10,v11}
        \fmf{photon,label=$\leftarrow$,label.dist=3}{v10,v00}
        \fmf{
          dbl_wiggly
        }{v01,v11}
        \fmffreeze
        \fmf{fermion,left=0.3,label=$a$,label.dist=3}{v00,v01}
        \fmf{fermion,left=0.3,label=$i$,label.dist=3}{v01,v00}
        \fmf{fermion,left=0.3,label=$b$,label.dist=3}{v10,v11}
        \fmf{fermion,left=0.3,label=$j$,label.dist=3}{v11,v10}
      \end{fmfgraph*}
      \end{fmffile}
    }
    \hspace*{1ex}
    +
    \hspace*{1ex}
    \diagramBox[1.0]{
      \begin{fmffile}{ACRPA3}
      \begin{fmfgraph*}(32,32)
        \fmfstraight
        \fmfset{arrow_len}{6}
        \fmfleft{v00,v01}
        \fmfright{v10,v11}
        \fmf{photon,label=$\leftarrow$,label.dist=3}{v11,v01}
        \fmf{
          dbl_wiggly
        }{v00,v10}
        \fmffreeze
        \fmf{fermion,left=0.3,label=$a$,label.dist=3}{v00,v01}
        \fmf{fermion,left=0.3,label=$i$,label.dist=3}{v01,v00}
        \fmf{fermion,left=0.3,label=$b$,label.dist=3}{v10,v11}
        \fmf{fermion,left=0.3,label=$j$,label.dist=3}{v11,v10}
      \end{fmfgraph*}
      \end{fmffile}
    }
    \hspace*{1ex}
    +
    \hspace*{1ex}
    \diagramBoxBorder[0.5]{0ex}{2ex}{
      \begin{fmffile}{ACRPA4}
      \begin{fmfgraph*}(32,64)
        \fmfkeep{ACRPA4}
        \fmfstraight
        \fmfset{arrow_len}{6}
        \fmfleft{v00,v01,v02}
        \fmfright{v10,v11,v12}
        \fmf{photon,label=$\leftarrow$,label.dist=3}{v11,v01}
        \fmf{
          dbl_wiggly,right=1.25
        }{v00,v12}
        \fmffreeze
        \fmf{fermion,left=0.3,label=$a$,label.dist=3}{v00,v01}
        \fmf{fermion,left=0.3,label=$i$,label.dist=3}{v01,v00}
        \fmf{fermion,left=0.3,label=$b$,label.dist=3}{v11,v12}
        \fmf{fermion,left=0.3,label=$j$,label.dist=3}{v12,v11}
      \end{fmfgraph*}
      \end{fmffile}
    }
    \hspace*{5ex}
  \right)
  \label{eqn:SOSEXRPACases}
\end{equation}
The diagrams are drawn such that the imaginary frequency $\nu$ goes from
right to left on the Coulomb interaction as indicated by the arrow.

For real valued spin-orbitals $\psi_p(\vec x)$ the Coulomb integrals
exhibit time reversal symmetry at each vertex such that
$
  V_{ij}^{ab} = V_{ib}^{aj} = V_{aj}^{ib} = V_{ab}^{ij}
$.
The same holds for the screened Coulomb integrals since the independent particle
polarizability $\vec X_0(\im\nu)$ is real valued. This simplifies
(\ref{eqn:SOSEXRPA}) to
\[
  E_{\rm c}^{\rm RPA} =
  -\frac12\int\frac{\d\nu}{2\pi}\,
    \sum_{ijab}
    V_{ab}^{ij} \overline W_{ji}^{ba}(\im\nu)f_{ia}(\im\nu)f_{jb}(\im\nu)\,,
  \qquad
  \textnormal{with }
  f_{ia}(\im\nu) = \frac{2(\eps_a-\eps_i)}{(\eps_a-\eps_i)^2+\nu^2}\,.
\]
In this form, the frequency dependent RPA energy expression bears resemblance to
the drCCD RPA expression
$
  \frac12\sum_{ijab} t_{ij}^{ab} V_{ab}^{ij}
$
and we can define the \emph{AC-SOSEX}\index{AC-SOSEX} by anti-symmetrizing the
Coulomb interaction $V$ in analogy to the drCCD SOSEX expression, arriving at
\begin{equation}
  E_{\rm c}^{\rm AC-SOSEX} = +\frac12\int\frac{\d\nu}{2\pi}\,
    \sum_{ijab}
    V_{ab}^{ji} \overline W_{ij}^{ab}(\im\nu)f_{ia}(\im\nu)f_{jb}(\im\nu)\,.
  \label{eqn:SOSEXAC}
\end{equation}

Above equation still requires $\mathcal O(N^4)$ of memory from the two
interactions $V_{ab}^{ij}$ and $\overline W_{ij}^{ab}$. We can, however,
transform it back into the position basis as discussed in Section
\ref{sec:MBPTpropagators}, giving
\[
  E_{\rm c}^{\rm AC-SOSEX} = -\frac12\int\frac{\d\nu}{2\pi}\,
  \Tr\left\{
    \vec P_{\rm x}^{\rm AC}(\im\nu) \overline{\vec W}(\im\nu)
  \right\}
\]
defining the imaginary frequency dependent exchange polarizability
for the AC-SOSEX
\begin{align}
  \nonumber
  {\vec P_{\rm x}^{\rm AC}}_{\vec x_1 \vec x_2}(\im\nu) =
    -\iint\d\vec x_3\,\d\vec x_4\,\frac1{|\vec r_3-\vec r_4|} &
    \sum_{ia}
      \psi^\ast_i(\vec x_4)\psi_i(\vec x_1)
      \psi^\ast_a(\vec x_1)\psi_a(\vec x_3)
      f_{ia}(\im\nu)
  \\ &
    \sum_{jb}
      \psi^\ast_j(\vec x_3)\psi_j(\vec x_2)
      \psi^\ast_b(\vec x_2)\psi_b(\vec x_4)
      f_{jb}(\im\nu)\,.
  \label{eqn:SOESX_PxAC}
\end{align}
Note that the exchange polarizability is defined negative since it contains
only one Fermion loop.
We can give a closed form for the coupling strength averaged screened
interaction finding a matrix function with the same Taylor expansion. This
yields the final expression for the AC-SOSEX:
\begin{equation}
  E_{\rm c}^{\rm AC-SOSEX} = +\frac12\int\frac{\d\nu}{2\pi}\,
  \Tr\left\{
    \vec P_{\rm x}^{\rm AC}(\im\nu)
    \Big(\vec X_0(\im\nu)\vec V\vec X_0(\im\nu)\Big)^{-1}
    \Big(
      \log\Big(\vec 1-\vec X_0(\im\nu)\vec V\Big)+\vec X_0(\im\nu)\vec V
    \Big)
  \right\}\,.
\end{equation}
The exchange polarizability is a quantity depending on two positions.
Thus, evaluating the AC-SOSEX as given above requires only $\mathcal O(N^2)$
of memory rather than $\mathcal O(N^4)$, greatly broadening the applicability
of the AC-SOSEX to larger systems.
However, calculating the exchange polarizability still requires
$\mathcal O(N^5)$ steps, which is equally time consuming as calculating
the SOSEX from the direct ring Coupled Cluster Doubles amplitudes.
In the Random Phase Approximation the respective polarizability analogous to
$\vec P_{\rm x}^{\rm AC}(\im\nu)$ simply factors into
$\vec X_0(\im\nu)\vec V\vec X_0(\im\nu)$,
allowing for an evaluation in only $\mathcal O(N^3)$ steps. This can
not be done for the AC-SOSEX since $\vec x_3$ and $\vec x_4$ occur in both
sums in (\ref{eqn:SOESX_PxAC}).
Reducing the memory consumption from $\mathcal O(N^4)$ to $\mathcal O(N^4)$ is
still an important improvement since it is easier to allocate more CPUs to a
calculation than it is to allocate more memory per CPU.

\section{Difference between drCCD SOSEX and AC-SOSEX}
\label{sec:SOSEX_Difference}
\begin{figure}
\begin{center}
  \hspace*{-2ex}
  \diagramBox{
    \begin{fmffile}{CCDmissingExchange}
    \begin{fmfgraph}(80,50)
      \fmfkeep{CCDmissingExchange}
      \fmfset{arrow_len}{6}
      \fmfstraight
      \fmfleft{v00,v01,v02}
      \fmfright{v140,v141,v142}
      \fmf{phantom,tension=4}{v00,v10} \fmf{photon}{v10,v50}
      \fmf{phantom}{v50,v90}
      \fmf{photon}{v90,v130} \fmf{phantom,tension=4}{v130,v140}
      \fmf{phantom,tension=4}{v01,v11} \fmf{phantom}{v11,v51}
      \fmf{photon}{v51,v91}
      \fmf{phantom}{v91,v131} \fmf{phantom,tension=4}{v131,v141}
      \fmf{phantom,tension=12}{v02,v12} \fmf{photon}{v12,v132}
      \fmf{phantom,tension=12}{v132,v142}
      \fmffreeze
      \fmf{fermion,left=0.2}{v12,v10,v12}
      \fmf{fermion,left=0.3}{v50,v51} \fmf{fermion}{v51,m} \fmf{plain}{m,v90}
      \fmf{fermion,right=0.3}{v90,v91} \fmf{fermion}{v91,m} \fmf{plain}{m,v50}
      \fmf{fermion,left=0.2}{v132,v130,v132}
    \end{fmfgraph}
    \end{fmffile}
  }
  \hspace*{-1ex}
  \qquad\qquad
    \diagramBoxBorder[0.5]{0ex}{2ex}{
      \fmfreuse{ACSOSEX4}
    }
  \qquad
\end{center}
\caption{
  Diagrams contained in the AC-SOSEX that are not part of the SOSEX based
  on the drCCD amplitudes. The factors from the coupling strength integration
  are omitted. The right diagram shows a particle turning into a hole and
  vice-versa, which has no correspondence in many-body perturbation theory.
}
\label{fig:SOSEXAcDifference}
\end{figure}
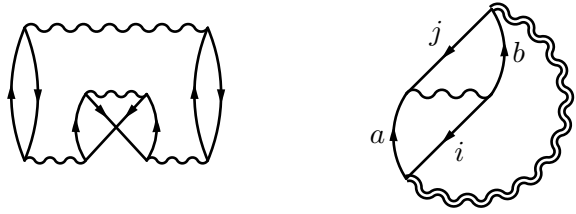

Despite the similarity of (\ref{eqn:SOSEXAC}) and the expression for the
drCCD SOSEX (\ref{eqn:SOSEXdrCCD}) they are only identical in second order
but not beyond. First, the drCCD amplitudes
$
  \diagramBox{\fmfreuse{CCDt}}
$
are constructed monotonous in time according
to (\ref{eqn:drCCDV})-(\ref{eqn:drCCDtVt}), which
guarantees that the closing Coulomb interaction in
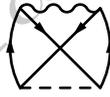
\[
  \diagramBox{\fmfreuse{drCCDx}}
\]
is indeed the last interaction in time. In contrast, the averaged screened
interaction $\overline{\vec W}(\im\nu)$ contains interactions that
reach both, in the past and in the future. Thus, the left diagram shown in
Figure \ref{fig:SOSEXAcDifference} is contained in the AC-SOSEX while it is
not contained in the drCCD SOSEX.
Furthermore, anti-symmetrizing the first and the last case in
(\ref{eqn:SOSEXRPACases}) yields terms that have no correspondence
in many-body perturbation theory. The AC-SOSEX introduces anti-symmetrization
by simply swapping $i$ and $j$ at the unscreened interaction $V$. For the last
term this yields for example
\[
  V_{ai}^{jb}\overline{W}_{bi}^{ja}\,.
\]
This term, however, contradicts the requirement of many-body perturbation
theory that upper
indices can only match lower indices and vice-versa since upper and lower
indices refer to creation and annihilation operators, respectively.
This term can therefore not be drawn diagrammatically in the usual manner such
that the upper/lower indices correspond to outgoing/incoming connections.
We can, however, draw the term respecting the propagation direction of particles
and holes. The resulting diagram is depicted on the right of Figure
\ref{fig:SOSEXAcDifference}. It exhibits a particle $a$ turning into a
hole $j$ at the left vertex of the Coulomb interaction and a hole $i$ turning
into a particle $b$ at the right vertex.
This diagram will be termed
\emph{swapped ladder diagram}\index{swapped ladder diagram} since it resembles
a particle-hole ladder diagram where $b$ and $j$ are swapped.
Employing the same notion of diagrams as in
(\ref{eqn:SOSEXRPACases}), the AC-SOSEX can be depicted diagrammatically by
\begin{equation}
  E_{\rm c}^{\rm AC-SOSEX} = +\frac12\,
  \left(
    \hspace*{4ex}
    \diagramBox[0.5]{
      \begin{fmffile}{ACSOSEX1}
      \begin{fmfgraph*}(32,64)
        \fmfkeep{ACRPA1}
        \fmfstraight
        \fmfset{arrow_len}{6}
        \fmfleft{v00,v01,v02}
        \fmfright{v10,v11,v12}
        \fmf{photon}{v11,v01}
        \fmf{
          dbl_wiggly,left=1.25,label=$\overline W$,label.side=right,label.dist=3
        }{v10,v02}
        \fmffreeze
        \fmf{fermion,left=0.3,label=$a$,label.dist=3}{v01,v02}
        \fmf{fermion,label=$i$,label.dist=3,label.side=left}{v02,v11}
        \fmf{fermion,right=0.3,label=$b$,label.dist=3}{v10,v11}
        \fmf{fermion,label=$j$,label.dist=3,label.side=right}{v01,v10}
      \end{fmfgraph*}
      \end{fmffile}
    }
    \hspace*{1ex}
    +
    \hspace*{1ex}
    \diagramBox[0.0]{
      \begin{fmffile}{ACSOSEX2}
      \begin{fmfgraph*}(32,32)
        \fmfstraight
        \fmfset{arrow_len}{6}
        \fmfleft{v00,v01}
        \fmfright{v10,v11}
        \fmf{photon}{v10,v00}
        \fmf{
          dbl_wiggly
        }{v01,v11}
        \fmffreeze
        \fmf{fermion,left=0.3,label=$a$,label.dist=3}{v00,v01}
        \fmf{plain}{v01,m1}
          \fmf{fermion,label=$i$,label.dist=3,label.side=left}{m1,v10}
        \fmf{fermion,right=0.3,label=$b$,label.dist=3}{v10,v11}
        \fmf{plain}{v11,m2}
          \fmf{fermion,label=$j$,label.dist=1.5,label.side=right}{m2,v00}
      \end{fmfgraph*}
      \end{fmffile}
    }
    \hspace*{1ex}
    +
    \hspace*{1ex}
    \diagramBox[1.0]{
      \begin{fmffile}{ACSOSEX3}
      \begin{fmfgraph*}(32,32)
        \fmfstraight
        \fmfset{arrow_len}{6}
        \fmfleft{v00,v01}
        \fmfright{v10,v11}
        \fmf{photon}{v11,v01}
        \fmf{
          dbl_wiggly
        }{v00,v10}
        \fmffreeze
        \fmf{fermion,left=0.3,label=$a$,label.dist=3}{v00,v01}
        \fmf{plain}{v01,m1}
          \fmf{fermion,label=$j$,label.dist=1.5,label.side=left}{m1,v10}
        \fmf{fermion,right=0.3,label=$b$,label.dist=3}{v10,v11}
        \fmf{plain}{v11,m2}
          \fmf{fermion,label=$i$,label.dist=3,label.side=right}{m2,v00}
      \end{fmfgraph*}
      \end{fmffile}
    }
    \hspace*{1ex}
    +
    \hspace*{1ex}
    \diagramBoxBorder[0.5]{0ex}{1ex}{
      \begin{fmffile}{ACSOSEX4}
      \begin{fmfgraph*}(32,64)
        \fmfkeep{ACSOSEX4}
        \fmfstraight
        \fmfset{arrow_len}{6}
        \fmfleft{v00,v01,v02}
        \fmfright{v10,v11,v12}
        \fmf{photon}{v11,v01}
        \fmf{
          dbl_wiggly,right=1.25
        }{v00,v12}
        \fmffreeze
        \fmf{fermion,left=0.3,label=$a$,label.dist=3}{v00,v01}
        \fmf{fermion,label=$i$,label.dist=3,label.side=left}{v11,v00}
        \fmf{fermion,right=0.3,label=$b$,label.dist=3,label.side=right}{v11,v12}
        \fmf{fermion,label=$j$,label.dist=3,label.side=right}{v12,v01}
      \end{fmfgraph*}
      \end{fmffile}
    }
    \hspace*{5ex}
  \right)\,.
  \label{eqn:SOSEXACCases}
\end{equation}

\subsection{Diamond C(A4)}
We can study the swapped ladder diagram for a small test system consisting of
diamond C(A4) with 128 states. It is based on a Hartree-Fock reference with
2$\times$2$\times$2 $k$-points in a primitive cell comprising 2 atoms with 4
electrons per atom.
Although the system is only coarsely described by 128 states it serves well as
a benchmark for individual diagrams beyond second order.
In this finite band gap system we can limit
the order of the AC-SOSEX diagrams from (\ref{eqn:SOSEXACCases})
solely to third order while still getting finite results. This excludes all
diagrams that are not contained in the drCCD SOSEX except the third order
swapped ladder diagram.
The AC-SOSEX expression from (\ref{eqn:SOSEXACCases}) expands in third order
to all possible permutations of the interaction times, analogous to Figure
\ref{fig:MP3Feyn}. Including the factor $1/3$ from the coupling strength
integration this gives
\begin{align}
  \nonumber
  {E_{\rm c}^{\rm AC-SOSEX}}^{(3)} = \frac1{2\cdot3} &\left(
    \diagramBox{
      \begin{fmffile}{MP3Exchange}
      \begin{fmfgraph*}(50,50)
        \fmfkeep{MP3Exchange}
        \fmfset{arrow_len}{6}
        \fmfstraight
        \fmfleft{v00,v01,v02}
        \fmfright{v20,v21,v22}
        \fmf{boson}{v00,v10,v20}
        \fmf{phantom}{v01,v11} \fmf{boson}{v11,v21}
        \fmf{boson}{v02,v12} \fmf{phantom}{v12,v22}
        \fmffreeze
        \fmf{fermion,left=0.25}{v00,v02}
        \fmf{fermion}{v02,v11}
        \fmf{fermion,right=0.25}{v11,v12}
        \fmf{fermion}{v12,v00}
        \fmf{fermion,left=0.25}{v20,v21,v20}
      \end{fmfgraph*}
      \end{fmffile}
    }
    +
    \diagramBox{
      \begin{fmffile}{MP3SwappedLadder}
      \begin{fmfgraph*}(50,50)
        \fmfkeep{MP3SwappedLadder}
        \fmfset{arrow_len}{6}
        \fmfstraight
        \fmfleft{v00,v02,v01}
        \fmfright{v20,v22,v21}
        \fmf{boson}{v00,v10,v20}
        \fmf{phantom}{v01,v11} \fmf{boson}{v11,v21}
        \fmf{boson}{v02,v12} \fmf{phantom}{v12,v22}
        \fmffreeze
        \fmf{fermion,left=0.25}{v00,v02}
        \fmf{fermion}{v11,v02}
        \fmf{fermion,right=0.25}{v12,v11}
        \fmf{fermion}{v12,v00}
        \fmf{fermion,left=0.25}{v20,v21,v20}
      \end{fmfgraph*}
      \end{fmffile}
    }
    \ +
    \diagramBox{
      \begin{fmffile}{MP3ExchangeUp}
      \begin{fmfgraph*}(50,50)
        \fmfkeep{MP3ExchangeUp}
        \fmfset{arrow_len}{6}
        \fmfstraight
        \fmfleft{v02,v01,v00}
        \fmfright{v22,v21,v20}
        \fmf{boson}{v00,v10,v20}
        \fmf{phantom}{v01,v11} \fmf{boson}{v11,v21}
        \fmf{boson}{v02,v12} \fmf{phantom}{v12,v22}
        \fmffreeze
        \fmf{fermion,left=0.25}{v02,v00}
        \fmf{fermion}{v11,v02}
        \fmf{fermion,right=0.25}{v12,v11}
        \fmf{fermion}{v00,v12}
        \fmf{fermion,left=0.25}{v21,v20,v21}
      \end{fmfgraph*}
      \end{fmffile}
    }
    +
  \right. \\[2ex]
  &\left.
    \hspace{1.75ex}
    \diagramBox{
      \begin{fmffile}{MP3Exchange2}
      \begin{fmfgraph*}(50,50)
        \fmfset{arrow_len}{6}
        \fmfstraight
        \fmfleft{v01,v00,v02}
        \fmfright{v21,v20,v22}
        \fmf{boson,tension=3}{v00,v10}
          \fmf{phantom,tension=12}{v10,c}
          \fmf{phantom,tension=4}{c,v20}
        \fmf{phantom}{v01,v11} \fmf{boson}{v11,v21}
        \fmf{boson}{v02,v12} \fmf{phantom}{v12,v22}
        \fmffreeze
        \fmf{boson}{v10,v20}
        \fmf{fermion,left=0.25}{v00,v02}
        \fmf{plain}{v02,m}
          \fmf{fermion}{m,v11}
        \fmf{fermion,right=0.125}{v11,c}
          \fmf{plain,right=0.125}{c,v12}
        \fmf{fermion}{v12,v00}
        \fmf{fermion,left=0.25}{v20,v21,v20}
      \end{fmfgraph*}
      \end{fmffile}
    }
    +
    \diagramBox{
      \begin{fmffile}{MP3SwappedLadder2}
      \begin{fmfgraph*}(50,50)
        \fmfset{arrow_len}{6}
        \fmfstraight
        \fmfleft{v01,v02,v00}
        \fmfright{v21,v22,v20}
        \fmf{boson}{v00,v10,v20}
        \fmf{phantom}{v01,v11} \fmf{boson}{v11,v21}
        \fmf{boson}{v02,v12} \fmf{phantom}{v12,v22}
        \fmffreeze
        \fmf{fermion,left=0.25}{v02,v00}
        \fmf{fermion}{v02,v11}
        \fmf{fermion,right=0.25}{v11,v12}
        \fmf{fermion}{v00,v12}
        \fmf{fermion,left=0.25}{v20,v21,v20}
      \end{fmfgraph*}
      \end{fmffile}
    }
    \ +
    \diagramBox{
      \begin{fmffile}{MP3ExchangeUp2}
      \begin{fmfgraph*}(50,50)
        \fmfset{arrow_len}{6}
        \fmfstraight
        \fmfleft{v02,v00,v01}
        \fmfright{v22,v20,v21}
        \fmf{boson,tension=3}{v00,v10}
          \fmf{phantom,tension=12}{v10,c}
          \fmf{phantom,tension=4}{c,v20}
        \fmf{phantom}{v01,v11} \fmf{boson}{v11,v21}
        \fmf{boson}{v02,v12} \fmf{phantom}{v12,v22}
        \fmffreeze
        \fmf{boson}{v10,v20}
        \fmf{fermion,left=0.25}{v02,v00}
        \fmf{plain}{m,v02}
          \fmf{fermion}{v11,m}
        \fmf{fermion,right=0.125}{c,v11}
          \fmf{plain,right=0.125}{v12,c}
        \fmf{fermion}{v00,v12}
        \fmf{fermion,left=0.25}{v20,v21,v20}
      \end{fmfgraph*}
      \end{fmffile}
    }
  \right)\,.
  \label{eqn:SOSEX3AC}
\end{align}
The lower diagrams are equivalent to the upper diagrams since each
can be continuously deformed into the respective upper diagram without changing
the order of the Coulomb interactions. The left two and the right two diagrams
are identical due to time reversal symmetry. This leaves two diagrams with
distinct energies to evaluate, the exchange diagram and the swapped ladder
diagram:
\begin{align}
  \label{eqn:DiamondExchange}
    \diagramBox{
      \begin{fmffile}{MP3ExchangeLabelled}
      \begin{fmfgraph*}(50,50)
        \fmfset{arrow_len}{6}
        \fmfstraight
        \fmfleft{v00,v01,v02}
        \fmfright{v20,v21,v22}
        \fmf{boson}{v00,v10,v20}
        \fmf{phantom}{v01,v11} \fmf{boson}{v11,v21}
        \fmf{boson}{v02,v12} \fmf{phantom}{v12,v22}
        \fmffreeze
        \fmf{fermion,left=0.25,label=$a$,label.dist=3,label.side=left}{v00,v02}
        \fmf{fermion,label=$j$,label.dist=3,label.side=right}{v02,v11}
        \fmf{fermion,right=0.25,label=$b$,label.dist=3,label.side=right}{v11,v12}
        \fmf{fermion,label=$i$,label.dist=3,label.side=left}{v12,v00}
        \fmf{fermion,left=0.25}{v20,v21,v20}
      \end{fmfgraph*}
      \end{fmffile}
    }
  &\ =
  (-1)^{2+3}\sum_{ijkabc}\frac{
    V_{ab}^{ji} V_{ik}^{ac} V_{jc}^{bk}
  }{
    (\eps_i+\eps_k-\eps_a-\eps_c)(\eps_i+\eps_j-\eps_a-\eps_b)
  } =
  -1.151\,\textnormal{m}E_h\,N
  \\
  \label{eqn:DiamondSwappedLadder}
    \diagramBox{
      \begin{fmffile}{MP3SwappedLadderLabelled}
      \begin{fmfgraph*}(50,50)
        \fmfset{arrow_len}{6}
        \fmfstraight
        \fmfleft{v00,v02,v01}
        \fmfright{v20,v22,v21}
        \fmf{boson}{v00,v10,v20}
        \fmf{phantom}{v01,v11} \fmf{boson}{v11,v21}
        \fmf{boson}{v02,v12} \fmf{phantom}{v12,v22}
        \fmffreeze
        \fmf{fermion,left=0.25,label=$a$,label.dist=3,label.side=left}{v00,v02}
        \fmf{fermion,label=$j$,label.dist=3,label.side=right}{v11,v02}
        \fmf{fermion,right=0.25,label=$b$,label.dist=3,label.side=right}{v12,v11}
        \fmf{fermion,label=$i$,label.dist=3,label.side=left}{v12,v00}
        \fmf{fermion,left=0.25}{v20,v21,v20}
      \end{fmfgraph*}
      \end{fmffile}
    }
  &\ =
  (-1)^{2+3}\sum_{ijkabc}\frac{
    V_{ai}^{jb} V_{ik}^{ac} V_{bc}^{jk}
  }{
    (\eps_i+\eps_k-\eps_a-\eps_c)(\eps_j+\eps_k-\eps_b-\eps_c)
  } =
  -1.141\,\textnormal{m}E_h\,N
\end{align}
Although the denominators differ, they yield almost the
same energy per electron. Thus, the AC-SOSEX energy for the test system in
third order hardly differs from the drCCD SOSEX energy in third order. The
latter is simply given by the exchange diagram in (\ref{eqn:DiamondExchange}):
\begin{align}
  \nonumber
  {E_{\rm c}^{\rm AC-SOSEX}}^{(3)} =&
    -\frac13\big(2\cdot1.151+1.141\big) = -1.148\,\textnormal{m}E_h\,N
  \\
  \nonumber
  {E_{\rm c}^{\rm SOSEX}}^{(3)} =&
    -1.151\,\textnormal{m}E_h\,N
\end{align}

In fourth order, the AC-SOSEX expression from (\ref{eqn:SOSEXACCases}) expands
to 24 possible permutations of the interaction times, 12 of which are distinct
Goldstone diagrams. This cancels the factor $1/2$ of (\ref{eqn:SOSEXACCases}).
The factor from the coupling strength integration is $1/4$, giving
\begin{align}
  \nonumber
  {E_{\rm c}^{\rm AC-SOSEX}}^{(4)} = \frac14 &\left(
    \diagramBox{
      \begin{fmffile}{MP4x1a}
      \begin{fmfgraph}(60,60)
        \fmfkeep{MP4x1a}
        \fmfset{arrow_len}{6}
        \fmfstraight
        \fmfleft{v00,v01,v02,v03}
        \fmfright{v30,v31,v32,v33}
        \fmf{boson}{v00,v30}
        \fmf{phantom}{v01,v21} \fmf{boson,tension=2}{v21,v31}
        \fmf{phantom}{v02,v12} \fmf{boson}{v12,v22} \fmf{phantom}{v22,v32}
        \fmf{boson,tension=2}{v03,v13} \fmf{phantom}{v13,v33}
        \fmffreeze
        \fmf{fermion,left=0.35}{v30,v31,v30}
        \fmf{fermion,left=0.35}{v21,v22,v21}
        \fmf{fermion,left=0.25}{v00,v03}
        \fmf{fermion}{v03,v12}
        \fmf{fermion,right=0.25}{v12,v13}
        \fmf{fermion}{v13,v00}
      \end{fmfgraph}
      \end{fmffile}
    }
    +\
    \diagramBox{
      \begin{fmffile}{MP4x1b}
      \begin{fmfgraph}(60,60)
        \fmfkeep{MP4x1b}
        \fmfset{arrow_len}{6}
        \fmfstraight
        \fmfleft{v00,v02,v01,v03}
        \fmfright{v30,v32,v31,v33}
        \fmf{boson}{v00,v30}
        \fmf{phantom}{v01,v21} \fmf{boson,tension=2}{v21,v31}
        \fmf{phantom}{v02,v12} \fmf{boson}{v12,v22} \fmf{phantom}{v22,v32}
        \fmf{boson,tension=2}{v03,v13} \fmf{phantom}{v13,v33}
        \fmffreeze
        \fmf{fermion,left=0.25}{v30,v31,v30}
        \fmf{fermion,left=0.35}{v21,v22,v21}
        \fmf{fermion,left=0.25}{v00,v03}
        \fmf{fermion}{v03,m1} \fmf{plain}{m1,v12}
        \fmf{fermion,right=0.25}{v12,v13}
        \fmf{plain}{v13,m2} \fmf{fermion}{m2,v00}
      \end{fmfgraph}
      \end{fmffile}
    }
    +
    \diagramBox{
      \begin{fmffile}{MP4x1c}
      \begin{fmfgraph}(60,60)
        \fmfkeep{MP4x1c}
        \fmfset{arrow_len}{6}
        \fmfstraight
        \fmfleft{v01,v00,v02,v03}
        \fmfright{v31,v30,v32,v33}
        \fmf{boson}{v00,v30}
        \fmf{phantom}{v01,v21} \fmf{boson,tension=2}{v21,v31}
        \fmf{phantom}{v02,v12} \fmf{boson}{v12,v22} \fmf{phantom}{v22,v32}
        \fmf{boson,tension=2}{v03,v13} \fmf{phantom}{v13,v33}
        \fmf{phantom,tension=6}{v00,l}
          \fmf{phantom,tension=21}{l,r} \fmf{phantom,tension=14}{r,v30}
        \fmffreeze
        \fmf{fermion,left=0.35}{v30,v31,v30}
        \fmf{fermion,left=0.125}{v21,l} \fmf{plain,left=0.125}{l,v22}
        \fmf{fermion,left=0.125}{v22,r} \fmf{plain,left=0.125}{r,v21}
        \fmf{fermion,left=0.25}{v00,v03}
        \fmf{fermion}{v03,v12}
        \fmf{fermion,right=0.25}{v12,v13}
        \fmf{fermion}{v13,v00}
      \end{fmfgraph}
      \end{fmffile}
    }
    +
  \right. \\[2ex]
  &\left.
    \hspace{1.75ex}
    \diagramBox{
      \begin{fmffile}{MP4x2a}
      \begin{fmfgraph}(60,60)
        \fmfkeep{MP4x2a}
        \fmfset{arrow_len}{6}
        \fmfstraight
        \fmfleft{v00,v01,v03,v02}
        \fmfright{v30,v31,v33,v32}
        \fmf{boson}{v00,v30}
        \fmf{phantom}{v01,v21} \fmf{boson,tension=2}{v21,v31}
        \fmf{phantom}{v02,v12} \fmf{boson}{v12,v22} \fmf{phantom}{v22,v32}
        \fmf{boson,tension=2}{v03,v13} \fmf{phantom}{v13,v33}
        \fmffreeze
        \fmf{fermion,left=0.35}{v30,v31,v30}
        \fmf{fermion,left=0.25}{v21,v22,v21}
        \fmf{fermion,left=0.25}{v00,v03}
        \fmf{fermion}{v12,v03}
        \fmf{fermion,right=0.25}{v13,v12}
        \fmf{fermion}{v13,v00}
      \end{fmfgraph}
      \end{fmffile}
    }
    +\
    \diagramBox{
      \begin{fmffile}{MP4x2b}
      \begin{fmfgraph}(60,60)
        \fmfkeep{MP4x2b}
        \fmfset{arrow_len}{6}
        \fmfstraight
        \fmfleft{v00,v02,v03,v01}
        \fmfright{v30,v32,v33,v31}
        \fmf{boson}{v00,v30}
        \fmf{phantom}{v01,v21} \fmf{boson,tension=2}{v21,v31}
        \fmf{phantom}{v02,v12} \fmf{boson}{v12,v22} \fmf{phantom}{v22,v32}
        \fmf{boson,tension=2}{v03,v13} \fmf{phantom}{v13,v33}
        \fmffreeze
        \fmf{fermion,left=0.2}{v30,v31,v30}
        \fmf{fermion,left=0.25}{v21,v22,v21}
        \fmf{fermion,left=0.25}{v00,v03}
        \fmf{fermion}{v03,v12}
        \fmf{fermion,right=0.25}{v12,v13}
        \fmf{fermion}{v13,v00}
      \end{fmfgraph}
      \end{fmffile}
    }
    +
    \diagramBox{
      \begin{fmffile}{MP4x2c}
      \begin{fmfgraph}(60,60)
        \fmfkeep{MP4x2c}
        \fmfset{arrow_len}{6}
        \fmfstraight
        \fmfleft{v01,v00,v03,v02}
        \fmfright{v31,v30,v33,v32}
        \fmf{boson}{v00,v30}
        \fmf{phantom}{v01,v21} \fmf{boson,tension=2}{v21,v31}
        \fmf{phantom}{v02,v12} \fmf{boson}{v12,v22} \fmf{phantom}{v22,v32}
        \fmf{boson,tension=2}{v03,v13} \fmf{phantom}{v13,v33}
        \fmffreeze
        \fmf{fermion,left=0.35}{v30,v31,v30}
        \fmf{fermion,left=0.2}{v21,v22,v21}
        \fmf{fermion,left=0.25}{v00,v03}
        \fmf{fermion}{v12,v03}
        \fmf{fermion,right=0.25}{v13,v12}
        \fmf{fermion}{v13,v00}
      \end{fmfgraph}
      \end{fmffile}
    }
    +\ t.r.
  \right)\,,
  \label{eqn:SOSEX4AC}
\end{align}
where $t.r.$ denotes the time reversed variants of the shown 6 diagrams.
Note that the diagrams are drawn such that the permutations of the
interaction times are evident and not according to minimal self intersection.
The diagrams can be evaluated by iterating the doubles amplitudes $t_{ij}^{ab}$
employing only a subset of the direct ring Coupled Cluster Doubles amplitude
equations. The following steps are for instance used to calculate the second
diagram in (\ref{eqn:SOSEX4AC}) indicating the employed part of the drCCD
amplitude equation above each equals sign:
\begin{align}
  \nonumber
  \diagramBoxBorder{1ex}{0ex}{
    \begin{fmffile}{CCDtLeftT0}
    \begin{fmfgraph*}(48,48)
      \fmfkeep{CCDtLeftT0}
      \fmfset{arrow_len}{6}
      \fmfstraight
      \fmfleft{v00,v01,v02,v03}
      \fmfright{v60,v61,v62,v63}
      \fmf{phantom,tension=4}{v02,v12} \fmf{dashes,label=$t^{(0)}$}{v12,v52}
        \fmf{phantom,tension=4}{v52,v62}
      \fmf{phantom}{v03,v23,v43,v63}
      \fmffreeze
      \fmf{phantom}{v03,v13,v23}
      \fmf{fermion,left=0.1}{v23,v12,v03}
      \fmf{fermion,left=0.1}{v63,v52,v43}
    \end{fmfgraph*}
    \end{fmffile}
  }
  \hspace*{-1ex}
  \quad
  {t_{ij}^{ab}}^{(0)}
    &\stackrel{(\ref{eqn:drCCDV})}{=}
    \frac{V_{ij}^{ab}}{\eps_i+\eps_j-\eps_a-\eps_b}
  \quad
  \hspace*{3ex}
  \diagramBox{
    \begin{fmffile}{CCDv0}
    \begin{fmfgraph}(48,48)
      \fmfset{arrow_len}{6}
      \fmfstraight
      \fmfleft{v00,v01,v02,v03}
      \fmfright{v60,v61,v62,v63}
      \fmf{phantom,tension=4}{v01,v11} \fmf{boson}{v11,v51}
        \fmf{phantom,tension=4}{v51,v61}
      \fmf{phantom}{v03,v23,v43,v63}
      \fmffreeze
      \fmf{fermion,left=0.1}{v23,v11,v03}
      \fmf{fermion,left=0.1}{v63,v51,v43}
    \end{fmfgraph}
    \end{fmffile}
  }
  \\
  \nonumber
  \diagramBoxBorder{1ex}{0ex}{
    \begin{fmffile}{CCDtLeftT1}
    \begin{fmfgraph*}(48,48)
      \fmfkeep{CCDtLeftT1}
      \fmfset{arrow_len}{6}
      \fmfstraight
      \fmfleft{v00,v01,v02,v03}
      \fmfright{v60,v61,v62,v63}
      \fmf{phantom,tension=4}{v02,v12} \fmf{dashes,label=$t^{(1)}$}{v12,v52}
        \fmf{phantom,tension=4}{v52,v62}
      \fmf{phantom}{v03,v23,v43,v63}
      \fmffreeze
      \fmf{phantom}{v03,v13,v23}
      \fmf{fermion,left=0.1}{v23,v12,v03}
      \fmf{fermion,left=0.1}{v63,v52,v43}
    \end{fmfgraph*}
    \end{fmffile}
  }
  \hspace*{-1ex}
  \quad
  {t_{ij}^{ab}}^{(1)}
    &\stackrel{(\ref{eqn:drCCDtVt})}{=}
    \frac{
      \sum_{klcd}{t_{ik}^{ac}}^{(0)}V_{cd}^{kl}{t_{lj}^{db}}^{(0)}
    }{
      \eps_i+\eps_j-\eps_a-\eps_b
    }
  \quad
  \diagramBox{
    \begin{fmffile}{CCDtvt0}
    \begin{fmfgraph*}(96,48)
      \fmfset{arrow_len}{6}
      \fmfstraight
      \fmfleft{v00,v01,v02,v03}
      \fmfright{v120,v121,v122,v123}
      \fmf{phantom}{v01,v41} \fmf{boson}{v41,v81} \fmf{phantom}{v81,v121}
      \fmf{phantom,tension=12}{v00,v10}
        \fmf{dashes,tension=4,label=$t^{(0)}$}{v10,v40}
        \fmf{phantom,tension=3}{v40,v80}
        \fmf{dashes,tension=4,label=$t^{(0)}$}{v80,v110}
        \fmf{phantom,tension=12}{v110,v120}
      \fmf{phantom}{v03,v23,v43,v63,v83,v103,v123}
      \fmffreeze
      \fmf{phantom}{v81,v91,v101,v111,v121}
      \fmf{phantom}{v103,v113,v123}
      \fmf{fermion,left=0.1}{v23,v10,v03}
      \fmf{fermion,left=0.1}{v123,v110,v103}
      \fmf{fermion,left=0.4,label=$c$,label.dist=4}{v40,v41}
        \fmf{fermion,left=0.4,label=$k$,label.dist=4}{v41,v40}
      \fmf{fermion,left=0.4,label=$d$,label.dist=4}{v80,v81}
        \fmf{fermion,left=0.4,label=$l$,label.dist=4}{v81,v80}
    \end{fmfgraph*}
    \end{fmffile}
  }
    \\
  \nonumber
  \diagramBox{\fmfreuse{MP4x1b}}\ =
  \diagramBox{
    \begin{fmffile}{drCCDxt1}
    \begin{fmfgraph*}(30,30)
      \fmfkeep{drCCDxt1}
      \fmfset{arrow_len}{6}
      \fmfstraight
      \fmfbottom{lb,rb}
      \fmftop{lt,rt}
      \fmf{dashes,label=$t^{(1)}$}{lb,rb}
      \fmf{boson}{lt,rt}
      \fmf{fermion,left=0.25}{lb,lt}
        \fmf{fermion}{lt,m1} \fmf{plain}{m1,rb}
      \fmf{fermion,right=0.25}{rb,rt}
        \fmf{fermion}{rt,m2} \fmf{plain}{m2,lb}
    \end{fmfgraph*}
    \end{fmffile}
  }
  \  &\stackrel{(\ref{eqn:SOSEXdrCCD})}{=}
    -\frac12\, \sum_{ijab} {t_{ij}^{ab}}^{(1)} V_{ab}^{ji}\,.
\end{align}
Evaluating the second row of diagrams in (\ref{eqn:SOSEX4AC}) requires two
additional parts of the doubles amplitude equation that are not part of the
drCCD amplitude equation:
\begin{align}
  \diagramBoxBorder{2ex}{0ex}{
    \fmfreuse{CCDtLeft}
  }
  \quad
  t_{ij}^{ab} &=
  \frac{-\sum_{kc}V_{ia}^{ck}t_{kj}^{cb}}{
    \eps_i+\eps_j-\eps_a-\eps_b
  }
  \quad
  \hspace*{4ex}
  \diagramBox{
    \begin{fmffile}{CCDswappedvt}
    \begin{fmfgraph*}(64,48)
      \fmfset{arrow_len}{6}
      \fmfstraight
      \fmfleft{v00,v01,v02,v03}
      \fmfright{v80,v81,v82,v83}
      \fmf{dots,tension=12}{v01,v11} \fmf{boson,tension=4}{v11,v41}
        \fmf{dots,tension=3}{v41,v81}
      \fmf{phantom,tension=3}{v00,v40} \fmf{dashes,tension=4}{v40,v70}
        \fmf{phantom,tension=12}{v70,v80}
      \fmf{dots}{v03,v23,v43,v63,v83}
      \fmffreeze
      \fmf{phantom}{v41,v51,v61,v71,v81}
      \fmf{phantom}{v63,v73,v83}
      \fmf{fermion}{v41,m1} \fmf{plain}{m1,v03}
      \fmf{plain}{v23,m2} \fmf{fermion}{m2,v11}
      \fmf{fermion,left=0.1}{v83,v70,v63}
      \fmf{fermion,left=0.1,label=$c$,label.dist=4}{v40,v11}
        \fmf{fermion,left=0.4,label=$k$,label.dist=4}{v41,v40}
      \fmfv{label=$a$,label.angle=90}{v03}
      \fmfv{label=$i$,label.angle=90}{v23}
      \fmfv{label=$b$,label.angle=90}{v63}
      \fmfv{label=$j$,label.angle=90}{v83}
      \fmfv{label=$t=0$,label.angle=0,label.dist=16}{v73}
      \fmfv{label=$t$,label.angle=0,label.dist=16}{v71}
    \end{fmfgraph*}
    \end{fmffile}
  }
  \label{eqn:CCDswappedvt}
  \\[2ex]
  \diagramBoxBorder{2ex}{0ex}{
    \fmfreuse{CCDtLeft}
  }
  \quad
  t_{ij}^{ab} &=
  \frac{-\sum_{klcd}t_{ik}^{ac}V_{cd}^{lk}t_{lj}^{db}}{
    \eps_i+\eps_j-\eps_a-\eps_b
  }
  \quad
  \diagramBox{
    \begin{fmffile}{CCDtxt}
    \begin{fmfgraph*}(96,48)
      \fmfset{arrow_len}{6}
      \fmfstraight
      \fmfleft{v00,v01,v02,v03}
      \fmfright{v120,v121,v122,v123}
      \fmf{dots}{v01,v41} \fmf{boson}{v41,v81} \fmf{dots}{v81,v121}
      \fmf{phantom,tension=12}{v00,v10} \fmf{dashes,tension=4}{v10,v40}
        \fmf{phantom,tension=3}{v40,v80} \fmf{dashes,tension=4}{v80,v110}
        \fmf{phantom,tension=12}{v110,v120}
      \fmf{dots}{v03,v23,v43,v63,v83,v103,v123}
      \fmffreeze
      \fmf{phantom}{v81,v91,v101,v111,v121}
      \fmf{phantom}{v103,v113,v123}
      \fmf{fermion,left=0.1}{v23,v10,v03}
      \fmf{fermion,left=0.1}{v123,v110,v103}
      \fmf{fermion,left=0.4,label=$c$,label.dist=4}{v40,v41}
        \fmf{plain}{v41,m1} \fmf{fermion,label=$l$,label.dist=4}{m1,v80}
      \fmf{fermion,right=0.4,label=$d$,label.dist=4}{v80,v81}
        \fmf{plain}{v81,m2} \fmf{fermion,label=$k$,label.dist=4}{m2,v40}
      \fmfv{label=$a$,label.angle=90}{v03}
      \fmfv{label=$i$,label.angle=90}{v23}
      \fmfv{label=$b$,label.angle=90}{v103}
      \fmfv{label=$j$,label.angle=90}{v123}
      \fmfv{label=$t=0$,label.angle=0,label.dist=16}{v113}
      \fmfv{label=$t$,label.angle=0,label.dist=16}{v111}
    \end{fmfgraph*}
    \end{fmffile}
  }
  \label{eqn:CCDtxt}
\end{align}
In the case of the swapped ladder amplitude equation (\ref{eqn:CCDswappedvt})
there is one additional hole $k$ but no additional loop, resulting in a
negative sign. In the exchange amplitude equation (\ref{eqn:CCDtxt})
there are two additional holes and one additional loop also giving a negative
sign. The latter is part of the full Coupled Cluster Doubles (CCD)
amplitude equations while the swapped ladder equation only occurs in the
AC-SOSEX. Table \ref{tab:SOSEXDiamondOrders} lists the energies per electron
of all diagrams of the AC-SOSEX in third and fourth order modulo time reversal
symmetry. The AC-SOSEX energy in fourth order is thus
\[
  {E_{\rm c}^{\rm AC-SOSEX}}^{(4)} =
  \frac12\big(0.212+0.105+0.208+0.208+0.103+0.209\big) =
  +0.522\,\textnormal{m}E_h\,N\,.
\]
The drCCD SOSEX energy is formed by the diagrams of the first row in
Table \ref{tab:SOSEXDiamondOrders} only, yielding a very similar result:
\[
  {E_{\rm c}^{\rm SOSEX}}^{(4)} =
    0.212+0.105+0.208 =
    +0.525\,\textnormal{m}E_h\,N\,.
\]

\begin{table}
\begin{center}
  \begin{tabular}{|c|c|ccc|}
    \hline
    \multirow{2}{*}{\textbf{C(A4)}} &
    \multirow{2}{*}{\textbf{3\textsuperscript{rd} order}} &
                    \multicolumn{3}{c|}{\textbf{4\textsuperscript{th} order}}
    \\
    & &  (a) & (b) & (c)
    \\\hline &&&& \\[-1ex]
    \textbf{SOSEX} & \ \ \diagramBox{\fmfreuse{MP3Exchange}}\ \ &
                       \ \ \diagramBox{\fmfreuse{MP4x1a}}\ \ &
                       \ \ \diagramBox{\fmfreuse{MP4x1b}}\ \ &
                       \ \ \diagramBox{\fmfreuse{MP4x1c}}
    \\[7ex]
    $E$ [m$E_h\,N$] & $-1.151$ & $+0.212$ & $+0.105$ & $+0.208$
    \\\hline &&&& \\[-1ex]
    \textbf{AC-SOSEX}& \diagramBox{\fmfreuse{MP3SwappedLadder}} &
                       \diagramBox{\fmfreuse{MP4x2a}} &
                       \diagramBox{\fmfreuse{MP4x2b}} &
                       \diagramBox{\fmfreuse{MP4x2c}}
    \\[7ex]
    $E$ [m$E_h\,N$] & $-1.141$ & $+0.208$   & $+0.103$ & $+0.209$
    \\ \hline
  \end{tabular}
\end{center}
\caption{
  Goldstone diagrams forming the third and the fourth order of the drCCD SOSEX
  and the AC-SOSEX correction. The drCCD SOSEX consists of diagrams of the
  first row only.
  A weighted average of the first and second row yields the AC-SOSEX energy.
}
\label{tab:SOSEXDiamondOrders}
\end{table}

The top row in Table \ref{tab:SOSEXDiamondOrders} shows diagrams where the last
interaction is anti-symmetrized while in the second row it is the second last
interaction that is anti-symmetrized.
The table indicates that the energy of the SOSEX or AC-SOSEX diagrams hardly
depends on which of the interactions is anti-symmetrized when using the swapped
ladder diagram instead of the proper ladder diagram.
Finally, the factor $1/n$ from the coupling strength integration simply
averages the energies from anti-symmetrizing each of the $n$ interactions,
as shown in (\ref{eqn:SOSEX3AC}) for third order and in (\ref{eqn:SOSEX4AC})
for fourth order.
Since the energies of the diagrams are very similar anti-symmetrizing
each of the $n$ interactions,
the average is also very similar to the energy of the diagrams where only
the last interaction is anti-symmetrized. The averaged energy forms the
AC-SOSEX energy while the latter forms the drCCD SOSEX originally proposed by
Freeman.
This argument holds at least up to fourth order in the diamond test system
and the truncation at fourth order describes the drCCD SOSEX already to an
accuracy of 5\%.

\subsection{Uniform electron gas}
In a metallic system it is not possible to truncate the AC-SOSEX expression at
any finite order beyond the second order since all of the described higher
order exchange diagrams diverge. It is, however, possible to study the
AC-SOSEX diagrams of (\ref{eqn:SOSEXACCases})
numerically for the uniform electron gas (UEG) as the prototypical metal.
In the UEG we let spin and momenta denote the states $i,j,a$ and $b$ occurring
in the diagrams. For the third and the fourth diagram of
(\ref{eqn:SOSEXACCases}) we choose for instance the following definition of
momenta:
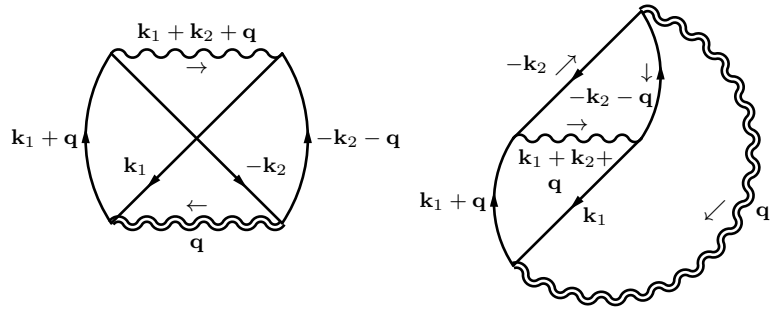
\[
  \diagramBoxBorder{3ex}{3ex}{
    \begin{fmffile}{ACSOSEX3Labeled}
      \scriptsize
      \begin{fmfgraph*}(64,64)
        \fmfkeep{ACSOSEX3Labeled}
        \fmfstraight
        \fmfset{arrow_len}{6}
        \fmfleft{v00,v01}
        \fmfright{v10,v11}
        \fmf{photon,label.side=right,label.dist=-10,label=$
          \begin{array}{c}
            \vec k_1+\vec k_2+\vec q \\[1ex]
            \rightarrow
          \end{array}
        $}{v11,v01}
        \fmf{dbl_wiggly,label.side=left,label.dist=-12,label=$
          \begin{array}{c}
            \leftarrow \\[1.2ex]
            \vec q
          \end{array}
        $}{v00,v10}
        \fmffreeze
        \fmf{fermion,left=0.3,label.dist=3,label=$
          \vec k_1+\vec q
        $}{v00,v01}
        \fmf{plain}{v01,m1}
          \fmf{fermion,label.side=left,label.dist=3,label=$
            -\vec k_2
          $}{m1,v10}
        \fmf{fermion,right=0.3,label.dist=3,label=$
          -\vec k_2-\vec q
        $}{v10,v11}
        \fmf{plain}{v11,m2}
          \fmf{fermion,label.side=right,label.dist=3,label=$
            \vec k_1
          $}{m2,v00}
      \end{fmfgraph*}
    \end{fmffile}
  }
  \qquad
  \hspace*{10ex}
  \diagramBoxBorder{0ex}{3ex}{
    \begin{fmffile}{ACSOSEX4Labeled}
    \scriptsize
    \begin{fmfgraph*}(48,96)
      \fmfstraight
      \fmfset{arrow_len}{6}
      \fmfleft{v00,v01,v02}
      \fmfright{v10,v11,v12}
      \fmf{photon,label.side=right,label.dist=-22,label=$
        \begin{array}{c}
          \rightarrow\\[1ex]
          \vec k_1+\vec k_2+\quad\\
          \vec q\quad\quad
        \end{array}
      $}{v11,v01}
      \fmf{dbl_wiggly,right=1.25,label.dist=-14,label=$
        \swarrow\quad\vec q
      $}{v00,v12}
      \fmffreeze
      \fmf{fermion,left=0.3,label.side=left,label.dist=3,label=$
        \vec k_1+\vec q
      $}{v00,v01}
      \fmf{fermion,label.dist=3,label.side=left,label=$
        \vec k_1
      $}{v11,v00}
      \fmf{fermion,right=0.3,label.side=left,label.dist=-2,label=$
        \begin{array}{r}\\[-0.5ex]\downarrow\\-\vec k_2-\vec q\end{array}
      $}{v11,v12}
      \fmf{fermion,label.dist=0.1,label.side=right,label=$
        -\vec k_2\nearrow
      $}{v12,v01}
    \end{fmfgraph*}
    \end{fmffile}
  }
\]
$\vec k_1$ and $-\vec k_2$ are required to be hole states while
$\vec k_1+\vec q$ and $-\vec k_2-\vec q$ must be particle states.
Thus, $\vec k_i$ must be below and $\vec k_i+\vec q$ must be above the Fermi
momentum $k_{\rm F}$ for $i\in\{1,2\}$. Note that in the swapped ladder diagram
on the right the momenta $-\vec k_2$ and $-\vec k_2-\vec q$ are defined
opposite to the propagation directions of the respective states, as indicated
by the arrows next to the labels. The spins of all four states must be the
same.
In the UEG the sum over the states is replaced by the sum over the spin and
integrals over all internal momenta according to (\ref{eqn:UEGSumOverStates}),
arriving at
\begin{align}
  \nonumber
  E_{\rm c}^{\rm AC-SOSEX} =&
    +\frac\Omega N \frac12 \int\frac{\d\nu}{2\pi}\,
    \int \frac{\d\vec q}{(2\pi)^3}\,
    \sum_\sigma\iint\limits_{|\vec k_i|<k_{\rm F}<|\vec k_i+\vec q|}
      \frac{\Omega^2 \d\vec k_1 \d\vec k_2}{(2\pi)^6}\,
    V(\vec k_1+\vec k_2+\vec q)
    \overline W(\vec q,\nu)\,
  \\
  \nonumber
  &
    \left(
      \frac1{
        (\Delta\eps_{\vec k_1,\vec q}-\im\nu)
        (\Delta\eps_{-\vec k_2,-\vec q}-\im\nu)
      } +
      \frac1{
        (\Delta\eps_{\vec k_1,\vec q}-\im\nu)
        (\Delta\eps_{-\vec k_2,-\vec q}+\im\nu)
      } +
    \right.
  \\
  &
  \hspace*{1.5ex}
    \left.
      \frac1{
        (\Delta\eps_{\vec k_1,\vec q}+\im\nu)
        (\Delta\eps_{-\vec k_2,-\vec q}-\im\nu)
      } +
      \frac1{
        (\Delta\eps_{\vec k_1,\vec q}+\im\nu)
        (\Delta\eps_{-\vec k_2,-\vec q}+\im\nu)
      }
    \right)\,,
  \label{eqn:SOSEX_ACUEG}
\end{align}
with $i\in\{1,2\}$ and the single particle excitation energy
$\Delta\eps_{\vec k_i,\vec q} = (\vec k_i+\vec q)^2/2 - \vec k_i^2/2$.
The same expression can be derived using imaginary frequency propagators
defined in (\ref{eqn:RPAUEGPropagatorImag}) and integrating out the two
additional imaginary frequencies analogous to the derivation of $\chi_0$ in
(\ref{eqn:RPAChiFreqInt}).

Integrating out $\vec k_1$ and $\vec k_2$ turns out to be a tedious task
for the above equation. Although there are closed expressions for the second
and the third case they are overly complicated.
For the swapped ladder diagrams in the first and the fourth case no
such expressions were found.
However, a straight-forward Monte-Carlo integration of $\vec k_1$ and
$\vec k_2$ has proven to be sufficiently accurate when sampling the momenta
$\vec k_i$ with a probability density function (PDF) given by
\[
  \PDF(\vec k_i) \propto
    \left\{
      \begin{array}{ll}
        \displaystyle
        \left|\frac1{\Delta\eps_{\vec k_i,\vec q}\pm\im\nu}\right| &
        \textnormal{for }|\vec k_i|<k_{\rm F}<|\vec k_i+\vec q|\,, \\[2ex]
        \displaystyle 0 & \textnormal{otherwise.}
      \end{array}
    \right.
\]
Figure \ref{fig:SOSEXUEG} and Table \ref{tab:SOSEX_UEG} show the resulting
AC-SOSEX energies as a function of density given by the Wigner-Seitz radius,
$r_s$. The uncertainties from the integrations are indicated by the error bars. 
For the Monte-Carlo integration of $\vec k_1$ and $\vec k_2$ a precision of 5
significant digits can be achieved with less than $30000$ samples for each $q$
and $\nu$, depending on momentum, frequency and density.
The error from the momentum and imaginary frequency integration is of
similar magnitude.

The differences between the two SOSEX variants are not as small as for an
isolating system but they are still below 3\% for the density range with
$r_s\leq10$. For lower densities no drCCD SOSEX reference values were found.

\section*{Summary}
The Random Phase Approximation (RPA) systematically overestimates the negative
correlation energy. This originates, at least partially, from violations of
the Pauli exclusion principle in the ring diagrams of the RPA such as
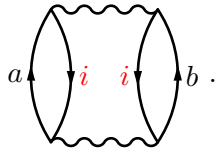
\[
  \diagramBox{\fmfreuse{Bubbles2Pauli}}\hspace*{3ex}.
\]
By the merit of Wick's theorem, violations of the Pauli exclusion
principle do not have to be considered as long as all contractions
of the occurring operators are included. The contractions correcting
these violations are those from the respective exchange diagrams where
the offending states are crossed by anti-symmetrizing an affected
Coulomb interaction.

Thus, the lowest order correction to the Random Phase Approximation
anti-symmetrizes one Coulomb interaction occurring in the ring diagrams.
If this interaction is the last interaction in time the respective
correction is termed Second Order Screened Exchange (SOSEX) containing the
following diagrams:
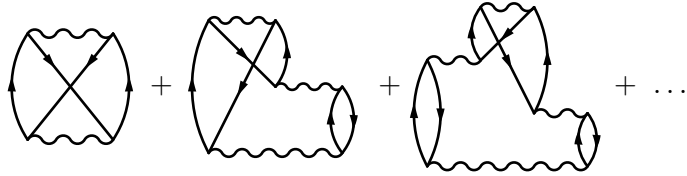
\[
  \diagramBox{
    \fmfreuse{Exchange2}
  }
  +
  \
  \diagramBox{
    \fmfreuse{BubblesExchange3}
  }
  \ +
  \diagramBoxBorder{0ex}{2ex}{
    \fmfreuse{BubblesExchange4}
  }
  +\ \ldots
\]
The SOSEX can
only be computed from the direct ring Coupled Cluster Doubles (drCCD)
amplitudes $\diagramBox{\fmfreuse{CCDt}}$ since monotonicity in time is
required to be able to anti-symmetrize only the last interaction. When
calculating the RPA using the drCCD amplitudes the SOSEX can be easily
computed with hardly any additional costs. However, calculating the drCCD
amplitudes requires $\mathcal O(N^4)$ of memory limiting the applicability of
RPA+SOSEX to rather small systems.

To overcome the limitations regarding memory consumption the
Adiabatic-Connection (AC) SOSEX can be used requiring only $\mathcal O(N^2)$
of memory. It yields very similar results compared to the SOSEX although
it contains terms that have no correspondence in many-body perturbation
theory where particles turn into holes and vice-versa herein called swapped
ladder diagrams.
This is due to the numerical oddity that it hardly matters which of the $n$
occurring interactions in $n$-th order are anti-symmetrized as long as
the swapped ladder diagrams are used rather than actual ladder diagrams.
In third order this means for instance:
\[
  \diagramBox{\fmfreuse{MP3Exchange}}
  \hspace*{1ex} \approx \hspace*{1ex} 
  \diagramBox{\fmfreuse{MP3SwappedLadder}}
  \hspace*{2ex} \approx \hspace*{2ex} 
  \diagramBox{\fmfreuse{MP3ExchangeUp}}
  \hspace*{1ex} ,
\]
such that
\[
  \underbrace{
    \frac 1 3 \left\{
      \hspace*{2ex}
      \diagramBoxBorder{0ex}{1ex}{\fmfreuse{MP3Exchange}}
      \hspace*{1ex} + \hspace*{1ex} 
      \diagramBox{\fmfreuse{MP3SwappedLadder}}
      \hspace*{2ex} + \hspace*{2ex}
      \diagramBox{\fmfreuse{MP3ExchangeUp}}
      \hspace*{1.5ex}
    \right\}
  }_{
    {E_{\rm c}^{\rm AC-SOSEX}}^{(3)}
  }
  \approx
  \underbrace{
    \hspace*{2ex} \diagramBoxBorder{0ex}{1ex}{\fmfreuse{MP3Exchange}}
  }_{
    {E_{\rm c}^{\rm SOSEX}}^{(3)}
  }\ .     
\]
Thus, the AC-SOSEX can be considered a recipe for imitating the SOSEX energy
while reducing the memory requirements to $\mathcal O(N^2)$.

Apart from technical convenience when having the drCCD amplitudes at
hand, there is however no reason to limit the considered exchange diagrams
to those where only the last interaction is anti-symmetrized. From an
\textit{ab-initio} point of view one should consider all diagrams where
one interaction is anti-symmetrized and where violations of the Pauli
principle can occur as the lowest order correction to the RPA.
This is discussed in the next chapter.



  \chapter{Adjacent Pairs Exchange}
\label{cha:APX}

The Second Order Screened Exchange (SOSEX) correction to the Random Phase
Approximation (RPA) arises from anti-symmetrizing only
the last Coulomb interaction of RPA's ring diagrams as shown in Figure
\ref{fig:SOSEXDiagrams}. When calculating the RPA using the direct ring
Coupled Cluster Doubles (drCCD) amplitudes $t_{ij}^{ab}$, as introduced by
\parencite{freeman_coupled-cluster_1977}, this comes at virtually no extra
costs and represents a natural choice for the lowest order correction to the
RPA.
However, when calculating the RPA in the frequency domain, which is more
efficient, there is a priori no reason to choose this particular class of
diagrams as the lowest order correction to the RPA. Furthermore, it is not
possible to evaluate the SOSEX diagrams directly since the last interaction
in time cannot be explicitly addressed in the frequency domain. The
AC-SOSEX approach, discussed in Section {\ref{sec:SOSEX_ACSOSEX}}, offers an
approximation but it contains swapped ladder diagrams that are not part of the
many-body perturbation expansion. So the question remains, which diagrams,
that are actually part of the many-body perturbation expansion, can be
efficiently evaluated in the frequency domain and offer a good
lowest order correction to the systematic error of the
Random Phase Approximation.

\section{Drivation of the Adjacent Pairs Exchange correction}
As discussed in the beginning of Chapter \ref{cha:SOSEX}, violations of the
Pauli exclusion principle in RPA's ring diagrams suggest to anti-symmetrize
the Coulomb interactions wherever such a violation can occur. In lowest
order it should be only one but not necessarily just the last Coulomb
interaction to be anti-symmetrized. This is done by cutting out one
Coulomb interaction including the adjacent pair bubbles from the RPA ring
diagrams and anti-symmetrizing this interaction if that can correct for a
violation of the Pauli exclusion principle.

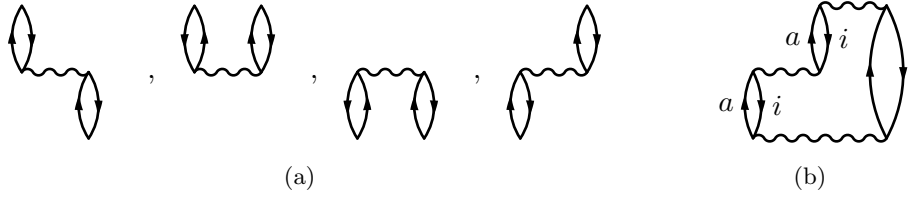
\begin{figure}
\begin{center}
  \subfigure[]{
    \diagramBoxBorder{0ex}{2ex}{
      \begin{fmffile}{Bubbles2UpDown}
      \begin{fmfgraph}(25,50)
        \fmfkeep{Bubbles2UpDown}
        \fmfset{arrow_len}{6}
        \fmfstraight
        \fmfleft{v11,v12,v13}
        \fmfright{v21,v22,v23}
        \fmffreeze
        \fmf{photon}{v12,v22}
        \fmf{fermion,left=0.3}{v12,v13,v12}
        \fmf{fermion,left=0.3}{v21,v22,v21}
      \end{fmfgraph}
      \end{fmffile}
    }
    ,
    \diagramBox{
      \begin{fmffile}{Bubbles2Up}
      \begin{fmfgraph}(25,50)
        \fmfkeep{Bubbles2Up}
        \fmfset{arrow_len}{6}
        \fmfstraight
        \fmfleft{v11,v12,v13}
        \fmfright{v21,v22,v23}
        \fmffreeze
        \fmf{photon}{v12,v22}
        \fmf{fermion,right=0.3}{v13,v12,v13}
        \fmf{fermion,left=0.3}{v23,v22,v23}
      \end{fmfgraph}
      \end{fmffile}
    }
    ,
    \diagramBox{
      \begin{fmffile}{Bubbles2Down}
      \begin{fmfgraph}(25,50)
        \fmfkeep{Bubbles2Down}
        \fmfset{arrow_len}{6}
        \fmfstraight
        \fmfleft{v21,v22,v23}
        \fmfright{v11,v12,v13}
        \fmffreeze
        \fmf{fermion,left=0.3}{v11,v12,v11}
        \fmf{fermion,right=0.3}{v21,v22,v21}
        \fmf{photon}{v12,v22}
      \end{fmfgraph}
      \end{fmffile}
    }
    ,
    \diagramBox{
      \begin{fmffile}{Bubbles2DownUp}
      \begin{fmfgraph}(25,50)
        \fmfkeep{Bubbles2DownUp}
        \fmfset{arrow_len}{6}
        \fmfstraight
        \fmfleft{v11,v12,v13}
        \fmfright{v21,v22,v23}
        \fmffreeze
        \fmf{photon}{v12,v22}
        \fmf{fermion,left=0.3}{v11,v12,v11}
        \fmf{fermion,left=0.3}{v22,v23,v22}
      \end{fmfgraph}
      \end{fmffile}
    }
    \label{sfg:APX_BubbleCases}
  }
  \qquad
  \subfigure[]{
    \diagramBoxBorder{0ex}{2ex}{
      \begin{fmffile}{Bubbles3Pauli}
      \begin{fmfgraph*}(50,50)
        \fmfset{arrow_len}{6}
        \fmfstraight
        \fmfleft{v00,v02,v01}
        \fmfright{v20,v22,v21}
        \fmf{boson}{v00,v10,v20}
        \fmf{phantom}{v01,v11} \fmf{boson}{v11,v21}
        \fmf{boson}{v02,v12} \fmf{phantom}{v12,v22}
        \fmffreeze
        \fmf{fermion,left=0.25,label=$a$,label.dist=4}{v00,v02}
        \fmf{fermion,left=0.25,label=$i$,label.dist=4}{v02,v00}
        \fmf{fermion,left=0.25,label=$i$,label.dist=4}{v11,v12}
        \fmf{fermion,left=0.25,label=$a$,label.dist=4}{v12,v11}
        \fmf{fermion,left=0.25}{v20,v21}
        \fmf{fermion,left=0.25}{v21,v20}
      \end{fmfgraph*}
      \end{fmffile}
    }
    \label{sfg:APX_Bubbles3Pauli}
  }
\end{center}
\caption{
  \subref{sfg:APX_BubbleCases} The possible time orders of a Coulomb
  interaction and its two adjacent pair bubbles.
  \subref{sfg:APX_Bubbles3Pauli} Anti-symmetrizing the first and the last case
  only cancels contributions $i$ and $a$ as indicated here for example.
  In general, these contributions do not violate the Pauli exclusion principle.
}
\end{figure}
\subsection{Two Sided Adjacent Pairs Exchange}
\begin{figure}
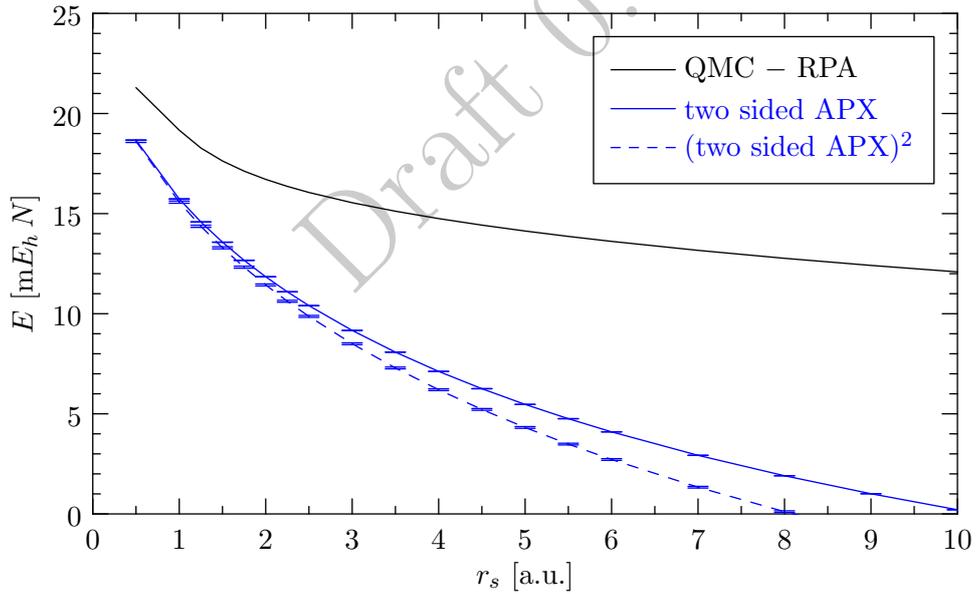

\begin{center}
\begin{asy}
  real[][] eps = input("EpsPara.dat").line().dimension(0,0);
  real[][] apx = input("AllApxPara.dat").line().dimension(0,0);
  real[][] apx2 = input("All2ApxPara.dat").line().dimension(0,0);

  sizeRatio(width=360);
  plotXY(eps, "QMC $-$ RPA");
  plotXYDY(apx, blue, "two sided APX");
  plotXYDY(apx2, blue+dashed, "(two sided APX)$^2$");
  axisXY(
    Label("$r_s$ [a.u.]",MidPoint), Label("$E$ [m$E_h\,N$]",MidPoint),
    (0,0), (10,25),
    yTicks=RightTicks(Step=5,n=5)
  );
  attach(legend(1,6,6,24,vskip=1),(10,25),12SW,UnFill);
\end{asy}
\end{center}
\caption{
  The Two sided Adjacent Pairs Exchange (2APX) energy per electron
  for the uniform electron gas compared to the error of the Random Phase
  Approximation with respect to Quantum Monte-Carlo (QMC)
  calculations by \parencite{ceperley_ground_1980}
  fitted by \parencite{perdew_self-interaction_1981}.
  Anti-symmetrizing more than one interaction in the RPA
  ring diagrams by the application of (\ref{eqn:APX_TwoSidedExchangeCases})
  worsens the accuracy with respect to Quantum Monte Carlo results.
}
\label{fig:APX_TwoSidedUegPara}
\end{figure}
Two adjacent pair bubbles have four possible time orders with respect
to the Coulomb interaction between them, shown in Figure
\ref{sfg:APX_BubbleCases}.
In the first and in the last case an anti-symmetrization of the contained
Coulomb interaction only cancels contributions where the same states $i$ and
$a$ occur in consecutive bubbles, as illustrated in Figure
\ref{sfg:APX_Bubbles3Pauli}.
In general, such contributions do not violate the Pauli exclusion principle.
Following this argument, we exclude these cases and study a correction to
the Random Phase Approximation where both remaining cases are anti-symmetrized
to
\begin{equation}
  \diagramBox{
    \begin{fmffile}{Exchange2Up}
    \begin{fmfgraph}(30,30)
      \fmfkeep{Exchange2Up}
      \fmfset{arrow_len}{6}
      \fmfstraight
      \fmfleft{v21,v22}
      \fmfright{v11,v12}
      \fmffreeze
      \fmf{fermion,right=0.3}{v11,v12}
      \fmf{fermion,left=0.3}{v21,v22}
      \fmf{plain}{v12,m} \fmf{fermion}{m,v21}
      \fmf{plain}{v22,m} \fmf{fermion}{m,v11}
      \fmf{photon}{v11,v21}
    \end{fmfgraph}
    \end{fmffile}
  }
  ,
  \diagramBox{
    \begin{fmffile}{Exchange2Down}
    \begin{fmfgraph}(30,30)
      \fmfkeep{Exchange2Down}
      \fmfset{arrow_len}{6}
      \fmfstraight
      \fmfleft{v21,v22}
      \fmfright{v11,v12}
      \fmffreeze
      \fmf{fermion,right=0.3}{v11,v12}
      \fmf{fermion,left=0.3}{v21,v22}
      \fmf{plain}{v11,m} \fmf{fermion}{v22,m}
      \fmf{plain}{v21,m} \fmf{fermion}{v12,m}
      \fmf{photon}{v12,v22}
    \end{fmfgraph}
    \end{fmffile}
  }
  ,
  \label{eqn:APX_TwoSidedExchangeCases}
\end{equation}
and inserted into the RPA ring diagrams. Since these diagrams are
polarization parts with two open vertices the memory
requirement for calculating this correction scales
like $\mathcal O(N^2)$ with the system size $N$. This is equivalent to
the memory requirement of the AC-SOSEX and considerably less than
$\mathcal O(N^4)$ of the conventional SOSEX, calculated from the
direct ring Coupled Cluster Doubles (drCCD) amplitudes.
We term this correction \emph{two sided Adjacent Pairs Exchange (2APX)} and
its expansion in terms of Goldstone diagrams is given by
\begin{align}
  \nonumber
  E_{\rm c}^{\rm 2APX} =&
  \hspace*{1ex}\diagramBox{\fmfreuse{Exchange2}}\hspace*{1ex} + \\[1ex]
  \nonumber
  &
  \ \hspace*{2ex}\diagramBox{\fmfreuse{MP3Exchange}}\hspace*{1ex} + 
  \hspace*{2ex}\diagramBox{\fmfreuse{MP4x1a}}\ +
  \hspace*{1ex}\diagramBox{\fmfreuse{MP4x2b}}\hspace*{2ex} +\ \ldots\ +
  \\[1ex]
  &\ \ t.r.\ +\ \ldots
  \label{eqn:APX_2APXDiagrams}
\end{align}
The second row contains only the right contribution of
(\ref{eqn:APX_TwoSidedExchangeCases}) and
$t.r.$ refers to diagrams emerging from the second row by time reversal,
containing only the left contribution of (\ref{eqn:APX_TwoSidedExchangeCases}).

The two sided Adjacent Pair Exchange correction differs from the
Second Order Screened Exchange correction already in third order.
Time reversal of any SOSEX diagram beyond second order gives a distinct
diagram and that is contained in the two sided APX.
Since third order is the lowest order where the two sided APX and SOSEX differ
and since this order is in general negative, the two sided APX correction
is expected to be less than the SOSEX correction. Evaluating the two sided APX
for the Uniform Electron Gas, as discussed in Section \ref{sec:APX_UEG},
confirms this expectation. As shown in Figure \ref{fig:APX_TwoSidedUegPara} the
two sided APX considerably underestimates the desired energy correction,
given by the difference of the RPA and Quantum Monte Carlo results.
At low densities, where $r_s>10$, it even becomes negative, actually worsening
the systematic error of RPA.

The two sided Adjacent Pairs Exchange correction seems the
most plausible lowest order correction to the Random Phase Approximation.
However, despite improving on the RPA in the UEG for densities with $r_s<10$,
the two sided APX does not offer a balanced correction to the RPA since
it always underestimates, but never overestimates the missing correlation
energy. One could include diagrams, where two, three or more of the RPA's
Coulomb interactions are anti-symmetrized, in the fashion discussed above,
as the next orders of the correction. These corrections are still forming a
ring and can thus be efficiently evaluated in the frequency domain, once the
two sided Adjacent Pairs Exchange polarization part $\vec P_{\rm x}^{\rm 2AXP}$
has been calculated. In terms of Feynman diagrams and propagator matrices,
the corrections with one or two Coulomb interactions anti-symmetrized
are given by
\begin{align}
  \nonumber
  E_{\rm c}^{\rm 2APX} =&
  \hspace*{-3ex}
  \diagramBox{
    \begin{fmffile}{ApxFeyn}
    \begin{fmfgraph*}(45,35)
      \fmfkeep{ApxFeyn}
      \fmfset{arrow_len}{6}
      \fmfsurroundn{v}{4}
      \fmf{plain,right=0.6}{v2,v4,v2}
      \fmfpoly{phantom,label=$\vec P_{\rm x}$}{v2,v4}
      \fmf{dbl_wiggly,left=1.3}{v2,v4}
    \end{fmfgraph*}
    \end{fmffile}
  }
  \\[1ex]
  =&
  \ \frac12\, \left(\frac12\,\vec P_{\rm x}\vec V +
    \vec P_{\rm x}\vec V\vec X_0\vec V +
    \vec P_{\rm x}\vec V\vec X_0\vec V\vec X_0\vec V + \ldots \right)
  \label{eqn:APX_TwoSidedApx}
  \\
  \nonumber
  E_{\rm c}^{({\rm 2APX}^2)} =&
  \hspace*{-0.6ex}
  \diagramBox{\fmfreuse{ApxFeyn}}
  \hspace*{1.6ex} +
  \diagramBox{
    \begin{fmffile}{Apx2Feyn}
    \begin{fmfgraph*}(45,45)
      \fmfset{arrow_len}{6}
      \fmfsurroundn{v}{8}
      \fmf{plain,right=0.6}{v8,v2,v8}
      \fmfpoly{phantom,label=$\vec P_{\rm x}$}{v8,v2}
      \fmf{dbl_wiggly}{v2,v4}
      \fmf{plain,right=0.6}{v4,v6,v4}
      \fmfpoly{phantom,label=$\vec P_{\rm x}$}{v4,v6}
      \fmf{dbl_wiggly}{v6,v8}
    \end{fmfgraph*}
    \end{fmffile}
  }
  \\
  =&
  \ E_{\rm c}^{\rm 2APX} +
  \frac1{2\cdot2}\,\Big(
    \vec P_{\rm x}\vec V + \vec P_{\rm x}\vec V\vec X_0\vec V +
    \vec P_{\rm x}\vec V\vec X_0\vec V\vec X_0\vec V + \ldots
  \Big)^2
\label{eqn:APX_TwoSidedApx2}
\end{align}
with
\[
  \diagramBox{
    \begin{fmffile}{Apx2Polarizability}
    \begin{fmfgraph*}(25,35)
      \fmfkeep{Apx2Polarizability}
      \fmfset{arrow_len}{6}
      \fmfsurroundn{v}{4}
      \fmf{plain,right=0.6}{v2,v4,v2}
      \fmfpoly{phantom,label=$\vec P_{\rm x}$}{v2,v4}
    \end{fmfgraph*}
    \end{fmffile}
  }
  =
  \hspace*{2ex}
  \diagramBox{\fmfreuse{Exchange2Up}}
  \hspace*{1ex}
  +
  \hspace*{1ex}
  \diagramBox{\fmfreuse{Exchange2Down}}
  \hspace*{1ex},
\]
and where the superscript of $\vec P_{\rm x}^{\rm 2APX}$, the trace,
the imaginary time integration - including the sign - and the imaginary time
arguments have been omitted for brevity. The RPA screened interaction
$
  \diagramBox{
    \begin{fmffile}{ScreenedWUnlabeled}
    \begin{fmfgraph}(26,10)
      \fmfstraight
      \fmfleft{v11}
      \fmfright{v21}
      \fmf{dbl_wiggly}{v11,v21}
    \end{fmfgraph}
    \end{fmffile}
  }
$ is given by (\ref{eqn:RPAScreenedW}).
All diagrams posses reflection symmetry but note that the symmetry factor
differs in the case where only one occurrence of $\vec P_{\rm x}^{\rm 2APX}$ is
inserted into the ring diagrams compared to other case. This is due to
a reflection symmetry introduced when closing $\vec P_{\rm x}^{\rm 2APX}$ with
only one Coulomb interaction $\vec V$.
Figure \ref{fig:APX_TwoSidedUegPara} shows $E_{\rm c}^{\rm 2APX}$ and
$E_{\rm c}^{(2APX)^2}$ according to (\ref{eqn:APX_TwoSidedApx}) and
(\ref{eqn:APX_TwoSidedApx2}). More insertions of $\vec P_{\rm x}^{\rm 2APX}$
into the ring diagrams of the Random Phase Approximation actually worsen the a
accuracy of the two sided APX with respect to Quantum Monte Carlo results. 
More than two insertions of $\vec P_{\rm x}^{\rm 2APX}$ into the ring diagrams
of the RPA offer no improvement either so none
of the two sided APX approximations ever overestimates the missing correlation
energy and thus none is balanced.

\begin{figure}
  \begin{center}
    \diagramBox{
      \begin{fmffile}{PauliNonAdjacent}
      \begin{fmfgraph*}(80,80)
        \fmfstraight
        \fmfset{arrow_ang}{25}
        \fmfset{arrow_len}{5pt}
        \fmfleft{v00,v01,v02,v03}
        \fmfright{v30,v31,v32,v33}
        \fmf{photon}{v00,v30}
        \fmf{photon}{v01,v11} \fmf{phantom}{v11,v21} \fmf{photon}{v21,v31}
        \fmf{photon}{v02,v12} \fmf{phantom}{v12,v22} \fmf{photon}{v22,v32}
        \fmf{photon}{v03,v33}
        \fmffreeze
        \fmf{fermion,left=0.3}{v00,v01,v00} \fmf{fermion,right=0.3}{v30,v31,v30}
        \fmf{
          fermion,left=0.3,label.dist=4,label=$\textcolor{red}{i}$
        }{v12,v11}
          \fmf{fermion,left=0.3}{v11,v12}
        \fmf{
          fermion,right=0.3,label.dist=1,label=$\textcolor{red}{i}$
        }{v22,v21}
          \fmf{fermion,right=0.3}{v21,v22}
        \fmf{fermion,left=0.3}{v02,v03,v02} \fmf{fermion,right=0.3}{v32,v33,v32}
      \end{fmfgraph*}
      \end{fmffile}
    }
    \quad
    $\rightarrow$
    \quad
    \diagramBox{
      \begin{fmffile}{PauliNonAdjacentExchange}
      \begin{fmfgraph*}(80,80)
        \fmfstraight
        \fmfset{arrow_ang}{25}
        \fmfset{arrow_len}{5pt}
        \fmfleft{v00,v01,v02,v03}
        \fmfright{v30,v31,v32,v33}
        \fmf{photon}{v00,v30}
        \fmf{photon}{v01,v11} \fmf{phantom}{v11,v21} \fmf{photon}{v21,v31}
        \fmf{photon}{v02,v12} \fmf{phantom}{v12,v22} \fmf{photon}{v22,v32}
        \fmf{photon}{v03,v33}
        \fmffreeze
        \fmf{fermion,left=0.3}{v00,v01,v00} \fmf{fermion,right=0.3}{v30,v31,v30}
        \fmf{fermion,left=0.3}{v11,v12}
          \fmf{
            fermion,label.dist=3,label.side=left,label=$\textcolor{blue}{i}$
          }{v12,vm1}
          \fmf{plain}{vm1,v21}
        \fmf{fermion,right=0.3}{v21,v22}
          \fmf{plain}{v22,vm2}
          \fmf{
            fermion,label.dist=1,label.side=left,label=$\textcolor{blue}{i}$
          }{vm2,v11}
        \fmf{fermion,left=0.3}{v02,v03,v02} \fmf{fermion,right=0.3}{v32,v33,v32}
      \end{fmfgraph*}
      \end{fmffile}
    }
  \end{center}
\caption{
  Exchange of pair bubbles that are non-adjacent
}
\label{fig:APX_NonAdjacentPauli}
\end{figure}
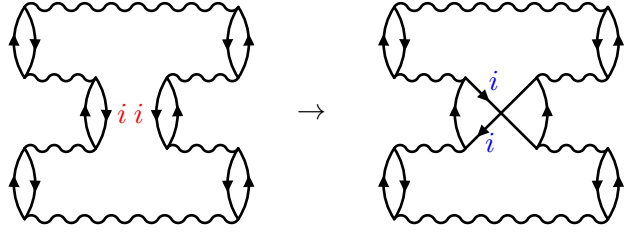

We could investigate more complex exchange processes, for
instance those correcting violations of the Pauli exclusion principle of
pair bubbles in the RPA that are not adjacent, as sketched in Figure
\ref{fig:APX_NonAdjacentPauli}. However, the exclusion principle is only
violated if the two effected pair bubbles propagate at overlapping times.
This would require a time order which can only be provided when evaluating
the Random Phase Approximation from the direct ring Coupled Cluster Doubles
(drCCD) amplitudes, loosing the advantage of the reduced memory requirements
of an RPA implementation in the frequency domain.

\subsection{Adjacent Pairs Exchange}
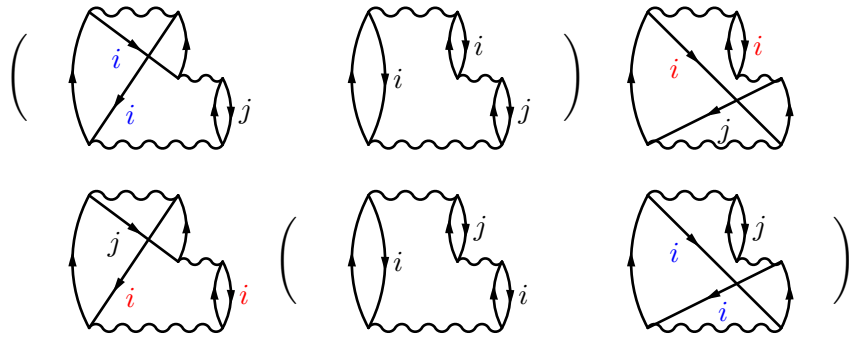
\begin{figure}
\begin{center}
  \begin{tabular}{ccccccc}
    \Bigg(&
    \diagramBox{
      \begin{fmffile}{MP3Exchangeiji}
      \begin{fmfgraph*}(50,50)
        \fmfset{arrow_len}{6}
        \fmfstraight
        \fmfleft{v00,v01,v02}
        \fmfright{v20,v21,v22}
        \fmf{boson}{v00,v20}
        \fmf{phantom}{v01,v11} \fmf{boson,tension=2}{v11,v21}
        \fmf{boson}{v02,v12} \fmf{phantom,tension=2}{v12,v22}
        \fmffreeze
        \fmf{plain,tension=2}{v12,m}
          \fmf{
            fermion,label.dist=3,label.side=left,label=$\textcolor{blue}{i}$
          }{m,v00}
        \fmf{fermion,left=0.25}{v00,v02}
        \fmf{
          fermion,label.dist=1.5,label.side=right,label=$\textcolor{blue}{i}$
        }{v02,v11}
        \fmf{fermion,right=0.25}{v11,v12}
        \fmf{fermion,left=0.25}{v20,v21}
        \fmf{
          fermion,label.dist=3,left=0.25,label=$j$
        }{v21,v20}
      \end{fmfgraph*}
      \end{fmffile}
    }
    &&
    \diagramBoxBorder{0ex}{2ex}{
      \begin{fmffile}{MP3Directiji}
      \begin{fmfgraph*}(50,50)
        \fmfset{arrow_len}{6}
        \fmfstraight
        \fmfleft{v00,v01,v02}
        \fmfright{v20,v21,v22}
        \fmf{boson}{v00,v20}
        \fmf{phantom}{v01,v11} \fmf{boson,tension=2}{v11,v21}
        \fmf{boson}{v02,v12} \fmf{phantom,tension=2}{v12,v22}
        \fmffreeze
        \fmf{fermion,left=0.25}{v00,v02}
        \fmf{
          fermion,left=0.25,label.dist=3,label=$i$
        }{v02,v00}
        \fmf{fermion,left=0.25}{v20,v21}
        \fmf{
          fermion,left=0.25,label.dist=3,label=$j$
        }{v21,v20}
        \fmf{fermion,left=0.25}{v11,v12}
        \fmf{
          fermion,left=0.25,label.dist=3,label=$i$
        }{v12,v11}
      \end{fmfgraph*}
      \end{fmffile}
    }
    &\Bigg)&
    \diagramBox{
      \begin{fmffile}{MP3ExchangeUpiji}
      \begin{fmfgraph*}(50,50)
        \fmfset{arrow_len}{6}
        \fmfstraight
        \fmfleft{v00,v01,v02}
        \fmfright{v20,v21,v22}
        \fmf{boson}{v00,v20}
        \fmf{phantom}{v01,v11} \fmf{boson,tension=2}{v11,v21}
        \fmf{boson}{v02,v12} \fmf{phantom,tension=2}{v12,v22}
        \fmffreeze
        \fmf{
          fermion,label.dist=3,label.side=left,label=$j$
        }{v21,v00}
        \fmf{fermion,left=0.25}{v00,v02}
        \fmf{
          fermion,label.dist=1.5,label.side=right,label=$\textcolor{red}{i}$
        }{v02,m}
          \fmf{plain,tension=2}{m,v20}
        \fmf{fermion,right=0.25}{v20,v21}
        \fmf{fermion,left=0.25}{v11,v12}
        \fmf{
          fermion,label.dist=3,left=0.25,label=$\textcolor{red}{i}$
        }{v12,v11}
      \end{fmfgraph*}
      \end{fmffile}
    }
    &
    \\[8ex]
    &
    \diagramBox{
      \begin{fmffile}{MP3Exchangeiij}
      \begin{fmfgraph*}(50,50)
        \fmfset{arrow_len}{6}
        \fmfstraight
        \fmfleft{v00,v01,v02}
        \fmfright{v20,v21,v22}
        \fmf{boson}{v00,v20}
        \fmf{phantom}{v01,v11} \fmf{boson,tension=2}{v11,v21}
        \fmf{boson}{v02,v12} \fmf{phantom,tension=2}{v12,v22}
        \fmffreeze
        \fmf{plain,tension=2}{v12,m}
          \fmf{
            fermion,label.dist=3,label.side=left,label=$\textcolor{red}{i}$
          }{m,v00}
        \fmf{fermion,left=0.25}{v00,v02}
        \fmf{
          fermion,label.dist=1.5,label.side=right,label=$j$
        }{v02,v11}
        \fmf{fermion,right=0.25}{v11,v12}
        \fmf{fermion,left=0.25}{v20,v21}
        \fmf{
          fermion,label.dist=3,left=0.25,label=$\textcolor{red}{i}$
        }{v21,v20}
      \end{fmfgraph*}
      \end{fmffile}
    }
    &\Bigg(&
    \diagramBoxBorder{0ex}{2ex}{
      \begin{fmffile}{MP3Directiij}
      \begin{fmfgraph*}(50,50)
        \fmfset{arrow_len}{6}
        \fmfstraight
        \fmfleft{v00,v01,v02}
        \fmfright{v20,v21,v22}
        \fmf{boson}{v00,v20}
        \fmf{phantom}{v01,v11} \fmf{boson,tension=2}{v11,v21}
        \fmf{boson}{v02,v12} \fmf{phantom,tension=2}{v12,v22}
        \fmffreeze
        \fmf{fermion,left=0.25}{v00,v02}
        \fmf{
          fermion,left=0.25,label.dist=3,label=$i$
        }{v02,v00}
        \fmf{fermion,left=0.25}{v20,v21}
        \fmf{
          fermion,left=0.25,label.dist=3,label=$i$
        }{v21,v20}
        \fmf{fermion,left=0.25}{v11,v12}
        \fmf{
          fermion,left=0.25,label.dist=3,label=$j$
        }{v12,v11}
      \end{fmfgraph*}
      \end{fmffile}
    }
    &&
    \diagramBox{
      \begin{fmffile}{MP3ExchangeUpiij}
      \begin{fmfgraph*}(50,50)
        \fmfset{arrow_len}{6}
        \fmfstraight
        \fmfleft{v00,v01,v02}
        \fmfright{v20,v21,v22}
        \fmf{boson}{v00,v20}
        \fmf{phantom}{v01,v11} \fmf{boson,tension=2}{v11,v21}
        \fmf{boson}{v02,v12} \fmf{phantom,tension=2}{v12,v22}
        \fmffreeze
        \fmf{
          fermion,label.dist=3,label.side=left,label=$\textcolor{blue}{i}$
        }{v21,v00}
        \fmf{fermion,left=0.25}{v00,v02}
        \fmf{
          fermion,label.dist=1.5,label.side=right,label=$\textcolor{blue}{i}$
        }{v02,m}
          \fmf{plain,tension=2}{m,v20}
        \fmf{fermion,right=0.25}{v20,v21}
        \fmf{fermion,left=0.25}{v11,v12}
        \fmf{
          fermion,label.dist=3,left=0.25,label=$j$
        }{v12,v11}
      \end{fmfgraph*}
      \end{fmffile}
    }
    &\Bigg)
  \end{tabular}
\end{center}
\caption{
  The two sided Adjacent Pairs Exchange correction introduces new violations
  of the Pauli exclusion principle, as shown here in third order in the
  case of two identical hole indices.
  One of the two exchange diagrams exactly cancels the offending contributions
  of the RPA diagram, indicated by parenthesis and blue index labels.
  The other exchange diagram, however, introduces new violations shown by
  red index labels.
}
\label{fig:APX_IntroducedViolations}
\end{figure}

The unbalanced performance of the two sided APX and its higher order variants
rises the question whether really all Coulomb interactions should be
anti-symmetrized if that can correct for violations of the Pauli exclusion
principle.
It turns out that the two sided APX, while indeed correcting for all
violations occurring in adjacent pairs, introduces new violations.
In third order this is most apparent and illustrated in Figure
\ref{fig:APX_IntroducedViolations}. The lower/upper row shows the case
where the lower/upper two pair bubbles of the RPA propagate in the same hole
state $i$.
The diagram shown with blue index labels exchanges these two propagators and
exactly cancels the offending contributions of the RPA. This is indicated
by parenthesis around the RPA and the respective exchange diagram.
The other diagram, however, introduces new violations to the Pauli exclusion
principle, shown by the red index labels.

The key issue in this case is that the leftmost and longest pair bubble of
the RPA diagram is exchanged in both diagrams of the two sided APX corrections.
While this guarantees that all possible violating contributions are
canceled, it always introduces new violations.
For a more balanced correction we therefore require that each pair bubble
is exchanged at most once. The simplest correction satisfying this requirement
is the (single sided)
\emph{Adjacent Pairs Exchange (APX)}\index{Adjacent Pairs Exchange (APX)}
correction, where only one case of (\ref{eqn:APX_TwoSidedExchangeCases}) is
contained.
For a system with time reflection symmetry it is irrelevant which of the
cases we choose and without loss of generality we choose to anti-symmetrize
adjacent pair bubbles in the third case of time orders shown in Figure
\ref{sfg:APX_BubbleCases}:
\begin{equation}
  \diagramBox{
    \begin{fmffile}{Bubbles2DownCentered}
    \begin{fmfgraph}(30,30)
      \fmfset{arrow_len}{6}
      \fmfstraight
      \fmfleft{v21,v22}
      \fmfright{v11,v12}
      \fmffreeze
      \fmf{fermion,right=0.3}{v11,v12,v11}
      \fmf{fermion,left=0.3}{v21,v22,v21}
      \fmf{photon}{v12,v22}
    \end{fmfgraph}
    \end{fmffile}
  }
  \mapsto
  \hspace*{1ex}
  \diagramBox{\fmfreuse{Exchange2Down}}
  \hspace*{1ex}.
  \label{eqn:APX_ExchangeCases}
\end{equation}
In terms of Feynman diagrams and the matrix notation of the propagators,
introduced in Section \ref{sec:RPAFreq}, the APX correction is then given by
\begin{align}
  E_{\rm c}^{\rm APX} =&
  \hspace*{-1ex}
  \diagramBox{\fmfreuse{ApxFeyn}}
  \hspace*{1.5ex}
  =
  \diagramBox{\fmfreuse{MP2x}}
  +
  \diagramBox{
    \begin{fmffile}{ApxVFeyn}
    \begin{fmfgraph*}(45,45)
      \fmfkeep{ApxVFeyn}
      \fmfset{arrow_len}{6}
      \fmfsurroundn{v}{8}
      \fmf{fermion,right=0.35}{v8,v2,v8}
      \fmf{boson}{v2,v4}
      \fmf{plain,right=0.6}{v4,v6,v4}
      \fmfpoly{phantom,label=$\vec P_{\rm x}$}{v4,v6}
      \fmf{boson}{v6,v8}
    \end{fmfgraph*}
    \end{fmffile}
  }
  +
  \diagramBox{
    \begin{fmffile}{ApxV2Feyn}
    \begin{fmfgraph*}(55,55)
      \fmfkeep{ApxV2Feyn}
      \fmfset{arrow_len}{6}
      \fmfsurroundn{v}{12}
      \fmf{fermion,right=0.35}{v2,v4,v2}
      \fmf{boson}{v4,v6}
      \fmf{plain,right=0.6}{v6,v8,v6}
      \fmfpoly{phantom,label=$\vec P_{\rm x}$}{v6,v8}
      \fmf{boson}{v8,v10}
      \fmf{fermion,right=0.35}{v10,v12,v10}
      \fmf{boson}{v12,v2}
    \end{fmfgraph*}
    \end{fmffile}
  }
  +\ \ldots
\label{eqn:APX_Feyn}
  \\
  \nonumber
  =&
  \,
  -\frac12\, \int\frac{\d\nu}{2\pi}\,
  \Tr\Big\{
    \vec P_{\rm x}^{\rm APX}\vec V +
    \vec P_{\rm x}^{\rm APX}\vec V\vec X_0\vec V +
    \vec P_{\rm x}^{\rm APX}\vec V\vec X_0\vec V
      \vec X_0\vec V + \ldots
  \Big\}
  \\
  \nonumber
  =&
  \,
  -\frac12\, \int\frac{\d\nu}{2\pi}\,
  \Tr\Big\{
    \vec P_{\rm x}^{\rm APX}\vec W
  \Big\}
\end{align}
with
\[
  \diagramBox{
    \fmfreuse{Apx2Polarizability}
  }
  =
  \hspace*{2ex}
  \diagramBox{\fmfreuse{Exchange2Down}}
  \hspace*{2ex}
  ,
\]
and where the imaginary time arguments as well as
the superscript of $\vec P_{\rm x}^{\rm APX}$ in the diagrams have
been omitted for brevity. All diagrams exhibit a single reflection symmetry
but note that, unlike in the two sided APX case, there is no additional
symmetry introduced when closing $\vec P_{\rm x}^{\rm APX}$ with one
Coulomb interaction $\vec V$ since the $\vec P_{\rm x}^{\rm APX}$ contains
only one of the two time orders contained in $\vec P_{\rm x}^{\rm 2APX}$.
Despite the ring form of the APX diagrams beyond second order, none of them
has a rotational symmetry in contrast to the respective RPA diagrams.
This simplifies the sum over all orders of the perturbation compared to the
RPA since all orders have the same
factor. We can use the infinite sum of a geometric series to give an
explicit form for the APX energy:
\begin{equation}
  E_{\rm c}^{\rm APX} = -\frac12\int_{-\infty}^\infty\frac{\d\nu}{2\pi}\,
  \Tr\left\{
    \vec P_{\rm x}^{\rm APX}(\im\nu)\vec V
    \Big(\vec 1-\vec X_0(\im\nu)\vec V\Big)^{-1}
  \right\}\,,
\label{eqn:APX_APX}
\end{equation}
where the imaginary frequency dependent exchange polarizability for the APX
contains only one of the four time orders contained in
$\vec P_{\rm x}^{\rm AC}$, which was given in (\ref{eqn:SOESX_PxAC}):
\begin{align}
  \nonumber
  {\vec P_{\rm x}^{\rm APX}}_{\vec x_1 \vec x_2}(\im\nu) =
    -\iint\d\vec x_3\,\d\vec x_4\,\frac1{|\vec r_3-\vec r_4|} &
    \sum_{ia}
      \psi^\ast_i(\vec x_4)\psi_i(\vec x_1)
      \psi^\ast_a(\vec x_1)\psi_a(\vec x_3)
      \frac1{\eps_a-\eps_i+\im\nu}
  \\ &
    \sum_{jb}
      \psi^\ast_j(\vec x_3)\psi_j(\vec x_2)
      \psi^\ast_b(\vec x_2)\psi_b(\vec x_4)
      \frac1{\eps_b-\eps_j-\im\nu}\,.
  \label{eqn:APX_Px}
\end{align}

    \section{APX for the uniform electron gas}
\label{sec:APX_UEG}
Evaluating the Adjacent Pairs Exchange (APX) energy for the
Uniform Electron Gas (UEG) is very similar to evaluating the AC-SOSEX
energy according to (\ref{eqn:SOSEX_ACUEG}). In contrast to the AC-SOSEX
expression, only one of the four time orders of two adjacent pair bubbles
are contained in the APX. In the chosen order, the Coulomb interaction of the
exchange polarizability $\vec P_{\rm x}^{\rm APX}$ occurs after both open
vertices of $\vec P_{\rm x}^{\rm APX}$\footnote{
In the given diagram, the time order of the
vertices is only relevant with respect to the Coulomb interaction. It is
therefore neither a Goldstone nor a Feynman diagram. The proper Feynman
diagram of APX is the leftmost diagram of (\ref{eqn:APX_Feyn}).
}
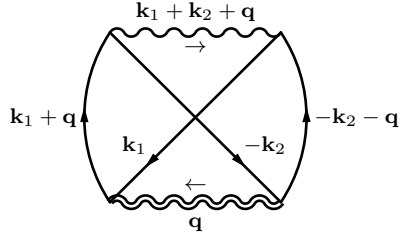
\[
  \diagramBoxBorder{3ex}{3ex}{
    \scriptsize
    \fmfreuse{ACSOSEX3Labeled}
  }\hspace{10ex}.
\]
Furthermore, the APX uses the screened interaction $W$, represented by
the double wiggly line, as given by (\ref{eqn:RPAScreenedW}) rather than the
coupling strength averaged screened interaction $\overline W$,
since all orders of the expansion have the same factor.
The APX correction to the Random Phase Approximation per electron is thus given
by
\begin{align}
  \nonumber
  E_{\rm c}^{\rm APX} =
    +\frac\Omega N \frac12 \int\frac{\d\nu}{2\pi}\,
    \int \frac{\d\vec q}{(2\pi)^3}\,
    \sum_\sigma\iint\limits_{|\vec k_i|<k_{\rm F}<|\vec k_i+\vec q|} &
      \frac{\Omega^2 \d\vec k_1 \d\vec k_2}{(2\pi)^6}\,
    V(\vec k_1+\vec k_2+\vec q)
    W(\vec q,\nu)\,
  \\
  &
      \frac1{
        (\Delta\eps_{\vec k_1,\vec q}+\im\nu)
        (\Delta\eps_{-\vec k_2,-\vec q}-\im\nu)
      }\,,
  \label{eqn:APX_APXUEG}
\end{align}
with $i\in\{1,2\}$ and the single particle excitation energy
$\Delta\eps_{\vec k_i,\vec q} = (\vec k_i+\vec q)^2/2 - \vec k_i^2/2$.
As in the case of the AC-SOSEX, the above expression can be evaluated
by a Monte-Carlo integration for $\vec k_1$ and $\vec k_2$ using
a probability density function (PDF) given by
\[
  \PDF(\vec k_i) \propto
    \left\{
      \begin{array}{ll}
        \displaystyle
        \left|\frac1{\Delta\eps_{\vec k_i,\vec q}\pm\im\nu}\right| &
        \textnormal{for }|\vec k_i|<k_{\rm F}<|\vec k_i+\vec q|\,, \\[2ex]
        \displaystyle 0 & \textnormal{otherwise.}
      \end{array}
    \right.
\]
The momentum $\vec q$ and the imaginary frequency $\nu$ can be integrated
using a Gauss--Kronrod rule, analogous to the evaluation of the Random Phase
Approximation. However, the asymptotic behavior of the APX energy for
large imaginary frequencies differs from that of the RPA and the AC-SOSEX,
such that a different variable transform for integrating the frequency tail
must be used. The asymptotic behavior for large frequencies
is discussed Subsection \ref{ssc:APX_UEGLargeFrequency}.

\begin{figure}[p]
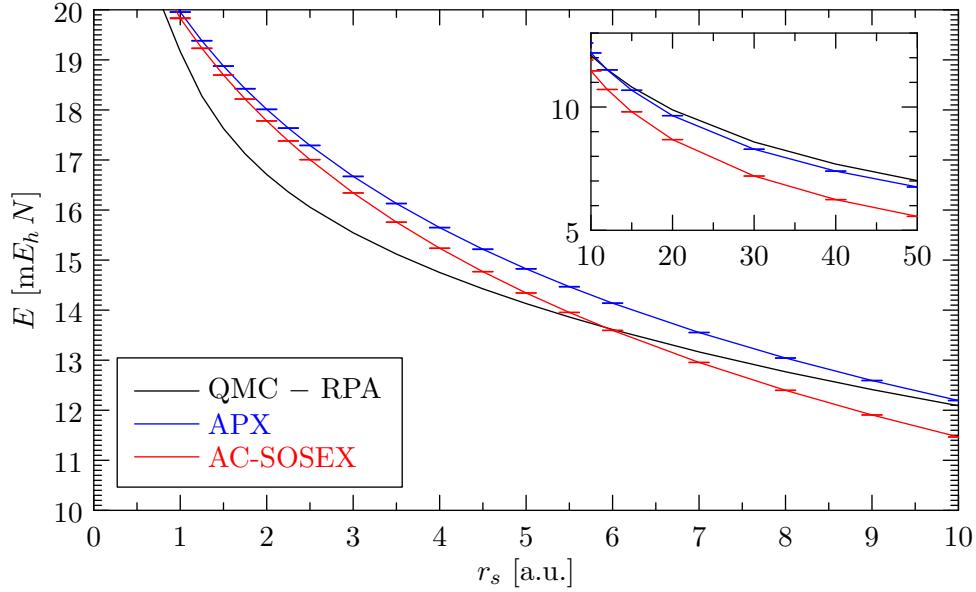

\begin{center}
\begin{asy}
  real[][] eps = input("EpsPara.dat").line().dimension(0,0);
  real[][] apx = input("ApxPara.dat").line().dimension(0,0);
  real[][] acSosex = input("AcSosexPara.dat").line().dimension(0,0);

  sizeRatio(width=360);
  plotXY(eps, "QMC $-$ RPA");
  plotXYDY(apx, blue, "APX");
  plotXYDY(acSosex, red, "AC-SOSEX");
  axisXY(
    Label("$r_s$ [a.u.]",MidPoint), Label("$E$ [m$E_h\,N$]",MidPoint),
    (0,10), (10,20),
    yTicks=RightTicks(Step=1,n=10)
  );
  attach(legend(1,6,6,24,vskip=1),(0,10),12NE,UnFill);

  picture insetPic;
  sizeRatio(insetPic, width=0.4*360);
  plotXY(insetPic, eps);
  plotXYDY(insetPic, apx, blue);
  plotXYDY(insetPic, acSosex, red);
  axisXY(insetPic, (10,5), (50,13), yTicks=RightTicks(Step=5,n=5));

  add(insetPic.fit(), (10,20), 12SW);
\end{asy}
\end{center}
\vspace*{-2ex}
\caption{
  The Adjacent Pairs Exchange (APX) energy per electron for the 
  uniform electron gas compared to the error of the Random Phase Approximation
  with respect to Quantum Monte-Carlo (QMC)
  calculations by \parencite{ceperley_ground_1980}
  fitted by \parencite{perdew_self-interaction_1981}.
  RPA+APX also fortuitously matches the QMC results but only at $r_s\approx10$.
  However, in the low density regime, where correlation is stronger, APX is
  considerably closer to the QMC results than AC-SOSEX.
  For results of the spin-polarized uniform electron gas see Subsection
  \ref{ssc:APX_UegFerromagnetic}.
}
\label{fig:APX_UegPara}
\end{figure}

For the lowest order of the APX, the accuracy of the numerical integrations can
be benchmarked against the MP2 exchange energy, which is independent of the
density and it is known \emph{analytically}
from the work of \parencite{onsager_integrals_1966}:
\[
  \frac{\log(2)}6 - 3\frac{\zeta(3)}{(2\pi)^2} \approx
  0.02417915891814441 E_h\,N\,.
\]
For the Monte-Carlo integration of $\vec k_1$ and $\vec k_2$ a precision of 5
significant digits can be achieved with less than $30000$ samples for each $q$
and $\nu$, depending on momentum, frequency and density.
The error from the momentum and imaginary frequency integration is of
similar magnitude.
Figure \ref{fig:APX_UegPara} and Table \ref{tab:APX_UegPara} show the resulting
APX energies as a function of density given by the Wigner-Seitz radius,
$r_s$. The uncertainties from the integrations are indicated by the error bars.
\begin{table}[p]
\begin{center}
\begin{tabular}{|c|r|rr|rr|}
\hline
&&&&& \\[-2.5ex]
  \multicolumn{1}{|c|}{$r_s$} &
    \multicolumn{1}{|c|}{$(E_{\rm c}-E_{\rm c}^{\rm RPA})$} &
    \multicolumn{2}{c|}{$E_{\rm c}^{\rm APX}$} &
    \multicolumn{2}{c|}{$E_{\rm c}^{\rm AC-SOSEX}$}
    \\
  \multicolumn{1}{|c|}{$[$a.u.$]$} &
    \multicolumn{1}{|c}{[m$E_h\,N$]} &
    \multicolumn{1}{|c}{[m$E_h\,N$]} &
    \multicolumn{1}{c|}{$\pm$} &
    \multicolumn{1}{|c}{[m$E_h\,N$]} &
    \multicolumn{1}{c|}{$\pm$}
     \\
&&&&& \\[-2.6ex]
\hline
&&&&& \\[-2.5ex]
 1 & 19.167 & 19.956 & 0.005 & 19.832 & 0.009 \\
 2 & 16.710 & 18.012 & 0.005 & 17.780 & 0.003 \\
 3 & 15.545 & 16.672 & 0.005 & 16.342 & 0.003 \\
 4 & 14.752 & 15.649 & 0.005 & 15.237 & 0.003 \\
 5 & 14.131 & 14.825 & 0.004 & 14.343 & 0.003 \\
 6 & 13.613 & 14.139 & 0.004 & 13.595 & 0.003 \\
 7 & 13.165 & 13.553 & 0.004 & 12.955 & 0.003 \\
 8 & 12.768 & 13.043 & 0.004 & 12.398 & 0.003 \\
 9 & 12.413 & 12.594 & 0.004 & 11.906 & 0.003 \\
10 & 12.090 & 12.194 & 0.004 & 11.466 & 0.003 \\
12 & 11.521 & 11.506 & 0.004 & 10.712 & 0.003 \\
15 & 10.810 & 10.680 & 0.004 &  9.806 & 0.003 \\
20 &  9.884 &  9.650 & 0.004 &  8.679 & 0.003 \\
30 &  8.582 &  8.289 & 0.004 &  7.201 & 0.003 \\
40 &  7.685 &  7.400 & 0.004 &  6.246 & 0.003 \\
50 &  7.014 &  6.757 & 0.004 &  5.564 & 0.003 \\
  \hline
\end{tabular}
\end{center}
\caption{
  Data of Figure \ref{fig:APX_UegPara} including low densities.
}
\label{tab:APX_UegPara}
\end{table}

      \subsection{Large momentum behavior}
\label{ssc:APX_UEGLargeMomentum}
In a solid or in a molecule, the finite resolution of the DFT or the
Hartree-Fock reference imposes an upper limit $G_{\rm max}$ on up to where
the exchange polarizability $\vec P_{\rm x}^{\rm APX}$ and the independent
particle polarizability $\vec X_0$ can be evaluated. Assuming that the
system is sufficiently homogeneous at that resolution the asymptotic
behavior for large momenta in the Uniform Electron Gas can be used to
extrapolate results retrieved at finite $G_{\rm max}$ to the limit
of an infinite basis set, where $G_{\rm max}\rightarrow\infty$.

We can write the APX correction in the UEG from (\ref{eqn:APX_APXUEG}) in
terms an exchange polarizability $P_{\rm x}^{\rm APX}(q, \im\nu)$, analogous
to the general APX expression (\ref{eqn:APX_APX}):
\begin{equation}
  E_{\rm c}^{\rm APX} =
    -\frac\Omega N \frac12 \int\frac{\d\nu}{2\pi}\,
    \int \frac{\d\vec q}{(2\pi)^3}\,
    P_{\rm x}^{\rm APX}(\vec q,\im\nu)
    V(\vec q)\Big(1-\chi_0(\vec q,\im\nu)V(\vec q)\Big)^{-1}\,,
\label{eqn:APX_ApxUEGPx}
\end{equation}
\begin{equation}
  P_{\rm x}^{\rm APX}(\vec q,\im\nu) =
    -\sum_\sigma\iint\limits_{|\vec k_i|<k_{\rm F}<|\vec k_i+\vec q|}
      \frac{\Omega^2 \d\vec k_1 \d\vec k_2}{(2\pi)^6}\,
    V(\vec k_1+\vec k_2+\vec q)\,
      \frac1{
        (\Delta\eps_{\vec k_1,\vec q}+\im\nu)
        (\Delta\eps_{-\vec k_2,-\vec q}-\im\nu)
      }\,.
\label{eqn:APX_PxUeg}
\end{equation}
To get an approximation $P_{\rm x}^{\rm APX}$ for large momenta $q$ we can
trivially integrate $\vec k_1$ and $\vec k_2$, since
$|\vec k_i|<k_{\rm F}\ll q$, getting
\begin{equation}
  P_{\rm x}^{\rm APX}(q,\im\nu) \sim
  \frac1{q^2}\,\frac1{(q^2/2+\im\nu)(q^2/2-\im\nu)}\,.
  \label{eqn:APX_PxLargeQ}
\end{equation}
Inserting this and the approximation of $\chi_0(q,\im\nu)V(q)$ for large $q$,
given in (\ref{eqn:RPA_ChiLargeQ}), into the APX energy expression,
we can integrate $\nu$ for a given, large momentum $q$:
\[
  \int\frac{\d\nu}{2\pi}\,
  \frac1{q^4}\,\frac1{(q^2/2+\im\nu)(q^2/2-\im\nu)}
  \left(1- \frac1{\left(q^2/2\right)^2 + \nu^2} \right)^{-1}
  = \frac1{q^4\sqrt{q^4-4}}\,.
\]
The missing energy $\Delta E_{\rm c}^{\rm APX}$ of the APX correction when
truncating the momentum integration at a finite but large momentum $G_{\rm max}$
is thus
\begin{equation}
\Delta E_{\rm c}^{\rm APX}(G_{\rm max}) \sim
  \int_{G_{\rm max}}^\infty \frac{q^2}{q^4\sqrt{q^4-4}}
  \sim a_3\,\frac1{G_{\rm max}^3} + a_7\,\frac1{G_{\rm max}^7}
  +\mathcal O\left(\frac1{G_{\rm max}^{11}}\right)\,,
\end{equation}
where we have expanded the integrand in the variable $u=1/q$ at $u=0$.
Hence, the APX energy and, similarly, the AC-SOSEX energy have the same
asymptotic behavior with respect to large momenta $\vec q$ as the Random Phase
Approximation. However, the convergence with respect to $G_{\rm max}$ is slower
since the exchange corrections are more short ranged, which
may require the use of more than the leading order term of the Taylor
expansion for an accurate extrapolation to the infinite basis set limit.

\subsection{Large imaginary frequency behavior}
\label{ssc:APX_UEGLargeFrequency}
\begin{figure}
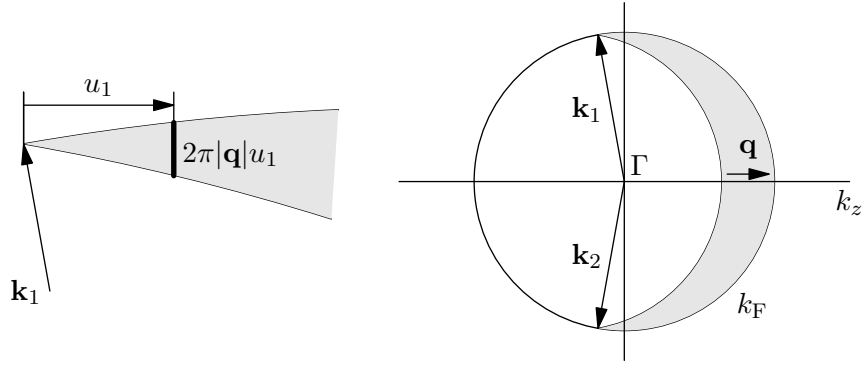

\begin{center}
\begin{asy}
    import graph;
    size(320,0);

    real dy=0.05;
    real dx=0.035;
    real q=0.35;
    real z=-q/2;
    real x=sqrt(1-z^2);
    pair gamma=(0,0);
    draw(circle(gamma,1),black);
    draw(arc((-q,0),(z,-x),(z,+x)),black);
    fill(
      arc(gamma,(z,-x),(z,+x))--arc((-q,0),(z,+x),(z,-x),CW)--cycle, lightgrey
    );
    draw(Label("$\vec q$",align=2N),(1-q+dx,dy)--(1-dx,dy),EndArrow);
    draw(Label("$\vec k_1$",align=W),gamma--(z,+x),EndArrow);
    draw(Label("$\vec k_2$",align=W),gamma--(z,-x),EndArrow);
    label("$\Gamma$",gamma,NE);
    label("$k_{\rm F}$",(1/sqrt(2),-1/sqrt(2)),SE);
    draw(Label("$k_z$",1,align=S), (-1.5,0)--(1.5,0));
    draw((0,-1.2)--(0,1.2));

    real phi=Degrees(asin(-z));
    real s=16;
    pair c=(-4,0.25);
    real b=0.75;
    draw(arc(c-s*(z,x),s,90+phi*(1-b),90+phi,CCW));
    draw(arc(c-s*(-z,x),s,90-phi,90-(1+b)*phi,CW));
    fill(
      arc(c-s*(z,x),s,90+phi*(1-b),90+phi,CCW)--
      arc(c-s*(-z,x),s,90-phi,90-(1+b)*phi,CW)--cycle,
      lightgrey
    );
    draw(Label("$\vec k_1$",0,align=W),(c-(z,x))--c,EndArrow);
    real l=1;
    real h=0.35;
    real x1=s*(sqrt(1-(-z+l/s)^2)-x);
    real x2=s*(sqrt(1-(-z-l/s)^2)-x);
    draw(c--(c+(0,h)));
    draw(c--(c+(0,h)));
    draw((c+(l,x2))--(c+(l,h)));
    draw(
      Label("$u_1$",align=N),
      (c+(0,h*0.75))--(c+(l,h*0.75)),
      EndArrow
    );
    draw(
      Label("$2\pi|\vec q|u_1$",0.4),
      (c+(l,x1))--(c+(l,x2)),linewidth(2.0)
    );
\end{asy}
\end{center}
\caption{
  Cross section of the set of momenta $\vec k_i$ such that
  $|\vec k_i|<k_{\rm F}<|\vec k_i+\vec q|$ for excitation momenta
  $\vec q\leq2$. The momentum of the Coulomb interaction
  $\vec k_1+\vec k_2+\vec q$ vanishes for the shown $\vec k_1$ and $\vec k_2$.
  The vicinity of the singular $\vec k_1$ is magnified
  on the left, showing the employed coordinates for the integration of this
  regions and a volume element in form of a ring with area $2\pi|\vec q|u_1$.
}
\label{fig:APX_DominantKs}
\end{figure}
For the frequency integration of metallic systems, such as the Uniform Electron
Gas, knowledge about the asymptotic behavior of the APX correction with
respect to large imaginary frequencies is important for choosing an appropriate
variable transform for the tail.

For large momenta $q$, we can use the approximation of $P_{\rm x}^{\rm APX}$ in
(\ref{eqn:APX_PxLargeQ}), gotten in the previous subsection.
Considering large frequencies $\nu$, we can still use the same approximation for
intermediate $q>2k_{\rm F}$ since the denominator of the propagator
$V(\vec k_1+\vec k_2+\vec q)$ is non-vanishing and the volume of the integration
region of $\vec k_1$ and $\vec k_2$ is independent of $q$. In this case the
integration of $\vec k_1$ and $\vec k_2$ merely averages the contributions.
For $q\leq2k_{\rm F}$, the volume of the integration region of $\vec k_1$ and
$\vec k_2$ depends on $q$ and there are also contributions where
the propagator $V(\vec k_1+\vec k_2+\vec q)$ becomes singular, as shown in
Figure \ref{fig:APX_DominantKs}.
In this case, we split the integration into two parts. In the first part
both, $\vec k_1$ and $\vec k_2$ are in the vicinity of the singular
contribution, shown for $\vec k_1$ on the left of Figure
\ref{fig:APX_DominantKs}. In the remaining part either $\vec k_1$, $\vec k_2$
or both are away from the singular contribution.
For the first part, we can transform the integration of $\vec k_1$ and
$\vec k_2$ into an integration of the respective distances $u_1$ and $u_2$
from the singular contribution in the direction of $\vec q$. In the remaining
part, we approximate the integral assuming that $\vec k_1+\vec k_2$ average out
and taking into consideration that the integration volume of each $\vec k_i$
scales like $q^3 + q^2$. We get
\begin{align}
  \nonumber
    \iint\limits_{|\vec k_i|<k_{\rm F}<|\vec k_i+\vec q|}
      \frac{\Omega^2 \d\vec k_1 \d\vec k_2}{(2\pi)^6}\,
    V(\vec k_1+\vec k_2+\vec q)
  &\sim
    \iint_0^{k_{\rm F}-q/2} \d u_1\d u_1\,
    \frac{(2\pi)^2q^2u_1u_2}{(u_1+u_2)^2}
    +
    \frac{(a_3q^3+a_2q^2)^2}{q^2}
  \\
    &\sim
    q^2 + \mathcal O\left(q^3\right)
  \,.
  \label{eqn:APX_PxIntegratedK}
\end{align}
Since $q\leq2$ and $\nu$ is large, we can
ignore the propagators of $P_{\rm x}^{\rm APX}$ containing the imaginary
frequency for the integration of $\vec k_1$ and $\vec k_2$.

\begin{figure}
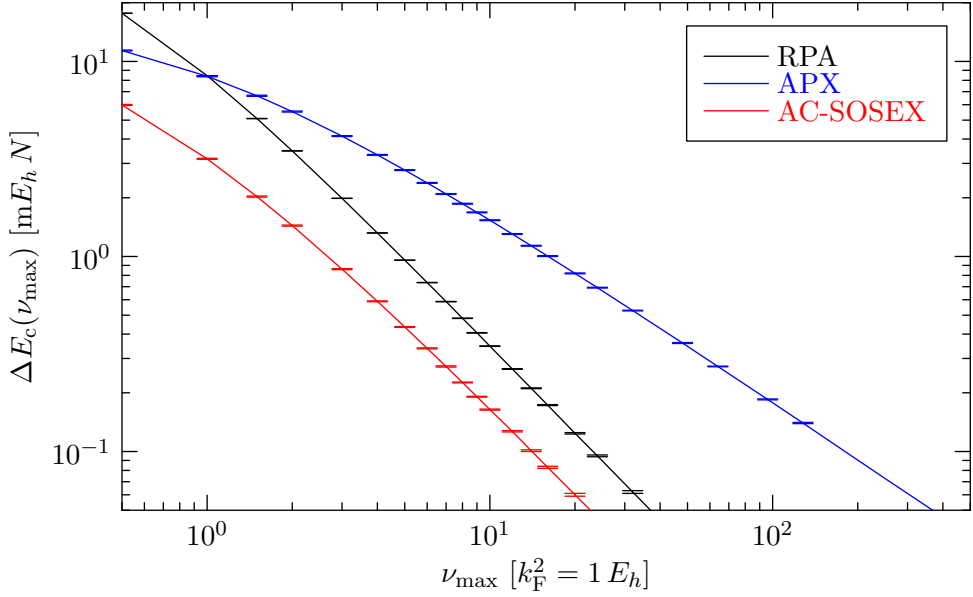

\begin{center}
\begin{asy}
  real[][] rpa = input("WTailRpaPara.dat").line().dimension(0,0);
  real[][] apx = input("WTailApxPara.dat").line().dimension(0,0);
  real[][] acSosex = input("WTailAcSosexPara.dat").line().dimension(0,0);

  sizeRatio(width=360);
  scale(Log,Log);
  plotXYDY(rpa, black, "RPA");
  plotXYDY(apx, blue, "APX");
  plotXYDY(acSosex, red, "AC-SOSEX");

  axisXY(
    Label("$\nu_{\rm max}$ [$k_{\rm F}^2 = 1\,E_h$]",MidPoint),
    Label("$\Delta E_{\rm c}(\nu_{\rm max})$ [m$E_h\,N$]",MidPoint),
    (0.5,5e-2), (500,20)
  );
  attach(legend(1,6,6,24,vskip=1),point(NE),12SW,UnFill);
\end{asy}
\end{center}
\caption{
  Numerical comparison of the missing correlation energy per electron for the
  Random Phase Approximation and its corrections when truncating
  the imaginary frequency integration at $\nu_{\rm max}$. $\nu_{\rm max}$
  is given in units of $k_{\rm F}^2$ and $k_{\rm F} = 1$\,a.u.
}
\label{fig:APX_WTailPara}
\end{figure}
We can now insert the respective approximation of $P_{\rm x}^{\rm APX}$
for small $q$, intermediate $q$ and large $q$ into the expression for
the APX energy (\ref{eqn:APX_ApxUEGPx}) and integrate $q$.
For $\chi_0$ we use the approximations (\ref{eqn:RPA_LargeFreqSmallQ}),
(\ref{eqn:RPA_LargeFreqIntermediateQ}) and (\ref{eqn:RPA_ChiLargeQ}) for
small, intermediate and large momenta $q$, respectively. The results are
long and apart from their expansion in $\nu$ of no particular interest. For
small and intermediate $q$ the integration of $q$ yields
$
  1/\nu^2 + \mathcal O\left(1/\nu^4\right)
$
and for large $q$ we get
$
  1/\nu^2 + \mathcal O\left(1/\nu^{5/2}\right)
$.
Finally, we can insert this expansion in the imaginary frequency integration
of the APX energy and estimate the missing energy per electron
$\Delta E_{\rm c}^{\rm APX}(\nu_{\rm max})$ when truncating the integration
at some finite but large imaginary frequency $\nu_{\rm max}$:
\begin{equation}
  \Delta E_{\rm c}^{\rm APX}(\nu_{\rm max} \sim \int_{\nu_{\rm max}}^\infty\d\nu
  \left(
    \frac1{\nu^2} + \mathcal O\left(\frac1{\nu^{5/2}}\right)
  \right)
  \sim \frac1{\nu_{\rm max}} +
    \mathcal O\left(\frac1{\nu_{\rm max}^{3/2}}\right)\,.
  \label{eqn:APX_LargeNu}
\end{equation}
The asymptotic behavior of the APX energy differs from that of the
Random Phase Approximation and from that of the AC-SOSEX, which can be
derived in an analogous fashion. This originates from the imaginary frequency
behavior of $P_{\rm x}^{\rm APX}$, containing only one of the four time orders
contained in the RPA and in the AC-SOSEX.
Figure \ref{fig:APX_WTailPara} compares the asymptotic behavior of RPA,
AC-SOSEX and APX for the Uniform Electron Gas numerically.

    \subsection{Spin-polarized Uniform Electron Gas}
\label{ssc:APX_UegFerromagnetic}
\begin{figure}[p]
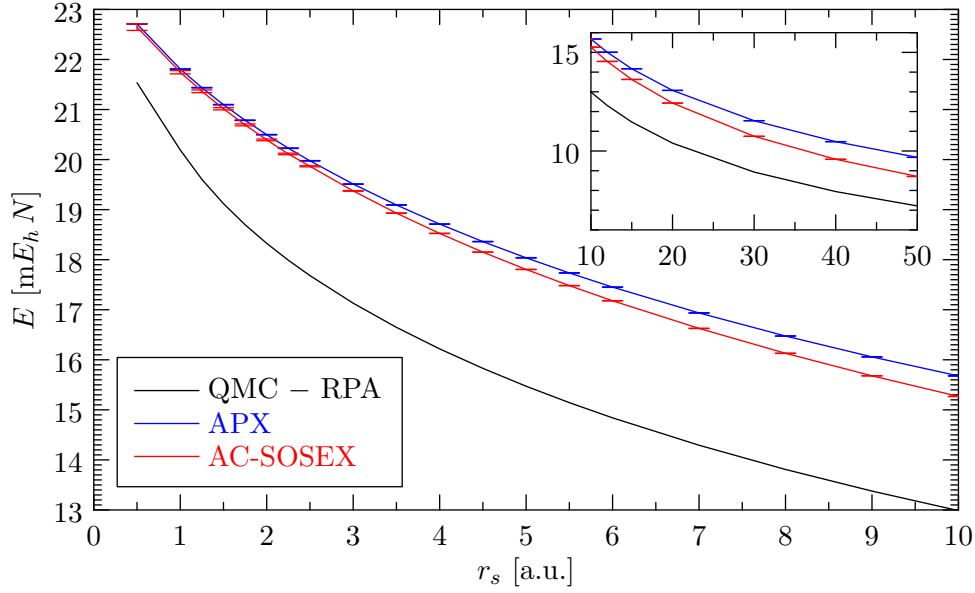

\begin{center}
\begin{asy}
  real[][] eps = input("EpsFerr.dat").line().dimension(0,0);
  real[][] apx = input("ApxFerr.dat").line().dimension(0,0);
  real[][] acSosex = input("AcSosexFerr.dat").line().dimension(0,0);

  sizeRatio(width=360);
  plotXY(eps, "QMC $-$ RPA");
  plotXYDY(apx, blue, "APX");
  plotXYDY(acSosex, red, "AC-SOSEX");
  axisXY(
    Label("$r_s$ [a.u.]",MidPoint), Label("$E$ [m$E_h\,N$]",MidPoint),
    (0,13), (10,23),
    yTicks=RightTicks(Step=1,n=10)
  );
  attach(legend(1,6,6,24,vskip=1),(0,13),12NE,UnFill);

  picture insetPic;
  sizeRatio(insetPic, width=0.4*360);
  plotXY(insetPic, eps);
  plotXYDY(insetPic, apx, blue);
  plotXYDY(insetPic, acSosex, red);
  axisXY(insetPic, (10,6), (50,16), yTicks=RightTicks(Step=5,n=5));

  add(insetPic.fit(), (10,23), 12SW);
\end{asy}
\end{center}
\vspace*{-2ex}
\caption{
  The Adjacent Pairs Exchange (APX) energy per electron for the
  spin-polarized uniform electron gas compared to the error of the Random Phase
  Approximation with respect to Quantum Monte-Carlo (QMC)
  calculations by \parencite{ceperley_ground_1980}
  fitted by \parencite{perdew_self-interaction_1981}.
  For low densities, anti-symmetrization of a ring diagram merely changes
  its sign, such that SOSEX or its variants simply remove the correlation
  energy already captured by the RPA.
}
\label{fig:APX_UegFerr}
\end{figure}
\begin{table}[p]
\begin{center}
\begin{tabular}{|c|r|rr|rr|}
\hline
&&&&& \\[-2.5ex]
  \multicolumn{1}{|c|}{$r_s$} &
    \multicolumn{1}{|c|}{$(E_{\rm c}-E_{\rm c}^{\rm RPA})$} &
    \multicolumn{2}{c|}{$E_{\rm c}^{\rm APX}$} &
    \multicolumn{2}{c|}{$E_{\rm c}^{\rm AC-SOSEX}$}
    \\
  \multicolumn{1}{|c|}{$[$a.u.$]$} &
    \multicolumn{1}{|c}{[m$E_h\,N$]} &
    \multicolumn{1}{|c}{[m$E_h\,N$]} &
    \multicolumn{1}{c|}{$\pm$} &
    \multicolumn{1}{|c}{[m$E_h\,N$]} &
    \multicolumn{1}{c|}{$\pm$}
     \\
&&&&& \\[-2.6ex]
\hline
&&&&& \\[-2.5ex]
 1 & 20.192 & 21.809 & 0.005 & 21.746 & 0.033 \\
 2 & 18.326 & 20.497 & 0.005 & 20.394 & 0.017 \\
 3 & 17.131 & 19.511 & 0.005 & 19.374 & 0.006 \\
 4 & 16.218 & 18.712 & 0.005 & 18.525 & 0.003 \\
 5 & 15.472 & 18.037 & 0.005 & 17.805 & 0.003 \\
 6 & 14.840 & 17.452 & 0.005 & 17.179 & 0.003 \\
 7 & 14.293 & 16.936 & 0.005 & 16.627 & 0.003 \\
 8 & 13.809 & 16.475 & 0.005 & 16.130 & 0.003 \\
 9 & 13.377 & 16.058 & 0.005 & 15.680 & 0.003 \\
10 & 12.987 & 15.678 & 0.005 & 15.268 & 0.003 \\
12 & 12.307 & 15.007 & 0.005 & 14.541 & 0.003 \\
15 & 11.469 & 14.169 & 0.004 & 13.628 & 0.003 \\
20 & 10.398 & 13.073 & 0.004 & 12.431 & 0.003 \\
30 &  8.932 & 11.536 & 0.004 & 10.745 & 0.003 \\
40 &  7.943 & 10.474 & 0.004 &  9.580 & 0.003 \\
50 &  7.215 &  9.678 & 0.004 &  8.710 & 0.003 \\
  \hline
\end{tabular}
\end{center}
\caption{
  Data of Figure \ref{fig:APX_UegFerr} including low densities.
}
\label{tab:APX_UegFerr}
\end{table}
We can readily evaluate the RPA and the Adjacent Pairs Exchange correction
for the spin-polarized Uniform Electron Gas using only one spin in the
sum over all spins $\sum_\sigma$, occurring in the expression of
$P_{\rm x}^{\rm APX}(\vec q,\im\nu)$ in (\ref{eqn:APX_PxUeg}) and
$\chi_0(\vec q,\im\nu)$
in (\ref{eqn:RPAChiUEG}). Considering that $k_{\rm F}$ also depends on
wheter the UEG is spin-polarized or not according to (\ref{eqn:RPA_UegKfOfRs}),
yields correlation energies of RPA+APX shown in
Figure \ref{fig:APX_UegFerr} and Table \ref{tab:APX_UegFerr}. For comparison,
the correlation energy of RPA+AC-SOSEX according to (\ref{eqn:SOSEX_ACUEG})
is also given.

Unlike in the non-spin-polarized case, neither of the SOSEX variants offers
a balanced correction to the Random Phase Approximation, overestimating
the missing energy for the entire range of densities. The accuracy with
respect to Quantum Monte Carlo results also worsens for low densities.
For low densities we can assume $\vec k_1+\vec k_2+\vec q \approx \vec q$
since $|\vec k_i|<k_{\rm F}\ll1$.
In this limit, two adjacent RPA bubbles only differ from
the respective exchange diagram in the Fermion sign and the additional
spin variable from having two loops instead of one:
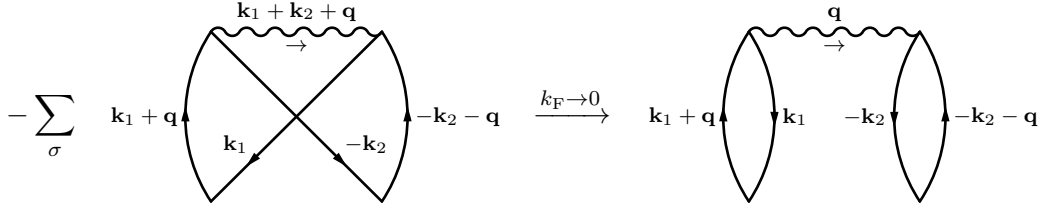
\[
  -\sum_\sigma
  \hspace*{8ex}
  \diagramBoxBorder{3ex}{3ex}{
    \begin{fmffile}{ACSOSEX3Open}
      \scriptsize
      \begin{fmfgraph*}(64,64)
        \fmfstraight
        \fmfset{arrow_len}{6}
        \fmfleft{v00,v01}
        \fmfright{v10,v11}
        \fmf{photon,label.side=right,label.dist=-10,label=$
          \begin{array}{c}
            \vec k_1+\vec k_2+\vec q \\[1ex]
            \rightarrow
          \end{array}
        $}{v11,v01}
        \fmffreeze
        \fmf{fermion,left=0.3,label.dist=3,label=$
          \vec k_1+\vec q
        $}{v00,v01}
        \fmf{plain}{v01,m1}
          \fmf{fermion,label.side=left,label.dist=3,label=$
            -\vec k_2
          $}{m1,v10}
        \fmf{fermion,right=0.3,label.dist=3,label=$
          -\vec k_2-\vec q
        $}{v10,v11}
        \fmf{plain}{v11,m2}
          \fmf{fermion,label.side=right,label.dist=3,label=$
            \vec k_1
          $}{m2,v00}
      \end{fmfgraph*}
    \end{fmffile}
  }
  \hspace*{10ex}
  \xrightarrow{k_{\rm F} \rightarrow 0}
  \hspace*{8ex}
    \diagramBoxBorder{3ex}{3ex}{
    \begin{fmffile}{RPA3Open}
      \scriptsize
      \begin{fmfgraph*}(64,64)
        \fmfstraight
        \fmfset{arrow_len}{6}
        \fmfleft{v00,v01}
        \fmfright{v10,v11}
        \fmf{photon,label.side=right,label.dist=-10,label=$
          \begin{array}{c}
            \vec q \\[1ex]
            \rightarrow
          \end{array}
        $}{v11,v01}
        \fmffreeze
        \fmf{fermion,left=0.3,label.dist=3,label=$
          \vec k_1+\vec q
        $}{v00,v01}
        \fmf{fermion,left=0.3,label.side=left,label.dist=3,label=$
          \vec k_1
        $}{v01,v00}
        \fmf{fermion,right=0.3,label.dist=3,label=$
          -\vec k_2-\vec q
        $}{v10,v11}
        \fmf{fermion,right=0.3,label.side=right,label.dist=3,label=$
          -\vec k_2
        $}{v11,v10}
      \end{fmfgraph*}
    \end{fmffile}
  }
  \hspace*{9ex}
\]
In the spin-polarized case, the exchange diagram entirely cancels
the two pair bubbles for densities low enough and as a consequence
RPA+SOSEX excatly cancels in the limit $k_{\rm F}\rightarrow0$.
RPA+APX becomes even positive, since APX contains more diagrams than RPA.
This occurs, however, only at very low densities beyond $r_s\approx88$.

    \section{APX for solids}
In solids or in molecules the Adjacent Pairs Exchange correction can be
evaluated in a similar fashion as the Random Phase Approximation.
Collecting positive and negative imaginary frequencies and rotating the
matrices cyclically converts all involved matrices into symmetric, real
valued matrices:
\[
  E_{\rm c}^{\rm APX} = -\frac12\int_0^\infty\frac{\d\nu}{2\pi}\,
  \Tr\left\{
    \vec V^\frac12
    \Big(
      \vec P_{\rm x}^{\rm APX}(\im\nu)+\vec P_{\rm x}^{\rm APX}(-\im\nu)
    \Big)\vec V^\frac12
    \Big(\vec 1-\vec V^\frac12\vec X_0(\im\nu)\vec V^\frac12\Big)^{-1}
  \right\}\,.
\]
This makes the evaluation of the inverse numerically more stable.
The independent particle polarizability $\vec X_0$ and the exchange
polarizability $\vec P_{\rm x}^{\rm APX}$ can be evaluated in imaginary
frequency according to (\ref{eqn:RPAChiImag}) and (\ref{eqn:APX_PxImagTime})
in $\mathcal O(N^3)$ and in $\mathcal O(N^5)$ steps, respectively.
The memory requirement for both cases is $\mathcal O(N^2)$.
Accepting a memory footprint of $\mathcal O(N^3)$, the evaluation of the 
APX and, similarly, of the AC-SOSEX correction can be improved to
$\mathcal O(N^4)$ by storing the intermediate tensor $\vec \Gamma(\im\nu)$
\[
  \vec \Gamma_{\vec x\vec x'\vec x''}(\im\nu) =
    \int_0^\infty\d\tau\,e^{+\im\nu\tau}
      {\vec G_0}_{\vec x\vec x'}(\im\tau)
      {\vec G_0}_{\vec x'\vec x''}(-\im\tau)
\]
and then evaluating the exchange polarizability in terms of $\vec \Gamma$:
\[
  {\vec P_{\rm x}^{\rm APX}}_{\vec x_1 \vec x_2}(\im\nu) =
    -\iint\d\vec x_3\,\d\vec x_4\,\frac1{|\vec r_3-\vec r_4|}
    \vec \Gamma_{\vec x_4\vec x_1\vec x_3}(+\im\nu)
    \vec \Gamma_{\vec x_3\vec x_2\vec x_4}(-\im\nu)\,.
\]

\subsection{$k$-points and $G_{\rm max}$ convergence}
\begin{figure}[t]
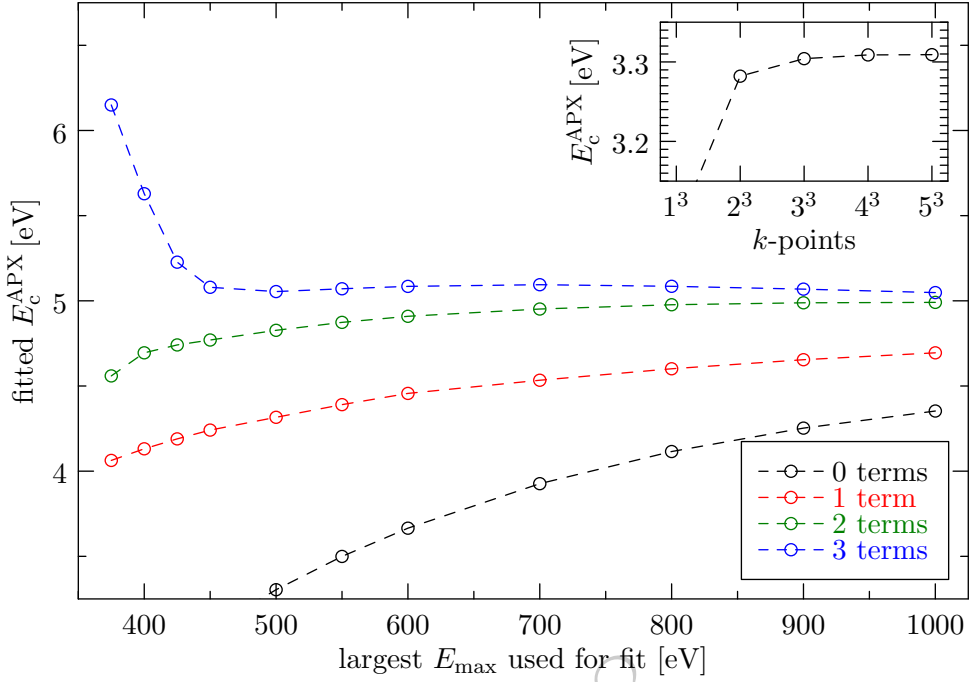

\begin{center}
\begin{asy}
  real[][] apx = input("EncutConvergence.dat").line().dimension(0,0);
  real[][] apx1 = input("EncutConvergence1.dat").line().dimension(0,0);
  real[][] apx2 = input("EncutConvergence2.dat").line().dimension(0,0);
  real[][] apx3 = input("EncutConvergence3.dat").line().dimension(0,0);

  sizeRatio(width=360,ratio=sqrt(2));
  plotXY(apx, black+dashed, "0 terms", marker(scale(0.8mm)*unitcircle,black));
  plotXY(apx1, red+dashed, "1 term", marker(scale(0.8mm)*unitcircle,red));
  plotXY(apx2, green*0.5+dashed, "2 terms", marker(scale(0.8mm)*unitcircle,green*0.5));
  plotXY(apx3, blue+dashed, "3 terms", marker(scale(0.8mm)*unitcircle,blue));
  axisXY(
    Label("largest $E_{\rm max}$ used for fit [eV]",MidPoint),
    Label("fitted $E_{\rm c}^{\rm APX}$\,[eV]",MidPoint),
    (350,3.25), (1025,6.75)
  );
  attach(legend(1,6,6,24,vskip=1),point(SE),10NW,UnFill);

  picture insetPic;
  // TODO: Housenumbers ;-)
  real[][] apxK = input("KPointsConvergence.dat").line().dimension(0,0);
  sizeRatio(insetPic, width=0.4*360);
  plotXY(insetPic, apxK, black+dashed,marker(scale(0.8mm)*unitcircle,black));
  axisXY(
    insetPic,
    xLabel=Label("$k$-points",MidPoint),
    yLabel=Label("$E_{\rm c}^{\rm APX}$\,[eV]",MidPoint),
    xTicks=LeftTicks("$
    (0.75,3.15), (5.25,3.35)
  );

  add(insetPic.fit(), (1025,6.75), 10SW);
\end{asy}
\end{center}
\caption{
  Basis set extrapolation of the APX energy for LiF with 10 electrons.
  The black points show the energies obtained at different cutoff energies
  $E_{\rm max}$.
  The colored points show a fit using one, two or three orders in the
  expansion of the APX energy with respect to large cutoff momenta.
  Each fit only uses data points with a cutoff energy below or equal to that
  of the fitted point.
  The inset shows the convergence of the APX energy with respect to different
  $k$ point meshes.
}
\label{fig:APX_LiFConvergence}
\end{figure}
In the diagrams of the Random Phase Approximation each Coulomb interaction
mediates the same momentum $q$ according to momentum conservation at each
vertex. This makes the RPA the most important contribution for small momenta
or, equivalently, at long distances. The diagrams of the Adjacent Pairs
Exchange correction contain one Coulomb interaction with a different momentum
than all others. Thus, the APX correction is more short ranged than the
Random Phase Approximation resulting in a faster $k$-point convergence, as
shown in the inset of Figure \ref{fig:APX_LiFConvergence}.


The downside of the short ranged nature of the APX is that the convergence of
the APX energy is slower with respect to the highest momentum $G_{\rm max}$
contained in $\vec X_0$ and $\vec P_{\rm x}^{\rm APX}$. Assuming that
the system is homogeneous at distances short enough, the asymptotic behavior
of the missing APX energy is equivalent to that of the Uniform Electron Gas,
shown in Subsection \ref{ssc:APX_UEGLargeMomentum}.
Figure \ref{fig:APX_LiFConvergence} shows
the convergence of the APX energy with respect to the employed energy cutoff
$E_{\rm max}=G_{\rm max}^2/2$. Note that the automatic finite basis set
extrapolation implemented in \VASP\ cannot be used for the APX correction for
two reasons. First, unlike $\vec X_0$, $\vec P_{\rm x}^{\rm APX}$ changes
with each considered cutoff momentum since the Coulomb kernel changes.
Recalculating the exchange polarizability is, however, too time consuming.
Second, it is often not sufficient to consider only the leading order
term of the asymptotic behavior. For an accurate extrapolation to
$G_{\rm max}\rightarrow\infty$ it may be necessary to include the second, or
higher order terms of the large momentum expansion for the Uniform Electron Gas.
The asymptotic behavior can then be fit to
\[
  \Delta E_{\rm c}^{\rm APX}(G_{\rm max}) \sim
  a_3\,\frac1{G_{\rm max}^3} +
  a_7\,\frac1{G_{\rm max}^7} +
  \mathcal O\left(\frac1{G_{\rm max}^{11}}\right)\,,
\]
where $\texttt{ENMAX}=G_{\rm max}^2/2$ is manually risen.
Only the first result line with the largest automatically chosen
cutoff should be used and the number of bands \texttt{NBANDS} should be
close to the maximum number of bands specified in the
\texttt{OUTCAR} file of the DFT or Hartree-Fock calculation for each
respective value of \texttt{ENCUT}. Note that \texttt{NBANDS} must be a
multiple of the number of cores employed for the APX calculation.
To avoid aliasing effects it is therefore advantageous to choose \texttt{ENMAX}
such that the reported maximum number of bands is as close as possible
to \texttt{NBANDS}.

\subsection{Frequency grid}
Directly evaluating (\ref{eqn:APX_Px}) would require $\mathcal O(N^8)$ steps.
However, $\vec P_{\rm x}^{\rm APX}$ can be calculated in imaginary time
using the imaginary time propagators given in (\ref{eqn:RPAGreenImag}).
Subsequently, it can be Fourier transformed numerically to imaginary frequency
on a non-equidistant grid,  analogous to calculating the independent
particle polarizability as proposed by \parencite{kaltak_low_2014}.
\begin{align}
  \nonumber
  {\vec P_{\rm x}^{\rm APX}}_{\vec x_1 \vec x_2}(\im\nu) =
    -\iint\d\vec x_3\,\d\vec x_4\,\frac1{|\vec r_3-\vec r_4|} &
    \int_0^\infty\d\tau_1\,e^{+\im\nu\tau_1}
      {\vec G_0}_{\vec x_3\vec x_1}(\im\tau_1)
      {\vec G_0}_{\vec x_1\vec x_4}(-\im\tau_1)
  \\ &
    \int_0^\infty\d\tau_2\,e^{-\im\nu\tau_2}
      {\vec G_0}_{\vec x_4\vec x_2}(\im\tau_2)
      {\vec G_0}_{\vec x_2\vec x_3}(-\im\tau_2)
  \label{eqn:APX_PxImagTime}
\end{align}
The imaginary time propagators can be calculated in $\mathcal O(N^3)$ steps.
If the number of samples for the two imaginary time and the final imaginary
frequency integration is independent of the system size, the evaluation
of the APX scales like $\mathcal O(N^4)$. The prefactor might, however, be
considerable.

\begin{figure}
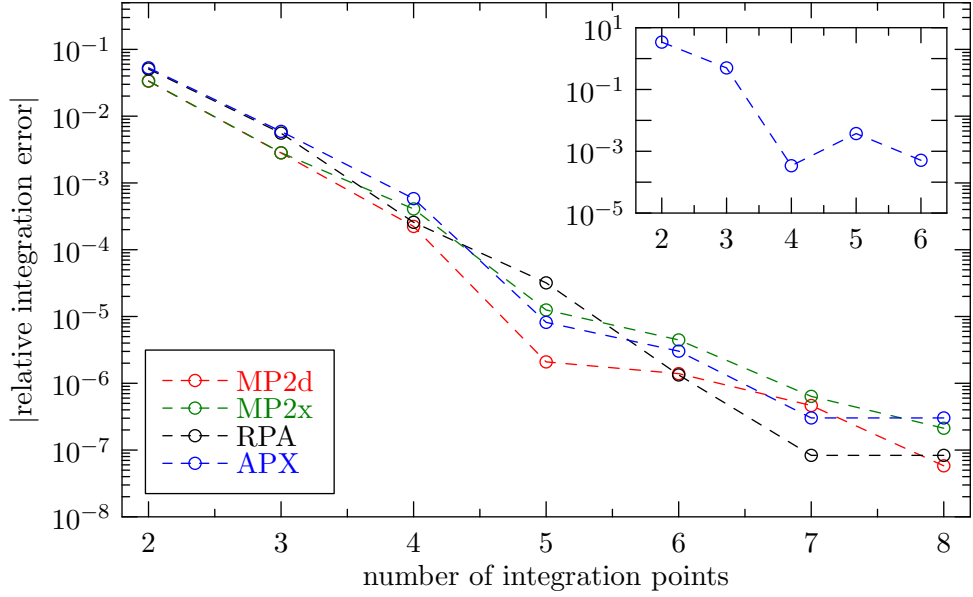

\begin{center}
\begin{asy}
  real[][] mp2d = input("FrequenciesErrorMp2d.dat").line().dimension(0,0);
  real[][] mp2x = input("FrequenciesErrorMp2x.dat").line().dimension(0,0);
  real[][] rpa = input("FrequenciesErrorRpa.dat").line().dimension(0,0);
  real[][] apx = input("FrequenciesErrorApx.dat").line().dimension(0,0);

  sizeRatio(width=360);
  scale(Linear,Log);
  plotXY(mp2d, red+dashed, "MP2d", marker(scale(0.8mm)*unitcircle,red));
  plotXY(mp2x, green*0.5+dashed, "MP2x",marker(scale(0.8mm)*unitcircle,green*0.5));
  plotXY(rpa, black+dashed, "RPA", marker(scale(0.8mm)*unitcircle,black));
  plotXY(apx, blue+dashed, "APX", marker(scale(0.8mm)*unitcircle,blue));
  axisXY(
    Label("number of integration points",MidPoint),
    Label("$|$relative integration error$|$",MidPoint),
    (1.8,1e-8), (8.2,0.5)
  );
  attach(legend(1,6,6,24,vskip=1),point(SW),12NE,UnFill);

  picture insetPic;
  real[][] apxDiff = input("FrequenciesErrorApxDiff.dat").line().dimension(0,0);
  sizeRatio(insetPic, width=0.4*360);
  scale(insetPic, Linear, Log);
  plotXY(insetPic, apxDiff, blue+dashed, marker(scale(0.8mm)*unitcircle,blue));
  axisXY(insetPic, (1.6,1e-5), (6.4,10), yTicks=RightTicks(N=2,n=1));

  add(insetPic.fit(), (8.2,-0.1), 12SW);
\end{asy}
\end{center}
\caption{
  Relative error of the imaginary frequency integration for different
  approximations of the correlation energy of LiF, depending on the
  number of samples. The system contains 10 electrons and<s 384 bands.
  The inset shows the error for the difference of the APX energy
  for different unit cell volumes, once at 15.4\,\AA$^3$ and once at
  15.9\,\AA$^3$.
  The error is below 1\,$\mu$eV beyond the shown number of integration points.
}
\label{fig:APX_LiFFrequencies}
\end{figure}
In the Random Phase Approximation the employed quadrature frequencies and
weights are determined from a fit to the function of the direct MP2 energy,
since it resembles the RPA energy function and its exact frequency integral is
known. In the case of the APX energy the respective lowest order function would
be the exchange MP2 energy
\[
  \frac12\int\frac{\d\nu}{2\pi}\,\frac{
    2(\eps_a-\eps_i)(\eps_b-\eps_j) + 2\nu^2
  }{
    \left((\eps_a-\eps_i)^2+\nu^2\right)
    \left((\eps_b-\eps_j)^2+\nu^2\right)
  }
  =
  \frac1{\eps_i+\eps_j-\eps_a-\eps_b}
\]
and the quadrature frequencies $\nu_k$ and weights $w_k$ could be fit
such that the dominant terms with $a=b$ and $i=j$
are best reproduced by the numeric integral
\begin{equation}
  \frac1{\eps_i+\eps_i-\eps_a-\eps_a}
  \approx
    \frac12 \sum_k w_k
    \frac{
      2(\eps_a-\eps_i)^2+2\nu_k^2
    }{
      \left((\eps_a-\eps_i)^2+\nu_k^2\right)^2
    }
\end{equation}
for all single particle excitation energies $\eps_a-\eps_i$. In general,
this yields a different grid of optimal frequencies and weights than the
grid obtained for the Random Phase Approximation according to
(\ref{eqn:RPA_NuFit}). However, the APX energy also contains the independent
particle polarizability in all orders beyond the second order, which
suggests that the a frequency grid, optimized for the exchange MP2 term, might
no longer be optimal for higher orders anyway.
Figure \ref{fig:APX_LiFFrequencies}
shows the convergence of the APX energy in LiF containing 10 electrons with
respect to the number of
imaginary frequency points, using a grid which is optimized for the
RPA rather than for the APX. The convergence hardly differs for the different
energies, justifying the use of the
RPA optimized grid for calculating $\vec P_{\rm x}^{\rm APX}$ and subsequently
the APX energy.
For metallic systemy, knowledge about asymptotic behavior of the APX with
respect to large imaginary frequencies is important since the quality
of the frequency grid degrades with vanishing band gap.
APX has a different behavior for large frequencies in the Uniform Electron Gas
than the Random Phase Approximation, as discussed in Subsection
\ref{ssc:APX_UEGLargeFrequency}. However, since APX requires a different
calculation setup with less $k$-points but higher plane wave cutoff momenta,
evaluating APX and RPA on different frequency grids in metallic systems does
not pose a considerable disadvantage.

\subsection{Lattice constants}
\begin{figure}[p]
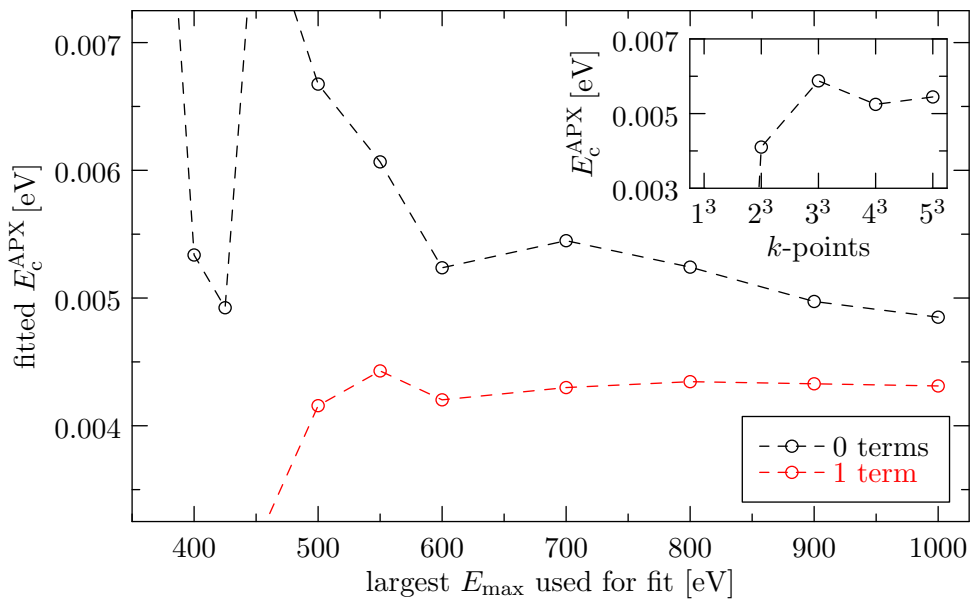

\begin{center}
\begin{asy}
  real[][] apx = input("EncutConvergenceDiff.dat").line().dimension(0,0);
  real[][] apx1 = input("EncutConvergenceDiff1.dat").line().dimension(0,0);

  sizeRatio(width=360);
  plotXY(apx, black+dashed, "0 terms", marker(scale(0.8mm)*unitcircle,black));
  plotXY(apx1, red+dashed, "1 term", marker(scale(0.8mm)*unitcircle,red));
  axisXY(
    Label("largest $E_{\rm max}$ used for fit [eV]",MidPoint),
    Label("fitted $E_{\rm c}^{\rm APX}$\,[eV]",MidPoint),
    (350,0.00325), (1025,0.00725)
  );
  attach(legend(1,6,6,24,vskip=1),point(SE),10NW,UnFill);

  picture insetPic;
  // TODO: converge k-point convergence data with respect to the cutoff
  real[][] apxK = input("KPointsConvergenceDiff.dat").line().dimension(0,0);
  sizeRatio(insetPic, width=0.4*360);
  plotXY(insetPic, apxK, black+dashed,marker(scale(0.8mm)*unitcircle,black));
  axisXY(
    insetPic,
    xLabel=Label("$k$-points",MidPoint),
    yLabel=Label("$E_{\rm c}^{\rm APX}$\,[eV]",MidPoint),
    xTicks=LeftTicks("$
    (0.75,0.003), (5.25,0.007)
  );
  add(insetPic.fit(), (1025,0.0073), 10SW);
\end{asy}
\end{center}
\caption{
  Basis set extrapolation of the APX energy difference at different volumes
  of LiF containing 10 electrons. The volumes of the primitive unit cell
  are 15.9\,\AA$^3$ and 15.4\,\AA$^3$.
  The black points show the energy differences obtained at different cutoff
  energies $E_{\rm max}$. Aliasing effects can be clearly seen.
  The red points show a fit using in the leading order of the
  expansion of the APX energy with respect to large cutoff momenta.
  Each fit only uses data points with a cutoff energy below or equal to that
  of the fitted point. Higher order fits offer no improvement since they
  largely cancel in the difference.
  The inset shows the convergence of the APX energy difference with respect to
  different $k$ point meshes.
}
\label{fig:APX_LiFDiffConvergence}
\end{figure}
\begin{table}[p]
\begin{center}
\begin{tabular}{|lc|rr|rr|rr|r|}
\hline
\multicolumn{2}{|c|}{\multirow{2}{*}{Solid}}& \multicolumn{7}{|c|}{} \\[-2.5ex]
  &&
    \multicolumn{7}{|c|}{$a_0$\,[\AA]}
  \\
&& \multicolumn{7}{|c|}{} \\[-2.6ex]
\cline{3-9}
&&&&&&&& \\[-2.5ex]
  &&
    \multicolumn{1}{|c}{RPA} &
    \multicolumn{1}{c}{\%} &
    \multicolumn{1}{|c}{SOSEX} &
    \multicolumn{1}{c}{\%} &
    \multicolumn{1}{|c}{APX} &
    \multicolumn{1}{c}{\%} &
    \multicolumn{1}{|c|}{exp.}
     \\
\hline
&&&&&&&& \\[-2.5ex]
C    &(A4) & 3.572 & 0.5 & 3.552 & $<$0.1 & 3.550 & $-0.1$ & 3.553 \\
LiH  &(B1) & 3.983 & 0.1 & 3.989 & 0.3 & 3.992 & 0.4 & 3.979 \\
LiF  &(B1) & 3.998 & 0.7 & 3.955 & $-$0.4 & 3.974 & $<$0.1 & 3.972 \\
LiCl &(B1) & 5.074 & $<$0.1 & \multicolumn{2}{|c|}{---} & --- & --- & 5.070 \\
NaF  &(B1) & 4.625 & 0.9 & \multicolumn{2}{|c|}{---} & --- & --- & 4.582 \\
MgO  &(B1) & 4.225 & 0.9 & \multicolumn{2}{|c|}{---} & --- & --- &  4.189 \\
GaP  &(B3) & 5.442 & $<$0.1 & \multicolumn{2}{|c|}{---} & --- & --- &  5.439 \\
\hline
\end{tabular}
\end{center}
\caption{
  Lattice constants from the Random Phase Approximation, SOSEX and the
  Adjacent Pairs Exchange correction compared to experiment. The
  experimental lattice constants were extrapolated to $T=0$\,K and to
  exclude the zero point energy (ZPE).
}
\label{tab:APX_LatticeConstants}
\end{table}
Calculating lattice constants beyond the Random Phase Approximation requires
sub-meV accuracy for the energy difference at different volumes of the
system under consideration.
In this case, the basis set extrapolation of the APX energy differences does
not require more than one term in the expansion of the large momentum behavior.
In fact, the quality of the extrapolation deteriorates when using more than one
term in the expansion since the higher order terms mostly cancel in the energy
difference. Figure \ref{fig:APX_LiFDiffConvergence} shows the basis set
convergence of the energy difference for LiF with 10 electrons. The
volumes of the primitive cells are 15.9\,\AA$^3$  and 15.4\,\AA$^3$.
The data contains
considerable aliasing effects since the number of bands must be a multiple
of the number of cores employed by the calculation, such that the number of
bands is not always equally close to the maximum number of plane waves
supported by the respective cutoff energy. All calculations were
conducted with a $3\times3\times3$ $k$-points mesh, providing sufficient
accuracy, as indicated in the inset of Figure \ref{fig:APX_LiFDiffConvergence}.

At each volume, the Adjacent Pairs Exchange
correction is calculated and converged with respect to the cutoff energy
using \VASP\ with a PBE-DFT reference. 
The Exact-Exchange+ RPA results were taken from the Birch Murnaghan fit of
previous converged RPA calculation, employing the same PAW potentials.
The resulting RPA+APX energy is then fit to a Birch-Murnaghan equation of state
to retrieve the RPA+APX lattice constants. The pressure derivative of the
bulk modulus was taken from ...
Table \ref{tab:APX_LatticeConstants} lists lattice constants from
RPA, SOSEX and APX in comparison with experiment, extrapolated to $T=0$\,K and
corrected to exclude the phononic zero point energy. 

\section{Summary and discussion}
The Random Phase Approximation exhibits a systematic error which origins, at
least partially, from violations of the Pauli exclusion principle. As discussed
in the beginning of Chapter \ref{cha:SOSEX}, violating contributions are
canceled by diagrams where the propagators of the offending states
are exchanged. In the lowest order only one pair of propagators is exchanged.

The Adjacent Pairs Exchange (APX) correction to the RPA constitutes the
largest set of Feynman diagrams correcting as many violations as possible
in lowest order without introducing new violations and still forming
one ring only:
\[
  E_{\rm c}^{\rm APX} =
  \diagramBox{\fmfreuse{MP2x}}
  +
  \hspace*{2ex}
  \diagramBox{\fmfreuse{ApxVFeyn}}
  \hspace*{1ex}
  +
  \hspace*{2ex}
  \diagramBox{\fmfreuse{ApxV2Feyn}}
  \hspace*{1ex}
  +\ \ldots
  \qquad
  \textnormal{with}
  \quad
  \diagramBox{
    \fmfreuse{Apx2Polarizability}
  }
  =
  \hspace*{2ex}
  \diagramBox{\fmfreuse{Exchange2Down}}
  \hspace*{2ex}
  .
\]
This makes the Random Phase Approximation plus the Adjacent Pairs Exchange
correction a purely \emph{ab-initio} choice of diagrams when the memory
scaling for their evaluation must not exceed $\mathcal O(N^2)$, since
the ring form of the APX allows its evaluation using only polarization parts
with 2 open ends and RPA is the most important class of diagrams for high
densities, becoming exact in the limit $r_s\rightarrow0$. This argument
relies on the prevalence of long ranged states making the RPA an accurate
approximation in the first place, which might not hold in the presence of more
localized states at the Fermi edge. However, in cases where the RPA offers a
good approximation, no test against experiment is required to argue
that APX offers the lowest order correction - which makes it
\emph{ab-initio}\footnote{
  Still, only a test against the uniform electron gas induced further
  examination of the two sided APX and the possibility of new violations
  introduced by it. Testing against systems - theoretical or not - is therefore
  necessary, of course.
}.

Table \ref{tab:APX_Approximations} compares the SOSEX, the AC-SOSEX and the APX
correction to the Random Phase Approximation in terms of contained Goldstone
diagrams, along with their respective computational costs in time and memory.
The diagrams of SOSEX constitute the largest set of
Goldstone, rather than Feynman, diagrams correcting as many violations as
possible under the same conditions than in the APX. In the case of SOSEX,
forming a ring is necessary for the applicability of the direct ring
Coupled Cluster Doubles amplitudes, which can be constructed in
$\mathcal O(N^5)$ steps, as discussed in Section \ref{sec:RPAdrCCD}.
Dropping the condition for the SOSEX diagrams to form a ring raises the time
complexity for the construction of the amplitudes to $\mathcal O(N^6)$.
Accepting this time complexity and the memory complexity of SOSEX, the full
Coupled Cluster Singles and Doubles (CCSD) approximation can be employed.
It includes considerably more diagrams, as indicated in Table
\ref{tab:APX_Approximations}, relieving the need of a correction to
the Random Phase Approximation in general. In cases, where CCSD is known
to be less accurate, as for instance in dissociation processes, the
distinguishable Coupled Cluster method may offer improvement at the same
computational costs.

\begin{table}[t]
\begin{center}
\begin{tabular}{|c|c|c|c|}
  \hline
  &&& \\[-2.5ex]
  \multirow{1}{*}{approximation} &
    \multirow{1}{*}{Goldstone diagrams} & time & memory \\
 &&& \\[-2.6ex]
  \hline
 &&& \\[-2.6ex]
  SOSEX &
    $
      \diagramBox{\fmfreuse{Exchange2}}
      +
      \
      \diagramBoxBorder{1ex}{1ex}{\fmfreuse{BubblesExchange3}}
      +
      \
      \diagramBoxBorder{1ex}{1ex}{\fmfreuse{BubblesExchange4}}
      \ +\ \ldots
    $
    & $\mathcal O(N^5)$ & $\mathcal O(N^4)$
  \\
  \hline
  AC-SOSEX &
    $
      \diagramBox{\fmfreuse{Exchange2}}
      +\displaystyle \frac13\left(
        \ldots + \hspace*{1ex} 
        \diagramBoxBorder{1ex}{1ex}{\fmfreuse{MP3SwappedLadder}}
        \hspace*{2ex} + \ldots
      \right)
      \ +\ \ldots
    $      
    & $\mathcal O(N^5)$ & $\mathcal O(N^2)$
  \\
  \hline
  APX &
   SOSEX +
      \diagramBoxBorder{1ex}{1ex}{\fmfreuse{MP4x2b}}
      +\ \ldots
    & $\mathcal O(N^5)$ & $\mathcal O(N^2)$
  \\
  \hline
  CCSD & RPA + APX +
    \hspace*{-1ex}
    \diagramBox{
      \begin{fmffile}{BubblesInter3}
      \begin{fmfgraph}(30,40)
        \fmfset{arrow_len}{6}
        \fmfstraight
        \fmfleft{v11,v12}
        \fmfright{v21,v22}
        \fmf{photon}{v11,v21}
        \fmf{photon}{v12,v22}
        \fmffreeze
        \fmf{fermion,left=0.3}{v11,v12} \fmf{fermion,left=0.15}{v12,vm1,v11}
        \fmf{fermion,left=0.15}{v21,vm2,v22} \fmf{fermion,left=0.3}{v22,v21}
        \fmf{photon,tension=0.8}{vm1,vm2}
      \end{fmfgraph}
      \end{fmffile}
    }
    \hspace*{-1ex}
    +
    \hspace*{-2ex}
    \diagramBoxBorder{1ex}{1ex}{
      \begin{fmffile}{BubblesLadder3}
      \begin{fmfgraph}(35,40)
        \fmfset{arrow_len}{6}
        \fmfstraight
        \fmfleft{v11,v12,v13}
        \fmfright{v31,v32,v33}
        \fmf{photon}{v11,v21} \fmf{phantom,tension=4}{v21,v31}
        \fmf{phantom}{v12,v22} \fmf{photon,tension=1.5}{v22,v32}
        \fmf{photon}{v13,v23} \fmf{phantom,tension=4}{v23,v33}
        \fmffreeze
        \fmf{fermion,left=0.3}{v11,v13,v11}
        \fmf{fermion,left=0.17}{v21,v22,v23,v32,v21}
      \end{fmfgraph}
      \end{fmffile}
    }
    \hspace*{-3ex}
    +\ \ldots
    & $\mathcal O(N^6)$ & $\mathcal O(N^4)$
  \\
  \hline
\end{tabular}
\end{center}
\caption{
  Comparison of different approximations beyond the Random Phase Approximation,
  showing the lowest order Goldstone diagrams introduced by the respective
  approximation.
  The AC-SOSEX is not derived within the same many-body perturbation theory
  framework as the other approximations. It can, however, be translated into
  Goldstone diagrams when including swapped ladder diagrams, as shown here
  in third order and which are discussed in Section \ref{sec:SOSEX_Difference}.
}
\label{tab:APX_Approximations}
\end{table}
Including exchange diagrams to cancel violating contributions in
the RPA diagrams offers an \emph{ab-initio} strategy for selecting the
lowest order diagrams of many-body perturbation theory, correcting for RPA's
systematic error.
In the framework of the Adiabatic Connection, the
unknown exchange-correlation kernel $\vec f^\lambda_{\rm xc}$, occurring in the
Dyson-like equation (\ref{eqn:RPAChiDyson}) of the polarizability
$\vec X_\lambda$, has to be included for a correction to the RPA.
The AC-SOSEX contains the exchange-correlation kernel in lowest order:
\[
  \vec X^{\rm AC-SOSEX}_\lambda =
    \vec X_0 \tilde{\vec f}^\lambda_{\rm xc} \vec X_0 +
    \vec X_0 \tilde{\vec f}^\lambda_{\rm xc} \vec X_0 \lambda\vec V\vec X_0 +
    \ldots\,,
\]
approximating the kernel by $\tilde{\vec f}^\lambda_{\rm xc}$, which is given
implicitly by
\[
  \vec X_0 \tilde{\vec f}^\lambda_{\rm xc}\vec X_0 =
  \lambda
  \left(
    \diagramBox[0.5]{
      \begin{fmffile}{fxc1}
      \begin{fmfgraph*}(32,64)
        \fmfstraight
        \fmfset{arrow_len}{6}
        \fmfleft{v00,v01,v02}
        \fmfright{v10,v11,v12}
        \fmf{photon}{v11,v01}
        \fmffreeze
        \fmf{fermion,left=0.3}{v01,v02}
        \fmf{fermion}{v02,v11}
        \fmf{fermion,right=0.3}{v10,v11}
        \fmf{fermion}{v01,v10}
      \end{fmfgraph*}
      \end{fmffile}
    }
    +
    \diagramBox[0.0]{
      \begin{fmffile}{fxc2}
      \begin{fmfgraph*}(32,32)
        \fmfstraight
        \fmfset{arrow_len}{6}
        \fmfleft{v00,v01}
        \fmfright{v10,v11}
        \fmf{photon}{v10,v00}
        \fmffreeze
        \fmf{fermion,left=0.3}{v00,v01}
        \fmf{plain}{v01,m1}
          \fmf{fermion}{m1,v10}
        \fmf{fermion,right=0.3}{v10,v11}
        \fmf{plain}{v11,m2}
          \fmf{fermion}{m2,v00}
      \end{fmfgraph*}
      \end{fmffile}
    }
    +
    \diagramBox[1.0]{
      \begin{fmffile}{fxc3}
      \begin{fmfgraph*}(32,32)
        \fmfstraight
        \fmfset{arrow_len}{6}
        \fmfleft{v00,v01}
        \fmfright{v10,v11}
        \fmf{photon}{v11,v01}
        \fmffreeze
        \fmf{fermion,left=0.3}{v00,v01}
        \fmf{plain}{v01,m1}
          \fmf{fermion}{m1,v10}
        \fmf{fermion,right=0.3}{v10,v11}
        \fmf{plain}{v11,m2}
          \fmf{fermion}{m2,v00}
      \end{fmfgraph*}
      \end{fmffile}
    }
    +
    \diagramBoxBorder[0.5]{0ex}{1ex}{
      \begin{fmffile}{fxc4}
      \begin{fmfgraph*}(32,64)
        \fmfkeep{ACSOSEX4}
        \fmfstraight
        \fmfset{arrow_len}{6}
        \fmfleft{v00,v01,v02}
        \fmfright{v10,v11,v12}
        \fmf{photon}{v11,v01}
        \fmffreeze
        \fmf{fermion,left=0.3}{v00,v01}
        \fmf{fermion}{v11,v00}
        \fmf{fermion,right=0.3}{v11,v12}
        \fmf{fermion}{v12,v01}
      \end{fmfgraph*}
      \end{fmffile}
    }
    \hspace*{.5ex}
  \right)\,.
\]
In the first and in the last diagram, particles turn into holes and vice versa
at the Coulomb interaction, which has no correspondence in many-body
perturbation theory.
This is, therefore, not an obvious choice for an approximation
in the author's opinion. It neither originates from time dependent DFT
nor from many-body perturbation theory, which is employed by AC-RPA after
all to calculate the independent particle polarizability $\vec X_0$. 
As discussed in Section \ref{sec:SOSEX_ACSOSEX}, the above approximation rather
comes from transforming the RPA energy expression into the particle/hole basis,
applying the analogy of SOSEX to exchange two indices, ignoring the time order,
and then transforming it back to the position basis.
However, the results are very similar to those of SOSEX for reasons discussed
in Section \ref{sec:SOSEX_Difference}, which are in no obvious connection to
its derivation. Owing to its similar results to SOSEX, the AC-SOSEX can be
considered a numerical recipe to approximate the SOSEX energy, requiring only
$\mathcal O(N^2)$ of memory.

The Adjacent Pairs Exchange corrections contains more diagrams than
the Second Order Screened Exchange correction since APX is the pendant of
SOSEX in terms of Feynman diagrams rather than Goldstone diagrams.
This does, however, not imply that APX is more accurate in any given situation.
After all, the surprisingly high accuracy of SOSEX or APX in case of the
uniform electron gas is certainly fortuitously and no theory, involving
only first and second order exchange diagrams can be expected to provide
this accuracy in general.
The spin-polarized uniform electron gas is an example of a system, where
RPA+SOSEX or RPA+APX can yield zero or even positive correlation energies in
the limit of dilute densities. For densities of real metals, the performance
in the spin-polarized case is better. APX misses an accurate correction
to RPA by less than 20\% and SOSEX is slightly better. It is assumed that
AC-SOSEX is close to SOSEX since no SOSEX calculations for the spin-polarized
UEG in the thermodynamic limit have been found. It is worth
mentioning that finite spin-polarized systems often exhibit a 
(quasi)-degenerate ground state, rendering many-body
perturbation theory, as discussed here and employed by most
implementations, inaccurate at best. (Quasi)-degenerate many-body perturbation
theory is discussed, for instance, in \parencite{shavitt_many-body_2009}.

APX and SOSEX are identical up to third order resulting only in small
differences. In the case of the non-spin-polarized uniform electron gas,
RPA+SOSEX matches Quamtum Monte Carlo correlation energies already at
$r_s\approx 5$, where RPA+APX still has an error of about than $0.7$\,m$E_h$
per electron, which is a relative error 5\%.
For lower densities with $r_s>8$, where correlation effects are more prevalent,
APX improves on SOSEX and its accuracy is never worse than $0.3$\,m$E_h$ per
electron, even up to $r_s=50$. Note, that the accuracy is given with respect
to the \parencite{perdew_self-interaction_1981} fit of the
\parencite{ceperley_ground_1980} QMC resutls.
The Adjacent Pairs Exchange correction can be computed in $\mathcal O(N^5)$
with a memory requirement scaling like $\mathcal O(N^2)$. This is equivalent
to the AC-SOSEX but considerably less memory demanding than SOSEX.
If a memory demand of $\mathcal O(N^3)$ is permissible, APX and AC-SOSEX
can also be evaluated in $\mathcal O(N^4)$ steps.
Both, APX and AC-SOSEX are implemented in \VASP\ and first applications
on lattice constants show excellent agreement with experiment.
A more extensive survey including atomization energies is still to be made.

Not all improvements beyond the Random Phase Approximation are based
on the inclusion of exchange processes. In the $GW$ approximation
the iterative scheme of RPA, used to build an effective interaction $W$ from
the Coulomb interaction $V$ according to (\ref{eqn:RPAScreenedW}), is extended
by an iterative scheme to improve the description of the propagator $G_0$
towards an effective propagator $G$ of the fully interacting system, referred
to as \emph{dressed}\index{dressed} propagator.
There is a whole family of such approximations, based on the work of
\parencite{hedin_new_1965}. The central quantity of interest in these
approximations is the dressed propagator, rather than connected diagrams,
as in the Goldstone approach of many-body perturbation theory, discussed
here. This makes a comparison between the two approaches hard, as there is
no consideration of symmetries and the propagators are approximated
by merely shifting the poles of the initial propagators $G_0$.
In most cases the dressed propagator is used to retrieve excitation
spectra but there are also total energy calculations within the $GW$
approximation, usually employing the formula of
\parencite{galitskii_application_1958}.
In Appendix \ref{apx:G0W0}, an alternative approach is suggested to evaluate
the total energy in the $G_0W_0$ approximation strictly within the many-body
perturbation theory discussed here. The total energies retrieved by this
approximation are expected to be more accurate than those of the RPA at
computational costs that should not exceed $\mathcal O(N^4)$ but
this remains to be tested in the future. Also, $G_0W_0$ is only the least
accurate approximation of the mentioned family of approximations, depending
on the reference such as DFT or HF and further, exchange effects are known to
be important not only to correct for violations of the exclusion principle,
so it is unclear, whether $G_0W_0$ total energies are a viable option to
RPA+APX or RPA+AC-SOSEX.

Finally, it is worth mentioning that most of time and memory requirement of
high accuracy methods, such as Coupled Cluster Singles and Doubles (CCSD)
strongly depends on the number of unoccupied states $\psi_a$, needed for a
convergent result. This number can be reduced without considerably sacrificing
the accuracy such that CCSD or even higher accuracy methods become feasible
for larger systems. One way of reducing the number of unoccupied states
is by means of natural orbitals, applicable in the case where large voids
are between the atoms or molecules, such as in atomization energy calculations
of solids. 
Another way of reducing the number of unoccupied states is by including
explicit correction already in the description of the unperturbed system.
This can be done by augmenting the Slater determinant, which is a product
of functions depending on one electron position only, by a set of functions
$f(\vec x_1,\vec x_2)$ explicitly depending on two electron positions,
hence the name $f_{12}$ methods.
In all cases there is still an underlying perturbation expansion to be
evaluated and the Adjacent Paris Correction as well as AC-SOSEX can also
profit from such a reduction of the number of unoccupied states.

\printbibliography

\begin{appendix}
\part{Appendices}
  \chapter{Total energies in $G_0W_0$ from connected diagrams}
\label{apx:G0W0}

Given the propagator $G_0$ in matrix form and in imaginary time according to
(\ref{eqn:RPAGreenImag})\footnote{
  Note that the propagator used here differs by a factor of $\im$ from
  the usual definition of the propagator, as discussed in the footnote of
  (\ref{eqn:MP2directGreens})
}
\[
  {\vec G_0}_{\vec x\vec x'}(\im\tau) =
  \left\{
  \begin{array}{ll}
    \displaystyle
    -\sum_i \psi_i(\vec x)\psi_i^\ast(\vec x') e^{-(\eps_i-\mu)\tau} &
    \textnormal{for }\tau\leq 0 \\[3ex]
    \displaystyle
    +\sum_a \psi_a(\vec x)\psi_a^\ast(\vec x') e^{-(\eps_a-\mu)\tau} &
    \textnormal{otherwise,}
  \end{array}
  \right.
\]
we can Fourier transform it to imaginary frequency $\vec G_0(\im\eta)$.
We also write the screened Coulomb interaction $\vec W_0$ in the Random Phase
Approximation (RPA) in the same form
\begin{equation}
  \begin{array}{ccc ccc ccc}
    \diagramBox{\fmfreuse{ScreenedW}}
    &=&
    \diagramBox{\fmfreuse{CoulombInteraction}}
    &+&
    \diagramBox{\fmfreuse{VChiV}}
    &+&
    \diagramBox{\fmfreuse{VChiVChiV}}
    &+&
    \ldots
    \\[4ex]
    \vec W_0(\im\nu)
    &=&
    \vec V
    &+&
    \vec V\vec X_0(\im\nu)\vec V
    &+&
    \vec V\vec X_0(\im\nu)\vec V\vec X_0(\im\nu)\vec V
    &+&
    \ldots
    \\[1ex]
    &=&
    \multicolumn{3}{c}{\vec V\,\Big(\vec 1-\vec X_0(\im\nu)\vec V\Big)^{-1}\,,}
  \end{array}
  \label{eqn:G0W0_W0}
\end{equation}
where $\vec X_0$ is the independent particle polarizability in imaginary
frequency as retrieved from a Fourier transform of (\ref{eqn:RPAChiImag}).
The \emph{irreducible self energy}\index{self energy} $\vec\Sigma$ can
then be approximated in the RPA by
\begin{equation}
  \begin{array}{ccl}
    \diagramBox{
      \begin{fmffile}{Sigma}
      \begin{fmfgraph*}(30,30)
        \fmfstraight
        \fmftop{v0}
        \fmfbottom{v2}
        \fmf{plain}{v0,v1,v2}
        \fmffreeze
        \fmfv{d.sh=circle,d.f=empty,d.si=20,l=$\Sigma_0$,l.d=0.}{v1}
      \end{fmfgraph*}
      \end{fmffile}
    }
    &=&
    \diagramBox{
      \begin{fmffile}{GVXV}
      \begin{fmfgraph*}(30,30)
        \fmfset{arrow_len}{6}
        \fmfsurroundn{v}{8}
        \fmf{fermion}{v6,v4}
        \fmf{fermion,right=0.3}{v8,v2,v8}
        \fmf{boson}{v2,v4}
        \fmf{boson}{v6,v8}
        \fmfv{label.angle=180,label=$\vec x'$}{v6}
        \fmfv{label.angle=180,label=$\vec x$}{v4}
      \end{fmfgraph*}
      \end{fmffile}
    }
    +
    \diagramBox{
      \begin{fmffile}{GVXVXV}
      \begin{fmfgraph}(40,40)
        \fmfset{arrow_len}{6}
        \fmfsurroundn{v}{12}
        \fmf{fermion}{v8,v6}
        \fmf{fermion,right=0.3}{v2,v4,v2}
        \fmf{fermion,right=0.3}{v10,v12,v10}
        \fmf{boson}{v12,v2}
        \fmf{boson}{v4,v6}
        \fmf{boson}{v8,v10}
      \end{fmfgraph}
      \end{fmffile}
    }
    +
    \ \ \ldots
    \\[5ex]
    &=&
    \hspace*{1ex}
    \diagramBox{
      \begin{fmffile}{GW}
      \begin{fmfgraph}(20,30)
        \fmfstraight
        \fmfset{arrow_len}{8}
        \fmfleft{v00,v01}
        \fmf{dbl_wiggly,right=0.9,label.side=right}{v00,v01}
        \fmf{fermion}{v00,v01}
      \end{fmfgraph}
      \end{fmffile}
    }
    -
    \hspace*{1ex}
    \diagramBox{
      \begin{fmffile}{GV}
      \begin{fmfgraph}(15,30)
        \fmfstraight
        \fmfset{arrow_len}{8}
        \fmfleft{v00,v01}
        \fmf{boson,right=0.9}{v00,v01}
        \fmf{fermion}{v00,v01}
      \end{fmfgraph}
      \end{fmffile}
    }
    \\[4ex]
    {\vec\Sigma_0}_{\vec x\vec x'}(\im\eta)
    &=&
    \displaystyle \int_{-\infty}^\infty\frac{\im\,\d\nu}{2\pi}\,
      {\vec G_0}_{\vec x\vec x'}(\im\eta+\im\nu)
      \Big({\vec W_0}_{\vec x\vec x'}(\im\nu) - \vec V_{\vec x\vec x'}\Big)
      \,.
  \end{array}
\end{equation}
Note that the matrices are multiplied elementwise and that we exclude the
exchange term $G_0V$ since it cancels with the effective interaction
in a Hartree-Fock reference as discussed in Section \ref{sec:MBPT_HartreeFock}.
In the $G_0W_0$ approximation the full propagator $\vec G$ is approximated
by inserting the above approximation to the irreducible self energy
$\vec\Sigma_0$ into the
propagator of the unperturbed system $\vec G_0$ arbitrarily many times
\vspace*{-3ex}
\begin{equation}
\begin{array}{c c ccccccc}
  \diagramBox{
    \begin{fmffile}{G}
    \begin{fmfgraph*}(15,30)
      \fmfset{arrow_len}{8}
      \fmfstraight
      \fmfbottom{v1}
      \fmftop{v2}
      \fmf{heavy}{v1,v2}
    \end{fmfgraph*}
    \end{fmffile}
  }
  &=&
  \diagramBox{
    \begin{fmffile}{G0}
    \begin{fmfgraph}(15,30)
      \fmfset{arrow_len}{6}
      \fmfstraight
      \fmfbottom{v1}
      \fmftop{v2}
      \fmf{fermion}{v1,v2}
    \end{fmfgraph}
    \end{fmffile}
  }
  &+&
  \diagramBox{
    \begin{fmffile}{G0SigmaG0}
    \begin{fmfgraph*}(15,60)
      \fmfstraight
      \fmfset{arrow_len}{6}
      \fmfbottom{v0}
      \fmftop{v2}
      \fmf{fermion}{v0,v1}
      \fmf{fermion}{v1,v2}
      \fmfv{d.sh=circle,d.f=empty,d.si=20,l=$\Sigma$,l.d=0.}{v1}
    \end{fmfgraph*}
    \end{fmffile}
  }
  &+&
  \diagramBox{
    \begin{fmffile}{G0SigmaG0SigmaG0}
    \begin{fmfgraph*}(15,90)
      \fmfstraight
      \fmfset{arrow_len}{6}
      \fmfbottom{v0}
      \fmftop{v3}
      \fmf{fermion}{v0,v1,v2,v3}
      \fmfv{d.sh=circle,d.f=empty,d.si=20,l=$\Sigma$,l.d=0.}{v1}
      \fmfv{d.sh=circle,d.f=empty,d.si=20,l=$\Sigma$,l.d=0.}{v2}
    \end{fmfgraph*}
    \end{fmffile}
  }
  &+&
  \ \ldots
  \\[10ex]
  \vec G &=&
  \vec G_0
  &+&
  \vec G_0\vec\Sigma_0\vec G_0
  &+&
  \vec G_0\vec\Sigma_0\vec G_0\vec\Sigma_0\vec G_0
  &+&
  \ \ldots\,,
  \label{eqn:G0W0_G}
\end{array}
\end{equation}
where we omit the imaginary frequency argument for brevity. Properties of
the interacting system can then be extracted from this approximation to the
full propagator $\vec G$. To get the total energy the formula of
\parencite{galitskii_application_1958}
can be used on the full imaginary time propagator $\vec G(\im\tau)$,
retrieved from an inverse Fourier transform of $\vec G(\im\eta)$ at
$\tau\rightarrow0$ from above:
\begin{equation}
  E = -\frac\im2 \int\d\vec x\,
    \lim_{\vec r'\rightarrow\vec r}\lim_{\tau\rightarrow 0^+}
    \left(\frac\der{\im\,\der\tau} - \im\,\hat h_\vec x\right)
    \vec G_{\vec r\alpha\,\vec r'\alpha}(\im\tau)\,,
\end{equation}
where $\hat h_\vec x$ is the single body Hamiltonian of the unperturbed
system acting on $\vec G_\vec x$. This can be interpreted as
cutting out one occurrence of $G_0$ by the differential operator and then
closing the remaining diagrams contained in $G$
\parencite{ziesche_high-density_2010}.

Here, an alternative way to the Galitskii-Migdal formula is proposed for
evaluating the total energy in $G_0W_0$, respecting the symmetries
of all connected diagrams occurring in this approximation, as discussed
in Subsection \ref{ssc:MBPT_FeynmanSymmetries}.
Figure \ref{fig:G0W0_totalE} shows all diagrams order by order, where each
row contains an order of the expansion of $\vec W_0$ contained in $\vec\Sigma_0$
according to (\ref{eqn:G0W0_W0}) and each column contains an order of
$\vec G$ according to (\ref{eqn:G0W0_G}). Note that the symmetries of the
first column differ from the symmetries of all other columns since closing
$\vec\Sigma_0$ with one $\vec G_0$ forms a bubble equivalent to $\vec X_0$.
The diagrams of the first column form the ring diagrams of the RPA. All the
remaining diagrams have rotational symmetry but no reflection symmetry and can
be summed to infinite order in the same fashion as it was done in the RPA in
(\ref{eqn:RPAReal}), arriving at
\begin{align}
  \nonumber
  E_{\rm c}^{G_0W_0} &=
    E_{\rm c}^{\rm RPA} +
    \im\int\frac{\im\,\d\eta}{2\pi}\Tr\left\{
      -\frac12\Big(\vec G_0(\im\eta)\Sigma_0(\im\eta)\Big)^2
      -\frac13\Big(\vec G_0(\im\eta)\Sigma_0(\im\eta)\Big)^3
      -\ldots
    \right\}
    \\[0.5ex]
  &=
    E_{\rm c}^{\rm RPA} -
    \int_{-\infty}^\infty\frac{\d\eta}{2\pi}\,
    \Tr\Big\{
      \log\Big(\vec 1-\vec G_0(\im\eta) \vec\Sigma_0(\im\eta)\Big)+
      \vec G_0(\im\eta)\vec\Sigma_0(\im\eta)
    \Big\}\,.
\end{align}

This approach still has to be tested and compared to previous $G_0W_0$
total energies, as for instance retrieved by \parencite{holm_total_2000}.
The two approaches do differ after all. Conventional $G_0W_0$ uses a
normalization factor for calculating $\vec G$ while this approach uses symmetry
factors.
An implementation for the uniform electron gas in the thermodynamic
limit is still ongoing.

\begin{figure}
\[
\begin{array}{ccc ccc ccc}
  E_{\rm c}^{G_0W_0}
    &=&
    E_{\rm c}^{\rm RPA}
    &-&
    \displaystyle \frac12\,\Big(
      \textcolor{red}{\vec G_0(\im\eta)}
      \textcolor{blue}{\vec \Sigma_0(\im\eta)}
    \Big)^2
    &-&
    \displaystyle \frac13\,\Big(
      \textcolor{red}{\vec G_0(\im\eta)}
      \textcolor{blue}{\vec \Sigma_0(\im\eta)}
    \Big)^3
    &-&
    \ldots
    \\[2ex]
    &=&
    \diagramBox{
      \begin{fmffile}{G0W012}
      \begin{fmfgraph*}(40,40)
        \fmfset{arrow_len}{6}
        \fmfsurroundn{v}{8}
        \fmf{boson,foreground=blue}{v2,v4}
        \fmf{boson,foreground=blue}{v6,v8}
        \fmf{fermion,right=0.25,foreground=blue}{v8,v2,v8}
        \fmf{fermion,right=0.25,foreground=blue}{v6,v4}
        \fmf{fermion,right=0.25,foreground=red}{v4,v6}
        \fmfv{label.angle=180,label=$G_0$}{v4}
        \fmfv{label=$\Sigma_0^{(2)}$}{v2}
        \fmf{dots}{v3,v7}
      \end{fmfgraph*}
      \end{fmffile}
    }
    &+&
    \diagramBox{
      \begin{fmffile}{G0W022}
        \begin{fmfgraph*}(60,60)
          \fmfkeep{G0W021}
          \fmfset{arrow_len}{6}
          \fmfsurroundn{o}{16}
          \fmf{boson,foreground=blue}{o2,i1}
            \fmf{
              fermion,right=0.4142,foreground=red,label.side=right,label=$G_0$
            }{i1,i2}
            \fmf{boson,foreground=blue}{i2,o8}
          \fmf{boson,foreground=blue}{o10,i3}
            \fmf{fermion,right=0.4142,foreground=red}{i3,i4}
            \fmf{boson,foreground=blue}{i4,o16}
          \fmffreeze
          \fmf{fermion,right=0.4142,foreground=blue}{i2,i3}
          \fmf{fermion,right=0.4142,foreground=blue}{i4,i1}
          \fmf{fermion,right=0.3,foreground=blue}{o16,o2,o16}
          \fmf{fermion,right=0.3,foreground=blue}{o10,o8,o10}
          \fmf{dots}{o5,o13}
          \fmfv{label=$\Sigma_0^{(2)}$}{o2}
        \end{fmfgraph*}
      \end{fmffile}
    }
    &+&
    \diagramBox{
      \begin{fmffile}{G0W032}
        \begin{fmfgraph*}(80,80)
          \fmfkeep{G0W031}
          \fmfset{arrow_len}{6}
          \fmfsurroundn{o}{24}
          \fmf{boson,foreground=blue}{o24,i1}\fmf{phantom,tension=0.366}{i1,o14}
          \fmf{boson,foreground=blue}{o2,i2} \fmf{phantom,tension=0.366}{i2,o12}
          \fmf{boson,foreground=blue}{o8,i3} \fmf{phantom,tension=0.366}{i3,o22}
          \fmf{boson,foreground=blue}{o10,i4}\fmf{phantom,tension=0.366}{i4,o20}
          \fmf{boson,foreground=blue}{o16,i5}\fmf{phantom,tension=0.366}{i5,o6}
          \fmf{boson,foreground=blue}{o18,i6}\fmf{phantom,tension=0.366}{i6,o4}
          \fmf{dots}{m,o5} \fmf{dots}{m,o13} \fmf{dots}{m,o21}
          \fmffreeze
          \fmf{fermion,right=0.268,foreground=blue}{i1,i2}
            \fmf{
              fermion,right=0.268,foreground=red,
              label.side=right,label.dist=2,label=$G_0$
            }{i2,i3}
            \fmf{fermion,right=0.268,foreground=blue}{i3,i4}
            \fmf{fermion,right=0.268,foreground=red}{i4,i5}
            \fmf{fermion,right=0.268,foreground=blue}{i5,i6}
            \fmf{fermion,right=0.268,foreground=red}{i6,i1}
          \fmf{fermion,right=0.35,foreground=blue}{o24,o2,o24}
          \fmf{fermion,right=0.35,foreground=blue}{o8,o10,o8}
          \fmf{fermion,right=0.35,foreground=blue}{o16,o18,o16}
          \fmfv{label.angle=45,label=$\Sigma_0^{(2)}$}{o2}
        \end{fmfgraph*}
      \end{fmffile}
    }
    &+&
    \ldots
    \\
    &+&
    \hspace*{-1ex}
    \diagramBox{
      \begin{fmffile}{G0W013}
      \begin{fmfgraph*}(45,45)
        \fmfset{arrow_len}{6}
        \fmfsurroundn{o}{12}
        \fmf{boson,foreground=blue}{o12,o2}
        \fmf{boson,foreground=blue}{o4,o6}
        \fmf{boson,foreground=blue}{o8,o10}
        \fmf{fermion,right=0.3,foreground=blue,label=$\Sigma_0^{(3)}$}{o2,o4}
          \fmf{fermion,right=0.3,foreground=blue}{o4,o2}
        \fmf{fermion,right=0.3,foreground=red}{o6,o8}
          \fmf{fermion,right=0.3,foreground=blue}{o8,o6}
        \fmf{fermion,right=0.3,foreground=blue}{o10,o12,o10}
        \fmf{dots}{o1,m}\fmf{dots}{o5,m}\fmf{dots}{o9,m}
        \fmfv{label.angle=180,label=$G_0$}{o6}
      \end{fmfgraph*}
      \end{fmffile}
    }
    \hspace*{-1ex}
    &+&
    \hspace*{-3ex}
    \diagramBox{
      \begin{fmffile}{G0W023}
      \begin{fmfgraph*}(120,120)
        \fmfset{arrow_len}{6}
        \fmfsurroundn{v}{8}
        \fmf{phantom,tension=2.0142}{v2	,p1}
          \fmf{phantom}{p1,c,p3}
          \fmf{phantom,tension=2.0142}{p3,v6}
        \fmf{phantom,tension=2.0142}{v4,p2}
          \fmf{phantom}{p2,c,p4}
          \fmf{phantom,tension=2.0142}{p4,v8}
        \fmffreeze
        \fmf{phantom,tension=1.732}{c,i1} \fmf{phantom}{i1,p1}
        \fmf{phantom,tension=1.732}{c,i2} \fmf{phantom}{i2,p2}
        \fmf{phantom,tension=1.732}{c,i3} \fmf{phantom}{i3,p3}
        \fmf{phantom,tension=1.732}{c,i4} \fmf{phantom}{i4,p4}
        \fmf{phantom,tension=1.732}{v1,o1} \fmf{phantom}{o1,p1}
        \fmf{phantom,tension=1.732}{v5,o2} \fmf{phantom}{o2,p2}
        \fmf{phantom,tension=1.732}{v5,o3} \fmf{phantom}{o3,p3}
        \fmf{phantom,tension=1.732}{v1,o4} \fmf{phantom}{o4,p4}
        \fmf{phantom,tension=1.366}{p1,m1}
          \fmf{phantom,tension=0.25}{m1,m4}
          \fmf{phantom,tension=1.366}{m4,p4}
        \fmf{phantom,tension=1.366}{p2,m2}
          \fmf{phantom,tension=0.25}{m2,m3}
          \fmf{phantom,tension=1.366}{m3,p3}
        \fmf{fermion,right=0.4142,tension=0,foreground=blue}{i4,i1}
        \fmf{
          fermion,right=0.4142,tension=0,foreground=red,
          label.side=right,label=$G_0$
        }{i1,i2}
        \fmf{fermion,right=0.4142,tension=0,foreground=blue}{i2,i3}
        \fmf{fermion,right=0.4142,tension=0,foreground=red}{i3,i4}
        \fmf{boson,tension=0,foreground=blue}{i1,m1}
          \fmf{boson,tension=0,foreground=blue}{o1,o4}
          \fmf{boson,tension=0,foreground=blue}{m4,i4}
        \fmf{
          fermion,right=0.35,tension=0,foreground=blue,label=$\Sigma_0^{(3)}$
        }{o1,m1}
          \fmf{fermion,right=0.35,tension=0,foreground=blue}{m1,o1}
        \fmf{fermion,right=0.35,tension=0,foreground=blue}{m4,o4,m4}
        \fmf{boson,tension=0,foreground=blue}{i2,m2}
          \fmf{boson,tension=0,foreground=blue}{o2,o3}
          \fmf{boson,tension=0,foreground=blue}{m3,i3}
        \fmf{fermion,right=0.35,tension=0,foreground=blue}{m2,o2,m2}
        \fmf{fermion,right=0.35,tension=0,foreground=blue}{m3,o3,m3}
        \fmf{phantom,tension=2}{v3,d1}
          \fmf{dots}{d1,d2}
          \fmf{phantom,tension=2}{d2,v7}
      \end{fmfgraph*}
      \end{fmffile}
    }
    \hspace*{-3ex}
    &+&
    \hspace*{-5ex}
    \diagramBox{
      \begin{fmffile}{G0W033}
      \begin{fmfgraph*}(120,120)
        \fmfset{arrow_len}{6}
        \fmfsurroundn{v}{12}
        \fmf{phantom}{v2,m1}
          \fmf{boson,foreground=blue}{m1,i1}
          \fmf{phantom}{i1,c,i4}
          \fmf{boson,foreground=blue}{i4,m4}
          \fmf{phantom}{m4,v8}
        \fmf{phantom}{v4,m2}
          \fmf{boson,foreground=blue}{m2,i2}
          \fmf{phantom}{i2,c,i5}
          \fmf{boson,foreground=blue}{i5,m5}
          \fmf{phantom}{m5,v10}
        \fmf{phantom}{v6,m3}
          \fmf{boson,foreground=blue}{m3,i3}
          \fmf{phantom}{i3,c,i6}
          \fmf{boson,foreground=blue}{i6,m6}
          \fmf{phantom}{m6,v12}
        \fmf{phantom}{v12,o6}
          \fmf{boson,foreground=blue}{o6,o1}
          \fmf{phantom}{o1,v2}
        \fmf{phantom}{v4,o2}
          \fmf{boson,foreground=blue}{o2,o3}
          \fmf{phantom}{o3,v6}
        \fmf{phantom}{v8,o4}
          \fmf{boson,foreground=blue}{o4,o5}
          \fmf{phantom}{o5,v10}
        \fmffreeze
        \fmf{fermion,right=0.268,foreground=blue}{i6,i1}
        \fmf{
          fermion,right=0.268,foreground=red,
          label.side=right,label.dist=3,label=$G_0$
        }{i1,i2}
        \fmf{fermion,right=0.268,foreground=blue}{i2,i3}
        \fmf{fermion,right=0.268,foreground=red}{i3,i4}
        \fmf{fermion,right=0.268,foreground=blue}{i4,i5}
        \fmf{fermion,right=0.268,foreground=red}{i5,i6}
        \fmf{fermion,right=0.35,foreground=blue}{m6,o6,m6}
          \fmf{fermion,right=0.35,foreground=blue,label=$\Sigma_0^{(3)}$}{o1,m1}
          \fmf{fermion,right=0.35,foreground=blue}{m1,o1}
        \fmf{fermion,right=0.35,foreground=blue}{m2,o2,m2}
          \fmf{fermion,right=0.35,foreground=blue}{o3,m3,o3}
        \fmf{fermion,right=0.35,foreground=blue}{m4,o4,m4}
          \fmf{fermion,right=0.35,foreground=blue}{o5,m5,o5}
        \fmf{dots}{c,d1} \fmf{phantom,tension=2}{d1,v3}
        \fmf{dots}{c,d2} \fmf{phantom,tension=2}{d2,v7}
        \fmf{dots}{c,d3} \fmf{phantom,tension=2}{d3,v11}
      \end{fmfgraph*}
      \end{fmffile}
    }
    \hspace*{-2ex}
    &+&
    \ldots
    \\
    &+&
    \vdots
    &+&
    \vdots
    &+&
    \vdots
    &+&
    \ddots
\end{array}
\]
\caption{
  Connected (closed) diagrams occurring in the $G_0W_0$ approximation of
  the correlation energy.
  The parts of the self energy are given in blue while the parts of the closing
  propagator are given in red.
  With each column the number of insertions of $\vec\Sigma_0\vec G_0$ into
  $\vec G$ increments, while the number of insertions of $\vec X_0\vec V$ into
  $\vec\Sigma_0$ increments with each row. The single Fermion loop in the
  center results in a negative Fermion sign of all diagrams beyond RPA.
  Imaginary units, the imaginary frequency integration and the trace are
  omitted for clarity.
}
\label{fig:G0W0_totalE}
\end{figure}

\end{appendix}


\end{document}